\documentclass[%
 aip,
 amsmath,amssymb,
 reprint,%
]{revtex4-1}

\usepackage{graphicx}% Include figure files
\usepackage{dcolumn}% Align table columns on decimal point
\usepackage{bm}% bold math
\usepackage[utf8]{inputenc}
\usepackage[T1]{fontenc}
\usepackage{mathptmx}
\usepackage{etoolbox}

\usepackage{mathtools}

\usepackage[shortlabels]{enumitem}
\usepackage{caption}
\usepackage{subcaption}
\usepackage{hyperref}

\usepackage{MnSymbol}

\usepackage{xcolor}
\newsavebox\MBox
\newcommand\Cline[2][red]{{\sbox\MBox{$#2$}%
  \rlap{\usebox\MBox}\color{#1}\rule[-1.2\dp\MBox]{\wd\MBox}{0.5pt}}}

\hypersetup{
    colorlinks = true,
    urlcolor   = black,
    citecolor  = black,
}

\newcommand{\RomanNumeralCaps}[1]

\makeatletter
\def\@email#1#2{%
 \endgroup
 \patchcmd{\titleblock@produce}
  {\frontmatter@RRAPformat}
  {\frontmatter@RRAPformat{\produce@RRAP{*#1\href{mailto:#2}{#2}}}\frontmatter@RRAPformat}
  {}{}
}%
\makeatother
\begin{document}

\preprint{AIP/123-QED}

\title{Recent developments and research needs in turbulence modeling of hypersonic flows}
% Force line breaks with \\
\author{Pratikkumar Raje}
 \email{praje@umich.edu}
\affiliation{%
Department of Aerospace Engineering, University of Michigan, 1320 Beal Ave, Ann Arbor, Michigan  48109, USA%\\This line break forced% with \\
}%

\author{Eric Parish}%
\affiliation{ 
Computational Data Science, Sandia National Laboratories, 7011 East Ave, Livermore, California 94550, USA %\\This line break forced with \textbackslash\textbackslash
}%

\author{Jean-Pierre Hickey}
% \homepage{http://www.Second.institution.edu/~Charlie.Author.}
\affiliation{%
Department of Mechanical and Mechatronics Engineering, University of Waterloo, 200 University Ave W, Waterloo, N2L 3G1, Ontario, Canada%\\This line break forced% with \\
}%

\author{Paola Cinnella}
% \homepage{http://www.Second.institution.edu/~Charlie.Author.}
\affiliation{%
Institut Jean Le Rond D'Alembert, Sorbonne Universit\'e, 4 Place Jussieu, Paris, 75005, France%\\This line break forced% with \\
}%

\author{Karthik Duraisamy}
% \homepage{http://www.Second.institution.edu/~Charlie.Author.}
\affiliation{%
Department of Aerospace Engineering, University of Michigan, 1320 Beal Ave, Ann Arbor, Michigan  48109, USA%\\This line break forced% with \\
}%

\date{}

\begin{abstract}
Hypersonic flow conditions pose exceptional challenges for Reynolds-Averaged Navier-Stokes (RANS) turbulence modeling. Critical phenomena include compressibility effects, shock/turbulent boundary layer interactions, turbulence-chemistry interaction in thermo-chemical non-equilibrium, and ablation-induced surface roughness and blowing effects. This comprehensive review synthesizes recent developments in adapting turbulence models to hypersonic applications, examining approaches ranging from empirical modifications to physics-based reformulations and novel data-driven methodologies. We provide a systematic evaluation of current RANS-based turbulence modeling capabilities, comparing eddy viscosity and Reynolds stress transport formulations in their ability to predict engineering quantities of interest such as separation characteristics and wall heat transfer. Our analysis encompasses the latest experimental and direct numerical simulation datasets for validation, specifically addressing two- and three-dimensional equilibrium turbulent boundary layers and shock/turbulent boundary layer interactions across both smooth and rough surfaces. 
 Key multi-physics considerations including catalysis and ablation phenomena along with the integration of conjugate heat transfer into a RANS solver for efficient design of a thermal protection system are also discussed. We conclude by identifying the critical gaps in the available validation databases and limitations of the existing turbulence models and suggest potential areas for future research to improve the fidelity of turbulence modeling in the hypersonic regime.
\end{abstract}

\maketitle

\section{\label{sec:level1}Introduction}

The development of hypersonic vehicles, including aircraft, missiles, glide vehicles, reusable launch vehicles, and spacecraft, is at the forefront of aerospace and defense research. Hypersonic vehicles are designed to operate at  Mach number greater than $5$ through the planetary atmosphere and they have the potential to transform military capabilities, space exploration, and commercial aviation. A hypersonic vehicle experiences a wide spectrum of flow regimes along its flight trajectory, ranging from subsonic to hypersonic speeds, continuum to free-molecular conditions, and laminar to turbulent flow behavior. These flow conditions are generally characterized using Mach (M: vehicle speed/sound speed), Knudsen (Kn: molecular mean free path/reference length), and Reynolds (Re: inertial/viscous forces) numbers. Moreover, high-temperature effects at hypersonic speeds lead to thermochemical non-equilibrium and result in widely varying Damk\"{o}hler numbers (Da: flow time/chemical reaction times). This review focuses on the modeling of hypersonic turbulent flows in the continuum regime as the vehicle design is generally based on such flow conditions because of the associated extreme aerothermal loads. The complex nature of these flows presents a formidable challenge for accurate physical modeling and numerical analysis and demands sophisticated techniques and significant computational resources.

\begin{figure*}
\centering
\includegraphics[width=0.65\linewidth, trim = 0mm 0mm 0mm 0mm]{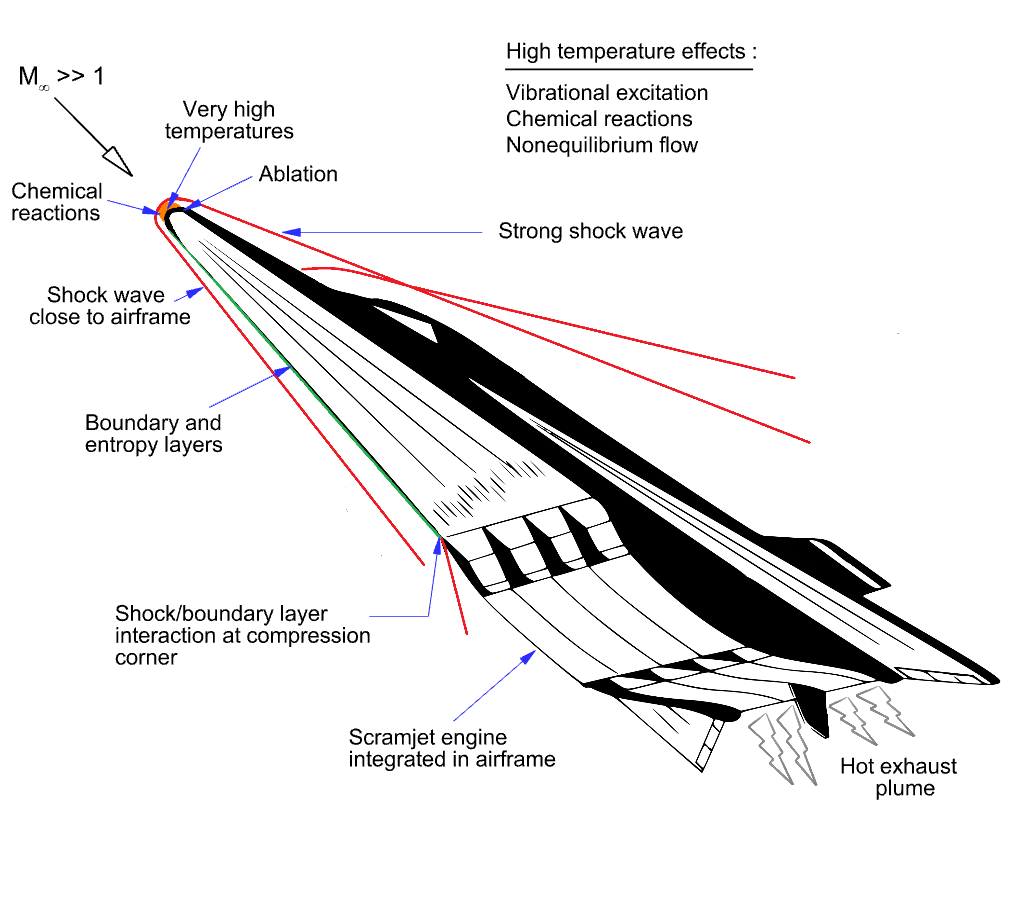}
\caption{Complex flow physics around a notional hypersonic flight system. Figure adapted from Ref.~\onlinecite{urzay2020physical} with permission from Javier Urzay. }
\label{fig:hypersonic_vehicle}
\end{figure*}

The complexity of modeling hypersonic turbulent flows relevant to practical applications arises from several unique physical phenomena that are not present or are significantly less pronounced at lower speeds. These include strong compressibility effects; the presence of strong shock waves that lie close to the vehicle body and the associated shock/turbulent boundary layer interactions i.e. SBLIs; significantly high temperatures due to extreme viscous dissipation within the turbulent boundary layers and strong shock waves leading to chemical and thermal non-equilibrium and the associated turbulence/chemistry interactions i.e. TCI; and surface reactions including ablation and the associated ablative particle-flow interaction (see Fig.~\ref{fig:hypersonic_vehicle}). Intense aerothermodynamic heating, unsteady pressure loads, and high skin-friction drag are the important global manifestations of the aforementioned complex physical phenomena crucial for engineering design considerations. Additionally, a better understanding of the flow-field in terms of locations of shock waves, expansion fans, turbulent boundary and free-shear layers, and separated regions including shock-induced recirculation bubbles, especially at off-design conditions, is critical for the successful and safe operation of hypersonic vehicles. For example, the shock wave from the engine cowl impinging on the pylon holding a dummy scramjet engine and shock waves impinging on the vertical tail caused severe heating and structural damage to the X-15 aircraft during its final hypersonic flight~\cite{anderson1989hypersonic}. Accurate prediction of these complex hypersonic flow-fields and the engineering quantities of interest (QoIs) is a significantly complex and challenging task for the Computational Fluid Dynamics (CFD) community.

%%%%%%%%%%%%%%%%%%%%%%%%%%%%%%%%%%%%%%%%%%%%%%%%%%%%%%
%%%%%%%%%%%%%%%%%%%%%%%%%%%%%%%%%%%%%%%%%%%%%%%%%%%%%%

The limitations of the present-day computational resources restrict the Direct Numerical Simulation (DNS)~\cite{moin1998direct} and/or Large eddy simulations (LES)~\cite{zhiyin2015large} to grid resolutions that cannot fully resolve the salient structures present in practical engineering applications~\cite{choi2012grid,yang2021grid}. A variety of methods like the wall-modeled LES (WMLES)~\cite{bose2018wall} and detached eddy simulations (DES)~\cite{spalart2009detached} are developed to efficiently handle the near-wall turbulence, however, Reynolds-averaged Navier-Stokes (RANS) remains the primary workhorse for numerical predictions of practical flows in the aerospace industry. In fact, RANS-based CFD plays an important role in obtaining certification from governing regulatory bodies. In the RANS approach, the mean flow field is computed, and the effect of turbulence on the mean flow is introduced using a turbulence model. RANS requires considerably coarser grid sizes than DNS and LES and is favored in the standard engineering design process because of significantly shorter turnaround times. Turbulence modeling is a crucially important aspect of the RANS-based CFD techniques as it considerably influences predictions for aerodynamic forces, heat transfer rates, and chemical reactions.

The modeling challenges posed by the hypersonic turbulent flows are rooted in the complexity of the strain field and the high rates of change to which the mean flow as well as the turbulence fields are subjected. Additionally, cold-wall conditions, characterized by the ratio of wall temperature to the adiabatic wall (or recovery) temperature i.e. $T_w/T_{aw}$, are common in hypersonic flight vehicles and wind tunnel experiments. This occurs because of transient heating of the vehicle/model, active cooling or ablation of the surface, and/or re-radiative cooling of the surface. Strong wall cooling increases the significance of compressibility effects, impacts near-wall turbulence characteristics, and may affect SBLI properties~\cite{sciacovelli2023priori}. Moreover, the surfaces of hypersonic vehicles, e.g. ballistic and maneuverable reentry vehicles, become rough, generally characterized by the dimensionless surface roughness height $k^+_s$, in key aerothermal heating areas due to the ablation loss of the heat shield. This leads to an increased heat transfer at the wall which in turn accelerates the ablation process. The ablation process leads to the thermal decomposition of the materials by pyrolysis reactions. The pyrolysis products are then outgassed into the boundary layer, resulting in the blowing phenomenon. Unlike surface roughness, blowing effects on the boundary layer decrease the heat load. Thus, the ablation process critically affects the aerothermal loads in the presence of roughness and blowing effects, and modeling these effects is a significantly challenging task~\cite{marchenay2022hypersonic}.

%Previous reviews on the status of turbulence modeling for hypersonic flows --- 
Marvin~\cite{marvin1983turbulence} and Marvin and Coakley~\cite{marvin1989turbulence} briefly discuss turbulence modeling for supersonic and hypersonic flows and evaluate the performance of several zero-, two-equation, and Reynolds stress transport models (RSTMs) against the experimental data for a number of hypersonic attached boundary layer and separated SBLI flows. Roy and Blottner~\cite{roy2006review} present a comprehensive review of the performance of a total of $18$ one- and two-equation Boussinesq hypothesis-based eddy viscosity models (EVMs) including Spalart-Allmaras (SA)~\cite{spalart1992one} and the variants of $k$-$\epsilon$ and $k$-$\omega$ models against the extended hypersonic SBLI experimental database of Settles and Dodson~\cite{settles1991} consisting of nine experiments on five different geometries. Marvin \textit{et al.}~\cite{marvin2013experimental} present a compilation of experimental validation database from $10$ experiments on six different geometries along with predictions obtained using $k$-$\epsilon$, $k$-$\omega$, and SA models. However, these works consider only two-dimensional (2D)/axisymmetric hypersonic wall-bounded flows. On the other hand, performance assessment of turbulence models for three-dimensional (3D) SBLIs is limited~\cite{bardina1992two,bardina1994three}.

Smits \textit{et al.}~\cite{smits2009current} reviews the status of basic research in computational methods for hypersonic turbulent flows including  RANS as well as DNS, LES, and RANS-LES hybrid approaches. Georgiadis \textit{et al.}~\cite{georgiadis2014status} provides an assessment of the ``state-of-practice'' in RANS with a focus on the two-equation $k$-$\epsilon$ and $k$-$\omega$ family of models for predicting hypersonic propulsion flowpaths including laminar-to-turbulent boundary layer transition, SBLIs, and modeling of the combustor and exhaust system in the context of a scramjet engine. A number of works~\cite{gnoffo2013uncertainty,brown2002turbulence} provide an assessment of the standard turbulence models in predicting the hypersonic zero-pressure-gradient (ZPG) turbulent boundary layers (TBLs) and SBLIs at different wall temperature conditions. Interestingly, there is a shift in the attention and recent works focus on evaluating the performance of these models and the compressibility corrections in predicting ZPG hypersonic TBLs at cold wall conditions~\cite{aiken2022assessment,huang2019assessment,rumsey2010compressibility}. The works discussed above generally evaluate the performance of Cebeci-Smith~\cite{smith1967numerical} and Baldwin-Lomax~\cite{baldwin1978thin} zero-equation models, SA one-equation model, and $k$-$\epsilon$ and $k$-$\omega$ model variants including Menter SST $k$-$\omega$~\cite{menter1994two} two-equation EVMs with and without the classical compressibility corrections. However, these works consider the hypersonic flow regime where calorically perfect gas models are appropriate. A recent work~\cite{sciacovelli2023priori} performs an \textit{a priori} assessment of the classical closures and compressibility corrections used with Menter SST $k$-$\omega$ model using DNS~\cite{passiatore2021finite,passiatore2022thermochemical} for ZPG TBLs at Mach numbers up to about $12.5$ with different levels of wall cooling including high enthalpy conditions with thermochemical non-equilibrium.

For hypersonic flows at high-enthalpy conditions, conversion of kinetic energy into internal energy can lead to sufficiently high temperatures to induce chemical reactions and/or thermal non-equilibrium effects that greatly influence surface forces and heat transfer  \citep{candler2019rate}. Depending on the flow Damk\"{o}hler number, i.e. the ratio of the flow characteristic time to
the characteristic times of chemical reactions or thermal relaxation, the flow can be considered as chemically (thermally) frozen, in chemical (thermal) equilibrium or in non-equilibrium conditions. The former case occurs when thermochemical processes are very slow compared to the flow residence time (Da $\rightarrow 0$), so that the advancement of such processes is essentially negligible; on the opposite side, when the reaction or thermal relaxation occurs on very fast time scales (Da $\rightarrow\infty$), the chemical composition and thermophysical flow properties change instantaneously as the flow conditions change. In many practical cases, the flow is characterized by small but non-negligible Damk\"{o}hler numbers, leading to nonequilibrium effects. In such cases, the governing equations are supplemented with additional transport equations describing the evolution of the chemical species/ions and/or the energies associated with the internal molecular degrees of freedom of polyatomic molecules \citep{anderson1989hypersonic,park1989nonequilibrium}, and most typically energies associated with vibrational modes. 
The Reynolds-averaged counterparts of the chemistry/vibrational energies equations involve additional unclosed terms. They are generally modeled using rough gradient approximations based on the introduction of constant nondimensional turbulent transport parameters (e.g. constant turbulent Schmidt numbers for the species and constant vibrational Prandtl numbers for the vibrational energy, which may be highly influential on the results \citep{georgiadis2014turbulence}.
More sophisticated, nonlinear representations have been attempted with some success \citep{bowersox2009extension,bowersox2010algebraic}.
Conducting experiments in these conditions is generally impractical or impossible, because of the extreme thermodynamic conditions and huge amounts of electric power needed to operate high-enthalpy wind tunnels \citep{urzay2021engineering}. On the other hand, numerical simulations are also very challenging, depending on the complexity of the thermochemical models in use, and on the range of physical space and time scales to be simulated. Only recently, DNS data are beginning to be available \citep{direnzo2021direct,passiatore2021finite,passiatore2022thermochemical}, providing valuable insights into the validity of modeling assumption.
The picture is further complicated if shock waves are present, promoting non-equilibrium conditions, for which experimental or high-fidelity data for turbulent high-enthalpy conditions are very scarce \citep{volpiani2021numerical,passiatore2023shock,di2024stagnation}.
%The analysis becomes particularly complex for high-enthalpy hypersonic flows involving turbulent flow regimes, such as those expected to occur in hypersonic flight at low altitudes.

This review paper aims to provide a comprehensive overview of the latest advancements in turbulence modeling for hypersonic flows. We limit our scope to the modeling of fully turbulent hypersonic boundary layer flows and do not consider transition modeling, which is also critically important in practical hypersonic flows. The performance of various turbulence modeling approaches for hypersonic flows, including zero-, one-, and two-equation eddy viscosity models, as well as RSTMs, Explicit Algebraic Reynolds Stress Models (EARSMs), and Non-Linear EVMs (NLEVMs) is discussed at length. This includes the generalization of low-speed models using compressibility corrections, heat flux modeling (e.g. variable turbulent Prandtl number models), etc. for hypersonic flows. The capabilities and limitations of these models in predicting key flow features and QoIs, such as shock-induced separations, wall pressure, surface heat transfer rates, and skin friction are highlighted. Furthermore, the ongoing challenges in hypersonic turbulence modeling, including the need for an improved physical understanding of compressibility effects on turbulence are discussed. We note here that the discussion does not involve turbulence models using wall functions, but includes the models where integration of the governing equations to the wall is performed.

The effects of thermochemical non-equilibrium relevant to high enthalpy flows expected during hypersonic flights on turbulence modeling are discussed. This includes the additional Favre-averaged species transport and vibrational energy equations and the modeling of unclosed terms therein. The available validation experimental dataset and DNS studies in the hypersonic regime including equilibrium TBLs and SBLIs over smooth and rough surfaces are provided. The emergence of machine learning and data-driven techniques has opened new avenues for turbulence modeling~\cite{duraisamy2019turbulence}. These approaches leverage high-fidelity simulation data and experimental results to improve model accuracy and generalizability. The role of machine learning and data-driven approaches in turbulence modeling for hypersonic flows and key insights in developing reliable data-driven models are discussed. Key multi-physics considerations are also discussed, including ablation and the conjugate heat transfer for efficient thermal protection system design. By examining the developments in the field, the paper seeks to highlight progress, identify persistent challenges, and suggest potential future research directions. As the field continues to evolve rapidly, driven by renewed interest in hypersonic flight technologies, a thorough understanding of the current state-of-the-art in turbulence modeling is essential for advancing the design and analysis capabilities for next-generation hypersonic vehicles.

The paper is organized as follows. Section \ref{sec2} provides the governing Favre-averaged equations for calorically and thermally perfect gases. Section \ref{sec3} gives a detailed discussion on turbulence modeling specifically for hypersonic flows. Eddy-viscosity models including linear EVMs (LEVMs), NLEVMs, and EARSMs along with the advanced RSTMs are covered. The performance of these modeling strategies in predicting attached and separated TBL flows is discussed in detail. Modeling for the thermo-chemical non-equilibrium effects in the hypersonic regime is also presented. Various modeling improvement strategies including ad-hoc fixes and physics-based corrections to the standard turbulence models for hypersonic flows to improve separation size and wall heat transfer predictions are discussed in Section \ref{sec4}. Critical modeling challenges pertinent to hypersonic flows are addressed in Section \ref{sec5}. This includes modeling in the presence of shock/turbulence interaction along with modeling for equilibrium hypersonic TBLs especially at cold wall conditions and SBLIs. Advanced topics like the modeling for surface roughness and blowing effects and thermochemical non-equilibrium effects relevant to hypersonic conditions are also discussed along with key insights into multiphysics considerations including ablation and conjugate heat transfer. Section \ref{sec6} provides the available validation datasets from the experiments and DNS studies performed to date in the hypersonic regime. The current limitations and unresolved issues in turbulence modeling for hypersonic flows are addressed in Section \ref{sec7} along with key insights on data-driven models. Finally, the paper concludes with a summary of key findings and perspectives on future directions in Section \ref{sec8}.

%%%%%%%%%%%%%%%%%%%%%%%%%%%%%%%%%%%%%%%%%%%%%%%%%%%%%%%%%%%%%%%%%%%%%%%%%%%%%%%%%

%%%%%%%%%%%%%%%%%%%%%%%%%%%%%%%%%%%%%%%%%%%%%%%%%%%%%%%%%%%%%%%%%%%%%%%%%%%%%%%%

\section{\label{sec2}Governing equations}

In this section, we limit our discussion to calorically and thermally perfect gases for simplicity. Also, nearly all the available hypersonic experimental databases are limited to low-enthalpy conditions (see Section \ref{sec6}), and turbulence model validation using perfect gas assumption is standard practice. However, temperatures in a hypersonic flow field can reach extremely high values inducing non-equilibrium effects including molecular dissociation, ionization, thermo-chemical reactions, radiation, and surface reactions such as ablation. The complex challenges of modeling these non-equilibrium effects and their profound implications for turbulence modeling will be addressed in Sections \ref{sec:reacteq} and \ref{sec:thermochemical}. In this review, our focus is confined to continuum flows, where the mean free path of gas molecules is small compared to the characteristic flowfield scale. The equations governing the hypersonic flows under these conditions are the Navier-Stokes (NS) equations describing the conservation of mass, momentum, and total energy supplemented by an equation of state. However, as we are interested in the mean behavior of these flows, it is customary to use Reynolds decomposition and express all the instantaneous quantities as the sum of mean and fluctuating parts. For the compressible flows, an established practice is to use Reynolds time-averaging for the density and pressure whereas Favre-averaging (also called density-weighted- or mass-averaging) for the velocity and energy (temperature) variables. The NS equations are then transformed into the Favre-averaged Navier-Stokes equations using the RANS formalism. Using the Einstein summation convention, these mean equations which account for density and temperature fluctuations together with the velocity and pressure fluctuations in a hypersonic flow field are given by \\
\vspace{0.2in}
%\noindent \textbf{Conservation of Mass}
\noindent \textit{Conservation of mass,}
\begin{equation}\label{favrecontinuity}
\dfrac{\partial \overline{\rho}}{\partial t} + \dfrac{\partial}{\partial x_i}(\overline{\rho} \widetilde{u}_i ) = 0
\end{equation}
%\noindent \textbf{Conservation of Momentum}
\noindent \textit{Conservation of momentum,}
\begin{equation}\label{favremom}
\dfrac{\partial}{\partial t} (\overline{\rho} \widetilde{u}_i) + \dfrac{\partial }{\partial x_j} (\overline{\rho} \widetilde{u}_j \widetilde{u}_i) +\Cline[blue]{\dfrac{\partial (-\tau_{ij}) }{\partial x_j} } = - \dfrac{\partial \overline{p}}{\partial x_i} + \dfrac{\partial \overline{\sigma}_{ij}}{\partial x_j}
\end{equation}
%\noindent \textbf{Conservation of total Energy}
\noindent \textit{Conservation of total energy,}
\begin{equation}\label{favreenergy}
\begin{split}
&\dfrac{\partial \overline{\rho} \widetilde{E}}{\partial t} + \dfrac{\partial}{\partial x_j} (\overline{\rho} \widetilde{H} \widetilde{u}_i )  + \Cline[blue]{ \dfrac{\partial (-\tau_{ij}) \widetilde{u}_i  }{\partial x_j} }  \\
&= \dfrac{\partial }{\partial x_j} \left[ - q_{L_{j}} - \Cline[blue]{ q_{T_{j}}  +  \overline{\sigma_{ji} u^{''}_i } - \overline{\rho u^{''}_j \frac{1}{2} u^{''}_i u^{''}_i } }  \right] + \dfrac{\partial (\overline{\sigma}_{ij} \widetilde{u}_i)  }{\partial x_j} 
\end{split}
\end{equation}
%\noindent \textbf{Equation of state}
\noindent \textit{Equation of state,}
\begin{equation}
\overline{p} = \overline{\rho} R \widetilde{T},
\end{equation}
where $\rho$ = density, $u$ = velocity, $p$ = pressure, $T$ = temperature, and $R$ = specific gas constant. Here, $\overline{(\cdot)}$, $\widetilde{(\cdot)}$, and $(\cdot)^{''}$ represent Reynolds time-averaged, Favre-averaged, and Favre fluctuating quantities, respectively. The averaged specific total energy and total enthalpy are given by
\begin{equation}
\begin{split}
\widetilde{E} = \widetilde{e} + \frac{1}{2} \widetilde{u}_i \widetilde{u}_i + k \\
\widetilde{H} = \widetilde{E} + \frac{\overline{p}}{\overline{\rho}} =  \widetilde{h} + \dfrac{1}{2} \widetilde{u_i} \widetilde{u}_i + k, \\
\end{split}
\end{equation}
where $k$ is the turbulence kinetic energy (TKE). The specific heats $C_p$ and $C_v$ are constant for a perfect gas and the averaged specific internal energy and enthalpy are given by
\begin{align}
\widetilde{e} = C_v \widetilde{T}, ~~ \widetilde{h} = C_p \widetilde{T}.
\end{align}
The averaged viscous stresses are described by,
\begin{equation}\label{eqn:viscous_stress}
\overline{\sigma}_{ij} = \widetilde{\mu} \left( \dfrac{\partial \widetilde{u}_i}{\partial x_j} + \dfrac{\partial \widetilde{u}_j}{\partial x_i}  - \dfrac{2}{3} \dfrac{\partial \widetilde{u}_k}{\partial x_k} \delta_{ij} \right)
\end{equation}
and the molecular viscosity is generally given by the Sutherland relation of the form $ \widetilde{\mu} = A \widetilde{T}^n/(B+\widetilde{T})$, where $n,A,$ and $B$ are constants that depend on the gas or Keyes law. The laminar heat flux is given by
\begin{equation}\label{eqn:laminar_heat_flux}
    q_{L_{j}} = - \dfrac{\widetilde{\mu} C_p}{Pr} \dfrac{\partial \widetilde{T}}{\partial x_j},
\end{equation}
and $Pr$ is the Prandtl number. Additional assumptions are made in the course of deriving the above equations, and these are the neglect of turbulent fluctuations of the dynamic viscosity, the thermal conductivity, and the specific heats.

All terms appearing in the Eqns.~(\ref{favrecontinuity})-(\ref{favreenergy}) are closed except for those resulting from turbulent fluctuations, represented by the blue underlined terms. These terms describe the average effect of turbulence on the mean flow. The blue underlined terms in Eqns.~(\ref{favremom}) and (\ref{favreenergy}) include the unclosed terms in the form of the Reynolds stresses $ \tau_{ij} = - \overline{\rho u^{''}_j u^{''}_i} $ and turbulent heat flux $q_{T_{j}} = \overline{\rho u^{''}_j h^{''}} $. The unclosed terms $ (D = \overline{\sigma_{ji} u^{''}_i}) $ and $ (T = \overline{\rho u^{''}_j \frac{1}{2} u^{''}_i u^{''}_i}) $ in Favre-averaged energy Eqn.~(\ref{favreenergy}) characterize the molecular diffusion and turbulent transport of TKE, respectively and they also appear in the Favre-averaged TKE transport equation (see Eqn.~(\ref{TKEeqn})). Notably, the gradients of Reynolds stresses in the Favre-averaged momentum Eqn.~(\ref{favremom}) play a significant role in determining the separation characteristics like separation location and length in case of flows involving shock-induced separations. On the other hand, the gradient of the turbulent heat-flux vector in the Favre-averaged total energy Eqn.~(\ref{favreenergy}) is crucial for the correct prediction of wall heat transfer rates. Hence, accurate representation of these turbulent quantities is key to correctly predicting the hypersonic flow field and the engineering QoIs. The goal of turbulence modeling is to develop appropriate closures for these unclosed terms by relating them to known mean flow quantities such as velocity and temperature.

%%%%%%%%%%%%%%%%%%%%%%%%%%%%%%%%%%%%%%%%%%%%%%%%%%%%%%%%%%%%%%%%%%%%%%%%%%%%%%%%%%%%%%%%%

Models for the Reynolds stresses   $ \tau_{ij} $ fall into two classes: ``eddy viscosity'' relations between the stresses and the mean velocity gradients at the same point in space; and models of the further unknowns in the exact partial differential transport equations for the individual Reynolds stress components. Eddy viscosity models are based on an analogy to laminar viscous transport given by Eqn.~(\ref{eqn:viscous_stress}) introducing eddy viscosity $\mu_T$ as an unknown and the exact formulation of $\mu_T$ is model dependent. The turbulent heat flux vector, on the other hand, is generally modeled in analogy to the laminar heat flux vector given by Eqn.~(\ref{eqn:laminar_heat_flux}) using a gradient diffusion hypothesis with a turbulent Prandtl number $Pr_t$ as an unknown model coefficient. The turbulent Prandtl number $Pr_t$ is typically set to a constant value, however, models that vary $Pr_t$ in a flow field using algebraic relations or partial differential equations also exist. The unknown molecular diffusion $D$ and turbulent transport $T$ (also known as triple velocity correlation) terms are also modeled using a gradient diffusion hypothesis albeit for the TKE which introduces unknown model coefficients the values of which depend on the specific turbulence model.

%%%%%%%%%%%%%%%%%%%%%%%%%%%%%%%%%%%%%%%%%%%%%%%%%%%%%%%%%%%%%%%%%%%%%%%%%%%%%%%%%%%%%%%%%

%%%%%%%%%%%%%%%%%%%%%%%%%%%%%%%%%%%%%%%%%%%%%%%%%%%%%%%%%%%%%%%%%%%%%%%%%%%%%%%%%%%%%%%%%

\section{Turbulence modeling for hypersonic flows}\label{sec3}

Historically,  turbulence models used to predict compressible flows have involved simple extensions of incompressible models using density-weighted averages. Variable density extensions of the existing incompressible turbulence models are obtained by invoking Morkovin's hypothesis. This hypothesis asserts that compressibility or Mach number only affects the turbulence, at least the velocity correlations, through variations in mean density $\overline{\rho}$ alone and that density and temperature fluctuations have negligible effect on the turbulence. Morkovin's hypothesis with density-weighted averaging effectively isolates the turbulence dynamics from the compressibility effects by focusing on how the turbulence would behave if the density were constant. It implies the structural similarity between compressible and incompressible turbulent flows and provides a foundation for applying incompressible modeling approaches to compressible flows.

DNS studies of Refs.~\onlinecite{duan2011assessment,huang2020simulation} have confirmed the validity of Morkovin's hypothesis for the mean velocity and mean temperature fields in the case of ZPG TBLs at hypersonic speeds (freestream Mach numbers ranging up to 14) with varying wall temperatures (wall-to-recovery temperature ratio between 0.18 and 1.0). Similar conclusions are drawn from the Mach $6.7$~\cite{horstman1972turbulent,owen1972structure,owen1975mean,mikulla1976turbulence}, Mach $7.2$~\cite{sahoo2009effects,baumgartner1997turbulence}, and Mach $11$~\cite{mcginley1994turbulence} experimental studies of ZPG flat plate TBLs. Furthermore, DNS studies of Ref.~\onlinecite{so1998morkovin} have also shown the applicability of the hypothesis for predicting the turbulence field itself for flat-plate TBLs up to a Mach number of at least 10 at adiabatic as well as cold wall conditions. However, the hypothesis is likely to become invalid under nonequilibrium conditions e.g. SBLIs. As the Mach numbers increase in the hypersonic regime, significant compressibility effects are introduced which are not adequately captured by the hypothesis. Additional terms such as those related to the dilatation dissipation, pressure work, pressure dilatation, etc. are generally added to the turbulence models or the incompressible model coefficients are modified to account for compressibility effects.

Most of the research efforts in turbulence modeling have focused on improving the Reynolds stress tensor closures. Hypersonic ZPG TBLs have unequal distribution of Reynolds normal stresses across the boundary layer thickness. Moreover, these TBLs interact with the shock waves in complex ways, leading to rapid distortion of the turbulence structure. A shock wave has strong directionality, i.e. it alters the flow quantities differently between the shock-normal and shock-transverse directions. This results in significant anisotropy in the Reynolds stresses. Reynolds stress tensor can be written in terms of its isotropic form and the deviations from isotropy as,
\begin{equation}
 \tau_{ij} = -  \overline{\rho u''_i u''_j} =  \underbrace{ \left( \tau_{ij} + \frac{2}{3} \overline{\rho} k\delta_{ij}\right)}_{deviatoric}  - \underbrace{\frac{2}{3} \overline{\rho}
 k\delta_{ij}}_{isotropic}.
\end{equation}
The departure from the isotropic state provides a measure of the Reynolds stress anisotropy state, i.e. deviation from the normal stress isotropy. The deviatoric or the anisotropy part of the Reynolds stress tensor i.e. $\tau^{dev}_{ij}$  can be equivalently expressed in terms of the Reynolds stress anisotropy tensor $ a_{ij}$ - a traceless and symmetric tensor - as, 
\begin{equation}
    a_{ij} =  \dfrac{- \tau^{dev}_{ij}}{\overline{\rho}  k}.
\end{equation}
The Reynolds stress tensor can be written as,
\begin{equation}
\tau_{ij} = - \overline{\rho} k \left( a_{ij} + \dfrac{2}{3} \delta_{ij} \right).
\label{eqn:Rij-aij-relation}
\end{equation}
Thus, the closure for Reynolds stress tensor amounts to constructing a representation for the anisotropy tensor $ a_{ij} $. The Reynolds stress tensor is linearly or non-linearly related to the mean strain-rate and/or vorticity tensors via $\mu_T$ in eddy-viscosity models, whereas transport equations for individual Reynolds stress components are solved in the case of Reynolds stress transport models. EVMs are generally the preferred choice while a very limited amount of work exists on RSTMs for hypersonic flow predictions.

The turbulent heat flux vector is another important unclosed term in the Favre-averaged total energy Eqn.~(\ref{favreenergy}). It is most commonly modeled by invoking the Reynolds analogy. A gradient diffusion approximation is generally used, similar to the eddy viscosity concept for momentum transfer. This is expressed as  
\begin{equation}
q_{T_{j}} = - \dfrac{\mu_T C_p}{Pr_t} \dfrac{\partial \widetilde{T}}{\partial x_j},
\end{equation}
where $ Pr_t $ is the turbulent Prandtl number. Morkovin's hypothesis, which assumes negligible total temperature fluctuations across an equilibrium TBL, leads to $Pr_t = 1$ for adiabatic wall conditions. Generally, a constant value of $0.9$ ($0.89$ in some works) for $Pr_t$ is used.

%---------------------------------------------------------------------------

\subsection{Modeling capabilities and limitations for hypersonic flows}

The most popular and widely used class of turbulence models is the eddy-viscosity models that introduce closures for $a_{ij}$ using the concept of eddy-viscosity $\mu_T$. A generalized expansion of $a_{ij}$ is given by the linear 
combination of the basis tensor $T_{ij}$ \citep{pope1975more}
\begin{equation}
    a_{ij} = \sum_{\lambda} \beta^{(\lambda)}  T^{(\lambda)}_{ij}.
    \label{eqn:aij-general}
\end{equation}
On dimensional grounds, $a_{ij}$ is assumed to be a function of ${S}^{\star}_{ij} = \tau S^{D}_{ij}$ and $\Omega^{\star}_{ij} = \tau \Omega_{ij}$, where $\tau = \epsilon/k = 1 / (C_{\mu} \omega)$ represents the turbulent time scale with $\epsilon,\omega$ being the dissipation rate of TKE and dissipation per unit TKE (or specific dissipation), respectively and $C_{\mu}$ is a model coefficient. Here, $S^{D}_{ij} = S_{ij} - (1/3) S_{kk} \delta_{ij}$ is the deviatoric part of the strain-rate tensor $S_{ij} = (1/2) (\partial \widetilde{u}_i/\partial x_j + \partial \widetilde{u}_j/\partial x_i)$ and $\Omega_{ij} = (1/2) (\partial \widetilde{u}_i/\partial x_j - \partial \widetilde{u}_j/\partial x_i)$ is the vorticity (or rotation) tensor. The scalar coefficients $\beta$'s may be functions of the invariants of ${S}^{\star}_{ij}$ and $\Omega^{\star}_{ij}$ and $\lambda$ is the number of linearly independent basis tensors.  It follows from the Cayley-Hamilton theorem that the number of independent invariants is two and $\lambda = 3$ for two-dimensional flows, whereas there are five independent invariants and $\lambda = 10$ for three-dimensional flows. Using Eqns.~(\ref{eqn:Rij-aij-relation}) and (\ref{eqn:aij-general}), the Reynolds stress tensor $\tau_{ij}$ can be expressed as 
\begin{equation}
    \tau_{ij} =  - \overline{\rho} k \sum_{\lambda} \beta^{(\lambda)}  T^{(\lambda)}_{ij} - \dfrac{2}{3} \overline{\rho} k \delta_{ij}.
    \label{eqn:Rij-general-final}
\end{equation}
Thus, the task of formulating a model for the Reynolds stress is reduced to that of determining $\beta$'s as the basis tensors $T_{ij}$ are known functions of ${S}^{\star}_{ij}$ and $\Omega^{\star}_{ij}$. In the EVM framework, eddy-viscosity is constructed from turbulence scalars, and a general expression for the eddy-viscosity hypothesis can be written as
\begin{equation}
\tau_{ij} = 2 \mu_T S^{D}_{ij} - \dfrac{2}{3} \overline{\rho} k \delta_{ij} - \overline{\rho} k a_{ij}^{(NL)}, 
 \label{eqn:Rij-general}
\end{equation}
where superscript ``\textit{NL}'' refers to the non-linear (quadratic and higher) contributions added to the linear part in the definition of $\tau_{ij}$. Eddy viscosity accounts for the momentum transfer and energy dissipation due to turbulent fluctuations. Unlike the molecular viscosity $\mu$, eddy viscosity $\mu_T$ is a hypothetical property of the flow that needs to be modeled. Depending on whether the non-linear contributions are considered or not in Eqn.~(\ref{eqn:Rij-general}), the EVMs can be categorized into Boussinesq hypothesis-based models (also called Linear Eddy-Viscosity Models or LEVMs) and Extended Boussinesq hypothesis-based models. Furthermore, $\mu_T$ is modeled in terms of length and velocity scales of turbulence based on dimensional grounds. The way these turbulence scales are determined defines the type of eddy viscosity model to be used. If both the scales are determined algebraically from mean flow data, the models are referred to as zero-equation or algebraic models. If the length scale is determined algebraically, but the velocity scale is determined from a field equation such as the TKE equation, or alternatively, a transport equation for some form of $\mu_T$ is directly solved, the model is referred to as a one-equation model. If both the scales are determined from field equations, the resulting model is called a two-equation model. Zero-, one-, and two-equations-based LEVMs exist but the extended Boussinesq hypothesis is generally used in a two-equation modeling framework.

%%%%%%%%%%%%%%%%%%%%%%%%%%%%%%%%%%%%%%%%%%%%%%%%%%%%%%%%%%%%%%%%%%%%%%%%%%%%%%%%%%%

\subsubsection{Boussinesq hypothesis-based models}

Boussinesq eddy-viscosity hypothesis~\cite{boussinesq1877essai} linearly relates the deviatoric part of the Reynolds stress tensor to the traceless mean strain rate tensor $S^{D}_{ij}$ via the eddy-viscosity $\mu_T$,
\begin{equation}
\tau_{ij}  = 2 \mu_T S^{D}_{ij} - \frac{2}{3} \overline{\rho} k \delta_{ij}.
\label{eqn:Boussinesq}
\end{equation}
It is the most commonly used approach to model the Reynolds stress tensor. However, it does not accurately represent the Reynolds stress tensor, and the key deficiencies of this constitutive relation influential in hypersonic flows predictions are :

\begin{itemize}
    \item It cannot correctly predict Reynolds stress anisotropy, a key feature in hypersonic SBLI flows. Application of the thin-shear-layer approximations to Eqn.~(\ref{eqn:Boussinesq}) yields purely isotropic normal stresses i.e. $\tau_{xx} = \tau_{yy} = \tau_{zz} = -(2/3) \overline{\rho} k$, which is inaccurate even for a non hypersonic ZPG TBL. %\EP{Isn't it inaccurate for a subsonic TBL as well?}
    \item Turbulent transport of momentum is determined by a single scalar $\mu_T$, so at most one Reynolds stress component (generally, the shear stress) can be represented accurately.   
    \item It is not a realizable model and breaks down in the presence of shock waves. As $\tau_{ij}$ is proportional to mean strain rates, it increases monotonically with mean velocity gradients thereby significantly overestimating Reynolds stresses at shock waves. Limiting functions or damping factors are generally used to prevent excessive turbulence production or dissipation rates in the vicinity of shock waves. 
\end{itemize}
Despite these shortcomings, this modeling framework remains the most popular group of models in aerospace-related CFD. This is because the extremely complex geometries and the associated meshing problems involved in such applications favor the simplicity and economy of these models for engineering predictions. 

Zero-equation models are the simplest and most economical models as they do not involve solving additional transport equations. These models directly model $\mu_T$ using theoretical/empirical algebraic relations. Baldwin–Lomax (BL) model is the stand-out zero-equation model in hypersonic applications involving flow separation, which formulates $\mu_T$ into a two-layer structure. The inner layer part is based on the mixing length model with a van Driest viscous damping correction while the outer layer uses a function of the maximum value of the mean vorticity and total velocity difference. It was initially developed based on the Cebeci–Smith (CS) model with modifications that avoid the necessity for calculating the edge of the boundary layer. BL model shows disparities in mean profiles with DNS, especially for the temperature, under diabatic (cold or heated wall) conditions for ZPG hypersonic TBLs~\cite{chen2024improved}. In general, it tends to overpredict separation regions and wall heat-transfer rates when applied to two-dimensional and axisymmetric SBLIs~\cite{dilley2001evaluation,hejranfar2012dual,horstman1987prediction}. It gives incorrect surface pressure and wall heat-transfer rates for three-dimensional crossing SBLIs \citep{narayanswami1993numerical,panaras2015turbulence,bardina1994three}.
Modifications especially for hypersonic flows in terms of employing well-established relations for compressible turbulent mean flows including the velocity transformation and algebraic temperature–velocity relation~\cite{chen2024improved}, adjusting the model coefficients to take into account compressibility and pressure gradient effects \cite{hejranfar2012dual,hu1995turbulent,kim1988hypersonic,panaras2015turbulence}, show significant improvements over the original model for a range of hypersonic flow applications including SBLIs.

%_______________________________________________________________________________________

%\subsubsection{One-equation models}

The use of zero-equation models to provide a length scale for the TKE equation has been explored for hypersonic applications. For example, the McDonald-Camarata model~\cite{mcdonald1968extended}  with TKE-equation failed to improve the solution provided by the zero-equation model for a Mach $7.4$ hypersonic inlet~\cite{ng1989turbulence}. One equation models like SA, Goldberg $R_T$~\cite{goldberg2000hypersonic}, and Menter's one-equation model~\cite{menter1997eddy} solve for $\mu_T$ (or some form of it) directly without dimensionally relating it with the turbulence scales. Variable density extensions of these models have been applied for hypersonic flow predictions. The standard SA model without any corrections generally underpredicts the separation size but may give good predictions for surface heat transfer rates~\cite{goldberg2000hypersonic,paciorri1997validation}. The $R_T$ model has been shown to perform better than the SA model in hypersonic SBLIs predicting the correct extent of the separation region and heat-transfer rates for a Mach $9.22$ flow over $38^\circ$ cooled ramp. Also, the $R_T$ model does not require the calculation of wall distance, unlike the SA model. However, the performance of this model has not yet been tested widely for hypersonic boundary layer applications. On the other hand, Menter's one-equation model, derived from the $k$-$\epsilon$ model, performs poorly in predicting the separation and heat transfer rate for hypersonic SBLIs~\cite{goldberg2000hypersonic}.

%_______________________________________________________________________________________

%_______________________________________________________________________________________

%\subsubsection{Two-equation models}

In two-equation turbulence models the transport equations are solved for two turbulent fields that are directly related to the length and velocity scales. Unlike the zero- and one-equation models, these models do not require the specification of empirical turbulence scales that must be adjusted in an ad-hoc fashion from one flow to the other. Hence, these models represent the simplest level of Reynolds stress closure that can be formulated in a geometry-independent fashion. The TKE is generally the standard choice for the velocity scale whereas $\epsilon$ or $\omega$ are widely used to form the length scale. The standard $k$-$\epsilon$ and $k$-$\omega$ models, however, show discrepancies for surface heat transfer for ZPG hypersonic TBLs, especially at cold wall conditions, while reasonably predicting skin-friction and mean flow profiles. These models significantly underpredict the separation while grossly overpredicting peak heat transfer and skin-friction for hypersonic SBLIs. The widely popular Menter SST $k$-$\omega$, a combination of $k$-$\omega$ and $k$-$\epsilon$ models, also performs poorly in predicting surface heat transfer for hypersonic TBL attached and separated flows. The performance of Menter SST model in predicting SBLIs greatly depends on the choice of sensitizing $\mu_T$ i.e. using vorticity or strain-rate invariants and production limiters.

Other choices for the auxiliary equations to determine velocity and length scale exist, for example, $q = \sqrt{k}$, enstrophy $\zeta$ i.e. the R.M.S. fluctuating vorticity ($\zeta \sim \omega^2$), and integral length scale $l$. Refs.~\onlinecite{nance1999turbulence,xiao2007modeling,xiao2007role} apply the $k$-$\zeta$ model~\cite{robinson1998further} to Mach $5$ and Mach $9.2$ SBLIs involving attached and separated flows. The model is free of damping and wall functions like the $k$-$\omega$ models. The model predictions for separation length agree well with experiments while skin friction is underpredicted and peak heat transfer is significantly overpredicted for these cases. Coakley's $q$-$\omega$~\cite{coakley1983turbulence} and Smith's $k$-$l$~\cite{smith1995prediction} models generally result in smaller separation and overprediction of heat transfer for hypersonic SBLIs. Such models with scale-determining equations other than $k$ and $\epsilon$ or $\omega$, however, have not been pursued to any great extent for hypersonic flows.

%-------------------------------------------------------------------------------------

\subsubsection{Extended Boussinesq hypothesis-based models}

This class of models includes non-linear contributions to represent the Reynolds stress tensor in addition to the linear part, as given by Eqn.~(\ref{eqn:Rij-general}). The higher order representation allows for a more general coupling between the mean-field and $\tau_{ij}$, compared to the models based on the Boussinesq hypothesis. Non-linear eddy-viscosity models (NLEVMs) and Explicit algebraic Reynolds stress models (EARSMs) fall under this category. Despite the formal functional equivalence between NLEVMs and EARSMs, fundamental differences exist in their origin, precisely in the way the expansion coefficients $\beta$'s are obtained. In the case of NLEVMs, the expansion coefficients are determined based on calibrations with experimental or numerical data, and on some physical consistency constraints. On the other hand, in the case of EARSMs, the expansion coefficients are derived from the full differential Reynolds stress equations under weak equilibrium assumption. These models are developed to improve the predictions of Reynolds stress fields by employing either quadratic, cubic, or higher-order expansions of the Reynolds stresses in terms of the strain and vorticity tensors. Also, the turbulent eddy viscosity is additionally sensitized to the mean strain and vorticity tensors. Generally, a quadratic expansion of the Reynolds stresses is found to be sufficient to predict anisotropy in normal Reynolds stresses~\cite{leschziner2006modelling}, while cubic and higher-order terms are needed to predict streamline curvature, including swirl effects. NLEVMs and EARSM are generally used with the family of $k$-$\epsilon$ or $k$-$\omega$ models by replacing the Boussinesq hypothesis. The model development generally involves a straightforward compressibility extension of the incompressible model forms and most of the models neglect compressibility corrections.

Application of the compressible form of NLEVMs to hypersonic flows is seriously limited in the literature. Goldberg \textit{et al.}~\cite{goldberg2000hypersonic} apply a cubic $k$-$\epsilon$ model wherein Reynolds stresses are modeled using the mean strain and vorticity relations from the quadratic model of Shih \textit{et al.}~\cite{shih1993realisable} with the cubic extension proposed by Lien and Leschziner~\cite{lien1996low} to predict several hypersonic SBLIs. The model predicts correct trends of wall heat transfer including the peak when applied to Mach $8.03$ curved compression surface, with experimental data by Holden~\cite{holden1992turbulent} (cooled wall). However, it significantly overpredicts surface heat transfer while underpredicting separation size with incorrect peak pressure location for Mach $9.22$ flow over a $38^\circ$ cooled ramp of Coleman and Stollery~\cite{coleman1972heat}. For a Mach $8.3$ flow in a wedge inlet configuration involving complex three-dimensional crossing SBLI, the model grossly overpredicts surface heat transfer while giving good surface pressure prediction. Zhang \textit{et al.}~\cite{zhang2022application} apply the cubic $k$-$\epsilon$ model of Craft \textit{et al.}~\cite{craft2000progress,craft1996development} with low-Reynolds-number effects to Mach $7.05$ SBLIs with fully attached and fully separated flows. The model predicts reasonably correct surface pressure distributions for these cases while significantly overpredicting peak heat transfer rates.

EARSMs can be considered as the approximations of Reynolds stress transport models. They are numerically simpler and more economical than Reynolds stress transport closures and provide higher accuracy compared to LEVM and are therefore gaining popularity. The modeling strategy followed in deriving an EARSM involves a mathematically sound transfer of desirable model properties inherent to the elaborate RSTMs to a less elaborate modeling framework. Thus, EARSMs can be considered as an intermediate level between RSTMs and LEVMs. Specifically, these models neglect the advection and diffusion terms in the exact transport equation for $ a_{ij} $ which yields an algebraic expression that can be written as,
\begin{equation}
- \dfrac{\tau_{ij}}{\overline{\rho} k} (P - \epsilon) = P_{ij} - \epsilon_{ij} + \Pi_{ij}
\end{equation}
Here, $P$ and $\epsilon$ represent the TKE production and dissipation, respectively. On the other hand, $P_{ij}$ represents the production of Reynolds stresses, $ \Pi_{ij}$ and $\epsilon_{ij}$, respectively, the pressure-strain correlation representing the mechanism of turbulence redistribution and relaxation, and the dissipation rate tensor. Depending on the choice of the models for $ \epsilon_{ij} $ and $ \Pi_{ij} $, the expansion coefficients $\beta$'s in Eqn.~(\ref{eqn:Rij-general-final}) can be obtained.

The EARSM of Wallin and Johansson~\cite{wallin2000explicit} (WJ-EARSM) is one of the few EARSMs developed for compressible flows that have been used to predict hypersonic flows. WJ-EARSM is based on a recalibrated compressible form of the general linear model of the Launder, Reece, and Rodi RSTM rapid pressure-strain rate model. It uses wall-damping functions derived from the van Driest damping function to obtain the correct near-wall limits of the anisotropies. The new near-wall treatment ensures realizability for the individual stress components. The model accounts for three-dimensionality in the mean flow description of the stress anisotropy. WJ-EARSM represents a fully explicit and self-consistent algebraic relation and consists of a quartic expansion of the Reynolds stresses for a general three-dimensional mean flow. The model gives substantially improved predictions for the separation in case of the Schulein~\cite{schulein2006skin} Mach $5$ impinging oblique SBLI flows with $10^\circ$ and $14^\circ$ shock generators compared to the standard $k$-$\omega$ and $k$-$\epsilon$ models using Boussinesq hypothesis~\cite{lindblad1998prediction,wallin2000explicit,vemula2020explicit}. Vemula and Sinha~\cite{vemula2020explicit} propose improvements to the WJ-EARSM model to correctly predict the amplification of Reynolds stresses across shocks of varying strengths using linear interaction analysis and DNS results for canonical shock/turbulence interactions (STIs) at a range of Mach numbers. Vemula and Sinha~\cite{vemula2020explicit} applied the shock-unsteadiness (SU) modification of Sinha \textit{et al.}~\cite{sinha2003modeling} to WJ-EARSM production term and obtained better predictions for surface pressure and skin-friction for Schulein's~\cite{schulein2006skin} $10^\circ$ and $14^\circ$ cases with separations whereas the SU modified and original model gave similar results for the $6^\circ$ case and predicted a small separation in contrast to the experiments. 

Raje and Sinha~\cite{raje2021anisotropic} use the EARSM by Rung et al.~\cite{rung1999assessment} to augment the standard SST $k$-$\omega$ model for high-speed flows. Rung \textit{et al.} EARSM is a realizable quadratic eddy viscosity model and is an improved version of the EARSM by Gatski and Speziale~\cite{gatski1993explicit}. It enjoys the desirable properties inherent to the elaborate second-moment closure of Speziale, Sarkar, and Gatski~\cite{speziale1991modelling}. The model is developed for compressible flows and it adopts the compressibility correction in the form of Sarkar’s pressure-dilatation model. The Rung EARSM with the SST model gives poor predictions for the separated Mach $8$ compression corner and Mach $5$ impinging oblique SBLI cases and the EARSM does not improve upon the standard SST model predictions. Raje and Sinha~\cite{raje2021anisotropic} propose modifications to the structure parameter $a_1$ (defined as the ratio between Reynolds shear stress and TKE) using the log-layer analysis of EARSM. The structure parameter $a_1$, which is generally taken to be equal to $0.31$ in the Menter SST $k$-$\omega$ model, is made a function of the mean velocity field in the form of the magnitude of the deformation rate tensor. Furthermore, to limit any unphysical high values of $a_1$ in the shock regions, a limiter to $a_1$ is proposed using an analysis of the SU model at a shock wave. The proposed $a_1$ is not a constant, unlike the standard SST model, but responds to the changes in strain rate and vorticity magnitude with an added effect of unsteady shock oscillations in response to incoming turbulent fluctuations. The eddy viscosity formulation is modified using the new $a_1$ definition and deformation rate tensor and the resulting model is called SUQ-SST model. The model is applicable to high-speed turbulent boundary layers including shock waves. The SUQ-SST model consistently predicts the separation shock location and separation bubble size and provides improved surface pressure and skin distributions compared to the baseline model for a range of hypersonic separated flow cases including compression corner and impinging oblique SBLI. However, the model overpredicts surface heat transfer for these cases.

%-------------------------------------------------------------------------------------

\subsubsection{Reynolds stress transport models}\label{sec:rstm}

Reynolds stress transport models (RSTMs), also called Differential Reynolds stress models, consist of transport equations for all the components of the Reynolds stress tensor without recourse to the eddy viscosity concept. They constitute the highest level of RANS-based turbulence models. These models have the natural potential to deal with the dynamics of inter-component energy transfer, and account for the effects of stresses increase or decrease due to curvature, acceleration or deceleration, swirling flow, and so on. In the case of three-dimensional mean flows, they require the solution of six highly coupled, non-linear partial differential equations and at least one equation for the turbulence dissipation rate. Thus, Reynolds stress transport models are mathematically more complex, pose greater numerical challenges, and require higher computational costs than the eddy viscosity models. The transport equations for the Reynolds stress tensor can be mathematically derived from the momentum equations. The Favre-averaged Reynolds stress tensor transport  equation can be written as 
\begin{equation} 
%\frac{D\overline{\rho} R_{ij}}{Dt}=
\frac{\partial }{\partial t } (\overline{\rho} \tau_{ij} ) +  \frac{\partial }{\partial x_k} (\overline{\rho}\widetilde{u_k} \tau_{ij}) =
P_{ij} + \Pi_{ij} - \epsilon_{ij} + M_{ij} + D_{ij},
\label{eqn:RSTM}
\end{equation}
where
\begin{equation}
\begin{split}
    P_{ij} = -\overline{\rho} \tau_{ik} \dfrac{\partial \widetilde{u_j}}{x_k}  -\overline{\rho} \tau_{jk} \dfrac{\partial \widetilde{u_i}}{x_k}; \\
    \Pi_{ij} = \overline{p' \left( \dfrac{\partial u''_i}{\partial x_j} + \dfrac{\partial u''_j}{\partial x_i}  \right) } = \phi_{ij} + \dfrac{2}{3} \phi_p \delta_{ij} + D^{(p)}_{ij}; \\
   \phi_{ij} = \overline{p' \left( \dfrac{\partial u''_i}{\partial x_j} + \dfrac{\partial u''_j}{\partial x_i} - \dfrac{\partial u''_k}{\partial x_k} \delta_{ij}  \right) }  ; \hspace{0.1in}  \phi_p = \overline{p' \dfrac{\partial u''_k}{\partial x_k}} \delta_{ij}; \\
    \epsilon_{ij} = \overline{\sigma'_{ik} \dfrac{\partial u''_j}{ \partial x_k} } + \overline{\sigma'_{jk} \dfrac{\partial u''_i}{\partial x_k} }; \\
    M_{ij} = \overline{u''_i} \left( \dfrac{\partial \overline{\sigma}_{jk}}{\partial x_k} - \dfrac{\partial \overline{p}}{\partial x_j} \right) + \overline{u''_j} \left( \dfrac{\partial \overline{\sigma}_{ik}}{\partial x_k} - \dfrac{\partial \overline{p}}{\partial x_i} \right); \\
 D_{ij} = D^{(u)}_{ij} + D^{(p)}_{ij} + D^{(\mu)}_{ij}  = - \dfrac{\partial}{\partial x_k} \left( \overline{\rho u''_i u''_j u''_k } \right) - \\ \dfrac{\partial}{\partial x_k} \left( \delta_{ik} \overline{p' u''_j} + \delta_{jk} \overline{p' u''_i} \right) +  \dfrac{\partial}{\partial x_k}
\left( \overline{\sigma'_{ik} u''_j + \sigma'_{jk} u''_i} \right).
\end{split}
\end{equation}
The terms on the R.H.S. of Eqn.~({\ref{eqn:RSTM}) represent the production, pressure-strain
redistribution, dissipation (destruction), turbulent mass flux contribution (direct compressibility effects), and turbulent transport (triple velocity correlation + pressure diffusion) and viscous diffusion, respectively. Apart from the production term, which is exact, all other terms need to be modeled. In most RSTMs, direct compressibility effects in terms of $M_{ij}$ and $\phi_p$ are neglected. The dissipation $\epsilon_{ij}$ is usually modeled using an algebraic approach, where a single scalar equation is solved for the dissipation rate of TKE ($\epsilon$ or $\omega$) and an algebraic expression distributes this scalar on the different components of $\epsilon_{ij}$, using an appropriate tensorial representation. For example, the destruction term is generally modeled using the Kolmogorov hypothesis of local isotropy as
\begin{equation}
    \epsilon_{ij} = \dfrac{2}{3} \overline{\rho} \epsilon \delta_{ij} = \dfrac{2}{3} C_\mu \overline{\rho} k \omega \delta_{ij}.
\end{equation}
The diffusion term is generally modeled using a gradient diffusion model while the processes of turbulent diffusion by pressure fluctuations are usually neglected. The redistribution tensor $\Pi_{ij}$ generally exerts a major influence on the model performance and has received the greatest amount of attention. Modeling strategies for $\Pi_{ij}$ typically consist of some that attempt to model $\Pi_{ij}$ as a whole and others that model the deviatoric part $\phi_{ij}$ and pressure diffusion $D^{(p)}_{ij}$ separately. The classical approach is to split the model for $\phi_{ij} $ into slow and rapid parts. The slow pressure-strain term is also referred to as the return-to-isotropy term and represents turbulence-turbulence interaction. The rapid pressure-strain term ($\phi^r_{ij}$) responds directly to the change in mean velocity gradients and is of the same form as of production term in Eq.~(\ref{eqn:RSTM}). It tends to redistribute the energy from the Reynolds stress component, where the production is significantly large compared to other components, consequently reducing the anisotropy in Reynolds stresses.

Wilcox~\cite{wilcox1998turbulence} apply the Wilcox Stress-$\omega$ model to a Mach $11$ SBLI with cold wall $T_w/T_{aw} = 0.2$. The Wilcox model by design is similar to the Launder-Reece-Rodi (LRR) model for the pressure strain correlation model but uses $\omega$ instead of $\epsilon$ for the scale-determining equation and does not use a wall-reflection term. The model predicts a smaller separation than experiments. For a Mach $7$, $35^\circ$ cylinder-flare configuration with $T_w/T_{aw} = 0.4$, the model still predicts a smaller separation and also significantly higher peak heat transfer rates compared to the experiments. The model results are similar to Wilcox 2006 $k$-$\omega$ model. The deficiencies in the predictions were attributed to the deficiencies in modeling the scale-determining equation and modeling of pressure-strain correlation. Refs.~\onlinecite{bosco2011investigation,frauholz2014investigation} apply the Speziale–Sarkar–Gatski/Launder–Reece–Rodi (SSG/LRR)-$\omega$ model by Eisfeld and Brodersen~\cite{eisfeld2005advanced} to Mach $6.35$ compression corner fully attached ($15^\circ$ ramp angle) and separated flows ($40^\circ$ ramp angle). The SSG/LRR-omega model is a combination of the two previously existing models: the Speziale, Sarkar, and Gatski (SSG) model using an $\epsilon$-based length-scale equation is employed in the far field and coupled to the $\omega$-based Launder, Reece, and Rodi (LRR) model in its modified Wilcox version for the near-wall region. The development of the SSG/LRR-omega model follows the ideas used by Menter for the SST $k$-$\omega$ model. A good agreement with experiments is obtained for the separation size along with the surface pressure distribution and peak heat transfer on the ramp for both cases. For a three-dimensional two-ramp Mach $7.5$ intake at slightly off-design conditions of Mach $7.7$, the SSG/LRR-$\omega$ model predicts the pressure coefficient and the Stanton number of the lower intake wall along the center line in good agreement with the experiments. The flow partially relaminarizes while expanding and turning inward into the interior engine section, which is predicted well by the model. Gerolymos \textit{et al.}~\cite{gerolymos2012term} assembles a working model by combing the different closures for the $\epsilon_{ij}$, $\phi_{ij}$, $D^{(u)}_{ij}$, and $D^{(p)}_{ij}$ terms and neglect the direct compressibility effects i.e. $M_{ij}$ and $\phi_p$. The model uses the Launder–Sharma modified dissipation rate $\epsilon$ transport equation but with a tensorial diffusion coefficient. Comparisons with Gerolymos and Vallet (GV RSM)~\cite{gerolymos2002wall} and the low-Reynolds number Reynolds stress model WNF (wall normal free)-LSS (Launder–Shima–Sharma)-HL (Hanjalic–Launder) RSM~\cite{gerolymos2004contribution} are done for Schulein's Mach $5$, $14^\circ$ fully separated case. The WNF-LSS-HL RSM underestimates the recirculation zone whereas the new and old GV models give improved predictions of the upstream influence but underestimate the height of the recirculation zone. All these models fail to predict the correct shape of wall friction in the reattachment flow region.

%------------------------------------------------------------------------------------

%------------------------------------------------------------------------------------

\subsubsection{Modeling of scale determining equations}

A majority of the widely used EVMs, e.g. family of $k$-$\epsilon$ and $k$-$\epsilon $ models, use TKE to represent the velocity scale. The value of TKE directly represents the strength of the turbulence in the flow. The TKE transport equation for compressible flows is given by
\begin{equation}\label{TKEeqn}
\begin{split}
 \dfrac{\partial \overline{\rho} k}{\partial t} +  \dfrac{\partial \overline{\rho} \widetilde{u}_j k}{\partial x_j} =  P_k - \overline{\rho} \epsilon + D + T + \Pi^t + M + \Pi^d,
 \end{split}
\end{equation}
where 
\begin{equation}
\begin{split}
 P_k =  \tau_{ij} \dfrac{\partial \widetilde{u}_i}{\partial x_j}; \overline{\rho} \epsilon =    \overline{\sigma'_{ji} \dfrac{\partial u''_i}{\partial x_j}} ; D =  \dfrac{\partial }{\partial x_j} \left( \overline{\sigma'_{ji} u''_i } \right); \\
 T = -  \dfrac{\partial }{\partial x_j} \left( \overline{\rho u''_j \frac{1}{2} u^{''}_i u''_i } \right); \Pi^t = - \dfrac{\partial }{\partial x_j} \left( \overline{p' u''_j} \right); \\
 M =   \overline{u''_{i}} \left( \dfrac{\partial \overline{\sigma_{ji}}}{\partial x_j} - \dfrac{\partial \overline{p} }{\partial x_i} \right); \Pi^d =  \overline{p' \dfrac{\partial u'' }{\partial x_i} }.
\end{split}
\end{equation}

%%%%%%%%%%%%%%%%%%%%%%%%%%%%%%%%%%%%%%%%%%%%%%%%%%%%%%%%%%%
	
	\begin{figure*}
		\centering
		\begin{subfigure}{0.49\textwidth}
			\centering
			\includegraphics[width=1.0\linewidth, trim = 0mm 0mm 0mm 0mm]{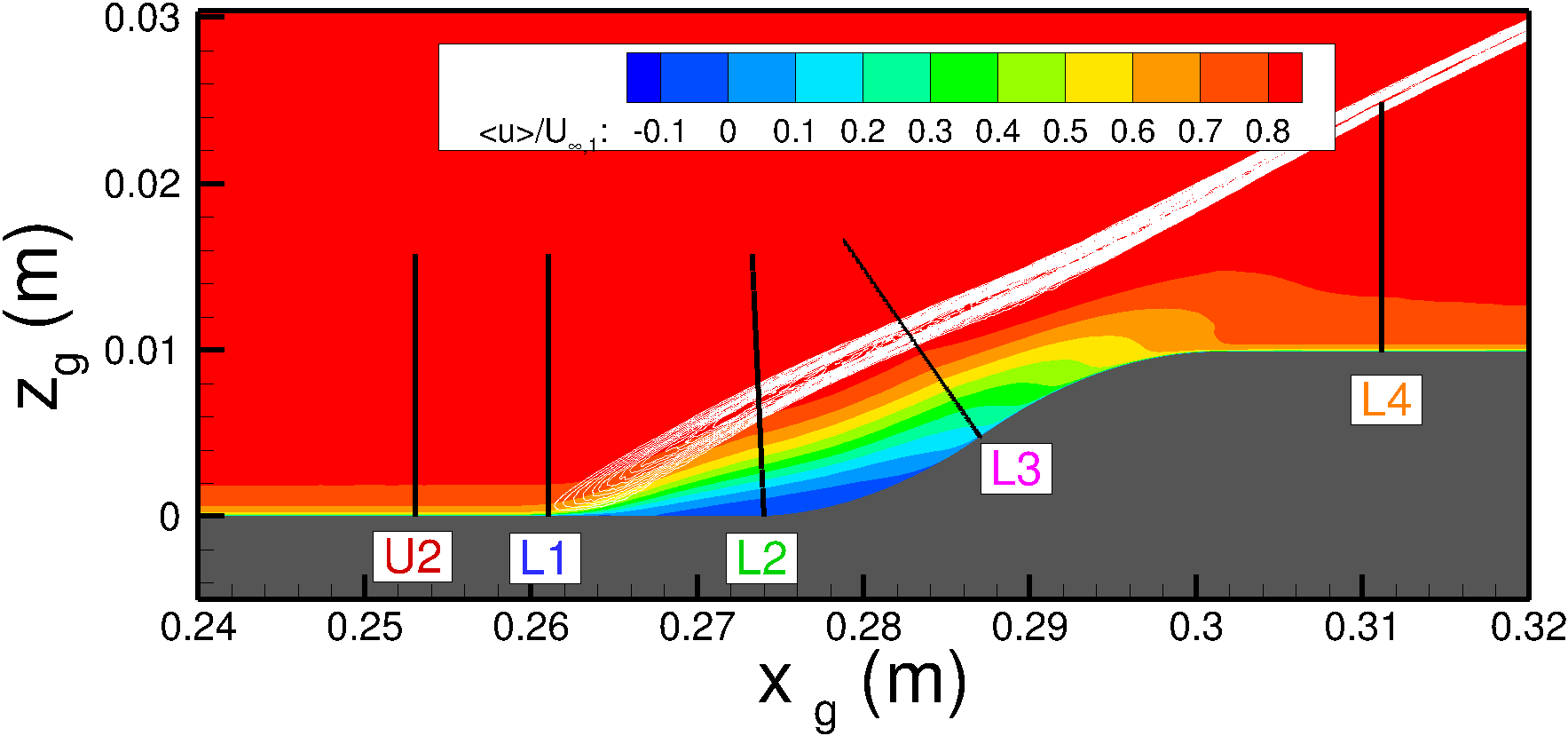}
			\caption{}
		\end{subfigure} 
		\begin{subfigure}{0.49\textwidth}
			\centering
			\includegraphics[width=0.9\linewidth, trim = 0mm 0mm 0mm 0mm]{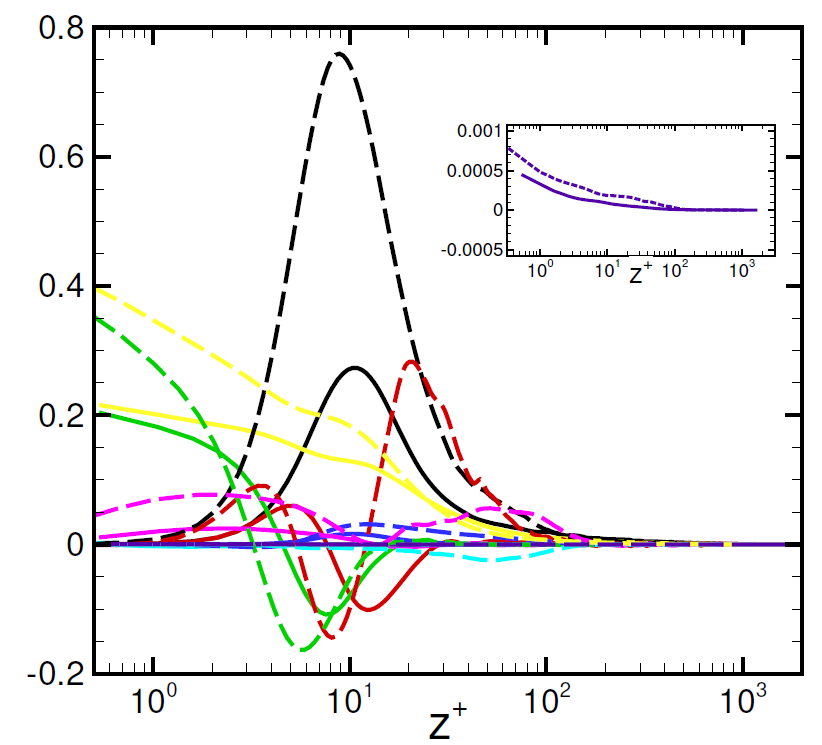}
			\caption{}
		\end{subfigure}	
  \caption{(a) Mean flow-field over the forward-facing geometry with wall curvature showing SBLI (reproduced from Ref.~\onlinecite{M5DNS_SBLI}) and (b) TKE budget : Solid lines represent the wall-normal variation of TKE budget terms at U$_2$ location and dashed lines represent corresponding quantities at L$_1$ location. Color code : Black - $P_k$, Yellow - $\epsilon_s$, Violet - $\epsilon_d$, Green - $D$, Red - $T$, Magenta - $\Pi^t$, Blue - $M$, and Cyan - $\Pi^d$ (data from the DNS of Ref.~\onlinecite{nicholson2024direct} taken from Ref.~\onlinecite{M5DNS_SBLI}). }
  \label{fig:DNS-TKE-budget}
	\end{figure*} 
	
%%%%%%%%%%%%%%%%%%%%%%%%%%%%%%%%%%%%%%%%%%%%%%%%%%%%%%%%%%%%%%%%

The terms on the L.H.S. in Eqn.~(\ref{TKEeqn}) represent the rate of change and transport by convection, respectively, and the terms on the R.H.S. represent Production $P_k$, Dissipation $ \overline{\rho} \epsilon$, Viscous Diffusion $D$, Turbulent Transport (triple velocity correlation term) $T$, Pressure Diffusion $\Pi^t$, mass flux contributions $M$, and pressure dilatation $\Pi^d$. The dissipation can be effectively split into solenoidal and dilatation parts i.e. $\overline{\rho} \epsilon = \overline{\rho}(\epsilon_s + \epsilon_d $) that represent fluctuating vorticity and divergence of fluctuating velocity, respectively. The solenoidal dissipation rate $\epsilon_s$ can be thought of as the dissipation due to the regular process of energy cascading to the smaller scales and in the absence of dilatational effects it can be considered to be equivalent to the ``incompressible'' dissipation rate. The dilatational dissipation $\epsilon_d$ (also referred to as compressible dissipation) is due to the non-divergent nature of the velocity fluctuations and it is an extra compressibility effect. The most commonly used approach to model unclosed terms in the TKE equation (\ref{TKEeqn}) is to generalize the low-speed closure approximations for the corresponding incompressible analogs. The terms $\epsilon_d$, $M$, and $\Pi^d$ are exactly zero for the incompressible flows and hence are generally neglected even for hypersonic flows. The pressure diffusion term $\Pi^t$ is also neglected in many works. On the other hand, the terms $D$ and $T$ are modeled using a gradient diffusion hypothesis as $( \widetilde{\mu} + \mu_T / \sigma_k) \partial k / \partial x_j$, where $\sigma_k$ is a model constant. However, these neglected terms can become significant under non-equilibrium conditions such as in the SBLI regions.

Figure~\ref{fig:DNS-TKE-budget}(a) shows the mean flow for a forward-facing geometry with wall curvature obtained using a recent DNS~\cite{nicholson2024direct} with an inflow Mach $4.9$ at mild cold wall conditions $T_w/T_{aw} = 0.91$ and an approximate friction Reynolds number of $1000$ immediately before the onset of wall curvature. The turbulent boundary layer fully separates at these geometric and flow conditions and SBLI is evident. Compared to the undisturbed boundary layer at location U$_2$, TKE budget at L$_1$ corresponding to the SBLI region shows that contributions due to all the terms on the R.H.S. of Eqn.~(\ref{TKEeqn}) increase. Production of TKE increases significantly and the terms $T$, $D$, $\Pi^t$, and $\epsilon_s$ become appreciable. Dilatation dissipation $\epsilon_d$ and pressure dilatation $\Pi^d$ remain comparatively smaller for this case. In the separated regions L$_2$ and L$_3$, the budget terms also become significant in the separated shear layer over the separation bubble in addition to post-shock regions. Thus, the usually neglected terms in the TKE transport equation can be expected to become considerable under strong nonequilibrium conditions. The turbulent fluctuations can become locally supersonic relative to the surrounding flow, likely creating local shocklets that may become the source of significant dilatational dissipation and entropy production.

The choice of length-scale determining equation plays a crucial role in determining the predictive capabilities of turbulence models. The most popular and widely used turbulence variable alongside TKE to construct the length-scale in the two-equation modeling framework and RSTMs is the turbulence dissipation rate $\epsilon$ or specific dissipation rate $\omega$. The choice between these variables significantly affects the near-wall treatment, wall and far-field boundary conditions, far-field behavior, and model performance. An $\epsilon$-based model is preferred away from the wall to avoid the high sensitivity to freestream turbulence observed in $\omega$-based models. However, the exact $\epsilon$-equation requires a lot of ad hoc modeling, substantially more than the TKE transport equation. Most of the terms in the $\epsilon$-equation are highly empirical and tuned for specific flow cases. As a result, there are numerous variants of the $k$-$\epsilon$ models, with major differences primarily in the near-wall modeling of the $\epsilon$-equation and the eddy-viscosity relation. Conversely, $\omega$-based models offer advantages near walls, allowing integration up to the wall without additional damping functions. However, they struggle in regions of strong SBLIs, where equilibrium conditions are disrupted. Both $\epsilon$- and $\omega$-based models exhibit deficiencies in non-equilibrium near-wall regions, which can lead to excessive length scale predictions in the separated flows, thereby significantly overestimating wall heat-transfer rates, particularly in reattachment regions. Moreover, advanced turbulence models such as NLEVMs, EARSMs, and RSTMs inherit these errors, which can undermine their superior theoretical foundation. Consequently, these advanced models may fail to improve upon standard two-equation models and, in some cases, even underperform in predicting basic wall-bounded flows.

%%%%% REACTIVE
\subsubsection{Favre-averaged equations for flows in thermochemical nonequilibrium}\label{sec:reacteq}

High-enthalpy hypersonic flows may incur chemical or thermal nonequilibrium effects, or both.
Proper modeling of such phenomena requires a precise picture of the intimate nature of molecules with their different modes of storing energies
(translational, rotational, vibrational, and electronic) and their quantum behavior \citep{park1989nonequilibrium,anderson1989hypersonic,colonna2001numerical}. Nevertheless, because of computational limitations, sophisticated ``state-to-state"
models involving detailed reaction kinetics (i.e. hundreds of species and thousands of intermediate reactions) can be used only for basic systems, and simplified thermochemical models are hence adopted in practice~\cite{park2001chemical}.

%%%%% CHEMICAL NONEQ
In the case of chemical non-equilibrium the fluid is modeled as a reacting mixture of chemical species, corresponding to the reactants and products of chemical reactions. The number of species to be considered depends on the chemical model adopted for air, which may undergo dissociation, recombination, and ionization of the component species, depending on the flight conditions and altitude (see Ref.~\onlinecite{anderson1989hypersonic}).
Chemistry models can vary in complexity, according to the number of chemical species and global or partial reactions used to describe the mixture. In complex configurations, air is often modelled as a five-species mixture of N$_2$, O$_2$, NO, O and N that undergoes seventeen dissociation/recombination reactions~\cite{park1989nonequilibrium}:
\begin{alignat}{3}
 \notag \text{R1}: & \qquad \text{N}_2 + \text{M} && \Longleftrightarrow 2\text{N} + \text{M} \\
 \notag \text{R2}: & \qquad \text{O}_2 + \text{M} && \Longleftrightarrow 2\text{O} + \text{M} \\
        \text{R3}: & \qquad \text{NO}  + \text{M} && \Longleftrightarrow  \text{N} + \text{O} + \text{M} \\
 \notag \text{R4}: & \qquad \text{N}_2 + \text{O} && \Longleftrightarrow  \text{NO}+ \text{N} \\
 \notag \text{R5}: & \qquad \text{NO}  + \text{O} && \Longleftrightarrow  \text{N} + \text{O}_2
\end{alignat}
being M any of the five species considered. The reader is referred to Ref.~\onlinecite{park1989nonequilibrium}
for details.

The flow dynamics is generally described by the conservation equations for the mass, momentum, and energy of the mixture, which still take the form (\ref{favrecontinuity})-(\ref{favreenergy}), and by supplementing them with transport equations for the chemical species. Because the sum of the masses of all species must equal the mass of the mixture, $N-1$ additional equations are needed, $N$ being the total number of species, with boundary conditions accounting for the catalytic or non-catalytic nature of the solid walls, i.e. for the presence of various degrees of chemical equilibrium between the flow and the wall.

%%%%%%%
In this paper, we are interested more specifically into the Favre-averaged form of the species transport equations:%
\begin{equation}
\frac{\partial \bar{\rho} \widetilde{Y}_{n}}{\partial t}+\frac{\partial\left(\bar{\rho} \widetilde{Y}_{n} \widetilde{u}_{j}\right)}{\partial x_{j}}= - \frac{\partial \overline{\rho Y_{n} u_{nj}^D}}{\partial x_{j}} + \overline{\dot{\omega}}_{n}+\frac{\partial}{\partial x_{j}} \overline{\rho u_{j}^{\prime \prime} Y_{n}^{\prime \prime}},
\end{equation}
$Y_n$ being the mass fraction, $u_{nj}^D$ the diffusion velocity and $\dot{\omega}_n$ the chemical production rate for the $n$-th species, respectively.
The unclosed term $\frac{\partial}{\partial x_{j}} \overline{\rho u_{j}^{\prime \prime} Y_{n}^{\prime \prime}}$ is the turbulent transport of chemical species.
Averages of the diffusion velocity contribution and of the source term are also unclosed. Passiatore \textit{et al.}~\cite{passiatore2021finite} found the former to be negligibly small in their DNS of a hypersonic boundary layer at Mach $10$, supporting the idea that it can removed from the equation.
The system of Favre-averaged mixture/species transport equations is supplemented with thermodynamic and transport models, as well as with models for the reaction rates  (see, e.g. Ref.~\onlinecite{sciacovelli2023priori}).
In addition, the unclosed terms arising from the Favre-averaging of the species equations must be modeled, as discussed in Section \ref{sec:thermochemical}.\\

%%%%%%%%%%% THERMAL NONEQ
If the flow residence time becomes comparable to the characteristic relaxation times of energy pools of species, thermal nonequilibrium effects must be taken into account.
In general,
the energies associated with the internal degrees of freedom of molecules may be
separated into rotational, vibrational, electronic and
nuclear contributions \citep{anderson1989hypersonic,park1989nonequilibrium}, in addition to the energy associated with the molecule translational motion. The internal energies are denoted as $e^{int}_m$ , where
the superscript indicates an internal mode and the subscript $m$ the internal mode ``energy pool''. The latter may represent individual species in the mixture, but if the coupling
between species is fast compared to the
coupling with other internal modes, then it may be acceptable to assign the same energy to all species. For this case,
subscript $m$ is dropped. To fix ideas, we will consider in the following the so-called two-temperature (2T) model proposed by Park~\cite{park1988two}, which assumes equilibrium between the translational and the rotational modes of all molecules and vibrational non equilibrium, with all diatomic species sharing however the same vibrational temperature.
%
%%%%%%%%%%%%%
Vibrational relaxation processes are then taken into account by adding a conservation equation for the vibrational energy $e_V$.

The Favre-averaged vibrational energy equation reads:
\begin{equation}
\begin{aligned}
    \frac{\partial \bar{\rho} \widetilde{e}_{V}}{\partial t}+\frac{\partial}{\partial x_{j}} \left(\bar{\rho} \widetilde{e}_{V} \tilde{u}_{j}\right) = \frac{\partial}{\partial x_{j}} \left(\overline{\rho u_{j}^{\prime \prime} e_{V}^{\prime \prime}}\right) \\ +  \frac{\partial}{\partial x_{j}} \left(-\overline{q_{V j}} + \sum_{m=1}^{\mathrm{NM}} \overline{\rho D_{m} \frac{\partial Y_{m}}{\partial x_{j}} e_{V m}}\right)\\
    +\sum_{m=1}^{\mathrm{NM}}\left(\overline{{Q_{\mathrm{TV} m}}} + \overline{\dot{\omega}_{m}e_{V m}} \right),
    \label{rans_vibrational}
\end{aligned}
\end{equation}
Here, $\text{NM}$ is the number of molecular species, $q_{Vj}$ the vibrational contribution of the heat flux and $e_{V_m}$ the vibrational energy per unit of volume of the $m$-th species.
Further, $Q_\text{TV}$ and $\dot{\omega}_m e_{Vm}$ are source terms denoting translational-vibrational energy exchanges and the vibrational energy variations due to chemical production/depletion, respectively. Equation (\ref{rans_vibrational}) contains the unclosed term $\frac{\partial}{\partial x_{j}} \left(\overline{\rho u_{j}^{\prime \prime} e_{V}^{\prime \prime}}\right)$, corresponding to the turbulent transport of vibrational energy. Furthermore, the averaged source terms, described by non-linear expressions, also need an appropriate closure model. These aspects are discussed in Section~\ref{sec:thermochemical}.

%----------------------------------------------------------------------------------------

\section{Modeling improvement strategies for hypersonic flows}\label{sec4} 

Baseline turbulence models that are direct extensions of incompressible forms are generally insufficient for accurately predicting many hypersonic flows, especially in the presence of shock waves. Several modeling treatments and fixes are available to improve their performance, especially in the framework of the industry-favored two-equation family of $k$-$\epsilon$ and $k$-$\omega$ models. We discuss key developments that have shown some success in predicting hypersonic flows in the two-equation modeling framework.

\subsection{Dilatation dissipation modeling}

Sarkar \textit{et al.}~\cite{sarkar1991analysis} and Zeman~\cite{zeman1990dilatation} propose models for the dilatational dissipation $\epsilon_d$ (also referred to as compressible dissipation) using the DNS analysis for homogeneous turbulence. Wilcox~\cite{wilcox1992dilatation} proposed an alternative form primarily applicable to the $k$-$\omega$ model and it has the same functional form as the Sarkar model. These approaches assume $\epsilon_d$ to be proportional to turbulent Mach number $M_T = \sqrt{2k}/\widetilde{a} $, where $\widetilde{a} = \sqrt{\gamma R \widetilde{T}}$ is the local speed of sound and $\gamma$ is the ratio of specific heats. The turbulent Mach number represents a ratio of the propagation of information by turbulence to acoustic propagation, with the turbulent kinetic energy providing a characteristic velocity scale at which turbulent fluctuations transfer information. These models are given by
\begin{equation}
    \epsilon = \epsilon_s (1 + F(M_T)),
\end{equation}
with
\begin{equation}
    F(M_T) = \begin{dcases}
        M^{2}_{T}~(\text{Ref.~\onlinecite{sarkar1991analysis}}), \\
        \dfrac{3}{4} \left[ 1 - e^{-\frac{1}{2} (\gamma+1)[(M_T - M_{T0})/a]^2} \right] \mathcal{H} (M_T - M_{T0}) \\ 
        (\text{Ref.~\onlinecite{zeman1990dilatation}}),  \\
        \dfrac{3}{2} \left[ M^{2}_{T} - M^{2}_{T0} \right] \mathcal{H} (M_T - M_{T0}) (\text{Ref.~\onlinecite{wilcox1992dilatation}}), 
    \end{dcases}
    \label{eqn:Cw_model}
\end{equation}
where $\mathcal{H} (\cdot)$ is the Heaviside function, $M_{T0} = 0.25$ for the Wilcox~\cite{wilcox1992dilatation} model and $M_{T0} = b \sqrt{2/(\gamma+1)}$ in the Zeman~\cite{zeman1990dilatation} model with $a = 0.66$, 
 $b=0.25$ for boundary layers and $a = 0.6$, $b=0.1$ for free shear flows. 
The effect of these corrections is to enhance the destruction of TKE when $M_T$ is large, resulting in an expected decrease in skin friction and wall heat transfer. 
Brown~\cite{brown2002turbulence} highlighted that the Sakar/Zeman/Wilcox corrections result in a significant reduction in wall shear stress (and heat transfer) for hypersonic boundary layers above Mach $5$ and proposed a further correction to improve predictions for hypersonic boundary layers. 

While dilatation-dissipation corrections have seen success in improving predictions for hypersonic flows, their underlying hypothesis remains in question. Wilcox~\cite{wilcox1998turbulence} notes that DNS results indicate that dilatation dissipation is small, and while recent DNS results~\cite{zhang2018direct} highlight that dilatation-dissipation effects become non-negligible at high enough Mach numbers, these Mach numbers are far in excess of those for which the Sakar/Zeman/Wilcox corrections were developed for and validated on.

\subsection{Pressure work modeling}

Hypersonic flows are generally characterized by large pressure gradients, especially in the regions of shock waves. The contribution due to the scalar product of the Favre-fluctuating velocity $\overline{u''}$ and the mean pressure gradient, i.e., pressure work, can thus become important. Turbulent mass flux $\overline{\rho' u'_i}$ is related to $\overline{u''}$ by
\begin{equation}
    \overline{u''} = - \dfrac{\overline{\rho' u'_i}}{\overline{\rho}},
\end{equation}
and this flux can be approximated with the standard gradient transport hypothesis. The pressure work is modeled as~\cite{grasso1993high,speziale1989preliminary}
   \begin{equation}
        \overline{u''_i} \dfrac{\partial \overline{p}}{\partial x_i} = \dfrac{\mu_T}{\overline{\rho}^2} \dfrac{1}{\sigma_p} \dfrac{\partial \overline{\rho}}{\partial x_j} \dfrac{\partial \overline{p}}{\partial x_j} \hspace{0.1in} \text{where} \hspace{0.1in} \sigma_p = 0.5.
    \end{equation}

\subsection{Pressure dilatation modeling}

The pressure dilatation is an explicit compressibility term in the TKE transport equation~(\ref{TKEeqn}) arising due to a non-divergent fluctuating velocity field. It refers to the work done due to simultaneous fluctuations in the volume of the fluid cell corresponding to the fluctuations in pressure. It can be either positive or negative and,  when negative, represents an extra dissipation. Sarkar~\cite{sarkar1992pressure} analyzed the governing equations for pressure fluctuations using homogeneous turbulence to obtain a formal expression for the pressure-dilatation from which a model is deduced using scaling arguments,
    \begin{equation}
        \overline{p' \dfrac{\partial u'' }{\partial x_i} } = \alpha_2 \tau_{ij} \dfrac{\partial \widetilde{u_i}}{\partial x_j} M_T + \alpha_3 \overline{\rho} \epsilon_s  M^{2}_{T},
    \end{equation}
where $\alpha_2 = 0.15$ and $\alpha_3 = 0.2$. The model has been calibrated using isotropic turbulence and homogeneous shear.

\subsection{Length scale correction}\label{sec:lengthscale}

Vuong and Coakley~\cite{vuong1987modeling} and Huang and Coakley~\cite{huang1993calculations} proposed a correction to remedy wall heat transfer overpredictions in the reattachment or shock impingement zone of SBLIs. The correction involves the use of an algebraic length scale which limits the length scale predicted by the two-equation turbulence models, which otherwise would become very large in these regions. For $k$-$\epsilon$ and $k$-$\omega$ models, the correction is given by
 \begin{equation}
   l = min \left( 2.5 y, \sqrt[3]{k} / \epsilon \right) = min \left( 2.5 y, \sqrt{k} / \omega \right)
 \end{equation}
where $l$ is the turbulent length scale which is taken to be the smaller of an algebraic expression $\kappa C_{\mu}^{-3/4} y = 2.5 y $ based on a von Karman constant of $\kappa = 0.41$ and the conventional length scale given by the two-equation models. Here, $y$ is the shortest distance from the wall. This relation is derived by bounding the turbulent viscosity by the $k$-$\epsilon$ or $k$-$\omega$ model by the turbulent viscosity as determined by Prandtl's model and by assuming Bradshaw's relation between shear stress and TKE and the logarithmic law for velocity to hold. Using this length scale, the value of $\epsilon$ or $\omega$ is recomputed and reset to be consistent with this value i.e. $\epsilon = \sqrt[3]{k} / l$ or $\omega = \sqrt{k}/l$. 
This length-scale correction has proven effective for improving heat transfer predictions in the reattachment region in SBLIs~\cite{huang1993calculations}. The correction, however, also impacts wall shear stress predictions and has seen minimal validation on this front.

Zhang \textit{et al.}~\cite{zhang2022application} recently proposed a new turbulent length scale correction for the $\epsilon$-equation taking inspiration from the length-scale modification discussed above. The correction is intended to control the growth of turbulence in the SBLI region. This involves the addition of an extra source term to the $\epsilon$-equation,
\begin{equation}
   S_{\epsilon cm} = H Y \lambda_l \dfrac{\epsilon^2}{k},
\end{equation}
where $ H = 3 $ is the suggested value based on numerical experimentation of a fully separated hypersonic SBLI flow. It is to be noted that the model predictions for surface pressure and heat transfer rate were found to be sensitive to the values of $H$. Here, $  \lambda_l = \mathrm{max} \left( \dfrac{k}{\epsilon} \dfrac{\partial \widetilde{u}_i}{\partial x_i}, -0.5\right) $ is designed to retain numerical stability and $ Y = \left( \mathrm{tanh}(2(\eta - 3)) - 1) \ \right)$, $\eta = \mathrm{max}(\widehat{S}, \widehat{\Omega})$ with $\widehat{S} = (k/\epsilon) \sqrt{2 S^{D}_{ij} S^{D}_{ij}} $ and $\widehat{\Omega} = (k/\epsilon) \sqrt{2 \Omega_{ij} \Omega_{ij}} $. This extra source term is specially designed for compressible flows and mainly remains effective in the near-wall or viscosity-dominated regions.

\subsection{Rapid compression correction}\label{sec:rapid_compression}

Vuong and Coakley~\cite{vuong1987modeling} and Huang and Coakley~\cite{huang1993calculations} proposed a modification for $k$-$\epsilon$ and $k$-$\omega$ models to improve predictions for flow separation in SBLIs. This modification was made to increase the size of computed separation-bubble regions by ensuring that the turbulent length scale does not change too quickly when undergoing rapid compression. The basic principle of this correction is that the product of the density and turbulent length scale $\overline{\rho} l$, where $l = k^{3/2}/ \epsilon = k^{1/2}/\omega$, should remain constant in a uniaxial compression, such as a shock wave, i.e.
\begin{equation}
   \dfrac{d \overline{\rho} l}{d t} = 0.
\end{equation}
Applying the continuity equation, $(1/\overline{\rho}) d \overline{\rho}/dt = - \partial \widetilde{u}_{k}/ \partial x_k$, to the above equation, the length scale equation can be written as
\begin{equation}
   \dfrac{1}{l} \dfrac{d l}{d t} = \dfrac{\partial \widetilde{u}_{k}}{\partial x_k} .
\end{equation}
Comparison between the above equation and the length scale equation derived from the $k$-$\epsilon$ or $k$-$\omega$ model equations under rapid dilatation gives appropriate values of the coefficients of the dilatation part in the source term of the $\epsilon$ or $\omega$ equation.} The corrected values of the dilatation coefficient for each model are
    \begin{equation}
        \alpha_{\epsilon} = 2 ; \alpha_{\omega} = 4/3.
    \end{equation}
The net effect of this correction is to decrease the turbulent length-scale in regions of rapid compression, or shock waves, which reduces $\mu_T$ and enhances separation.

\subsection{Shock-unsteadiness correction}
\label{subsection:SUcorrection}
%Sinha et al.~
Sinha \textit{et al.}~\cite{sinha2003modeling} propose modifications to the standard $k$-$\epsilon$ and $k$-$\omega$ models for high-speed shock-dominated flows to remedy the problem of excessive production obtained using standard models at shock waves. The modification is based on the physics of canonical STI and represents the damping effect of an unsteady shock oscillation due to interaction with the upstream vortical turbulence. The modeling approach replaces the TKE production term $P_{k} = \tau_{ij} \partial \tilde{u_i} / \partial x_j$ in the standard $k$-$\epsilon$ and $k$-$\omega$ models with a SU modified form in the vicinity of shock waves,
\begin{equation}
\begin{split}
  P_{k}^{SU} =  - \frac{2}{3} \overline{\rho} k S_{ii} (1 - b'_1) \\ \hspace{0.1in} \text{with} \hspace{0.1in} b'_1 = max[0, 0.4(1 - e^{1 - M_{1n}})].  
\end{split}
\label{eqn:SU-Prod}
\end{equation}
The model parameter $b'_1$ represents the damping effect caused by the coupling between the shock unsteadiness and the upstream velocity fluctuations. Here, $M_{1n}$ is the upstream shock-normal Mach number and it brings in the physical effect of the shock strength. The switching between the original TKE production and SU modified production term is done using a shock-detecting $\mathrm{tanh}$ function depending on the mean dilatation. Veera and Sinha~\cite{veera2009modeling} modify the original SU model with the added effect of entropy fluctuations representative of hypersonic turbulent boundary layers. However, the effect of this model correction on hypersonic SBLIs has not been tested. The SU model is not Galilean invariant through the use of mean flow Mach number as a model parameter.

In recent works~\cite{rathi2024simulation}, the original SU model parameter $b'_1$ is approximated in terms of a shock function $\psi$ given by
\begin{equation}
b'_{1} =  0.4 \left(1 - \dfrac{\sqrt{(6 \psi - 1)}}{5} \right). 
\label{eqn:b1p_psi}
\end{equation}
The function $\psi$ identifies the regions of the shock wave in a flow and brings in the history effect of the shock in the downstream flow. It is computed using a transport equation given by
\begin{equation}
    \dfrac{\partial \overline{\rho} \psi }{\partial t} + \dfrac{\partial (\overline{\rho} \widetilde{u}_i \psi)}{\partial x_i} = \overline{\rho} \psi S_{ii} - \dfrac{\overline{\rho} c (\psi - \psi_0) }{L},
    \label{eqn:psi}
\end{equation}
where $c$ is the local speed of sound, $L = a \Delta (1+b(\psi_o - \psi))$ is the characteristic relaxation length with $\Delta$ is the representative grid size and the constants $a = 3$, $b = 15$, and $\psi_o = 1$. The function $\psi$ is set to $1$ in the undisturbed boundary layer and takes a value of $1/r$ at shock waves with $r$ being the local density ratio. A shock sensor is also devised based on $\psi$ such that the SU model is only active in the vicinity of a shock wave.

\subsection{Diffusion term corrections}

Catris and Aupoix~\cite{catris2000density} propose corrections to the diffusion terms of the $k$-$\epsilon$, $k$-$\omega$, $k$-$l$, and SA models to account for the density variations and make these models consistent with the logarithmic law for high-speed compressible ZPG TBLs. These corrections are based on the analysis of the prediction of the logarithmic region of a compressible boundary layer by these Boussinesq hypothesis-based one- and two-equation models. The developments of these corrections assume equilibrium TBL and Bradshaw's assumption of shear stress proportional to TKE in a TBL. 
Their analysis shows that the diffused quantity should become $\overline{\rho} k$ in the TKE transport equation for $k$, while for the length scale determining equations, the diffused quantities should be $\sqrt{\overline{\rho}} \epsilon$, $\sqrt{\overline{\rho}} \omega$, and $l$ and $\overline{\rho} q^2$ for the transported quantities $ \overline{\rho} \epsilon$, $ \omega$, and $l/\sqrt{\overline{\rho}}$, respectively. Similarly, in the SA model, the diffused quantity becomes $\sqrt{\overline{\rho}} \nu_T$ for the transported quantity $ \overline{\rho} \nu_T$. This results in additional terms linked to density variations in the transport equations of these models in their classical form.

More recently, Pecnik and Patel~\cite{Pecnik2017scaling} investigated variable-property scaling for non-adiabatic, low-Mach turbulent duct flows and obtained a new density scaling of the diffusion terms~\cite{otero2018turbulence} that carries similarities with that of Catrix and Aupoix~\cite{catris2000density}.

All of the above-mentioned corrections consist of rescaling the diffusion terms to account for local variations of the mean density. Such a phenomenon is not peculiar to compressible flows, and can also be found in incompressible variable-property flows. Hasan \textit{et al.}~\cite{hasan2023incorporating} discussed a velocity transformation based on the supposed universality of the total shear stress (sum of the viscous and turbulent stress) in the near-wall region of attached flows to infer modifications to the turbulence model that account for genuine compressibility effects (outward shift of the turbulent stresses in compressible flow cases).

%%%%%%%%%%%%%%%%%%%%%%%%%%%%%%%%%%%%%%%%%%%%%%%%%%%%%%%%%%%%%%%%%%%

\subsection{$\omega$-equation correction}

Danis and Durbin~\cite{danis2022compressibility} proposed a compressibility correction for two-equation $k$-$\omega$ models which explicitly takes the effects of high Mach numbers and wall cooling in account. The compressibility correction is heuristic and aims to improve the predictions of the $k$-$\omega$ models that inaccurately predict high values of eddy viscosity and low mean velocities, as wall cooling increases. The correct profiles of eddy viscosity are achieved by modifying the closure coefficients of the production and destruction terms in the $\omega$-equation while the $k$-equation remains unchanged.

The baseline production and destruction terms in the $\omega$-equation are multiplied with a compressibility correction function $f$ given by
\begin{equation}\label{eqn:danis_comp_correc}
    f = (1 - \phi) \exp{-c_1 M^{*2}_{\tau} - c_2 B^{*}_{q}}) + \phi \exp{c_3 M^{*}_{\tau} + c_4 B^{*}_{q}},
\end{equation}
where
\begin{equation}
\begin{split}
        c_1 = 60, c_2 = 5, c_3 = \dfrac{15}{4} \exp{\left( -\dfrac{M}{10} \right)}, \\
        c_4 = - \dfrac{21}{4} \exp{\left(\dfrac{M}{5} \right)}, \phi = \tanh{\left( \dfrac{3}{4} M \right) }.
\end{split}
\end{equation}
Here, $M$ is the local Mach number, friction Mach number $M^{*}_{\tau} = \widetilde{u}^*/a$ with $\widetilde{u}^*$ being the function of the total local stress $\tau$ given by $ \widetilde{u}^* = \sqrt{\tau/\overline{\rho}}$ and the heat flux parameter $B^{*}_{q} = q/(\overline{\rho} C_p \widetilde{T} \widetilde{u}^*)$ with $q$ being the magnitude of the local heat flux components $q_{x_j}$ that are constructed from the boundary-layer approximations to the enthalpy equation and represent the effects of the molecular and turbulent diffusion of the mean enthalpy as well as the dissipative heating. 

The form of $f$ is based on extensive numerical experiments. The first term in Eqn.~(\ref{eqn:danis_comp_correc}) is active below the log layer and it controls the $y$ intercept of the log law while the second term is responsible for the log-law slope. The correction function $f$ is deactivated i.e $f \rightarrow 1$ at low speed as $M_\infty \rightarrow 0$ and $Tw/T_r \rightarrow 1$. The correction is, however, not Galilean invariant as it depends on the local Mach number $M$. 

The correction is validated using the DNS database~\cite{zhang2018direct,huang2020simulation} for ZPG cases for a range of Mach numbers between $2$ and $14$ and wall-to-recovery temperature ratios ($T_w/T_r$) between $0.18$ and $1$ as well as Mach $5$ favorable and adverse pressure gradient cases~\cite{nicholson2021simulation,nicholson2021fav} with $T_w/T_r = 0.91$. The correction collapses the eddy viscosity profiles on the incompressible data giving significantly improved velocity, temperature, skin friction, and wall heat transfer coefficient profiles than the baseline $k$-$\omega$ models.

%%%%%%%%%%%%%%%%%%%%%%%%%%%%%%%%%%%%%%%%%%%%%%%%%%%%%%
\begin{figure*}
\centering
\includegraphics[width=1.0\linewidth, trim = 0mm 0mm 0mm 0mm]{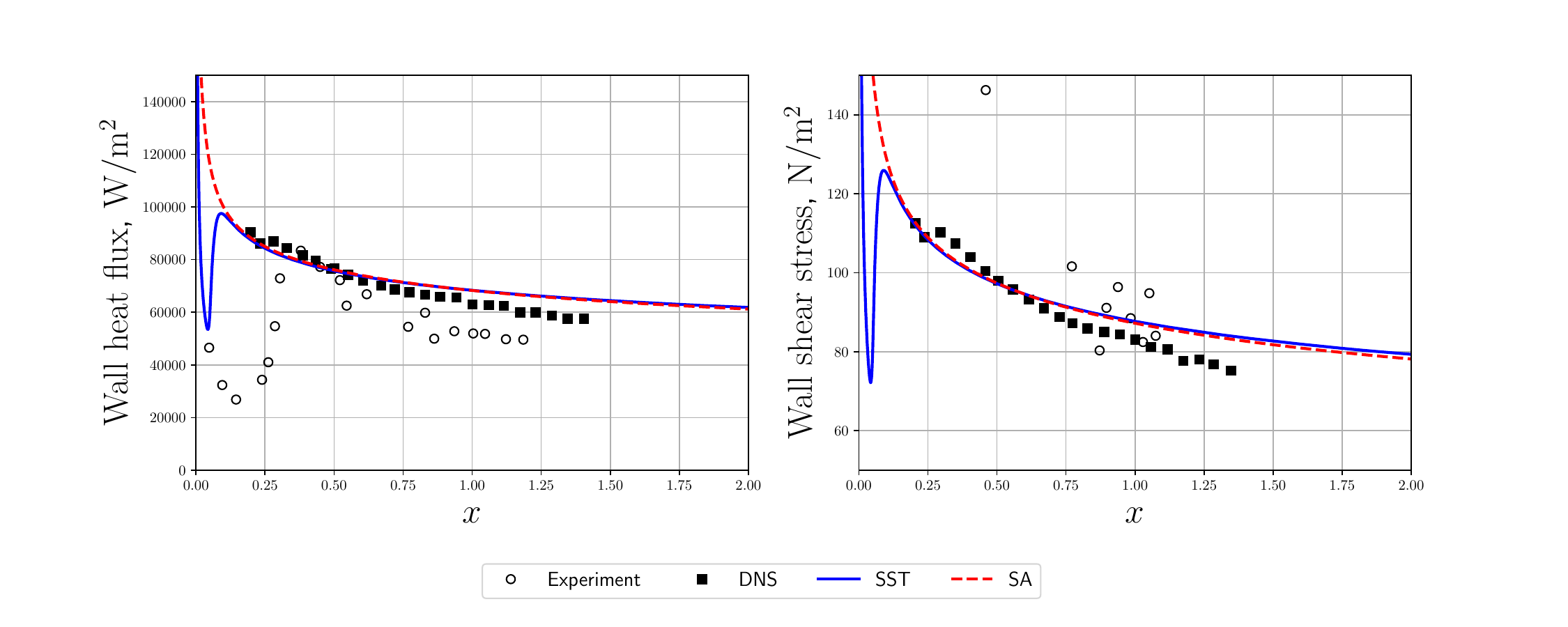}
\caption{Wall heat flux (left) and shear stress (right) predictions for a Mach 11 hypersonic boundary layer with $T_w/T_{r} = 0.20$. The DNS data are from Ref.~\onlinecite{huang2022direct} and the experimental data are from CUBRC~\cite{gnoffo2011uncertainty,gnoffo2013uncertainty}. }
\label{fig:m11_flat_plate}
\end{figure*} 
%%%%%%%%%%%%%%%%%%%%%%%%%%%%%%%%%%%%%%%%%%%%%%%%%%%%%%

\subsection{Variable $Pr_t$ modeling}\label{sec:var_prt}

The constant value of $Pr_t$  used in the standard gradient-diffusion heat flux model generally results in good predictions of wall heat transfer for ZPG hypersonic TBLs at adiabatic wall conditions. However, inaccurate predictions are obtained at the practically relevant cold wall conditions and the standard turbulence models generally overpredict the reattachment point heat transfer rates in the case of SBLIs. These inaccurate predictions can be attributed to a constant $Pr_t$ assumption along with other reasons like inaccurate length scale characterizations. DNS studies reveal that $Pr_t$ is not constant and varies in the near-wall region even though the average value across the TBL thickness is about $0.9-1$ for highly cooled equilibrium TBLs. The strong Reynolds analogy breaks down especially at the non-adiabatic wall conditions~\cite{maeder2001direct,duan2010direct,duan2011direct,liang2012dns,liang2013dns} and in SBLIs~\cite{Ni24,zhang2018direct}, and we can expect a stronger near-wall variation of $Pr_t$ with significant deviations from the nominally used value. Near-wall variation of $Pr_t$ is also true for air out of chemical equilibrium at high enthalpy conditions~\cite{passiatore2021finite}. Several modeling strategies that attempt to account for the variation of $Pr_T$ in the flow field have thus been developed, and some of these key works are discussed below.

\vspace{0.1in}

\noindent \textit{Variable $Pr_t$ model of Roy and Sinha }

\vspace{0.1in}

Roy and Sinha~\cite{roy2019turbulent} propose a variable $Pr_t$ model based on the conservation of total enthalpy fluctuations across an unsteady oscillating shock wave interacting with incoming turbulence consisting of vorticity and temperature fluctuations. The model development based on Linear Interaction analysis (LIA) data and uses the SU model of Sinha \textit{et al.}~\cite{sinha2003modeling}. The model is given by
\begin{equation}
\begin{split}
    Pr_t &= \dfrac{3/4}{1+b_1(r-1)}  \\ \text{with} \hspace{0.1in} b_1 &= 0.4 + 0.6 \left( \dfrac{6 \psi - 1}{5 \psi} \right)^{5.2} + 0.2 (\gamma -1) M^{2}_{1}.
    \label{eqn:varPrT-Roy}        
\end{split}
\end{equation}
Here, $M_1$ is the shock upstream mean flow Mach number, and $\psi$ is a shock function computed using a transport equation to identify the regions of the shock wave in a flow. The model is not Galilean invariant because of the use of $M_1$ as a parameter. In the earlier version of the model~\cite{roy2018variable}, which neglects the shock-upstream temperature fluctuations while considering only the incoming vortical turbulence, the term depending on $M_1$ is effectively zero, and only the first two terms in Eqn.~(\ref{eqn:varPrT-Roy}) for $b_1$ remain.

\vspace{0.1in}

\noindent \textit{Variable $Pr_t$ model of Xiao et al }

\vspace{0.1in}

Xiao \textit{et al.}~\cite{xiao2007role} model the transport equations for enthalpy variance $ \widetilde{h''^{2}} $ and its dissipation rate $\epsilon_h$ derived from the exact energy equation thereby accounting for the compressibility and dissipation terms. The modeled equations are free of damping and wall functions and are tensorially consistent and invariant under Galilean transformation. The modeled equations are
\begin{equation}
\begin{split}
\dfrac{\partial }{\partial t} (\overline{\rho} \widetilde{h''^{2}}/2)  +  \dfrac{\partial }{\partial x_j} (\overline{\rho} \widetilde{u}_j \widetilde{h}''^{2}/2) = \\ \dfrac{\partial }{\partial x_j} \left[ \overline{\rho} (\gamma \alpha + \alpha_t C_{h,2}) \dfrac{\partial }{\partial x_j} (\widetilde{h''^{2}}/2)  \right]  + \\ 2 \mu \gamma S_{ij} \left[ \dfrac{\partial }{\partial x_j} (q_{T,i} / \overline{\rho}) + \dfrac{\partial }{\partial x_i} (q_{T,j} / \overline{\rho}) \right] -  \dfrac{4}{3} \mu \gamma S_{kk} \dfrac{\partial }{\partial x_j} (q_{t,j} / \overline{\rho}) \\ - (\gamma - 1) \overline{\rho} \widetilde{u}_j \widetilde{h''^{2}} S_{kk}  -  q_{T,i} \dfrac{\partial \widetilde{h}}{\partial x_i} + 2 C_{h,4} \gamma \mu \sqrt{\widetilde{h''^{2}}} \zeta - \gamma \overline{\rho} \epsilon_h
\end{split}  
\end{equation}
\begin{equation}
\begin{split}
\dfrac{\partial }{\partial t} (\overline{\rho} \epsilon_h)  +  \dfrac{\partial }{\partial x_j} (\overline{\rho} \widetilde{u}_j \epsilon_h) = \overline{\rho} \epsilon_h \left( C_{h,5} a_{jk} + \dfrac{\delta_{jk}}{3} \right) \dfrac{\partial \widetilde{u}_j }{\partial x_k} + \\ C_{h,6} \overline{\rho} k \dfrac{\partial \sqrt{\widetilde{h''^{2}}} }{\partial x_j} \dfrac{\partial \widetilde{h}}{\partial x_j}  +  \dfrac{\partial }{\partial x_j} \left[ \overline{\rho} (\gamma \alpha + \alpha_t C_{h,7}) \dfrac{\partial \epsilon_h }{\partial x_j}  \right] + \\ C_{h,8} \dfrac{q_{T,j}}{\tau_h} \dfrac{\partial \widetilde{h}}{\partial x_j} - \gamma \overline{\rho} \epsilon_h \left[ \dfrac{C_{h,9}}{\tau_h} \dfrac{C_{h,10}}{\tau_k}  \right] + \\ C_{h,11} \epsilon_h \left[ \dfrac{D \overline{\rho}}{Dt} + \dfrac{\overline{\rho}}{\overline{p}} max \left(\dfrac{D \overline{p}}{Dt},0.0 \right) \right],
\end{split}  
\end{equation}
%Here, $b_{ij} = \tau_{ij} / (\overline{\rho} k) + (2/3) \delta_{ij}$ 
where $D/Dt$ represents the substantial or material derivative and $\tau_k = \overline{\rho} k/(\mu \zeta)$. The turbulent Prandtl number is given in terms of $\mu_T$ and turbulent thermal diffusivity $\alpha_T$ i.e. $Pr_t = \mu_T / (\overline{\rho} \alpha_T)$ and $\alpha_T$ is modeled as 
\begin{equation}
    \alpha_T = 0.5 \left( C_h k \tau_h + \mu_T / (0.89 \overline{\rho}) \right), \hspace{0.05in} \text{where} \hspace{0.05in}  \tau_h = \widetilde{h''^2} /  \epsilon_h. 
\end{equation}
The modeling of $\alpha_T$ is based on experiments in simple shear flows which showed that the appropriate timescale for temperature fluctuations is proportional to the arithmetic average of $\tau_h$ and $\tau_k$. The model constants are : $C_h = 0.0648, C_{h,2} = 0.5,  C_{h,4} = -0.4,  C_{h,5} = -0.05,  C_{h,6} =  -0.12,  C_{h,7} = 1.45,  C_{h,8} =  0.7597,  C_{h,9} = 0.87,  C_{h,10} =  0.25,$ and $C_{h,11} =  0.575$.

%-------------------------------------------------------------------------------------

%%%%%%%%%%%%%%%%%%%%%%%%%%%%%%%%%%%%%%%%%%%%%%%%%%%%%%%%%%%%%%%%%%%%%%%%%%%%%%%%%%%%%%%%%%%

\section{Modeling challenges pertinent to hypersonic flows}\label{sec5}

\subsection{Flat plate boundary layers}
Many one- and two-equation models struggle to accurately predict wall quantities of interest (wall shear stress and wall heat flux) for ZPG flat plate TBLs at hypersonic speeds~\cite{aiken2022assessment,rumsey2010compressibility,danis2022compressibility,BaPaJo24};  particularly for cold wall boundary layers at high Mach and Reynolds numbers~\cite{rumsey2010compressibility}. In general, RANS models tend to over-predict both wall shear stress and wall heat flux at high Reynolds numbers. Figure~\ref{fig:m11_flat_plate} illustrates this challenge by showing predictions of two standard RANS models --- the Menter SST model and the Spalart--Allmaras model --- on a Mach $11$ ZPG hypersonic TBL at a wall temperature-to-recovery ratio of $T_w/T_{r} = 0.20$, for which comparative DNS data are available from Ref.~\onlinecite{huang2022direct} and experimental data from the CUBRC facility~\cite{gnoffo2011uncertainty,gnoffo2013uncertainty}. For both turbulence models, wall shear stress and heat flux predictions show discrepancies compared to the experiment and DNS that increase as the boundary layer spatially evolves.

Various compressibility corrections have been assessed (see Refs.~\onlinecite{rumsey2010compressibility,aiken2022assessment}) and proposed (see Refs.~\onlinecite{danis2022compressibility,BaPaJo24}) to address these issues. Aiken \textit{et al.}~\cite{aiken2022assessment} compared the predictions of the SA, SST, and BSL models to DNS data~\cite{zhang2018direct} for ZPG, cold-wall boundary layers for Mach numbers ranging from 2 to 8 with wall-to-recovery-temperature ratios of $0.18$ to 1.0 and concluded that no model accurately predicts QoIs at the wall across this range of conditions. They further concluded that the Zeman compressibility correction tends to improve accuracy for wall QoIs, but does so at the expense of worse wall-normal temperature profiles. Rumsey~\cite{rumsey2010compressibility} studies the performance of the $k-\omega$ model over a range of Mach numbers and
wall-to-recovery-temperature ratios, and similarly shows degraded performance at high Mach number, cold-wall conditions. Rumsey notes an improved performance of algebraic models like Baldwin-Lomax over standard one- and two-equation models, and further shows that dilatation-dissipation corrections tend to bring $k$-$\omega$ predictions of wall QoIs more inline with the Baldwin-Lomax model. Barone \textit{et al.}~\cite{BaPaJo24} propose several corrections to the SA model to improve performance in ZPG TBLs including a modified near-wall viscous damping function (informed by hypersonic boundary layer DNS datasets) and a normal Reynolds stress correction. Of note, Ref.~\onlinecite{BaPaJo24} highlights that the wall-normal Reynolds stress has an appreciable impact on the wall-normal momentum balance for high-speed boundary layers; this aspect is neglected in traditional RANS models. Specifically, under the assumption that the wall-normal Reynolds stress scales with the freestream dynamic pressure, it is straightforward to show that the ratio of the wall-normal Reynolds stress to pressure scales with the Mach number squared, i.e., 
$$\frac{\tau_{22}}{p} \sim M_{\infty}^2.$$
For high Mach number flows, the contribution of the wall-normal Reynolds stress can be non-negligible; Barone confirmed this result with DNS. 
The resulting SA model gives improved performance for high Reynolds number boundary layers. While the aforementioned advancements have improved predictions, many are ad-hoc and, as Rumsey~\cite{rumsey2010compressibility} concludes, there is a clear need for an increased understanding of zero pressure gradient hypersonic boundary layers.

%%%%%%%%%%%%%%%%%%%%%%%%%%%%%%%%%%%%%%%%%%%%%%%%%%%%%%%%%%

\begin{figure*}
\centering
\includegraphics[width=0.9\linewidth, trim = 0mm 0mm 0mm 0mm]{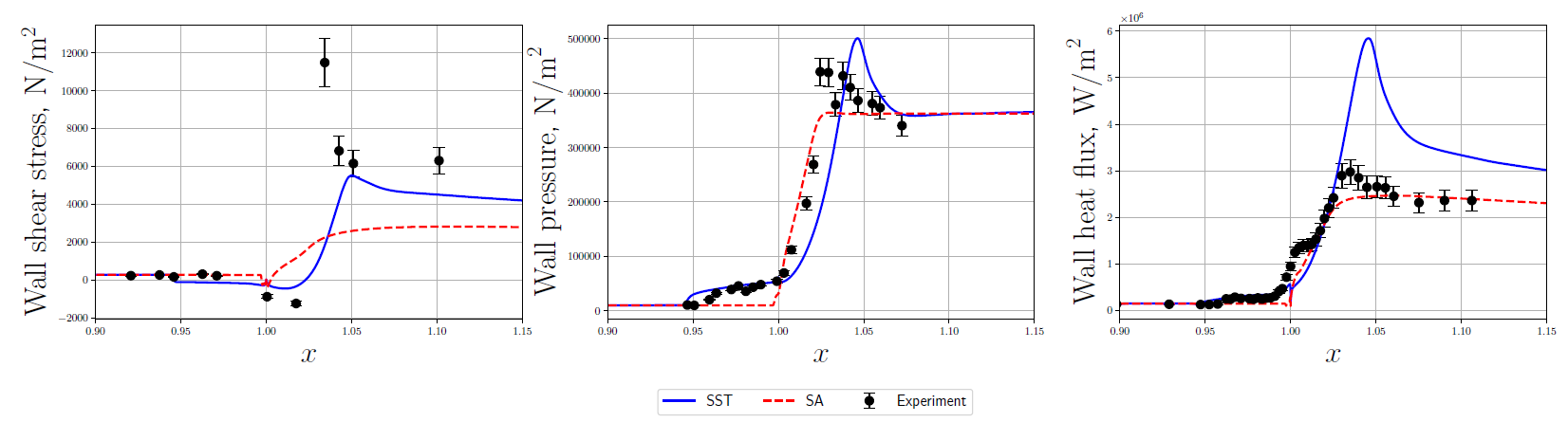}
\caption{RANS model predictions for wall shear stress (left), wall pressure (center), and wall heat flux (right) for a Mach 8, 33$^{\circ}$ compression ramp with $T_w/T_{t_e} = 0.29$.}
\label{fig:holden_m8}
\end{figure*}

%%%%%%%%%%%%%%%%%%%%%%%%%%%%%%%%%%%%%%%%%%%%%%%%%%%%%%%%%%

%-------------------------------------------------------------------------------------

\subsection{Shock/turbulence interaction and compressibility effects}

Shock waves, often embedded within a fully turbulent flow field, are an inevitable feature of high-speed applications. The interaction between turbulence and an oscillatory, corrugated, and sometimes broken shock wave causes a turbulence amplification and an increased anisotropy, which leads to increased dissipation and acoustic energy dispersion downstream of the shock. The physics of the shock turbulence interaction (STI) needs to be accurately modeled in the context of RANS. This is especially true for many applied problems, such as dual-bell nozzles, where incorrect modeling of the turbulence amplification directly impacts post-shock mixing effectiveness and thus the size of the recirculation regions characteristic under certain operating conditions. This, in turn, can influence the overall shock structure of the problem and have a first-order effect on the solution.  The modeling of turbulence amplification in STI was first studied by  \citet{Ribner_1954}
through linear interaction analysis (LIA); an approach that can capture many of the correct features of the STI problems, especially when the turbulent Mach number becomes small \citep{Ryu_Livescu_2014}. STI was later re-examined using rapid distortion theory (RDT) \citep{Lele_1992,Zank_Zhou_Matthaeus_Rice_2002}
but this approach leads to an over-amplification at high-Mach numbers as a result of the non-homogeneous compression of the shock due to unsteadiness and curvature, which is not modeled by RDT \citep{Jacquin_Cambon_Blin_1993}.
Despite their useful insight, analytical models generally provide an incomplete description of the turbulence amplification through shocks for direct use in RANS modeling. 

Although LIA has been used for turbulence modeling (see e.g. 
 \citet{Griffond_Soulard_2012})
these analytical solutions require compatibility conditions that limit their applicability to weak turbulence. For greater generalizability, ad-hoc STI models have been integrated into RANS solvers to correctly capture the TKE amplification across the shock. These ad-hoc models are necessary to address two important, yet related, problems when using the RANS equations for STI: (i) resolution of the shock structure, and (ii) shock unsteadiness and curvature. 

The shock thickness in a viscous fluid scales with the mean free path of gas molecules \citep{Howarth1953}
therefore, it represents a near discontinuity as observed from the relevant scales for typical RANS simulations. The turbulence causes instantaneous corrugation and unsteadiness of the shock structure. Thus, in a time-averaged sense, the shock will be significantly thicker than the instantaneous and local shock thickness. Despite its increased thickness, the shock still remains thin with respect to the typical scale of resolution of a RANS simulation. Therefore, high-speed RANS simulations will inevitably under-resolve the shock structure, which is problematic for most two-equation models. The TKE production in the classical turbulence transport equations (in standard two-equations models), takes the general form:
\begin{equation}
P_k=\tau_{ij} \frac{\partial u_i}{ \partial x_j},
    \label{eq:prod}
    % I may want to change this with respect to eddy viscosity instead
\end{equation}
where the production term is directly proportional to the velocity gradient and Reynolds stresses. As the shock thickness is effectively underresolved in applied, three-dimensional STI problems, the TKE production will effectively be tied to the mesh resolution at the shock; a significant problem for grid convergence. When the turbulence is high, thus a very diffusive shock, some models can reach grid convergence with sufficient resolution \citep{Schwarzkopf_Livescu_Baltzer_Gore_Ristorcelli_2016}.
However, this is not the case for most STI models.

A number of works \citep{Ryu_Livescu_2014,Chen_Donzis_2019,Lacombe_Roy_Sinha_Karl_Hickey_2021}
have proposed a scaling to approximate the time-averaged shock thickness, which could be used to bound the TKE production. The integration of the ensemble-averaged shock thickness has been one of the core ideas behind some STI modeling paradigms \citep{Braun_Gore_2018}. In the hypersonic regime, the time-averaged shock thickness becomes small, even sub-Kolmogorov, as the turbulent Mach numbers are typically smaller.  \citet{Ryu_Livescu_2014}
showed that the LIA solutions are regained when the viscous and nonlinear effects become small across the shock when the laminar shock thickness over the Kolmogorov length scale tends toward zero; a condition which is met in most hypersonic flows. Therefore, the challenges of vortical turbulence interacting with a planar shock in the hypersonic regime are simpler as the solution will tend towards turbulence amplification predicted by LIA.

Despite the under-resolution of the shock thickness, most turbulence models over-predict the TKE amplification. To address this over-amplification and embed stronger physical modeling of the shock corrugation and unsteadiness,  \citet{Sinha_Mahesh_Candler_2003}
, proposed a correction to the turbulence transport equations, discussed in greater depth in section \ref{subsection:SUcorrection}, which has become widely used and extended by others  \citep{Zhang_2017}. 
Other works impose an upper bound on the production term to mitigate unphysical TKE production through the shock 
\citep{prasad2023turbulence}.
\citet{Zhang_Gao_Jiang_Lee_2017} 
relied on a combination of shock sensors with locally applied damping functions to the TKE production term in $k-\omega$ SST to better match experimentally observed results, this model was recently extended to use a vorticity-based instead of a shear-based production term definition near the shock
\citep{Tian_Gao_Jiang_Lee_2023}.

The two-equation models cannot directly account for the important anisotropy generated through STI, which plays a critical role in post-shock mixing, energy redistribution, and turbulence decay. To address the lack of anisotropy in two-equation models, RSTMs for STI have been proposed. Similar to the TKE, the Reynolds stresses are over-predicted using conventional RSTM models. A physics-based modification to RSTM was proposed by \citet{Vemula_Sinha_2017}
that accounts for the damping effect of the shock unsteadiness and showed some grid convergence properties.  A number of works have integrated STI models within the transport equations for the Reynolds stresses by relying on two-length scale models which have shown good agreement with LIA
\citep{Schwarzkopf_Livescu_Baltzer_Gore_Ristorcelli_2016, Braun_Gore_2018}
but require a significant grid resolution.
\citet{Karl_Hickey_Lacombe_2019}
proposed a simple, but empirically derived correction method that adjusts the TKE jump, decay, and anisotropy; the generalizability of which, especially for oblique shocks has not been shown.

%-------------------------------------------------------------------------------------

\subsection{Shock/boundary layer interaction and wall heat-transfer}\label{sec:SBLI}

SBLIs occur when a turbulent boundary layer intersects with a shock wave and result in large peaks in wall heat flux, wall shear stress, and wall pressure. RANS modeling of SBLIs at hypersonic speeds is extremely difficult as the flow is largely anisotropic and standard modeling assumptions such as the Boussinesq relationship and constant turbulent Prandtl number break down. RANS predictions of SBLI flows are subject to erroneous predictions for flow separation, wall heat flux, and wall shear stress. Figure~\ref{fig:holden_m8} illustrates this issue by presenting predictions from the Menter SST and Spalart--Allmaras RANS models for an SBLI emerging from a Mach 8, 33$^{\circ}$ compression ramp with $T_w/T_{t_e} = 0.29$ ($T_{t_e}$ is the free-stream stagnation temperature) and comparing them to experimental data from Holden~\cite{holden2010experimental}. The model predictions deviate significantly from each other and illustrate the challenges in modeling SBLI flows. The SST model yields reasonable predictions for peak wall pressure but drastically over-predicts wall heat flux. The SA model fails to predict flow separation, under-predicts peak wall heat transfer and pressure, and massively under-predicts wall shear stress.

As seen above, two significant challenges RANS models encounter for SBLIs are (1) accurate prediction of flow separation and (2) the joint prediction of wall heat flux and wall shear stress. 
With respect to the prediction of flow separation, standard two-equation models without stress limiters, along with the Spalart--Allmaras model, tend to under-predict flow separation~\cite{wilcox1998turbulence}. A variety of fixes have been devised to address this issue. These fixes include, but are not limited to, rapid compression corrections (see Sec.~\ref{sec:rapid_compression}), stress limiters~\cite{menter1994two,KaWi95}, and shock unsteadiness models (Sec.~\ref{subsection:SUcorrection}). In general, these methods act by decreasing the eddy viscosity in regions near the SBLI, which in turn enhances separation. However, these corrections are typically calibrated for a flow regime of interest and a specific turbulence model; these models require further development for greater generalizability~\cite{wilcox1998turbulence}. Further, different turbulence models are subject to different systematic deficiencies. As an example, the 2003 variants of the SST model, which use the strain invariant instead of the vorticity invariant in the denominator of the eddy viscosity, systematically over-predict the size of the separation region in hypersonic SBLIs~\cite{raje2021anisotropic,parish2024report}. Similar results exist with, e.g., the RNG $k$-$\epsilon$ model~\cite{huang1994turbulence}. A recent collaborative effort on RANS modeling of the hypersonic flared cone --- using 11 different commercial, institutional, and open-source codes --- revealed a large variation in the onset of flow separation~\cite{Hoste2024}. Thus, while progress has been made in predicting flow separation in SBLIs, challenges remain.

With respect to the joint prediction of wall heat flux and wall shear stress, standard two-equation models tend to massively over-predict wall heat flux near the point of reattachment. This over-prediction is typically worse for $k$-$\epsilon$-based models than for $k$-$\omega$-based models~\cite{coakley1994turbulence,wilcox1998turbulence}. Similar to ZPG TBLs, the over-prediction in peak heat flux appears to worsen as the Reynolds number grows~\cite{parish2024report}. Length-scale corrections (Sec.~\ref{sec:lengthscale}) and variable turbulent Prandtl number models (Sec.~\ref{sec:var_prt}) have shown promise for addressing this issue. Huang \textit{et al.}~\cite{huang1993turbulence} show that length-scale corrections result in a dramatic improvement in wall heat flux predictions for a range of SBLI flows. It is worth highlighting, however, that length-scale corrections additionally reduce wall shear stress; this impact was not assessed in Ref.~\onlinecite{huang1993turbulence}. The variable turbulent Prandtl number of Roy and Sinha (see Sec.\ref{sec:var_prt}) similarly shows promise for addressing over-prediction in wall heat flux. Ref.~\onlinecite{SuSi19} shows that an extension of the model for hypersonic flows results in improved predictions for wall heat flux on a variety of hypersonic compression ramps and impinging SBLI cases. Unlike length-scale limiting, the variable turbulent Prandtl number model decouples the turbulent thermal diffusivity from the eddy viscosity. This allows for a reduction in wall heat flux with a minimal impact on the wall shear stress, and vice versa. This approach is supported by experimental data, which suggests that the rate at which wall shear stress rises through an SBLI decouples from the rate at which wall heat flux rises as shock strength increases~\cite{schulein2006skin}. Lastly, we remark that the SA model does not result in the same type of over-predictions in wall heat flux in SBLIs as two-equation models, as seen above. However, the improved performance of the SA model for wall heat flux is at the expense of an under-prediction in wall shear stress~\cite{parish2024report}.

%-------------------------------------------------------------------------------------
\subsection{Surface roughness and blowing effects}\label{sec:roughness}

%% Roughness/blowing -- 

The development of roughness patterns and gasification during the ablation process (see Section \ref{sec:ablation}) significantly alters the boundary layer state and affects SBLI properties. Modeling the transpired turbulent boundary layers over rough walls is a challenging task. Roughness effects tend to increase peak surface heat transfer compared to smooth wall conditions~\cite{prince2005experiments,marchenay2022hypersonic}. On the other hand, blowing creates a cooler film in the near wall region and provides additional protection against the convective heat flux, referred to as the blockage effect of blowing, thus reducing the total heat load~\cite{holden2008review}.

%% roughness -- consequences and modeling

The distribution of Reynolds normal and shear stresses across a TBL is significantly affected due to the roughness effects~\cite{sahoo2009effects,peltier2016crosshatch}. However, logarithmic behavior is preserved for the mean velocity and the outer region is independent of the wall conditions as per Townsend's hypothesis. A downward shift in the streamwise velocity is observed in the log region compared to ZPG TBLs over a smooth surface~\cite{berg1977surface,sahoo2009effects,peltier2016crosshatch}. The decreased momentum in the incoming TBL associated with the roughness effects before interaction with a shock wave results in an earlier separation and a delayed reattachment compared to a smooth wall case for identical flow conditions and shock strengths~\cite{prince2005experiments}. In the RANS framework, the discrete element method has been widely used to describe the interaction between well-defined distributed roughness patterns and turbulent flow. This method involves directly including some roughness corrective terms in the form of sources and sinks into the RANS or boundary layer equations~\cite{finson1980effect}. However, it requires modification of the original system of RANS equations and thus it is not desirable. 

A classical and the most commonly used method in industrial applications to describe the effect of any given roughness is the equivalent sand grain approach. Equivalent sand grain roughness is defined as the height of Nikuradse's sand grain roughness that would be required to produce the same velocity defect. In the fully rough regime, it is characterized by the roughness parameter (also called roughness Reynolds numbers) $k^+_s$, i.e. dimensionless equivalent sand grain height $k_s$ 
 \begin{equation}
     k^+_s = \dfrac{ \overline{\rho}_w k_s u_\tau }{\mu_w}.
 \end{equation}
Unlike the discrete element method, this method is not intrusive to an existing code and thus appears to be the most suitable for industrial purposes. It involves modification of the boundary conditions of a turbulence model using the sand grain roughness height as a new parameter to recover the expected law of the wall. The velocity shift is achieved by imposing an artificial increase of the eddy viscosity to increase turbulence levels at the wall. Several roughness corrections are available in the literature; however, they are developed for incompressible flows~\cite{aupoix2015roughness}, and their application to hypersonic flows is limited. The extension of these corrections to compressible flows is based on the observations of Goddard~\cite{goddard1959effect,lin1982turbulence} which imply that the effect of roughness for compressible flow can be formulated in a fashion similar to that of the incompressible case if the pertinent parameters are evaluated based on local flow properties at the wall. These roughness corrections generally result in an over-prediction of the wall heat flux with a constant $Pr_T$ value. A correction to $Pr_T$ based on a large database of low-speed turbulent boundary layers over rough surfaces obtained using the discrete element method~\cite{aupoix2015improved} is found to improve wall heat flux predictions. Olazabal \textit{et al.}~\cite{olazabal2017study,olazabal2019roughness} applied the dynamic and thermal corrections of Aupoix~\cite{aupoix2015roughness,aupoix2015improved} to the SST $k$-$\omega$ model to compute  $7^\circ$ half-angle sharp and blunt cones at Mach $10$ and $6^\circ$ half-angle slender cone at Mach $11.3$. Reasonable comparisons between the experiments and model predictions are obtained for skin friction and wall heat transfer, however, discrepancies remain. On the other hand, Marchenay \textit{et al.}~\cite{marchenay2022hypersonic} applied the Aupoix corrections to the SST $k$-$\omega$ model along with Sarkar and Zeman compressibility corrections and obtained excellent predictions for skin friction and Stanton number for a hemispherical model and $6^\circ$ slender cone configuration at Mach $11$ for which the standard SST model largely overestimated the surface heat transfer.

%% blowing -- consequences and modeling

Similar to wall roughness effects, wall blowing also tends to decrease the viscous sublayer thickness and alters the velocity profile in the inertial region. Experimental investigations on the effects of blowing are generally motivated by turbulent flow control and transpiration cooling for thermal protection systems (TPS). The classical RANS models show an over-prediction of near-wall turbulence production, especially for the SA model, which results in a faster growing boundary layer over the blowing region \cite{Bukva2021}. Blowing effects are introduced by altering the slope of the mean velocity profile in the inertial region through a blowing parameter $v_{w}^{+} = v_w/u_\tau$, i.e. wall-normal velocity at the wall (also called blowing velocity). Wilcox's correction~\cite{wilcox1988reassessment} to the $k$-$\omega$ model to account for the blowing effects has the same effect as the roughness corrections. It consists of a reduction of the specific dissipation rate wall condition given by
\begin{equation}
    \omega_{w}^{+} = \dfrac{25}{v_{w}^{+}(1+5 v_{w}^{+})}.
\end{equation}
However, the correction has not been tested for hypersonic flows.

%%Combined effects of roughness/blowing -- modeling

The roughness and blowing corrections discussed above cannot be employed together since they play on the same wall boundary conditions. Recent works of Marchenay \textit{et al}.~\cite{marchenay2021modeling,marchenay2022hypersonic} model the combined effects of surface roughness and blowing due to the ablation process for $k$-$\omega$ based models. The modeling strategy relies upon the characterization of the velocity profile in the inertial region and the identification of the separate contributions of blowing and surface roughness on the velocity
shift $\Delta U^+$ of the logarithmic law,
\begin{equation}
    u^{+} = \dfrac{1}{\kappa} \ln y^+ + C - \Delta U^+|_{app}(k_{s}^{+},v_{w}^{+}).
    \end{equation}
Further, the apparent velocity shift is decomposed as
\begin{equation}
\Delta U^+|_{app}(k_{s}^{+},v_{w}^{+}) = U^{+}_{r}(k_{s}^{+}) + U^{+}_{rb}(k_{s}^{+},v_{w}^{+}),
\end{equation}
where $U^{+}_{r}$ is the velocity shift given by standard roughness corrections and $U^{+}_{rb} $ depicts the effect of the interaction between roughness elements and blowing on the skin friction. The apparent velocity shift is imposed by modifying the wall-boundary conditions for $k$ and $\omega$ in the following manner,
\begin{equation}
    k_w = k^+_w u^2_\tau\hspace{0.1in} \text{and} \hspace{0.1in} \omega_w = \overline{\rho} \omega^+_w u^2_\tau / \mu,
\end{equation}
where
\begin{equation}
    \begin{split}
        k^+_w &= max (0,  k^+_0), \\
        k^+_0 &= \dfrac{1}{\sqrt{\beta^*}} \tanh \left[ \left( \dfrac{\ln (k^+_s/30)}{\ln(8)} + \right. \right. \\ &
    \left. \left. 0.5 [ 1 - \tanh (k^+_s/100) ] \tanh (k^+_s/75) \right)  \right], \\
        \omega^+_w &= \dfrac{400000}{k^{+4}_s} \left( \tanh \dfrac{10000}{3k^{+3}_s} \right)^{-1} + \dfrac{70}{k^+_s} \left[ 1 - \exp \left( - k^+_s / 300   \right)  \right]. 
    \end{split}
\end{equation}
The equivalent sand grain height, $ k_s$, is given by
\begin{equation}
    \begin{split}
        k^+_s &= \max(k_1,k_2), k_1 = k^+_s ( 1 + f v^+_w), \\ k_2 &= \exp \left[ \dfrac{\kappa (b_1 - C)}{1 - \kappa b_2}  \right] ( 1 + f v^+_w)^{1/(1 - \kappa b_2)} \\
        f &= 5.9 [1 + \tanh ( k^+_s - 7) ] + 4.2,
    \end{split}
\end{equation}
where $\kappa = 0.4$, $ C= 0.55$, and the constants $b_1$ and $b_2$ are given by
\begin{equation}
 \begin{split}
     k^+_s < 3.5 &: \hspace{0.1in} b_1 = 5.5 \hspace{0.1in} b_2 = 1/\kappa, \\
     3.5 \leq k^+_s < 7 &: \hspace{0.1in} b_1 = 6.59 \hspace{0.1in} b_2 = 1.52, \\
     7 \leq k^+_s < 14 &: \hspace{0.1in} b_1 = 9.58 \hspace{0.1in} b_2 = 0, \\
     14 \leq k^+_s < 68 &: \hspace{0.1in} b_1 = 11.5 \hspace{0.1in} b_2 = -0.7, \\
     68 \leq k^+_s &: \hspace{0.1in} b_1 = 8.48 \hspace{0.1in} b_2 = 0.
 \end{split}   
\end{equation}
The equivalent sand grain height estimation is difficult and relies on empirical roughness correlations. A correction to $Pr_T$ is also proposed to account for the blowing thermal effects on smooth and rough surfaces. The turbulent mixing increase is represented by the roughness function $\Delta U^+$, the effect of surface topology or the geometry of roughness above the meltdown surface is parameterized using the corrected wetted surface ratio $S_{corr}$, and the correction is imposed only in the roughness sublayer via the mean physical height $k_m$ of the rough surface.%%%%

A contribution that accounts for the effects as mentioned earlier is added to the constant turbulent Prandtl number used for smooth walls $Pr_{T,s}$
\begin{equation}
    \begin{split}
        Pr_T &= Pr_{T,s} + \Delta Pr_T, \Delta Pr_T =  \left( A \Delta U^{+2} + B \Delta U^+ \right) e^{-y/k_m} \\ 
        A &= ( 0.0155 - 0.0035 S_{corr} ) [ 1 - e^{-12(S_{corr} - 1) } ], \\ B &= -0.08 + 0.25  e^{-10(S_{corr} - 1)} \\
        \Delta U^+ &= \Delta U^+_r + \Delta U^+_{rb}, \\ \Delta U^+_r &= \dfrac{1}{\kappa} \ln ( k^+_s ) + C - b_1 - b_2 \ln ( k^+_s ), \\ \Delta U^+_{rb} &=  \dfrac{1}{\kappa} \ln (1 + f_T v^+_w ) \\ 
        f_T &= 8 [1 + \tanh ( k^+_s - 7) ], 
    \end{split}
\end{equation}
where the parameters $k_m$ and $S_{corr}$ need to be estimated. The modification is designed to recover the turbulent Prandtl number's constant value above the roughness sublayer. These dynamic and thermal corrections when applied to the SST $k$-$\omega$ model along with the Zeman compressibility correction resulted in improved predictions for skin friction and Stanton number for a $10.5^\circ$ slender cone configuration at Mach $11$ for different blowing rates $F = (\overline{\rho}_w v_w) / (\overline{\rho_e} \widetilde{u_e}) $~(see Ref.~\onlinecite{marchenay2022hypersonic}). 
%%%%%%%%%%%%%%%%%%%%%%%%%%%%%%%%%%%%%%%%%%%%%%%%%%%%%%%%%%%%%%%

%-------------------------------------------------------------------------------------

%-------------------------------------------------------------------------------------

%-------------------------------------------------------------------------------------

\subsection{Thermochemical non-equilibrium effects}\label{sec:thermochemical}
% 
%%%%%%%%
When thermochemical non-equilibrium effects are at stake, the governing equations for the gas are supplemented with additional transport equations describing the evolution of the chemical species (chemical non-equilibrium) and/or the modal energy(ies), see Section \ref{sec:reacteq}. 

Favre-averaging of such equations introduces additional unclosed terms, which will be examined in the following.

We focus first on the unclosed terms in the species transport equations, and specifically on the turbulent mass transport term $\overline{\rho u_{j}^{\prime \prime} Y_{n}^{\prime \prime}}$. The latter is usually modeled by introducing a ``turbulent'' Schmidt number, such that:
\begin{equation}
      \overline{\rho u_{j}^{\prime \prime} Y_{n}^{\prime \prime}} \simeq -\overline{\rho}D_{t,n}\frac{\partial \widetilde{Y}_n}{\partial x_j}=-\frac{\mu_t}{Sc_{t,n}}\frac{\partial \widetilde{Y}_n}{\partial x_j},
       \label{eq:turbfick}
\end{equation}
where $D_{t,n}$ is a turbulent diffusivity coefficient for the $n$th species, and $Sc_{t,n}=\frac{\mu_t}{D_t}$ is the corresponding turbulent Schmidt number, representing the ratio of turbulent transport of momentum to turbulent transport of mass. Early CFD modeling of the turbulent mass transport of a passive scalar in incompressible flow used $Sc_t=0.7$, since “there is wall-authenticated evidence that the time–mean concentration profile is appreciable “fatter” than that of velocity” \citep{spalding1971concentration}. On the other hand, Launder~\cite{launder2005heat} pointed out that $Sc_t$ shows the value of $0.9$ for turbulence near the wall. Quoting from these results, the values $0.7$ or $0.9$ have been widely used to model turbulent mass diffusion in CFD codes.  However, in the review paper of Tominaga and Stathopoulos~\cite{tominaga2007turbulent} the  turbulent Schmidt number providing the best fit
to the experimental observations was found to vary in a range as large as $0.2 \le Sc_t \le 1.3$, depending on the flow problem. 
For compressible flows, experimental studies~\cite{marquardt2020experimental} have shown that $Sc_t$ differs significantly from unity: in shock-free flows, the turbulent Schmidt number is found to be in the range $0.5\le Sc_t\le 1.5$, and even larger variations are observed in case of SBLIs. In addition to not having a well-defined value for all flows, $Sc_t$ is also seen to vary across the flow, which invalidates the use of the linear constitutive relation (\ref{eq:turbfick}) to model turbulent mass transport.

Sciacovelli \textit{et al.}~\cite{sciacovelli2023priori}  conducted \emph{a priori} tests of the validity of the constant turbulent Schmidt number assumption for hypersonic boundary layers at Mach 10 and 12.48, under chemical nonequilibrium and adiabatic wall and thermochemical nonequilibrium and cooled wall conditions. For nearly parallel flows, the ``exact'' species Schmidt numbers can be computed from DNS data as:
\begin{align}
\quad Sc_{t,n}&=\frac{\overline{\rho u^{\prime \prime} v^{\prime \prime}} \partial \widetilde{Y}_{n} / \partial y}{\overline{\rho v^{\prime \prime} Y_{n}^{\prime \prime}} \partial \widetilde{u} / \partial y}.
\end{align}
The results showed that, even for such relatively simple flows, $Sc_{t,n}$ is not constant but instead, like the turbulent Prandtl number, it exhibits a bump in the logarithmic region, more pronounced for the cooled wall case, and is below the average value in the outer region. The value of 0.9, often used in the literature for linear turbulent mass flux models, is only representative of the average across the boundary layer. The profiles for different chemical species were not found to be much different from each other, but their values and the position of the bump in the logarithmic region were clearly dependent on $M_\infty$ and the wall conditions.\\

The influence of species mass fraction and temperature fluctuations on the species production rates, i.e. the intensity of turbulence-chemistry interactions (TCI), is evaluated through the quantity:
\begin{equation}\label{eq:tci}
\dot{\omega}_n^{I} = \left[\overline{\dot{\omega}_{n}\left(T, \rho_{n}\right)}-\dot{\omega}_{n}(\overline{T}, \overline{\rho_{n}})\right] \frac{\overline{\mu}}{\overline{\rho} \, \overline{\tau}_w},
\end{equation}
where $\overline{\dot{\omega}_n(T,\rho_n)}$ is a source term that represents the production or depletion of the $n$-th species in the mixture due to chemical reactions.
This parameter represents the chemical production due to turbulent fluctuations, leading to $\overline{\dot{\omega}_n(T,\rho_n)}\ne\dot{\omega}_n(\overline{T},\overline{\rho_n})$ due to the strong nonlinearity of the chemical source term $\dot{\omega}_n$.
A simplification often adopted in the RANS framework is to assume $\overline{\dot{\omega}_n(T,\rho_n}) \approx \dot{\omega}_n(\overline{T},\overline{\rho_n})$, which is an acceptable approximation only if the turbulence-chemistry interactions are limited. Less simplistic models exist for combustion applications \citep{xiang2020turbulence}, based on the Eddy Dissipation Concept (EDC):
\begin{equation}
    \overline{\dot{\omega}_n} \approx \gamma^{*} \dot{\omega}_n(\overline{T},\overline{\rho_n}),
\end{equation}
where $\gamma^{*}  \approx 9.7 (\nu \varepsilon / k^2)^{\frac{3}{4}}$ represents the fine-scale structure volume fraction, i.e. the fraction of the volume in which chemical reactions take place which is assumed to be in the region where the turbulent kinetic energy is quasi-steady.
Direct extraction of $\dot{\omega}_n^{I}$ from DNS data shows that the indicator takes large values close to the wall for the adiabatic case M10C, and in the buffer region for the wall-cooled case M12, where turbulent fluctuations are significant.

\emph{A priori} tests for the above-mentioned hypersonic boundary layers showed that the assumption $\overline{\dot{\omega}_n(T,\rho_n}) \approx \dot{\omega}_n(\overline{T},\overline{\rho_n})$ is reasonably accurate for adiabatic wall conditions, but underestimates the DNS value in the cooled case. On the contrary, the EDC model does not prove to be satisfactory, mainly due to the isotropic assumption in the formulation of $\gamma^{*}$ causing the model to fail close to the wall. Further research is required on closures for this term, by leveraging, for instance, presumed PDF models typically used in turbulent combustion applications \citep{bray2006finite}.

The behavior of unclosed terms in the averaged vibrational energy equation was investigated in the DNS study of Passiatore \textit{et al.}~\cite{passiatore2022thermochemical}. They showed that the vibrational turbulent heat flux $\overline{\rho u_j''e''_v}$ is a prominent term in the vibrational energy budget, and should be taken into account in RANS models. A simple gradient model, relying on the introduction of a vibrational turbulent Prandtl number $Pr_{t}^\text{V}$ was introduced to model such term, by analogy with the rototranslational turbulent heat flux~\cite{baranwal2020vibrational}. For nearly parallel flows, an ``exact'' value can be extracted from DNS data:
\begin{equation}
    Pr_{t}^\text{V}=\frac{\overline{\rho u^{\prime \prime} v^{\prime \prime}} \partial \widetilde{e}_{\mathrm{V}} / \partial y}{\overline{\rho v^{\prime \prime} e_{\mathrm{V}}^{\prime \prime}} \partial \widetilde{u} / \partial y},
\end{equation}
such that the model reads:
\begin{equation}
    \overline{\rho u_{j}^{\prime \prime} e_{V}^{\prime \prime}} = \frac{\mu_t}{Pr_{t}^\text{V}}\frac{\partial \widetilde{e}_V}{\partial x_j},
\end{equation}
where $Pr_t^\textbf{V}$ exhibits a behavior similar to that of $Pr_t$ across the boundary layer height, with a bump in the logarithmic region. In this case, a value of $0.9$ is reasonably well recovered only in the outer region.

Finally, the Favre-averaged source term in the vibrational energy equation also needs attention. Such a term measures the intensity of the interaction between turbulence and thermal relaxation, which can be considered as weak when $\overline{Q_{\mathrm{TV}}\left(T, T_{\mathrm{V}}, \rho, p, Y_n\right)}\approx Q_{\mathrm{TV}}\left(\widetilde{T}, \widetilde{T_{\mathrm{V}}}, \overline{\rho}, \overline{p}, \widetilde{Y_n}\right)$, where $\overline{Q_{\mathrm{TV}}}$ is a source term accounting for vibrational energy
production/depletion due to translational-vibrational energy transfers. Deviation from such behavior can be measured using the indicator \citep{passiatore2022thermochemical}:
\begin{equation}
    Q_{\mathrm{TV}}^I=\frac{\overline{Q_{\mathrm{TV}}\left(T, T_{\mathrm{V}}, \rho, p, Y_n\right)}-Q_{\mathrm{TV}}\left(\widetilde{T}, \widetilde{T_{\mathrm{V}}}, \overline{\rho}, \overline{p}, \widetilde{Y_n}\right)}{\overline{Q_{\mathrm{TV}}^\text{max}}},
\end{equation}
 with $\overline{Q_{\mathrm{TV}}^\text{max}}$ the maximum wall-normal value at the selected station.
Differences up to $\approx 15\%$ in the inner layer, indicating a strong interaction between turbulence and thermal non-equilibrium, were observed for the thermochemical nonequilibrium boundary layer at Mach 12.48.
Although a few models do exist for the chemistry source term $\dot{\omega}$,  no models are found in the literature for $\overline{Q_{\mathrm{TV}}}$, which is potentially critical for RANS modeling of hypersonic flows out of thermal equilibrium.

%-------------------------------------------------------------------------------------

\subsection{Multi-physics considerations }

\subsubsection{Catalysis and ablation phenomena}\label{sec:ablation}
Catalysis and ablation are physical phenomena that occur at the interface between the hypersonic flow and the wall, typically as part of  thermal protection system (TPS) used to reduce the heat transfer from the fluid to the vehicle. 

Catalysis is a chemical gas/surface interaction, and it relates to a series of chemo-physical phenomena such as diffusion of reactants from the gas to the surface, chemical reactions at the interface, and diffusion of the products into the gas. In hypersonic flow models, catalysis is generally modeled by introducing a recombination coefficient (designated $\gamma=M_{rec}/M_{imp}$), corresponding to the ratio of recombining atoms $M_{rec}$ over atoms that impinge on a wall $M_{imp}$.
Such a ratio can be measured experimentally, but large uncertainties on its value exist \cite{longo2004modelling}. 

Ablation of aerospace materials is a widely employed TPS solution used to protect vehicles from radiative heating. Convective heat-transfer rates
are nearly zero because the outward flow of ablative products impedes the conductive heat
flux to the wall. The product gas of ablation forms a layer,
called ablation-product layer, which prevents the hot shock layer gas from reaching the wall \citep{longo2004modelling}.
The ablation-product layer absorbs a portion of the radiative flux directed toward the wall. Lightweight ablative materials are characterized by high porosities, with pores being the site of chemical reactions \citep{mansour2024flow}. Turbulence models then largely borrow from pre-existing studies on incompressible flows on porous materials. Specific models for hypersonic flows are an open research topic. 

Finally, ablation induces geometrical modifications of the surface and generates surface roughness, which plays a major role in transition and turbulence \citep{schneider2010hypersonic}. This is another open topic (see Section \ref{sec:roughness}) that needs attention for the development of improved models.

\subsubsection{Conjugate heat transfer}
Strong aerothermal heating of high-enthalpy flows represents a key consideration for thermal management systems in hypersonic vehicles. The predicted thermal loading characteristics dictate the material and TPS, which in turn has direct implications on vehicle weight and design. Both maximum and total integrated thermal loads are important design parameters but their accurate prediction requires a two-way coupled, conjugate heat transfer (CHT)  evaluation of the heat flux on the vehicle.

The main contribution to the heat flux at the solid-fluid interface is governed by the temperature gradient at the wall, which in turn depends on the near wall turbulence and heavily on the temperature of the wall. Whereas the heat flux shows a linear dependence on wall temperature for a perfect gas in the laminar regime  \citep{Yang_Wang_Gao_2022},
the effects of temperature are much more complex with turbulent flows for which temperature scaling has been recently proposed for high-speed, non-adiabatic boundary layer flows\citep{Cheng_Fu_2024}.
The spatially and temporally varying wall temperature depends not only on the aerothermal heating and chemical kinetics of the gas but also on internal heat sources (propulsion system, combustion chamber, electronic equipment, etc.) and external heat transfer through radiation or other shock boundary layer interaction-induced heating. Furthermore, the wall temperature is influenced by the heat conduction within the vehicle structure, the convective and radiative losses, ablative effects, and active thermal protection system.  The aggregate of these multiphysics effects will govern the time-varying heat flux in the vehicle.

The integration of CHT demands many algorithmic and computational changes to classical CFD solvers, readers can refer to a review of the recent studies on the topic \citep{Lewis_Hickey_2023}.
As the accuracy of the near-wall temperature and velocity gradients is central to CHT computations (especially in the turbulent regime), the compounding modeling challenges of high-speed near-wall flow, discussed earlier, take on even greater importance. Within the context of RANS, two modeling parameters are particularly important:  turbulent Prandtl number \citep{Yoder_2016}
and the modeling of thermal fluctuations
\citep{Yang_Iacovides_Craft_Apsley_2021}.
Within the context of higher fidelity simulations such as LES, the wall-modeling in the context of CHT is a particularly relevant topic. A good number of works have looked at wall-modeled LES (WMLES) with CHT in lower speed application 
\citep{Maheu_Moureau_Domingo_2012,Gelain2021}
very few works have started to address this problem for high-speed flows~\citep{Muller_Dutta_Boisvert_Oefelein_2024,muller2024investigation}.

%%%%%%%%%%%%%%%%%%%%%%%%%%%%%%%%%%%%%%%%%%%%%%%%%%%%%%%%%%%%%%%%%%%%%%%

%%%%%%%%%%%%%%%%%%%%%%%%%%%%%%%%%%%%%%%%%%%%%%%%%%%%

%\section{Model Validation : Available datasets}\label{sec6} %% and Uncertainty quantification

\section{Model Validation : Available datasets, Model performance, and Industrial applicability }\label{sec6} %% and Uncertainty quantification

%%%%%%%%%%%%%%%%%%%%%%%%%%%%%%%%%%%%%%%%%%%%%%%%%%%%%%%%%%%%%%%%%%%%%%%%%%%%%%%%%%%%%%%%%%%%%
%\textbf{Experiments - FP }

\begin{table*}
\caption{\label{tab:experiments-fp}Experiments -- ZPG hypersonic TBL : Flat plate - FP; Nozzle Wall - NW; Wind tunnel wall - WTW; Cone/Ogive Cylinder - COC; Hollow cylinder - HC; Sharp circular cone - SCC; $^\ostar$ high-enthalpy case; $^\filledstar$ ASCII datafile available on the NASA Turbulence Modeling Resource website~\cite{Hypersonic_SBLI_Expt}.}
\begin{ruledtabular}
\begin{tabular}{llccccc}
 %&\multicolumn{2}{c}{$D_{4h}^1$}&\multicolumn{2}{c}{$D_{4h}^5$}\\
 Configuration & Experiments& $M_e$    & $T_w/T_{aw}$
 &$H_{e}$ &$Re_e$/m  & Quantities
		\\ \vspace{0.1in}
		&  &  &     & (MJ/kg) & $\times 10^6$ & reported \\ \hline 
%\vspace{0.1in}
 %%%%%%%%%%%%%%%%%%%%%%%%%%%%%%%%%%%%%%%%%%%%%%%%%%%%%%%%%%%%%%%%%%%%%%%%%%%%%%%%%%%%%%	
        & Winkler \& Cha~(Ref.~\onlinecite{winkler1959investigation}) \& & $4.98-5.21$    & $0.58-0.83$  & $0.35-0.5$ & $9.3-12.6$  & $P_w$,$C_f$,$Q_w$   \\
        &  Winkler~(Ref.~\onlinecite{winkler1961investigation}) &   &  &  &   &    \\
        & Danberg~(Refs.~\onlinecite{danberg1964characteristics,danberg1967characteristics}) & $6.7$   & $0.41$ \& $0.87$ & $0.5$ \& $0.5$  & $8$ \& $19$ & $P_w$,$Q_w$     \\ %BL surveys of $P_t$, $T_t$, and 
    FP   & Young~(Ref.~\onlinecite{young1965experimental}) &  $4.91$    & $0.54$ \& $1$ & $0.335$ \& $0.64$ & $17.2$ \& $45.2$  & $P_w$,$C_f$,$Q_w$    \\  %BL surveys of $P_t$, $T_t$, and 
           % & \cite{heronimus1966hypersonic} & $5.4 - 11.7$ &  Tw/Tt $0.09-0.3$  & $1.04-2.4$ & $1-200$ & $C_f$,$Q_w$   \\
           % & (data taken from \cite{cary1970summary}) &  &  (assuming $Tw = 311$K \& $Pr_T = 0.89$)  &  &  &    \\
            & Neal~(Ref.~\onlinecite{neal1966study}) & $6.8$ & $0.5$ &  $0.62$   & $2-16$  &  $C_f$,$Q_w$  \\
        & Wallace~(Ref.~\onlinecite{wallace1967hypersonic})  & $7.4-10.7$  & $0.16-0.33$  &  $1-2.12$ & $29.5-265.7$ &  $P_w$,$C_f$,$Q_w$   \\
         & Hopkins~(Ref.~\onlinecite{hopkins1969summary}) & $6.5$ & $0.32-0.51$  & $0.69-1.09$ &$1.7-4.07$  & $C_f$,$Q_w$   \\
        & Voisinet \& Lee~(Ref.~\onlinecite{voisinet1972measurements}) & $4.9$  & $0.25-1$  &  $0.4$ & $9-25$ &  $C_f$,$Q_w$   \\ % BL surveys of $P_t$, $T_t$, and 
         & Holden~(Ref.~\onlinecite{holden1972shock})  & $7.4-12$   & $0.16-0.34$  & $0.99-2.07$  & $15.1-191.8$ & $P_w$,$C_f$,$Q_w$   \\
         & Watson~(Ref.~\onlinecite{watson1973measurements}) & $9.5-10.1$    & $1$  & $0.3$ & $7.769-34.544$  & $P_w$,$C_f$,$Q_w$    \\  %BL profiles of $P_t$, $T_t$, and
         & Goyne~(Ref.~\onlinecite{goyne2003skin})$^\ostar$ & $5.3-6.7$  & $0.022-0.11$  & $3.2-13.1$ & $0.43-16.1$ & $P_w$,$C_f$,$Q_w$    \\
 		& Schulein~(Ref.~\onlinecite{schulein2006skin})$^\filledstar$ & $5$   & $0.8$ & $0.41$  & $37$ & $P_w$,$C_f$,$Q_w$  \\
%            &  & &   &  &     &  & $U,M,T,\overline{\rho}$ profiles at various $x$-locations \\
		& Holden~(Ref.~\onlinecite{holden2010experimental})$^\filledstar$  & $8-11.4$   & $0.195-0.325$  & $1.02-1.71$  & $36-205.2$ & $P_w$,$C_f$,$Q_w$   \\ 
           % & \cite{keener1972measurements}, \cite{hopkins1972hypersonic} & $5.9-7.8$ & $0.28-0.51$  & $0.69-1.15$ &$2-38$  & $C_f$,$Q_w$   \\
           % & (FP at angles of attack) &  &   &  &  &   \\
            %& \cite{suraweera2006reynolds} & $4.2-5.9$  &  & $4.8-9.5$ & -- & $C_f$,$Q_w$    \\
%            &  &  &  &  &   &  &      \\
%            &  &  &  &  &   &  &      \\
%            &  &  &  &  &   &  &      \\
             \hline
 %%%%%%%%%%%%%%%%%%%%%%%%%%%%%%%%%%%%%%%%%%%%%%%%%%%%%%%%%%%%%%%%%%%%%%%%%%%%%%%%%%%%%% 
		& Hill~(Ref.~\onlinecite{hill1959turbulent}) & $8.25-10.06$ & $0.47-0.52$  & $0.76-0.83$ & $5.7-9.4$ &  $C_f$,$Q_w$   \\
	NW	& Wallace~(Ref.~\onlinecite{wallace1967hypersonic})  & $6.6-8$ & $0.1-0.33$  &  $1-3.2$ & $1.2-36.1$ &  $P_w$,$C_f$,$Q_w$   \\
 & Lee~(Ref.~\onlinecite{lee1969velocity}) & $4.91$  & $0.52$  &  $0.61$ & $5.06$ &  $P_w$,$C_f$,$Q_w$   \\ 
    & Backx~(Refs.~\onlinecite{backx1973measurements,backx1974experimental}) \& & $15$ \& $19.8$    & *($T_w/T_{t_e} : 0.15$ \& $0.11$)  & $1.7$ \& $2.3$ & $27.9$ \& $11.2$  & $P_w$,$Q_w$   \\  
        & Backx \& Richards~(Ref.~\onlinecite{backx1976high}) &   &  &  &  &   \\  
 \hline
  %%%%%%%%%%%%%%%%%%%%%%%%%%%%%%%%%%%%%%%%%%%%%%%%%%%%%%%%%%%%%%%%%%%%%%%%%%%%%%%%%%%%%% 
 %       & \cite{matting1961turbulent} & $6.7, 9.9$  & $---$    &   &  & $P_w$,$Q_w$ \\ 
 %           & \cite{laderman1974mean} & $5.3$   & --  & -- & --  & --   \\ 
  WTW           & Hopkins~(Ref.~\onlinecite{hopkins1969summary}) & $7.4$ & $0.32-0.49$  & $0.7-1.05$  &$0.63-4.01$  & $C_f$,$Q_w$   \\
\hline  %BL profiles and 
 %%%%%%%%%%%%%%%%%%%%%%%%%%%%%%%%%%%%%%%%%%%%%%%%%%%%%%%%%%%%%%%%%%%%%%%%%%%%%%%%%%%%%% 
		& Horstman \& Owen~(Ref.~\onlinecite{horstman1972turbulent}) \& & $7.2$  & $0.49$  & $0.67$  & $10.9$  & $P_w$,$C_f$,$Q_w$    \\	 %BL surveys of $P_t$, $T_t$, and 	
	COC	& Owen \& Horstman~(Ref.~\onlinecite{owen1972structure}) &    &  &   &  &    \\
 \hline
  %%%%%%%%%%%%%%%%%%%%%%%%%%%%%%%%%%%%%%%%%%%%%%%%%%%%%%%%%%%%%%%%%%%%%%%%%%%%%%%%%%%%%%    
       HC  & Samuels~(Ref.~\onlinecite{samuels1967experimental})  & $6$  & $0.55$ \& $0.49$ & $0.49$ \& $0.54$  & $32.6$ \& $28.6$  & $P_w$,$C_f$,$Q_w$ \\  % BL surveys of $P_t$, $T_t$, and  
       & Murray~(Ref.~\onlinecite{murray2013experimental}) & $8.9$    & $0.27$  &  $1.2$ & $48$ & $P_w$,$Q_w$ \\ \hline
 %%%%%%%%%%%%%%%%%%%%%%%%%%%%%%%%%%%%%%%%%%%%%%%%%%%%%%%%%%%%%%%%%%%%%%%%%%%%%%%%%%%%%% 
	SCC	& Kimmel~(Ref.\onlinecite{kimmel1993experimental,kimmel1997effect})  & $7.93$    & $0.46$   & $0.725$ & $6.6$ & $P_w$,$Q_w$    \\ % ($7^\circ$ HAgC) $T_w/T_{t_e} = 0.42$  T0 = 722 Tw 303.24 Taw 655.12 
	%	& Chien 1974 ($5^\circ$ HAgC)   & $7.15$ & *$T_w/T_{t_e} : 0.11-0.41$ &  & $24.2-36.4$ &    \\		
	%	& Rumsey et al. ($7.5^\circ$ HAgC)   & $5.15$ &  &   & $31$  &    \\
%		& Stainback et al. ($10^\circ$ HAgC) & $6$  &  &   &  &    \\
%          & Hillier et al. ($7^\circ$ HAgC) & $9.16$  &  &   &  &    \\
%          & Holden et al. ($6^\circ$ HAgC) & $13.04$  &  &   &  &    \\
\end{tabular}
\end{ruledtabular}
\end{table*}

%%%%%%%%%%%%%%%%%%%%%%%%%%%%%%%%%%%%%%%%%%%%%%%%%%%%%%%%%%%%%%%%%%%%%%%%%%%%%%%%%%%%%%%%%%%%%%%%

%%%%%%%%%%%%%%%%%%%%%%%%%%%%%%%%%%%%%%%%%%%%%%%%%%%%%%%%%%%%%%%%%%%%%%%%%%%%%%%%%%%%%%%%%%%%%
%\textbf{DNS - FP }

\begin{table*}
\caption{\label{tab:dns-fp} DNS -- ZPG hypersonic TBLs; $^\filledstar$ ASCII datafile available on the NASA Turbulence Modeling Resource website~\cite{Hypersonic_FP_DNS}. }
\begin{ruledtabular}
\begin{tabular}{llccccc}
 %&\multicolumn{2}{c}{$D_{4h}^1$}&\multicolumn{2}{c}{$D_{4h}^5$}\\
 		Condition & DNS& $M_e$    &   $T_w/T_{aw}$
  &$Re_\theta$ 
		\\ %\vspace{0.1in}
		  & &  &    & $\times 10^3$  \\\hline 	%\vspace{0.05in}
 %%%%%%%%%%%%%%%%%%%%%%%%%%%%%%%%%%%%%%%%%%%%%%%%%%%%%%%%%%%%%%%%%%%%%%%%%%%%%%%%%%%%%%	
		 & Maeder \textit{et al.}~(Ref.~\onlinecite{maeder2001direct}) & $6$ & $1$   & $2.945$  \\
          & Xin-Liang \textit{et al.}~(Ref.~\onlinecite{xin2006direct}) & $6$ & $0.94$  & $109.5$  \\
          & Martin~(Ref.~\onlinecite{martin2007direct}) & $4.97$ \& $5.98$ & $0.95$  & $6.23$ \& $8.43$  \\
          & Duan \textit{et al.}~(Ref.~\onlinecite{duan2010direct}) & $5$ & $0.18-1$   & $1.28-4.84$  \\
          & Duan \textit{et al.}~(Ref.~\onlinecite{duan2011direct}) & $4.9-11.93$ & $0.97-1.05$    & $1.51-11.3$  \\
          & Liang~(Ref.~\onlinecite{liang2012dns}) \& & $8$ & $0.15$ \& $0.77$   & $22$ \& $78$  \\
          & Liang \& Li~(Ref.~\onlinecite{liang2013dns}) &  &    &   \\
   Low enthalpy &  Chu \textit{et al.}~(Ref.~\onlinecite{chu2013effect}) & $4.9$ & $0.5-1.5$   & $3.48-7.61$  \\
          & Duan \textit{et al.}~(Ref.~\onlinecite{duan2016pressure}) & $5.86$ & $0.76$   & $9.45$  \\
          & Zhang \textit{et al.}~(Ref.~\onlinecite{zhang2017effect}) & $5.86$ & $0.25$ \& $0.76$  & $2.121$ \& $9.455$  \\
          & Zhang \textit{et al.}~(Ref.~\onlinecite{zhang2018direct})$^\filledstar$ & $5.84-13.64$ & $0.18-0.76$   & $2.12-14.4$  \\
          & Huang \textit{et al.}~(Ref.~\onlinecite{huang2020simulation}) : & $10.9$ \& $13.64$ & $0.2$ \& $0.18$   & $9.08-14.258$  \\
         & (similar to CUBRC M11 FP experiments) &   &   &   \\
          &  Sciacovelli \textit{et al.}~(Ref.~\onlinecite{sciacovelli2020numerical}) & $6$ & $0.72$    & $5.72$ \\ %%tinf 78  Tw 422.5 
          & Nicholson \textit{et al.}~(Ref.~\onlinecite{nicholson2021simulation}) & $4.9$ & $0.91$  & $17.406$  \\
          & Huang \textit{et al.}~(Ref.~\onlinecite{huang2022direct}) & $4.9-13.4$ & $0.18-0.91$   & $1.465-26.762$  \\
          & Dang \textit{et al.}~(Ref.~\onlinecite{dang2022direct}) & $10$ & $0.2$    & $24.8$  \\
          & Aultman \textit{et al.}~(Ref.~\onlinecite{aultman2024asymptotic}) & $5.84$ & $0.25$    & $2$  \\
          \hline
          & Duan \textit{et al.}~(Ref.~\onlinecite{duan2011direct4}) & $9-9.4$ & $0.12-0.13$  & $2.45-3.06$  \\
   High enthalpy       & Passiatore \textit{et al.}~(Ref.~\onlinecite{passiatore2021finite}) \& & $10$ & $1$    & $7.15$  \\
  & Sciacovelli \textit{et al.}~(Ref.~\onlinecite{sciacovelli2023priori}) &  &     &  \\
          & Passiatore \textit{et al.}~(Ref.~\onlinecite{passiatore2022thermochemical}) & $12.48$ & $0.1$ & $18.66$  \\ %tinf 594.3  Tw 1800 Taw 17255.54 
\end{tabular}
\end{ruledtabular}
\end{table*}

%%%%%%%%%%%%%%%%%%%%%%%%%%%%%%%%%%%%%%%%%%%%%%%%%%%%%%%%%%%%%%%%%%%%%%%%%%%%%%%%%%%%%%%%%%%%%%%%

%%%%%%%%%%%%%%%%%%%%%%%%%%%%%%%%%%%%%%%%%%%%%%%%%%%%%%%%%%%%%%%%%%%%%%%%%%%%%%%%%%%%%%%%%%%%%
%\textbf{Two-dimensional planar SBLI }

\begin{table*}
\caption{\label{tab:SBLI-2D} Experiments -- Two-dimensional planar SBLI : Compression Corner - CC; Impinging Shock - IS; $\theta$ = flow deflection (or shock generator) angle; $^\filledstar$ ASCII datafile available on the NASA Turbulence Modeling Resource website~\cite{Hypersonic_SBLI_Expt}. }
\begin{ruledtabular}
\begin{tabular}{llccccccc}
 %&\multicolumn{2}{c}{$D_{4h}^1$}&\multicolumn{2}{c}{$D_{4h}^5$}\\
		Configuration & Experiments& $M_e$   & $\theta$ (degree)  & $T_w/T_{aw}$
 &$H_{e}$ &$Re_e$/m  & Quantities
		\\ %\vspace{0.1in}
		& & &    &  & (MJ/kg) & $\times 10^6$ & reported \\\hline 	%\vspace{0.05in}
 %%%%%%%%%%%%%%%%%%%%%%%%%%%%%%%%%%%%%%%%%%%%%%%%%%%%%%%%%%%%%%%%%%%%%%%%%%%%%%%%%%%%%%	
%%%%%%%%%%%%%%%%%%%%%%%%%%%%%%%%%%%%%%%%%%%%%%%%%%%%%%%%%%%%%%%%%%%%%%%%%%%%%%%%%%%%%%	
		& Coleman \& Stollery~(Ref.~\onlinecite{coleman1972heat})$^\filledstar$ & $9$ & $15$-$38$  &  $0.28$ & $1.1$ & $47$ & $P_w$,$Q_w$  \\
        & Holden~(Ref.~\onlinecite{holden1972shock}) & $6.25-14.9$  & $27-36$ & $0.12-0.9$  & $0.75-3.01$ & $15.4-101.7$  &  $P_w$,$C_f$,$Q_w$ \\ % Tw 303.33 - 663 Taw 2503.176 - 736.44 
   	CC	& Appels~(Ref.~\onlinecite{appels1973turbulent}) & $11.7$   & $30-42$ & $0.15$  & $2.23$ & $36.6$   &  $P_w$,$Q_w$ \\
        & Coet~(Ref.~\onlinecite{coet1993experiments})  & $5$  & $35$ & $0.24$ \& $0.72$  & $0.5$ & $40$  &  $P_w$,$Q_w$ \\
		& Holden~(Ref.~\onlinecite{holden2010experimental})$^\filledstar$ & $8$  & $27-36$   & $0.34$ & $0.97$ & $145.7$ & $P_w$,$C_f$,$Q_w$  \\
        &  & $11.3$ & $36$  & $0.2$ & $1.63$ & $36$ & $P_w$,$C_f$,$Q_w$  \\ %tin 61, tw 300 
        \hline
 %%%%%%%%%%%%%%%%%%%%%%%%%%%%%%%%%%%%%%%%%%%%%%%%%%%%%%%%%%%%%%%%%%%%%%%%%%%%%%%%%%%%%%
            
 %%%%%%%%%%%%%%%%%%%%%%%%%%%%%%%%%%%%%%%%%%%%%%%%%%%%%%%%%%%%%%%%%%%%%%%%%%%%%%%%%%%%%% 
		& Schulein~(Ref.~\onlinecite{schulein2006skin})$^\filledstar$ & $5$   & $6-14$ & $0.8$  & $0.41$ & $39.1$ & $P_w$,$C_f$,$Q_w$ \& \\
%            &  & &   &  &   &   &  & $U,M,T,\overline{\rho}$ profiles at various $x$-locations \\
    IS		& Kussoy \& Horstman~(Ref.~\onlinecite{kussoy1991documentation})$^\filledstar$ & $8.18$    &  $5-11$ & $0.29$  & $1.1$  & $4.87$ & $P_w$,$Q_w$ \\	
     %      & Kussoy and Horstman & $7$   &  $15$ &   &   &  & $P_w$,$Q_w$ \\	
%           &  & &   &  &   &   &  & $U,P,T,\overline{\rho}$ profiles at various $x$-locations \\
		& Holden~(Ref.~\onlinecite{holden2010experimental})$^\filledstar$ & $11.4$  & $20$ & $0.2$ & $1.71$ & $40.4$ & $P_w$,$C_f$,$Q_w$  \\ %tin 62.8 tw 300
        & Holden~(Ref.~\onlinecite{holden1972shock})  & $6.25-14.9$  & $12.5-19.8$ & $0.2-0.9$  & $0.75-3.01$ & $15.4-101.7$  &  $P_w$,$C_f$,$Q_w$ \\ % Tw/Ttinf $0.13-0.39$
\end{tabular}
\end{ruledtabular}
\end{table*}

%%%%%%%%%%%%%%%%%%%%%%%%%%%%%%%%%%%%%%%%%%%%%%%%%%%%%%%%%%%%%%%%%%%%%%%%%%%%%%%%%%%%%%%%%%%%%%%%

%\textbf{Axisymmetric SBLI }

%%%%%%%%%%%%%%%%%%%%%%%%%%%%%%%%%%%%%%%%%%%%%%%%%%%%%%%%%%%%%%%%%%%%%%%%%%%%%%%%%%%%%%%%%%%%%

\begin{table*}
\caption{\label{tab:sbli-axisymmetric} Experiments -- Axisymmetric SBLI : Cone/Cylinder/Flare - CCF; Hollow Cylinder/Flare - HCF; Cone/Flare - CF; Hollow Cylinder/Cowl - HCC; hemisphere cylinder/flare - HemCF; Cone/ogive-cylinder - COC; Ogive Cylinder/Flare - OCFl; Ogive Cylinder/Fin - OCFi; $\theta$ = flow deflection (or shock generator) angle; Half-angle cone - HAC; Flare Angle - FA; $^\filledstar$ ASCII datafile available on the NASA Turbulence Modeling Resource website~\cite{Hypersonic_SBLI_Expt}. }
\begin{ruledtabular}
\begin{tabular}{llccccccc}
 %&\multicolumn{2}{c}{$D_{4h}^1$}&\multicolumn{2}{c}{$D_{4h}^5$}\\
		Configuration & Experiments& $M_e$  & $\theta$ (degree)  & $T_w/T_{aw}$
 &$H_{e}$ &$Re_\infty$/m  & Quantities
		\\ %\vspace{0.1in}
		& & &   &    & (MJ/kg) & $\times 10^6$ & reported \\\hline 	%\vspace{0.05in}
 %%%%%%%%%%%%%%%%%%%%%%%%%%%%%%%%%%%%%%%%%%%%%%%%%%%%%%%%%%%%%%%%%%%%%%%%%%%%%%%%%%%%%%	
%%%%%%%%%%%%%%%%%%%%%%%%%%%%%%%%%%%%%%%%%%%%%%%%%%%%%%%%%%%%%%%%%%%%%%%%%%%%%%%%%%%%%%	
	CCF	& Holden~(Ref.~\onlinecite{WaMaHo08})$^\filledstar$ :  & $6.5-7.2$  &  $7$ HAC; $33$ FA   & $0.11-0.16$ & $2.2-3.4$ &  $1.5-20$ & $P_w$,$Q_w$ \\
    	& (HiFire I) &  &   &  &  &   &  \\ %(sharp and blunted nose tip)
       
 	    & Coleman~(Ref.~\onlinecite{coleman1973study,coleman1973hypersonic,coleman1974incipient}) & $9.22$ & $10$ HAC ;   &  $0.29$ & $1.07$ & $47$ & $P_w$,$Q_w$  \\
        
      &  &  & $15-40$ FA  &   &  &  &   \\ \hline
 %%%%%%%%%%%%%%%%%%%%%%%%%%%%%%%%%%%%%%%%%%%%%%%%%%%%%%%%%%%%%%%%%%%%%%%%%%%%%%%%%%%%%% 
%		& Williams  & $9$  &  $36$ &  -- &  -- &  -- & $P_w$,$Q_w$  \\
				& Coleman~(Ref.~\onlinecite{coleman1973study,coleman1973hypersonic,coleman1974incipient}) & $9.22$ & $15$-$40$  &  $0.29$ & $1.07$ & $47$ & $P_w$,$Q_w$  \\	
	HCF	& Murray~(Ref.~\onlinecite{murray2013experimental})$^\filledstar$ & $8.9$    & $36$  & $0.27$   &  $1.2$ & $48$ & $P_w$,$Q_w$ \\	
		& Holden~(Ref.~\onlinecite{holden2014measurements,holden2018measurements}) & $5-8$  & $36$  &  $0.13-0.25$ &  $1.27-2.54$ & $4-22.3$ & $P_w$,$Q_w$ \\
        \hline
 %%%%%%%%%%%%%%%%%%%%%%%%%%%%%%%%%%%%%%%%%%%%%%%%%%%%%%%%%%%%%%%%%%%%%%%%%%%%%%%%%%%%%% 
		& Holden~(Ref.~\onlinecite{holden1991studies})$^\filledstar$  & $11-15.4$  &  $6-7$ HAC ; & $0.15-0.21$ &  $1.7-2.6$ &  $5-15$ & $P_w$,$Q_w$  \\	
  		&   &   &  $30-42$ FA & &  &  &   \\	
CF & Holden~(Ref.~\onlinecite{holden2014measurements,holden2018measurements})  & $4.96-8.21$  &  $7$ HAC; $40$ FA  &  $0.12-0.67$ &  $0.47-2.38$ & $4.4-49$ & $P_w$,$Q_w$ \\
  
  &  Running~(Ref.~\onlinecite{running2019hypersonic}) &  $6.14$ & $7$ HAC; $34-43$ FA & $0.15$ &  $0.9$ & $7.6-20$ & $Q_w$ \\ %T0 490, Tw 297
        \hline
 %%%%%%%%%%%%%%%%%%%%%%%%%%%%%%%%%%%%%%%%%%%%%%%%%%%%%%%%%%%%%%%%%%%%%%%%%%%%%%%%%%%%%% 
	HCC	& Murray~(Ref.~\onlinecite{murray2013experimental})$^\filledstar$ & $8.9$   & $4.7$ \& $10$  & $0.27$   &  $1.2$ & $48$ & $P_w$,$Q_w$ \\	
          \hline
 %%%%%%%%%%%%%%%%%%%%%%%%%%%%%%%%%%%%%%%%%%%%%%%%%%%%%%%%%%%%%%%%%%%%%%%%%%%%%%%%%%%%%% 
	HemCF	& Coleman~(Ref.~\onlinecite{coleman1973study,coleman1973hypersonic,coleman1974incipient})  & $9.22$  & $15$-$40$  &  $0.29$ & $1.07$ & $47$ & $P_w$,$Q_w$  \\		
          \hline    
  %%%%%%%%%%%%%%%%%%%%%%%%%%%%%%%%%%%%%%%%%%%%%%%%%%%%%%%%%%%%%%%%%%%%%%%%%%%%%%%%%%%%%%
	 COC & Kussoy and Horstmann~(Ref.~\onlinecite{kussoy1975experimental}) & $7.2$  & $7.5$ \& $15$  &  $0.45$ & $0.7$ & $7.1$ \& $7.7$ & $P_w$,$C_f$,$Q_w$  \\		
          \hline                   
 %%%%%%%%%%%%%%%%%%%%%%%%%%%%%%%%%%%%%%%%%%%%%%%%%%%%%%%%%%%%%%%%%%%%%%%%%%%%%%%%%%%%%% 
	OCFl & Kussoy~(Ref.~\onlinecite{kussoy1989documentation})$^\filledstar$ & $7$  &  $20-35$  & $0.385$ &  $0.89$ &  $5.8$ & $P_w$,$Q_w$  \\ \hline %- sharp nose\hline  
  %%%%%%%%%%%%%%%%%%%%%%%%%%%%%%%%%%%%%%%%%%%%%%%%%%%%%%%%%%%%%%%%%%%%%%%%%%%%%%%%%%%%%%
	OCFi	& Kussoy~(Ref.~\onlinecite{kussoy1989documentation})$^\filledstar$ & $7$  &  $10-20$  & $0.385$ &  $0.89$ &  $5.8$ & $P_w$,$Q_w$ %- half-angles
\end{tabular}
\end{ruledtabular}
\end{table*}

%\textbf{Three-dimensional SBLI}

%%%%%%%%%%%%%%%%%%%%%%%%%%%%%%%%%%%%%%%%%%%%%%%%%%%%%%%%%%%%%%%%%%%%%%%%%%%%%%%%%%%%%%%%%%%%%

\begin{table*}
	\caption{\label{tab:sbli-3d} Three-dimensional SBLI :  Single-fin SBLI - SFSBLI; crossing SBLI (double fin geometry) - CSBLI; $\theta$ = flow deflection angle; $^\filledstar$ ASCII datafile available on the NASA Turbulence Modeling Resource website~\cite{Hypersonic_SBLI_Expt}. } %Wedge on flat plate - WoP;
\begin{ruledtabular}
\begin{tabular}{llccccccc}
 %&\multicolumn{2}{c}{$D_{4h}^1$}&\multicolumn{2}{c}{$D_{4h}^5$}\\
		Configuration & Experiments& $M_e$   & $\theta$ (degree)  & $T_w/T_{aw}$
 &$H_{e}$ &$Re_e$/m  & Quantities
		\\ %\vspace{0.1in}
		& & &     &  & (MJ/kg) & $\times 10^6$ & reported \\\hline 	%\vspace{0.05in}
 
 %%%%%%%%%%%%%%%%%%%%%%%%%%%%%%%%%%%%%%%%%%%%%%%%%%%%%%%%%%%%%%%%%%%%%%%%%%%%%%%%%%%%%%
%      WoP   & Borovoy~(Ref.~\onlinecite{borovoy2009laminar,borovoy20123d}) & $5$ \& $6$  & $10-20$ & $0.62$ \& $0.59$  & $0.53$ \& $0.57$  & $10.9$ \& $7.7$  & $P_w$,$Q_w$ \\  \hline
%      sharp to 4 mm FP LE radius  &  &  &  &  &   & \\
 %%%%%%%%%%%%%%%%%%%%%%%%%%%%%%%%%%%%%%%%%%%%%%%%%%%%%%%%%%%%%%%%%%%%%%%%%%%%%%%%%%%%%%
 %%%%%%%%%%%%%%%%%%%%%%%%%%%%%%%%%%%%%%%%%%%%%%%%%%%%%%%%%%%%%%%%%%%%%%%%%%%%%%%%%%%%%%	
		& Kussoy~(Ref.~\onlinecite{kussoy1991documentation})$^\filledstar$ & $8.18$  & $5-15$ &  $0.29$ & $1.1$ & $4.87$ &  $P_w$,$C_f$,$Q_w$  \\
%            &  & &  &   &   &  &  Surface Streamline Angles, Flow Field Yaw Angles, \\
%            &  & &  &   &   &  &  Flow Field Pitot Pressure \\
            & Rodi~(Ref.~\onlinecite{rodi1991experimental,rodi1992experimental,rodi1995behavior})$^\filledstar$ & $4.9$   & $6-16$ & $0.8$  & $0.42$  & $38.1$ & $P_w$,$Q_w$ \\ %,surface flow visualization
    %        & \cite{rodi1992experimental}, &   &  &   &   &  &  \\
    %        & \cite{rodi1995behavior} &   &  &   &   &  &  \\
    SFSBLI	& Borovoy~(Ref.~\onlinecite{borovoy2009laminar,borovoy20123d,borovoy2016entropy}) & $5$ \& $6$  & $10-20$ & $0.62$ \& $0.59$  & $0.53$ \& $0.57$  & $10.9$ \& $7.7$  & $P_w$,$Q_w$ \\
%	SFS	& Borovoy~(Ref.~\onlinecite{borovoy2009laminar,borovoy20123d,borovoy2016entropy})  & $5-8$ & $10-20$ & Ptin Ttinf 530,560,690 & $0.53-0.69$  & $22-84.4$  & $P_w$,$Q_w$ \\ %,surface flow visualization
		& Law~(Ref.~\onlinecite{law1975three})$^\filledstar$ & $5.85$  & $6-16$ &  $0.51$ & $0.6$ & $32.8$ \& $98.4$ & $P_w$,$Q_w$ \\ %Streamwise and transverse 
 %           &  &    &  &   &   &  & profiles at various $x$-locations \\
           & Holden~(Ref.~\onlinecite{holden1984experimental}) & $11.3$ & $5-12.5$ & $0.2$  &  $1.66$ & $31.6$   & $P_w$,$Q_w$ \\   
           & Schulein~(Ref.~\onlinecite{schulein2001documentation,schulein2006skin}) & $5$ & $2-27$ & $0.8$  &  $0.41$ & $37$   & $P_w$,$C_f$,$Q_w$ \\  
         \hline
 %%%%%%%%%%%%%%%%%%%%%%%%%%%%%%%%%%%%%%%%%%%%%%%%%%%%%%%%%%%%%%%%%%%%%%%%%%%%%%%%%%%%%% 
		& Kussoy~(Ref.~\onlinecite{kussoy1993hypersonic})$^\filledstar$ & $8.28$  & $10,15$ &  $0.269$ & $1.2$ & $5.3$ &  $P_w$,$C_f$,$Q_w$  \\ %Streamwise and transverse
%            &  & &    &   &   &  & Surface Streamline Angles, Flow Field Yaw Angles,  \\
%            &  & &    &   &   &  &  Flow Field Pitot Pressure \\
%            &  &   &  &   &   &  & profiles at various $x$-locations \\
      CSBLI & Schulein~(Ref.~\onlinecite{schulein2001documentation,schulein2006skin}) & $5$  & $8-23$  & $0.8$  &  $0.41$ & $37$   & $P_w$,$Q_w$\\ % , surface streamlines - symmetric fin angles
      & Borovoy~(Ref.~\onlinecite{borovoy2009laminar,borovoy20123d,borovoy2016entropy}) & $5$ \& $6$  & $15$ & $0.62$ \& $0.59$  & $0.53$ \& $0.57$  & $10.9$ \& $7.7$  & $P_w$,$Q_w$ \\
 %        & Borovoy~(Ref.~\onlinecite{borovoy2009laminar,borovoy20123d,borovoy2016entropy}  & $5-8$ & $15$ & Ttinf 530,560,690 & $0.53-0.69$  & ReL 27,19.2,7  & $P_w$,$Q_w$ \\ %,surface flow visualization
 %      &  &  &  &  &   &   & \\ 
 %%%%%%%%%%%%%%%%%%%%%%%%%%%%%%%%%%%%%%%%%%%%%%%%%%%%%%%%%%%%%%%%%%%%%%%%%%%%%%%%%%%%%% 
\end{tabular}
\end{ruledtabular}
\end{table*}

%%%%%%%%%%%%%%%%%%%%%%%%%%%%%%%%%%%%%%%%%%%%%%%%%%%%%%%%%%%%%%%%%%%%%%%%%%%%%%%%%%%%%%%%%%%%%%%%

%%%%%%%%%%%%%%%%%%%%%%%%%%%%%%%%%%%%%%%%%%%%%%%%%%%%%%%%%%%%%%%%%%%%%%%%%%%%%%%%%%%%%%%%%%%%%
%\textbf{DNS SBLI }

\begin{table*}
	\caption{\label{tab:dns-sbli} DNS - SBLI : Impinging shock - IS; Compression corner - CC; BCC - Blunt Circular Cone; Cone/Flare - CF; Backward-facing curved wall - BFCW; Forward-facing curved wall - FFCW; $\theta$ - Flow deflection angle; Ramp angle - RA; Half-angle cone - HAC; Flare angle - FA; Sweep angle - SWA; * $Re_\theta$ based on $\mu_w$; na - not available; $^\filledstar$ ASCII datafile available on the NASA Turbulence Modeling Resource website~\cite{M5DNS_SBLI}.}
\begin{ruledtabular}
\begin{tabular}{llcccc}
 %&\multicolumn{2}{c}{$D_{4h}^1$}&\multicolumn{2}{c}{$D_{4h}^5$}\\
		Configuration & DNS & $M_e$  &   $T_w/T_{aw}$
 &$\theta$(degree)  &$Re_\theta \times 10^3 $  
		\\ 
	\hline 	%\vspace{0.05in}
 %%%%%%%%%%%%%%%%%%%%%%%%%%%%%%%%%%%%%%%%%%%%%%%%%%%%%%%%%%%%%%%%%%%%%%%%%%%%%%%%%%%%%%	
 %&  &  &  &   &  &  &  &  \\
 IS : (based on & Volpiani \textit{et al.}~(Ref.~\onlinecite{volpiani2020effects}) &  $5$  & $0.8$ \& $1.9$ & $6-14$ &  $3.8$ \& $5.4$    \\ 
  Schulein~\cite{schulein2006skin} experiments) & &   &  &  &   \\ 
 CC : (based on  & Priebe \& Martin~(Ref.~\onlinecite{priebe2021turbulence}) &  $7.2$   & $0.53$ & $8$ & $3.3$   \\ 
 Bookey~\cite{bookey2005new} experiments) &      &  &  &  &    \\ 
  IS & Volpiani~(Ref.~\onlinecite{volpiani2021numerical}) (3D simulation) &  $5.6$  & $0.2$  & $6$ & na $(Re_\tau = 190)$    \\ %%tinf 2275 taw 15116.92 tw 3042
  Swept CC  & Zhang \textit{et al.}~(Ref.~\onlinecite{zhang2022direct}) (3D simulation) &  $6$  & $0.5$ & $34$ RA; $45$ SWA & na $(Re_e/m = 10^6)$    \\ %tw 294 taw 585
    CC  & Dang \textit{ et al.}~(Ref.~\onlinecite{dang2022direct}) &  $6$  & $0.5$ & $34$  & $4.9$    \\ 
    BCC  & Dang \textit{ et al.}~(Ref.~\onlinecite{dang2022direct}) &  $6$  & $0.61$ & $7$ HAC  & $1.8$    \\ 
  CC  & Guo \textit{et al.}~(Ref.~\onlinecite{guo2023amplification}) &  $6$  & $0.75$ & $30$  & $6.62$    \\   %Tw 616 Taw 822.8
   CF : (based on  & Fulin \textit{et al.}~(Ref.~\onlinecite{fulin2023hypersonic}) &  $6$   & $0.62$ & $7$ HAC; $34$ FA & $2.5$    \\ 
 Running~\cite{running2019hypersonic} experiments) &      &  &  &  &    \\ 
 CC  & Di Renzo \textit{et al.}~(Ref.~\onlinecite{di2024stagnation}) &  $5-6$   & $0.4-0.714$ & $15$ & $(0.3-0.65)^*$   \\ 
  BFCW \& FFCW  & Nicholson \textit{et al.}~(Ref.~\onlinecite{nicholson2024direct})$^\filledstar$ &  $4.9$   & $0.91$ & $-$ & $16.1-22.4$   \\ 
\end{tabular}
\end{ruledtabular}
\end{table*}

%%%%%%%%%%%%%%%%%%%%%%%%%%%%%%%%%%%%%%%%%%%%%%%%%%%%%%%%%%%%%%%%%%%%%%%%%%%%%%%%%%%%%%%%%%%%%%%%

%%%%%%%%%%%%%%%%%%%%%%%%%%%%%%%%%%%%%%%%%%%%%%%%%%%%%%%%%%%%%%%%%%%%%%%%%%%%%%%%%%%%%%%%%%%%%
%\textbf{Rough surface experiments }

\begin{table*}
	\caption{\label{tab:exp-rough} Experimental studies on surface roughness effects : FP - Flat plate; CC - Compression Corner; na - not available.}
\begin{ruledtabular}
\begin{tabular}{llcccccc}
 %&\multicolumn{2}{c}{$D_{4h}^1$}&\multicolumn{2}{c}{$D_{4h}^5$}\\
		Configuration & Experiment & $M_e$    &   $T_w/T_{aw}$
 &$H_{e}$ &$Re_e$/m  & Quantities
		\\ \vspace{0.1in}
		& &     &   & (MJ/kg) & $\times 10^6$ & reported \\\hline 	%\vspace{0.05in}
 %%%%%%%%%%%%%%%%%%%%%%%%%%%%%%%%%%%%%%%%%%%%%%%%%%%%%%%%%%%%%%%%%%%%%%%%%%%%%%%%%%%%%%	
 %&  &  &  &   &  &  &  &  \\
 Flat plate (milled spanwise V-grooves) & Young~(Ref.~\onlinecite{young1965experimental}) &  $4.9$    & $0.5-1$ & $0.34-0.625$   & $16.4-45.8$  &  $P_w$,$C_f$,$Q_w$    \\ %BL surveys of $P_t$, $T_t$, and
Slender cones, & Holden~(Ref.~\onlinecite{holden1982experimental,holden1989studies})  & $6-13$   & *($T_w/T_{t_e} : 0.2-0.4$) & $0.69-2$ & $15-125.6$ & $P_w$,$C_f$,$Q_w$   \\
biconic \& hemispherical nosetips 	    &   &  &  &  &  \\
(sand grain \& patterned roughnesses)	    &  &   &  &  &  &  \\
Rough cylinder with $15^\circ,20^\circ$ flare  & Babinsky \& Edwards~(Ref.~\onlinecite{babinsky1997large})  & $5$   & na   & $0.4$ & $13$ & $P_w$,$Q_w$  \\
 (sawtooth and cavity shape roughnesses)	    &  &   &  &  &  &  \\
 $25^\circ-38^\circ$ CC  & Prince~(Ref.~\onlinecite{prince2005experiments})  & $8.2$    &  $0.24$ & $1.3$ & $9.3$ & $P_w$,$Q_w$  \\ %ptin 10.9 Ttin 1290 Tw 283 Taw 1169.93
 (sand grains on FP \& ramp)	   &  &   &  &  &  &  \\ % - $k_s \approx 0.3$mm 
 Sphere-cone and hemisphere  & Hollis~(Ref.~\onlinecite{hollis2014distributed})  & $6$   & $0.7$   & $0.5$ & $9.95-27.4$ & $Q_w$  \\
 (sand-grain and patterned roughnesses)	    &  &   &  &  &  &  \\
  Sphere cone & Wilder~(Ref.~\onlinecite{wilder2019rough})  & $9-9.7$   & $0.09-0.33$   & $0.78-0.86$ & $31$ & $Q_w$  \\
 (sand-grain and patterned roughnesses)	    &  &   &  &  &  &  \\
   Flat plate & Forsyth~(Ref.~\onlinecite{forsyth2024experimental})  & $4.9$   & $0.74-0.85$   & $0.38-0.44$ & $22.3-37.3$ & $Q_w$  \\
 (patterned hemispheres, 	    &  &   &  &  &  &  \\
 pseudomesh roughness,     &  &   &  &  &  &  \\
 and their superposition)	    &  &   &  &  &  &  \\
\end{tabular}
\end{ruledtabular}
\end{table*}

%%%%%%%%%%%%%%%%%%%%%%%%%%%%%%%%%%%%%%%%%%%%%%%%%%%%%%%%%%%%%%%%%%%%%%%%%%%%%%%%%%%%%%%%%%%%%%%%

%%%%%%%%%%%%%%%%%%%%%%%%%%%%%%%%%%%%%%%%%%%%%%%%%%%%%%%%%%%%%%%%%%%%%%%%%%%%%%%%%
Experiments and DNS play an important role in the development and validation of turbulence models by providing data on the flow physics required to substantiate modeling assumptions. Comprehensive model validation requires assessment across a broad parameter space, encompassing diverse Mach numbers, Reynolds numbers, wall temperature and surface conditions, and geometric complexities. This extensive validation helps ensure the model's robustness and applicability to a wide range of operating conditions for a given design requirement. However, there is a serious lack of wind tunnel, flight, and high-fidelity simulation data in the hypersonic regime compared to the lower speeds. 

%Moreover, the existing experimental validation database at $ Ma \gtrsim 5$ is exclusively confined to low enthalpy conditions, whereas a few DNS works consider high enthalpy conditions.

%%%%%%%%%%%%%%%%%%%%%%%%%%%%%%%%%%%%%%%%%%%%%%%%%%%%%%%%%%%%%%%%%%%%%%%%%%%%%%%%%%%%%%%

\subsection{Available datasets}\label{sec:datasets}

Tables~\ref{tab:experiments-fp}-\ref{tab:dns-sbli} present the available experiments and DNS for TBL flows in the hypersonic regime for smooth as well as rough surfaces, and it excludes the experimental works that have not reported wall heat transfer. The experiments are generally carried out using air, $N_2$, or $He$ as the working fluid. In the data, the subscript `e' indicates quantities at the boundary layer edge or the freestream, and unit Reynolds number $Re_e = \overline{\rho}_e \widetilde{u}_e/\mu_e$ with $\mu_e$ being the dynamic viscosity calculated using Sutherlands or Keyes law. The adiabatic wall temperature $T_{aw}$ (or equivalently the recovery temperature $T_r$) is given by
\begin{equation}
\begin{split}
    T_{aw} &= \widetilde{T}_e \left( 1 + \dfrac{(\gamma-1)}{2} Pr_T M^{2}_{e} \right) \\ &=  \widetilde{T}_{t_e} \left( 1 + \dfrac{(\gamma-1)}{2} M^{2}_{e} \right)^{-1} \left( 1 + \dfrac{(\gamma-1)}{2} Pr_t M^{2}_{e} \right),         
\end{split}
\end{equation}
where $\widetilde{T}_{t_e}$ is the total temperature at the edge of the boundary layer and unless specified otherwise $Pr_t$ is taken to be $0.9$. Total enthalpy per unit mass is given by $H_e = C_p \widetilde{T}_e + 0.5 \widetilde{u_{e}^{2}} = C_P \widetilde{T}_{t_e} $. Note that, in some cases, wall heat transfer ($Q_w$) is given in terms of Stanton number $S_t$
\begin{equation}
    S_t = \dfrac{Q_w}{\overline{\rho} \widetilde{u}_e C_p (T_w - T_{aw})}.
\end{equation}
All the available experimental data excluding the experiments of Goyne \textit{et al.}~\cite{goyne2003skin} correspond to low enthalpy conditions, whereas a few DNS studies cover high enthalpy conditions. It is important to note that during a hypersonic flight, total (stagnation) enthalpies as large as $100$ MJ/kg and higher can be experienced~\cite{collen2021development}.

The validation database for ZPG hypersonic smooth wall TBLs consists of flow over flat plates, wind tunnel walls, nozzle walls, and flow along model geometries with axisymmetric forebodies including hollow cylinders and sharp circular cones at different wall temperature conditions. The available experimental database is given in Table~\ref{tab:experiments-fp}. Roy and Blottner~\cite{roy2006review} present a large number of available equilibrium hypersonic TBL experiments, however, many of these experiments were performed to establish the accuracy of the compressible engineering correlations (CECs) theories; a number of them are not well documented with many missing wall heat transfer measurements and they are not suitable for turbulence model validation studies.

On the other hand, Table~\ref{tab:dns-fp} gives the available DNS studies for flat plate ZPG hypersonic TBLs. A majority of the DNS studies are based on perfect gas simulations while the works of Refs.~\onlinecite{duan2011direct4,passiatore2022thermochemical,passiatore2021finite} consider real gas models with stagnation enthalpies $\approx 20$ MJ/kg. Refs.~\onlinecite{maeder2001direct,xin2006direct,martin2007direct,duan2010direct,duan2011direct,chu2013effect} consider adiabatic wall conditions whereas the works in Refs.~\onlinecite{liang2012dns,liang2013dns,chu2013effect,duan2016pressure,zhang2017effect,zhang2018direct,huang2020simulation,nicholson2021simulation} study the cold wall ZPG TBLs. Ref.~\onlinecite{chu2013effect} is the only work to consider heated wall condition at Mach $8$ with $T_w/T_{aw} = 1.5$. A few DNS works study TCI~\cite{duan2011assessment} while some works~\cite{duan2011direct,duan2012study,duan2011direct4} study the effects of radiation on hypersonic TBLs.

Compressible Engineering Correlations (CECs) are generally used for flat plate ZPG TBL comparisons of skin friction and wall heat transfer. The commonly used CECs include van Driest II~\cite{van1956problem}, White–Christoph~\cite{white1972simple}, and Spalding-Chi~\cite{spalding1964drag} relations for skin friction estimation in terms of an equivalent incompressible TBL (see Appendix \ref{appendixA}). The correlations of van Driest II and White–Christoph generally give similar results for adiabatic as well as strongly cooled walls, but results using Spalding–Chi can differ significantly~\cite{rumsey2010compressibility}. Generally, these CECs give good results for adiabatic and moderately cold wall temperature ratios ($T_w/T_{aw}$). However, the accuracy of the results differs significantly depending on the Mach number, wall temperature ratio, and Reynolds numbers~\cite{hopkins1971evaluation,holden1972shock,goyne2003skin,huang2022direct}. Reynolds analogy factor $R_{af}$ is classically used to predict wall heat flux from $C_f$ obtained using the CECs,  
 \begin{equation}
     S_t = \dfrac{1}{2} C_f R_{af}.
 \end{equation}
The commonly used approximation for $R_{af}$ is $R_{af} = Pr^{-2/3}$ (with $Pr = 0.72$), $R_{af} = Pr_{t}^{-1}$ as per van Driest, and $R_{af} =1$. The experimental data display a rather wide range of values for $R_{af}$ ranging from $0.75$ to $1.3 $.

A limited number of experimental studies are available for hypersonic SBLIs. Generally, complex configurations are broken down into simpler geometries to isolate different kinds of SBLIs. Experiments and DNS/LES studies are then performed on these canonical geometric configurations. These simplified geometries include 2D planar configurations consisting of compression corner and oblique shockwave impingement, axisymmetric configurations consisting of cone-cylinder-flare, cone-flare, etc. as given in Tables~\ref{tab:SBLI-2D} and \ref{tab:sbli-axisymmetric}. Table~\ref{tab:sbli-3d} gives 3D SBLIs used for the model validation studies. The test geometry for the 3D configurations consists of one or two sharp or blunt fins that act as shock generators mounted on a flat plate. The two basic test models are the single fin and the symmetric double fin geometries. Surface flow visualization, surface streamline angles, flow field yaw angles, and mean profiles at various streamwise locations are available for these experiments in addition to the surface pressure, skin friction, and wall heat transfer measurements. 

DNS studies on SBLIs are limited (see Table \ref{tab:dns-sbli}) with most studies performing perfect gas simulations while only two works use a real gas model. LES studies are also limited with recent examples including the wall-resolved LES of BoLT-2 vehicle at Mach $6.11$ flow conditions ($T_w/T_{aw} = 1 \& 0.215$, $Re/m = 12.9\times10^6$) in the descent phase of the trajectory~\cite{bhagwandin2023wall}; compression corner SBLIs at Mach $7.16$ ($T_w/T_{aw} = 0.52$, $Re_\theta = 3300$) and Mach $9.05$ TBLs with $33^\circ$  and $34^\circ$ ramp angles ($T_w/T_{aw} = 0.33$, $Re_\theta = 8000$)~\cite{helm2022large}, respectively; Mach $10$ axisymmetric $34^\circ$ cylinder-flare configuration ($T_w/T_{aw} \approx 0.3$, $Re_\theta \approx 8000$)~\cite{bhagwandin2021shock}; a hypersonic cold wall flat plate TBL at Mach $6$ and $8$ and several wall temperature ratios~\cite{kianvashrad2021large}; $31^\circ-34^\circ$ compression ramps and $33^\circ$ axisymmetric cylinder flare at Mach $10$ ($T_w/T_{aw} \approx 0.3$, $Re_\theta \approx 8000$) \cite{bhagwandin2019shock}; and Mach $5$ SBLI generated by a $23^\circ$ single fin~\cite{fang2015large,fang2017investigation}. 

%Some of these DNS and LES studies~\cite{volpiani2021numerical,priebe2021turbulence,fulin2023hypersonic,bhagwandin2021shock} also attempt to compare the high-fidelity simulation results with the available experimental data at similar geometric and flow conditions. However, the comparisons show discrepancies and these studies remain unsubstantiated.   

%Here, we do not consider transitional SBLIs where the oncoming BL before the interaction is laminar.

Hypersonic TBL experiments on rough wall conditions with surface heat transfer measurements are also limited as presented in Table \ref{tab:exp-rough}. The test models use sand grain roughness and patterned roughness surfaces for these studies. Several experiments~\cite{berg1977surface,sahoo2009effects,peltier2016crosshatch,neeb2015experimental} study the boundary layer characteristics in the presence of different roughness elements. These experiments verify the validity of the law of wall and wake for a rough wall which is given by
\begin{equation}\label{eqn:van-Driest-u-rough}
\dfrac{u_{VD}}{u_\tau} =  \dfrac{1}{\kappa} log \left( \dfrac{y u_{\tau}}{\nu_w} \right) + C - H(k_{s}^{+}) + \dfrac{2 \Pi}{\kappa} sin^{2} \left( \dfrac{\pi}{2} \dfrac{y}{\delta} \right) . 
\end{equation}
The function $H(k_{s}^{+})$ gives the downward shift of the log region due to the roughness effect.

%%%%%%%%%%%%%%%%%%%%%%%%%%%%%%%%%%%%%%%%%%%%%%%%%%%%%%%%%%%%%%%%%%%%%%%%%%%%%%%%%%%%%%%

\subsection{Model performance}\label{sec:performance}

A total of $15$ turbulence models are assessed against the experimental database introduced in Section~\ref{sec:datasets} in predicting hypersonic SBLIs (see Table~\ref{tab:model-notations}). Popular EVMs including the zero-equation Baldwin-Lomax, one-equation Spalart-Allmaras, two-equation $k$-$\epsilon$ and $k$-$\omega$ models including Menter SST are considered along with two cubic $k$-$\epsilon$ NLEVMs, WJ-EARSM, and Wilcox Stress-$\omega$ and SSG/LRR-$\omega$ RSTMs that have been tested against the available hypersonic validation cases. Evaluation of three versions of the $k$-$\epsilon$ model and two versions of the Wilcox $k$-$\omega$ model which have considerable validation history at lower speeds is shown. Typical errors in model predictions for the separation length, heat transfer near the reattachment point, and peak wall-shear stress values are considered to evaluate the model performance. Generally, the errors in model predictions increases with an increase in shock strength or free stream Mach number or flow deflection angle, and typical errors in model predictions for fully separated SBLI flows are shown in Table~\ref{tab:model_performance}. All the models show significant errors in model predictions except the Goldberg cubic $k$-$\epsilon$ model~\cite{goldberg2000hypersonic}, which has been tested against a very limited validation database and requires further validation.

Similar to the $k$-$\epsilon$ and Wilcox $k$-$\omega$ models, the widely popular SA and Menter SST models also have several variants, e.g. SA model with Catris-Aupoix correction~\cite{catris2000density} and SST model with strain-rate magnitude instead of vorticity magnitude in the $\mu_T$ definition~\cite{menter2003ten}, and the different variants can show significantly different performance in hypersonic applications. Additionally, the performance of the Menter SST model in fully separated SBLI flows is sensitive to the values of the structure parameter $a_1$ in the $\mu_T$ definition, type of production term, and production limiter value~\cite{georgiadis2013recalibration,raje2021anisotropic}. 

It is important to note that the models can show significantly smaller separation than the experiments but yield good peak heat transfer rate predictions near the reattachment point~\cite{goldberg2000hypersonic,gnoffo2011uncertainty} or produce good separation length predictions while significantly overpredicting peak wall heat transfer~\cite{nance1999turbulence}. For the models that significantly overpredict the shock-induced separation bubble sizes, the separation point can move far upstream near the inlet of the computational domain resulting in failure to get convergence of the mean flow~\cite{gnoffo2011uncertainty}. Note that the simulation results may vary as a result of different implementations of the turbulence model within a code or between codes~\cite{gnoffo2011uncertainty,gnoffo2013uncertainty}. Also, note that the choice of initial guess can be a key determinant for a nonlinear solver to fall into the different basins of attraction, resulting in multiple machine-zero converged solutions on the same grid for hypersonic applications involving SBLIs, similar to the observations of Kamenetskiy \emph{et al.}~\cite{kamenetskiy2014numerical}. 

Table~\ref{tab:comp_corr} presents the popular compressibility corrections that are generally used with the two-equation $k$-$\epsilon$ and $k$-$\omega$ models including Menter SST. A combination of these corrections may be required to correctly predict the separation bubble size and wall heat transfer near the reattachment location. For example, Refs.~\onlinecite{huang1993turbulence,coakley1994turbulence} applied the length scale and rapid compression corrections to the Launder-Sharma $k$-$\epsilon$ and Wilcox 1988 $k$-$\omega$ models to get improvements over the baseline models without compressibility corrections in separation length predictions from as high as $75$\% underprediction and peak heat transfer from as high as $170$\% overprediction for two-dimensional and axisymmetric SBLI cases~\cite{coleman1972heat,kussoy1975experimental,kussoy1989documentation}. However, these modifications in many cases tend to overcorrect in the boundary layer and yield heat transfer and skin friction values that are too low~\cite{coakley1994turbulence,zhu2020analysis}. Also, discrepancies in the heating rate and skin friction distribution remain, particularly within the recirculation bubble and post-reattachment location, respectively. It is important to note that these model corrections may very well give poor predictions for flow situations that have not yet been validated. The Catris-Aupoix correction applied to the SA and $k$-$\omega$ models including Menter SST does not give significant improvements in the predictions of hypersonic TBLs and fully separated SBLI flows~\cite{catris2000density,guohua2012assessment,BaPaJo24}.

%%%%%%%%%%%%%%%%%%%%%%%%%%%%%%%%%%%%%%%%%%%%%%%%%%%%%%%%%%%%%%%%%%%%%%%%%%%%%%%%%%%%%%%%%%%%%
%\textbf{Turb Models }

\begin{table}
\caption{\label{tab:model-notations}Turbulence models assessed using the hypersonic validation database with notation.}
\begin{ruledtabular}
\begin{tabular}{llc}
 %&\multicolumn{2}{c}{$D_{4h}^1$}&\multicolumn{2}{c}{$D_{4h}^5$}\\
 Model type & Turbulence Model & Notation \\ \hline 
%\vspace{0.1in}
 %%%%%%%%%%%%%%%%%%%%%%%%%%%%%%%%%%%%%%%%%%%%%%%%%%%%%%%%%%%%%%%%%%%%%%%%%%%%%%%%%%%%%%	
 Zero-equation         &  Baldwin-Lomax~(Ref.~\onlinecite{baldwin1978thin}) &  BL  \\
 EVM & & \\
             \hline 
 %%%%%%%%%%%%%%%%%%%%%%%%%%%%%%%%%%%%%%%%%%%%%%%%%%%%%%%%%%%%%%%%%%%%%%%%%%%%%%%%%%%%%% 
One-equation		& Spalart-Allmaras~(Ref.~\onlinecite{spalart1992one}) & SA   \\ 
EVM & & \\
%		& Goldberg~(Ref.~\onlinecite{goldberg2000hypersonic})  & UG  \\ 
%   & Menter~(Ref.~\onlinecite{}) & MTR     \\  
 \hline
  %%%%%%%%%%%%%%%%%%%%%%%%%%%%%%%%%%%%%%%%%%%%%%%%%%%%%%%%%%%%%%%%%%%%%%%%%%%%%%%%%%%%%
   & $k$-$\epsilon$ Jones-Launder ~(Ref.~\onlinecite{jones1972prediction}) & $k \epsilon$JL  \\
  & $k$-$\epsilon$ Launder-Sharma ~(Ref.~\onlinecite{launder1974application}) & $k \epsilon$LS  \\
%    & $k$-$\epsilon$ Chien~(Ref.~\onlinecite{}) & $k \epsilon$CH  \\
%    & $k$-$\epsilon$ Nagano and Hishida~(Ref.~\onlinecite{}) & $k \epsilon$NH  \\
    & $k$-$\epsilon$ Rodi~(Ref.~\onlinecite{rodi1991experience}) & $k \epsilon$R  \\
%   & $k$-$\epsilon$ So~(Ref.~\onlinecite{}) & $k \epsilon$SO  \\
%     & $k$-$\epsilon$ Huang-Coakley~(Ref.~\onlinecite{}) & $k \epsilon$HC  \\
Two-equation      & $k$-$\omega$ Wilcox (1988)~(Ref.~\onlinecite{wilcox1988reassessment}) & $k \omega$88 \\
EVM     & $k$-$\omega$ Wilcox (2006)~(Ref.~\onlinecite{wilcox2008formulation}) & $k \omega$2006\\
   & $k$-$\omega$ Menter SST~(Ref.~\onlinecite{menter1994two}) & SST\\
 %  & $k$-$\omega$ Menter Baseline~(Ref.~\onlinecite{}) & BSL\\
%    & $k$-$l$ Smith~(Ref.~\onlinecite{}) & $kl$\\
    & $k$-$\zeta$ Robinson–Hassan~(Ref.~\onlinecite{robinson1998further}) & $k \zeta$\\
    & $q$-$\omega$ Coakley~(Ref.~\onlinecite{coakley1983turbulence}) & $q \omega$\\
\hline  
 %%%%%%%%%%%%%%%%%%%%%%%%%%%%%%%%%%%%%%%%%%%%%%%%%%%%%%%%%%%%%%%%%%%%%%%%%%%%%%%%%%%%%% 
NLEVM		& Goldberg cubic $k$-$\epsilon$~(Ref.~\onlinecite{goldberg2000hypersonic}) & Cubic-$k \epsilon$-UG      \\	 %BL surveys of $P_t$, $T_t$, and 	
		& Craft \emph{et al.} $k$-$\epsilon$~(Refs.~\onlinecite{craft2000progress} \& \onlinecite{craft1996development}) & Cubic-$k \epsilon$-CLS      \\
 \hline
  %%%%%%%%%%%%%%%%%%%%%%%%%%%%%%%%%%%%%%%%%%%%%%%%%%%%%%%%%%%%%%%%%%%%%%%%%%%%%%%%%%%%%%    
  EARSM         & Wallin-Johansson~(Ref.~\onlinecite{wallin2000explicit})  & WJ-EARSM  \\   \hline
 %%%%%%%%%%%%%%%%%%%%%%%%%%%%%%%%%%%%%%%%%%%%%%%%%%%%%%%%%%%%%%%%%%%%%%%%%%%%%%%%%%%%%% 
	RSTM  &  Wilcox Stress-$\omega$ RSTM (Ref.~\onlinecite{wilcox1998turbulence})  & Stress-$\omega$      \\ 
    & Eisfeld-Brodersen RSTM (Ref.~\onlinecite{eisfeld2005advanced})  & SSG/LRR-$\omega$      \\ 
\end{tabular}
\end{ruledtabular}
\end{table}

%%%%%%%%%%%%%%%%%%%%%%%%%%%%%%%%%%%%%%%%%%%%%%%%%%%%%%%%%%%%%%%%%%%%%%%%%%%%%%%%%%%%%%%%%%%%%%%%

%%%%%%%%%%%%%%%%%%%%%%%%%%%%%%%%%%%%%%%%%%%%%%%%%%%%%%%%%%%%%%%%%%%%%%%%%%%%%%%%%%%%%%%%%%%%%
%\textbf{Turb Models performance }

\begin{table*}
\caption{\label{tab:model_performance}Typical performance of popular turbulence models for two-dimensional and axisymmetric SBLIs against available experiments with error in model predictions for fully separated flows: Separation length - SL; Peak wall shear stress - $\tau_{w_p}$; Peak wall heat transfer - $Q_{w_p}$; not available - na.}
\begin{ruledtabular}
\begin{tabular}{lllcccc}
 %&\multicolumn{2}{c}{$D_{4h}^1$}&\multicolumn{2}{c}{$D_{4h}^5$}\\
 Turbulence & \multicolumn{2}{c}{Validation Experiments }    & \multicolumn{3}{c}{Typical model performance with prediction error } & Validation 
		\\ 
  Model & Two-dimensional  & Axisymmetric  & SL  & $\tau_{w_p}$ &  $Q_{w_p}$  &  References
		\\  \hline 
%\vspace{0.1in}
 %%%%%%%%%%%%%%%%%%%%%%%%%%%%%%%%%%%%%%%%%%%%%%%%%%%%%%%%%%%%%%%%%%%%%%%%%%%%%%%%%%%%%%	
   BL & Refs.~\onlinecite{coleman1972heat},\onlinecite{holden1984experimental},\onlinecite{holden2010experimental}  & Refs.~\onlinecite{coleman1973hypersonic},\onlinecite{holden1991studies}   & overprediction  & underprediction & underprediction   & Refs.~\onlinecite{vuong1987modeling},\onlinecite{horstman1987prediction},\onlinecite{marvin1989turbulence},\onlinecite{gnoffo2011uncertainty}     \\         
   
   & & & $150$\% & $15$\%  & $10$\% &  \\ \hline
   
 SA &  Refs.~\onlinecite{coleman1972heat},\onlinecite{kussoy1991documentation},\onlinecite{murray2007three} & Refs.~\onlinecite{kussoy1975experimental},\onlinecite{kussoy1989documentation},   & underprediction & underprediction & underprediction  &  Refs.~\onlinecite{paciorri1997validation},\onlinecite{paciorri1998exploring},\onlinecite{goldberg2000hypersonic},   \\ 
 
  &   & \onlinecite{delery1996shock},\onlinecite{holden1991studies}    &  $50$\% & $50$\% & $25$\%   &  \onlinecite{zhang2010turbulence},\onlinecite{brown2011shock},\onlinecite{gnoffo2011uncertainty},\onlinecite{guohua2012assessment},\onlinecite{pasha2012simulation},    \\ 

 & & &  &  &  & \onlinecite{gnoffo2013uncertainty},\onlinecite{brown2013hypersonic} \\ \hline
 
  $k \epsilon$JL & Refs.~\onlinecite{coleman1972heat},\onlinecite{elfstrom1972turbulent},\onlinecite{kussoy1991documentation},\onlinecite{settles1991}   & Refs.~\onlinecite{kussoy1975experimental},\onlinecite{kussoy1989documentation},\onlinecite{settles1991},\onlinecite{holden1991studies}    & underprediction & na & overprediction   & Refs.~\onlinecite{coakley1992turbulence},\onlinecite{horstman1992hypersonic},\onlinecite{huang1993turbulence},\onlinecite{roy2006review}      \\  
  
   & & & $50$\% & na & $65$\% &  \\ \hline 
   
 $k \epsilon$LS & Refs.~\onlinecite{coleman1972heat},\onlinecite{elfstrom1972turbulent},\onlinecite{schulein2006skin}   &  Refs.~\onlinecite{kussoy1975experimental},\onlinecite{kussoy1989documentation}    & underprediction & overprediction & overprediction & Refs.~\onlinecite{coakley1994turbulence},\onlinecite{roy2006review},\onlinecite{raje2021thesis} \\  
 & & & $75$\% & $25$\% & $200$\% &  \\ \hline 

 $k \epsilon$R    & Refs.~\onlinecite{coleman1972heat},\onlinecite{elfstrom1972turbulent},\onlinecite{kussoy1991documentation}  &  Refs.~\onlinecite{kussoy1975experimental},\onlinecite{kussoy1989documentation},\onlinecite{settles1991},\onlinecite{holden1991studies}   &  overprediction & na & overprediction & Refs.~\onlinecite{horstman1992hypersonic},\onlinecite{roy2006review} \\ 
 & & & $75$\% & na & $35$\% &  \\ \hline 
 
 $k \omega$88   & Refs.~\onlinecite{coleman1972heat},\onlinecite{elfstrom1972turbulent},\onlinecite{kussoy1975experimental}, & Refs.~\onlinecite{kussoy1975experimental},\onlinecite{kussoy1989documentation},\onlinecite{holden1991studies} & underprediction & overprediction  & overprediction  & Refs.~\onlinecite{coakley1992turbulence},\onlinecite{huang1993turbulence},\onlinecite{coakley1994turbulence},\onlinecite{brown2002turbulence},\\ 
 
 & \onlinecite{schulein2006skin} &  & $75$\% & $35$\%  & $50$\%  & \onlinecite{roy2006review},\onlinecite{raje2021thesis},\onlinecite{roy2018variable},\onlinecite{roy2019turbulent},\onlinecite{pasha2012simulation} \\
 
 \hline 
 
  $k \omega$2006  &  Refs.~\onlinecite{kussoy1991documentation},\onlinecite{holden2010experimental},\onlinecite{murray2007three}, & Refs.~\onlinecite{holden1991studies},\onlinecite{kussoy1991documentation}   & underprediction & underprediction  & overprediction & Refs.~\onlinecite{zhang2010turbulence},\onlinecite{gnoffo2011uncertainty},\onlinecite{jie2011stress},\onlinecite{brown2011shock}, \\ 
  
  & \onlinecite{schulein2006skin} & & $7.5$\% & $45$\% & $70$\% & \onlinecite{brown2013hypersonic},\onlinecite{gnoffo2013uncertainty},\onlinecite{li2021bayesian} \\ \hline 
 
SST  & Refs.~\onlinecite{coleman1972heat},\onlinecite{kussoy1991documentation},\onlinecite{schulein2006skin},  & Refs.~\onlinecite{kussoy1975experimental},\onlinecite{kussoy1989documentation}      & underprediction & good match & overprediction & Refs.~\onlinecite{steelant2002effect},\onlinecite{coratekin2004performance},\onlinecite{olsen2005lag},\\ 

& \onlinecite{murray2007three},\onlinecite{holden2010experimental}  &       &  $90$\%  &  $2$\%  & $200$\%    & \onlinecite{zhang2010turbulence},\onlinecite{georgiadis2013recalibration},\onlinecite{gnoffo2011uncertainty},\onlinecite{guohua2012assessment},\onlinecite{gnoffo2013uncertainty}    \\ 

&   &       &  &    & & \onlinecite{zhang2015turbulence},\onlinecite{raje2021anisotropic}    \\ \hline

$k \zeta$ & Refs.~\onlinecite{coleman1972heat},\onlinecite{kussoy1993hypersonic},\onlinecite{schulein2006skin}  &   na    & good match & underprediction  & overprediction & Refs.~\onlinecite{nance1999turbulence},\onlinecite{xiao2007role} \\ 

& & & $2$\% & $15$\% & $250$\% &  \\ \hline
                    
$q \omega$ & Ref.~\onlinecite{coleman1972heat}  &  Refs.~\onlinecite{kussoy1975experimental},\onlinecite{kussoy1989documentation}     &  underprediction & na & overprediction & Refs.~\onlinecite{vuong1987modeling},\onlinecite{marvin1989turbulence},\onlinecite{coakley1992turbulence},\onlinecite{roy2006review} \\ 

& & & $60$\% & na & $100$\% &  \\ \hline

Cubic-$k \epsilon$-UG & Refs.~\onlinecite{coleman1972heat},\onlinecite{holden1992turbulent}  &  na     & good match  &  na  & good match & Ref.~\onlinecite{goldberg2000hypersonic}   \\ 

& & & $1$\% & na & $2$\% &  \\ \hline

Cubic-$k \epsilon$-CLS & Refs.~\onlinecite{coleman1972heat},\onlinecite{elfstrom1972turbulent},\onlinecite{kussoy1975experimental}, & Ref.~\onlinecite{kussoy1989documentation} &  good match   & na  & overprediction    & Refs.~\onlinecite{zhang2021turbulence},\onlinecite{zhang2022application}   \\ 

& \onlinecite{kussoy1991documentation},\onlinecite{schulein2006skin} &  & $1$\% & na & $100$\% &  \\ \hline

WJ-EARSM & Refs.~\onlinecite{coleman1972heat},\onlinecite{schulein2006skin}  & na      & underprediction   & underprediction   & overprediction  & Refs.~\onlinecite{wallin2000explicit},\onlinecite{coratekin2004performance},\onlinecite{vemula2020explicit}    \\ 

& &  & $20$\% & $30$\% & $90$\% &  \\ \hline

Stress-$\omega$ & Ref.~\onlinecite{holden1978study}   & Ref.~\onlinecite{kussoy1989documentation}  &  underprediction & na   &  overprediction & Ref.~\onlinecite{wilcox1998turbulence} \\ 

& & & $35$\% & na & $50$\% &  \\ \hline

SSG/LRR-$\omega$ & Ref.~\onlinecite{bosco2011investigation}  &  na     & underprediction & na   & good match & Refs.~\onlinecite{bosco2011investigation},\onlinecite{frauholz2014investigation}   \\
& & & $20$\% & na & $1$\% &   \\
\end{tabular}
\end{ruledtabular}
\end{table*}

%%%%%%%%%%%%%%%%%%%%%%%%%%%%%%%%%%%%%%%%%%%%%%%%%%%%%%%%%%%%%%%%%%%%%%%%%%%%%%%%%%%%%%%%%%%%%%%%

%%%%%%%%%%%%%%%%%%%%%%%%%%%%%%%%%%%%%%%%%%%%%%%%%%%%%%%%%%%%%%%%%%%%%%%%%%%%%%%%%%%%%%%%%%%%%
%\textbf{Model corrections }

\begin{table*}
\caption{\label{tab:comp_corr}Popular compressibility corrections assessed using the available hypersonic validation experimental and DNS database : Separation length - SL; Peak wall heat transfer - $Q_{w_p}$.}
\begin{ruledtabular}
\begin{tabular}{lllll}
 %&\multicolumn{2}{c}{$D_{4h}^1$}&\multicolumn{2}{c}{$D_{4h}^5$}\\
 Compressibility correction & Intended improvement & Key consequences & Validation & Validation \\
   &  & in TBLs and SBLIs & Experiments/DNS & References  \\
  \hline 
%\vspace{0.1in}
 %%%%%%%%%%%%%%%%%%%%%%%%%%%%%%%%%%%%%%%%%%%%%%%%%%%%%%%%%%%%%%%%%%%%%%%%%%%%%%%%%%%%%%	
Dilatation dissipation  & Reduction in shear &  Increase in SL, & Refs.~\onlinecite{coleman1972heat},\onlinecite{elfstrom1972turbulent},\onlinecite{kussoy1975experimental},\onlinecite{kussoy1989documentation}, & Refs.~\onlinecite{coakley1992turbulence},\onlinecite{grasso1993high},\onlinecite{coakley1994turbulence},\onlinecite{steelant2002effect},\onlinecite{brown2002turbulence},  \\

(Sarkar~\cite{sarkar1991analysis},Zeman~\cite{zeman1990dilatation},Wilcox~\cite{wilcox1992dilatation})      &  layer spreading rate &  Overcorrection in $C_f$ & \onlinecite{kussoy1991documentation},\onlinecite{schulein2006skin},\onlinecite{holden2010experimental},\onlinecite{huang2019assessment},\onlinecite{sciacovelli2020numerical},\onlinecite{sciacovelli2020numerical}, & \onlinecite{zhang2010turbulence},\onlinecite{guohua2012assessment},\onlinecite{gnoffo2011uncertainty},\onlinecite{gnoffo2013uncertainty},\onlinecite{li2021bayesian},\onlinecite{aiken2022assessment},  \\

    &   & in TBLs  & \onlinecite{passiatore2022thermochemical},\onlinecite{huang2022direct} & \onlinecite{barone2022internal},\onlinecite{sciacovelli2023priori} \\
                
                & & & &  \\

%%%%%%%%%%%%%%%%%%%%%%%%%%%%%%%%%%%%%%%%%%%%%%%%%%%%%%%%%%%%%%%%%%%%%%%%%%%%%%%%%%%%%%  
                
Length scale  &  Reduction of & Reduction in $Q_{w_p}$ & Refs.~\onlinecite{coleman1972heat},\onlinecite{elfstrom1972turbulent},\onlinecite{watson1973measurements},\onlinecite{kussoy1975experimental}, & Refs.~\onlinecite{vuong1987modeling},\onlinecite{marvin1989turbulence},\onlinecite{coakley1992turbulence},\onlinecite{grasso1993high},\onlinecite{coakley1994turbulence},   \\

 correction~(Huang and Coakley~\cite{huang1993calculations})  & reattachment  & in separated flows & \onlinecite{kussoy1989documentation},\onlinecite{kussoy1991documentation} & \onlinecite{smith1996prediction},\onlinecite{brown2002turbulence},\onlinecite{coratekin2004performance},\onlinecite{zhang2010turbulence}  \\
 
 & heat transfer &  & &  \\
 
 & & & \\

%%%%%%%%%%%%%%%%%%%%%%%%%%%%%%%%%%%%%%%%%%%%%%%%%%%%%%%%%%%%%%%%%%%%%%%%%%%%%%%%%%%%%%
 
Rapid compression    & Increase of & Increase in SL & Refs.~\onlinecite{coleman1972heat},\onlinecite{coleman1973hypersonic},\onlinecite{kussoy1975experimental},\onlinecite{holden1984experimental}, & Refs.~\onlinecite{vuong1987modeling},\onlinecite{horstman1987prediction},\onlinecite{marvin1989turbulence},\onlinecite{coakley1992turbulence},\onlinecite{coakley1994turbulence}, \\

  correction~(Huang and Coakley~\cite{huang1993calculations})   &  separation length  &  & \onlinecite{kussoy1989documentation}   & \onlinecite{coratekin2004performance},\onlinecite{zhang2010turbulence} \\
  
  & & & &  \\

%%%%%%%%%%%%%%%%%%%%%%%%%%%%%%%%%%%%%%%%%%%%%%%%%%%%%%%%%%%%%%%%%%%%%%%%%%%%%%%%%%%%%%
  
Shock-unsteadiness    &  Increase of & Increase in SL, & Refs.~\onlinecite{schulein2006skin},\onlinecite{holden2010experimental}  & Refs.~\onlinecite{pasha2012simulation},\onlinecite{raje2021anisotropic},\onlinecite{raje2021thesis},\onlinecite{roy2018variable},\onlinecite{roy2019turbulent}\\

 correction (Sinha \emph{et al.}~\cite{sinha2003modeling})   &  separation length & Overcorrection in $C_f$ & & \\
 & & in separated flows & &  \\

  & & & &  \\

%%%%%%%%%%%%%%%%%%%%%%%%%%%%%%%%%%%%%%%%%%%%%%%%%%%%%%%%%%%%%%%%%%%%%%%%%%%%%%%%%%%%%%
 
Catris \& Aupoix diffusion   & Correct log-law & No significant & Refs.~\onlinecite{winkler1959investigation}, \onlinecite{sciacovelli2020numerical},\onlinecite{sciacovelli2020numerical},\onlinecite{passiatore2022thermochemical}, & Refs.~\onlinecite{catris2000density},\onlinecite{guohua2012assessment},\onlinecite{sciacovelli2023priori},\onlinecite{barone2024data} \\

term correction~\cite{catris2000density}   &  in compressible TBLs & improvements in TBLs  & \onlinecite{holden2010experimental},\onlinecite{holden2013measurements},\onlinecite{WaMaHo08},\onlinecite{huang2022direct} & \\
 & & and separated flows & &  \\
\end{tabular}
\end{ruledtabular}
\end{table*}

%%%%%%%%%%%%%%%%%%%%%%%%%%%%%%%%%%%%%%%%%%%%%%%%%%%%%%%%%%%%%%%%%%%%%%%%%%%%%%%%%%%%%%%%%%%%%%%%

%%%%%%%%%%%%%%%%%%%%%%%%%%%%%%%%%%%%%%%%%%%%%%%%%%%%%%%%%%%%%%%%%%%%%%%%%%%%%%%%%%%%%%%%

\subsection{Industrial applicability}\label{sec:efficiency}

The computational cost of turbulence models generally increases with the number of transport equations solved. The cost per iteration on identical grids and processors rises following the order: zero-equation, one-equation, two-equation, NLEVMs/EARSMs, and RSTMs. This progression reflects increasing model complexity, with each additional transport equation and/or non-linear terms in the stress-strain constitutive relationship adding to the computational burden. For most industrial applications, linear EVMs provide the optimal balance between accuracy and robustness.

The typical CPU time required to reach convergence for the different turbulence models can be used to assess the computational efficiency of these models. In comparison to the zero-equation turbulence models, one-equation models can require about $1.1-1.5$x, two-equation linear EVMs about $1.5-2$x, NLEVMs/EARSMs about $2-4$x, and RSTMs about $4-6$x more CPU time on identical grids and CPU cores to get converged solutions. The CPU time for a specific turbulence model depends on the solver efficiency and grid quality. Hypersonic flows have been shown to be extremely grid-sensitive. Strategies relying on adaptive or static mesh refinement and aligning the grid with the shock waves could strongly improve the model performance in terms of memory and CPU time while at the same time maintaining the accuracy of the results \cite{frauholz2014investigation}.

Zero-equation models like Bardin-Lomax and Cebeci-Smith can struggle to converge while showing massive and unsteady separation for hypersonic SBLIs for which one-equation and two-equation linear EVMs yield results~\cite{gnoffo2013uncertainty}. The one-equation Spalart-Allmaras and two-equation $k$-$\epsilon$ and $k$-$\omega$ linear EVMs including Menter SST $k$-$\omega$ are the commonly used model family for industrial flow simulations. The Spalart-Allmaras model can require fewer time steps for the mean flow to reach convergence compared to the two-equation $k$-$\epsilon$ and $k$-$\omega$ models and is a popular choice in the two- and three-dimensional simulations of a scramjet engine \cite{pecnik2012reynolds,barth2015effects,dai2022numerical} and reentry capsule \cite{reddy2009hypersonic} using thermochemical non-equilibrium gas models.

NLEVMs and EARSMs can offer improved accuracy over linear two-equation EVMs with a modest increase in computational cost and a similar general numerical behavior \cite{wallin2000explicit,zhang2022application}. For example, Zhang~\cite{zhang2021turbulence} reported that the use of Craft-Launder-Suga cubic $k$-$\epsilon$ NLEVM only increased CPU time by about $6$ \% at most compared with the Menter SST $k$-$\omega$ and Launder-Sharma $k$-$\epsilon$ model with Yap length-scale correction tested for the hypersonic experimental cases of Refs.~\onlinecite{coleman1972heat,elfstrom1972turbulent,kussoy1975experimental,kussoy1989documentation,kussoy1991documentation,schulein2006skin}. Raje and Sinha~\cite{raje2021anisotropic} reported that SUQ-SST EARSM required about $6$-$7$ \% extra computational time per time-step than the Menter SST $k$-$\omega$ model with validation studies conducted against the experiments of Refs.~\onlinecite{schulein2006skin,holden2013measurements}. Similar observations are noted for the WJ-EARSM compared to the Wilcox $k$-$\omega$ model for Schulein cases~\cite{wallin2000explicit}. On the other hand, RSTMs typically require more CPU time and memory than EVMs and are prone to convergence difficulties on complex applications and non-optimal grids. Bosco~\cite{bosco2011reynolds} reported that although the SSG/LRR-$\omega$ RSTM required approximately the same
number of iterations as the Menter SST $k$-$\omega$ and Wilcox $k$-$\omega$ models for a hypersonic compression corner case, it needed about $2$x the CPU time of the linear two-equation EVMs on identical grids and CPU cores. The Wilcox Stress-$\omega$ RSTM can require about $2-3$x longer than the time needed for the Wilcox $k$-$\omega$ model\cite{wilcox1998turbulence}. Moreover, the performance of the RSTMs in hypersonic flow applications involving SBLIs does not warrant the abandonment of eddy-viscosity models in industry.

%%%%%%%%%%%%%%%%%%%%%%%%%%%%%%%%%%%%%%%%%%%%%%%%%%%%%%%%%%%%%%%%%%%%%%%

%%%%%%%%%%%%%%%%%%%%%%%%%%%%%%%%%%%%%%%%%%%%%%%%%%%%%%%%%%%%%%%%%%%%%%%

\section{Challenges and Future Directions}\label{sec7}

Accurate modeling and prediction of hypersonic TBL flows is one of the most difficult yet important problems in the aerothermal design of hypersonic vehicles. Equilibrium TBLs at cold wall conditions and the regions of SBLIs and separated flows present the greatest challenge for turbulence modeling. In these flows, turbulent as well as thermo-chemical non-equilibrium and compressibility effects are of major importance. Popular turbulence models like Baldwin-Lomax, Spalart-Allmaras, $k$-$\epsilon$, $k$-$\omega$, Menter SST $k$-$\omega$, etc., as well as advanced models like NLEVMs, EARSMs, and RSTMs, were primarily developed for incompressible or moderately compressible flows. These models are generally insensitive to compressibility effects and are tuned for non-hypersonic smooth wall turbulent boundary layers with conventional rates of heat transfer. The straightforward compressible extensions of these models generally fail to capture the intricate phenomena present in hypersonic regimes at different wall conditions. Over the years, a major focus of turbulence modeling has been the adaptation and extension of the traditional low-speed models to better handle the high-speed conditions relevant to hypersonic flows.

%-------------------------------------------------------------------------------------

\subsection{Current Limitations and Unresolved Issues}

%% experimental and DNS data 
Turbulence modeling for hypersonic flows faces several critical limitations and unresolved issues, as discussed below, due to the extreme conditions and complex flow physics encountered at high Mach numbers. One of the main limitations, which echoes some of the previous recommendations by Roy and Blottner~\cite{roy2006review} from 2006, is the dearth of experimental data in the hypersonic regime. Additionally, many of the available experimental works reported in Roy and Blottner~\cite{roy2006review}, especially on ZPG hypersonic TBLs, are poorly documented, including inadequate specifications of the boundary conditions and geometry details required to set up a CFD simulation for accurate one-on-one comparisons. Experiments generally do not include detailed measurements of Reynolds stresses and mean flow surveys are generally limited. DNS and wall-resolved LES works are also significantly limited in Reynolds numbers and geometric sophistication. Moreover, attempts to match DNS with experiments at similar conditions show disagreements for ZPG TBLs as well as SBLI characteristics~\cite{volpiani2020effects}.  The scarcity of experimental and high-fidelity data makes it challenging to develop and calibrate turbulence models specifically for hypersonic flows, leading to uncertainty in their accuracy and reliability. Developing high-quality validation datasets and benchmarks for hypersonic turbulence models remains a critical issue.

%%%
Popular turbulence models generally do not correctly represent hypersonic cold-wall boundary layers. Reliable and accurate prediction of shock-induced separations and the associated wall pressure loads, surface heat-transfer rates, and skin-friction at different wall conditions remains an unsolved problem. Standard turbulence models generally lack sensitivity to adverse pressure gradients, often resulting in overestimated shear stresses and delayed flow separation predictions. Compressibility terms related to pressure fluctuations and dilatation dissipation are generally neglected due to a lack of information on these terms for wall-bounded flows even at lower speeds. The compressibility corrections in the form of dilatation dissipation and pressure dilatation of Sarkar or Zeman designed with some elegant theoretical framework, particularly with the compressible isotropic turbulence, are originally intended for improving mixing predictions in free shear layers. Their application to TBL flows with SBLIs is expected to adequately simulate the free shear layers over the separation bubbles. These compressibility corrections along with Wilcox's compressibility correction have been shown to improve model predictions for some hypersonic TBL flows involving SBLIs. However, improved separation length prediction generally results in significantly lower shear stress while the overpredicted peak heat transfer predictions remain largely unaffected. These corrections are also found to deteriorate model predictions for attached equilibrium TBLs in many cases. As a result, these compressibility corrections are not widely accepted. Further, the validity of the arguments underpinning dilatation-dissipation models are not substantiated by experimental or DNS data~\cite{wilcox1998turbulence}.

%%% length-scale equation and compressibility corrections
The length-scale determining equation remains a major limitation for the two-equation models including LEVMs, NLEVMs, and EARSMs. Several ad-hoc fixes are proposed as discussed in Section \ref{sec4} to improve length scale predictions, especially aimed at improving heat transfer predictions. Other corrections including modification to the production terms, diffusion terms, and turbulent Prandtl number developed to date to improve SBLI predictions show promising results in some cases, however, model predictions generally depend on the flow conditions and geometry configurations. 

In SBLIs, the average flow structure also results from shock motions with larger amplitude at very low frequencies, associated with large-scale bubble breathing. The compressible extensions of the standard turbulence models developed for low-speed flows cannot take into account the complex physical mechanisms driving the average shock position, separation length, surface pressure, skin friction coefficient, and wall heat transfer distributions. Attempts to incorporate the effects of small-scale high-frequency unsteady oscillations of shock waves upon interaction with upstream turbulence using LIA results based on canonical STI problems~\cite{sinha2003modeling} have shown improvements over the standard models for a number of hypersonic test cases. However, these models lack a rigorous theoretical foundation as the physics of SBLIs and canonical STIs is considerably different and the scaling of turbulence quantities across a shock can differ significantly in the presence of wall effects. Studies using these models also do not offer an evolution of turbulence quantities like TKE or TKE budgets in comparison to DNS or experiments.

%% Turbulent heat flux
%%ADDED by Paola*****
This review demonstrates that current turbulence models, while adequately predicting wall pressure distributions and separation zones, consistently overestimate heat flux at the wall surface. This discrepancy in heat flux prediction remains a significant challenge for accurate thermal analysis in hypersonic applications.
 Of note, while many modeling efforts have addressed the formulation of the constitutive equation for the Reynolds stresses and of the auxiliary transport equations, only a few studies have attempted to improve the turbulent heat flux formulation, which mostly relies on the over-simplistic constant turbulent Prandtl number hypothesis. 
Advanced turbulent heat flux models have been introduced with some success for incompressible, variable density flows \cite{daly1970transport,abe2001towards,shin2008elliptic} and, recently, for hypersonic flows \cite{roy2019turbulent}. Furthermore, data-driven turbulent heat flux models have been developed to improve the prediction of heat transfer effects in jet flows \cite{weatheritt2020data}. Extension of such approaches to hypersonic flows should be pursued.\\

%% thermo-chemical non-equilibrium
For flows with thermochemical non-equilibrium effects, only very rough models based on constant Schmidt or vibrational Prandtl numbers have been proposed. The validity of such models has only recently been assessed for simple hypersonic flow configurations such as flat plate boundary layers, showing a need for improvement. Furthermore, turbulence/chemistry and turbulence/vibrational energy interactions can be significant, especially in flows with shocks,  making the approximation of evaluating chemical and vibrational source terms as a function of averaged flow properties inaccurate. Direct application of interaction models developed for combustion problems has proved unfruitful, pointing out the necessity of developing specific models for hypersonic non-equilibrium flows.
%%***END PAOLA ADDITIONS-----

%%% surface roughness
The ablation process creates a complex interaction between the surface and TBL that must be properly modeled. Integration of rough wall effects into existing turbulence models should be a focus for improving overall prediction capabilities. Accurate modeling of blowing/surface roughness interactions in TBLs is necessary to correctly predict the aerothermal loads and efficiently design the ablative heat shield. Standard turbulence models fail to accurately estimate the velocity shift in the logarithmic law caused by roughness and blowing interactions. The development of sophisticated turbulence models that can accurately capture the combined effects of surface roughness/blowing in hypersonic conditions is an ongoing area of research.

%% advanced modeling 

%{\color{red}****Advanced modeling (Prof Karthik)}

%-------------------------------------------------------------------------------------

%-------------------------------------------------------------------------------------

\subsection{Role of Machine Learning and Data-Driven Approaches}

Data-driven  modeling has emerged as a promising technique to improve RANS models.  These methods combine high-fidelity and/or experimental data with more traditional phenomenological arguments to improve the predictive accuracy of turbulence models. 
There is a large body of work focused on data-driven turbulence modeling~\cite{duraisamy2019turbulence,zhao2020rans,duraisamy2021perspectives} with the goal of developing improved augmentations in the transport equations for turbulence scalars, or directly for the Reynolds stress tensor, heat flux models, etc. 
These model augmentations are described over an appropriate feature space.
Critical steps in this process include careful data preparation, choice of non-dimensional features, choice of learning algorithms, cross-validation strategies, and uncertainty quantification to assess model reliability across different flow regimes.
%%%

Data-driven models can be classified according to the representation adopted for the learned corrective terms and to the training procedure.
The main families of approaches can be identified as:
\begin{enumerate}
\item models using data-assimilation to identify corrective fields (to the RANS transport equations and/or to the Reynolds stresses), which are subsequently parametrized as functions of a set of flow features to allow predictions of new flows. For instance, the field inversion and machine learning (FIML) approach~\citep{parish2016paradigm} belongs to this family;
\item models that learn \emph{a priori} corrections of the Reynolds stresses and other terms from full fields of high-fidelity data~\citep{ling2016reynolds,weatheritt2016novel,schmelzer2020discovery,cherroud2022sparse}. The learned models are subsequently propagated through a CFD and assessed \emph{a posteriori} against test data;
\item model-consistent approaches, which use a coupled training strategy with the CFD solver in the loop, e.g., learn the corrective fields via backpropagation of errors through the RANS solver or via gradient-free algorithms~\citep{zhao2020rans,duraisamy2021perspectives,saidi2022}.
\end{enumerate}

Models of the first and second families involve tensor or vector representations of the corrective fields, and specifically the Reynolds tensor and, in some cases, the turbulent heat flux vector. The function coefficients of the representations are then learned from data using black-box models (e.g. neural networks, random forests \citep{ling2016reynolds,kaandorp2020data}) or symbolic regression (or open-box) techniques. The latter include gene expression programming \citep{weatheritt2016novel} and symbolic identification \citep{schmelzer2020discovery,cherroud2022sparse}.
Finally, both black-box and open-box models can be found in the third family \citep{holland2019towards,holland2019field,zhao2020rans,srivastava2021generalizable,saidi2022}.
%\textcolor{blue}{Should we put the equations? Seems like too much!}}
While much of the early work on machine learning augmented turbulence modeling involved \emph{a priori} learning (first and second family), model consistency through inference~\cite{duraisamy2021perspectives} is increasingly being adopted over the past few years by most practicioners. This step ensures that the learned augmentations are consistent with the RANS modeling environment. Validation against independent datasets is crucial to ensure generalizability, especially given the limited availability of hypersonic flow data. 
A more detailed overview of data-driven techniques for turbulence modeling can be found in Cinnella~\cite{cinnella2024data} and in Duraisamy~\cite{duraisamy2025book}.

While still an emerging area, data-driven methods show potential for enhancing model performance across a range of conditions relevant to hypersonic vehicle design, provided challenges in data scarcity and extrapolation to extreme conditions can be overcome.
Refs.~\onlinecite{huang2017high,wang2019prediction} demonstrate the use of machine learning  to improve Reynolds stress predictions in hypersonic boundary layers. The approach was seen to significantly improve the accuracy of normal stress predictions. However, only \textit{a priori} studies were performed, and the impact of the updated Reynolds stress predictions on the mean flow field was not assessed. In a similar vein, Parish \textit{et al.}~\cite{PaChMi23} developed a tensor basis neural network trained on data from both subsonic and hypersonic flows, and showed that the resulting model improves wall shear stress and wall heat flux predictions in hypersonic TBL flows. Jordan~\cite{Jo23} developed a machine learning approach utilizing for predicting an improved structure factor in the SST turbulence model, which plays an important role in hypersonic SBLIs. Jordan utilizes high-fidelity data to extract an ``exact" structure factor, and then develops an equation learning neural network to learn the structure factor. The resulting model is seen to improve predictions for flow separation and peak heat transfer on a range of SBLI cases. 

Parish \textit{et al.}~\cite{PaChJo24} utilized field inversion~\citep{singh2017machine,parish2016paradigm} to derive model consistent corrections and machine learning  to develop a variable turbulent Prandtl number model to address over-prediction in peak heat flux in cold-wall hypersonic SBLIs. The approach resulted in a model that gave consistent improvements over baseline RANS models across a suite of SBLI cases. 
Chowdhary \textit{et al.}~\cite{chowdhary2022calibrating} used surrogate-based Bayesian inference to calibrate the parameters of the SST $k-\omega$ model, using shock-tunnel data for the HIFiRE-1 vehicle. The authors found that the primary issues in estimating the SST model parameters are the limited information content of the heat flux and pressure measurements and the large model-form error encountered in a certain part of the flow.
Barone \textit{et al.}~\cite{BaPaJo24} utilized recent DNS simulation data of hypersonic boundary layers to develop modifications to the SA model to improve performance on hypersonic TBLs. The corrections include the development of a new near-wall damping function based on a novel eddy viscosity transformation as well as a normal stress correction to improve predictions for the wall-normal Reynolds stress. These corrections improved model performance for wall heat flux and wall shear stress in hypersonic SBLIs at sufficient Reynolds numbers. Zafar \textit{et al.}~\cite{zafar2024data}  developed a data-driven turbulence modeling approach for cold-wall boundary layers that utilizes an ensemble Kalman filter to train improved models for the Reynolds stress and turbulent Prandtl number model with mixed degree of success.

%\subsection{Integration of multi-physics modeling (e.g., chemical reactions, radiation)} 

Given the difficulty of obtaining abundant experimental datasets, the production and sharing of numerical ones should be encouraged and pursued, for instance via opendata repositories such as \url{https://blastnet.github.io} (mostly powered by the combustion research community for now). Furthermore, data-driven research should  focus more on model-consistent (or CFD-in-the-loop) training methods, which have the potential of learning from sparse and noisy data. With sufficient data available, data-driven turbulent modeling approaches can facilitate the development of novel closures for turbulent heat or mass fluxes, vibrational energy fluxes, roughness and ablation models and much more.

\subsubsection{Towards reliable data-driven models}
Machine learning augmentations are generally developed as a functional map over a feature space. The underlying feature selection must balance both classical machine learning criteria and domain-specific requirements. From a classical machine learning perspective, features should provide unique and complementary information to prevent overfitting, even when dealing with correlated features. A parsimonious feature set is crucial, as it helps minimize overfitting risk, improves computational efficiency, and simplifies model interpretation. Features must also demonstrate robustness by being resistant to noise and small data variations, ensuring model stability and generalizability. 
Feature sets have been proposed by various authors \cite{ling2015evaluation,wu2018data-driven}, mostly in the context of low speed flows. More effort is required to develop principles and frameworks to systematically derive the features. For hypersonic flows, these are required not only to augment the Reynolds tensor, but also the unclosed terms in the energy equation, and in any additional transport equations,e.g., for reacting species. Of course, the computational cost of calculating the features must be considered, especially when working with large datasets. Approaches favoring sparsity, as well as feature selection strategies based on information theory \cite{beneddine2023nonlinear} may help extracting optimal feature sets.

Domain-specific requirements add another layer of complexity to feature selection. Features must be locally non-dimensionalized to work effectively across different flow configurations. The selected features should respect important physical properties such as rotational, reflectional, and frame-invariance. Interpretability is another crucial aspect - features that are easily interpretable help avoid unphysical behavior and enable asymptotic analysis.

Generalization remains one of the most significant challenges in data-driven turbulence modeling. Model consistency is crucial for generalization, requiring careful consideration of the implicit relationship between state variables and augmentation values. Generalizable results in physical space can be achieved through careful interpolation over a well-designed feature space. However, this presents an important trade-off: while more features allow for more unique physical description, a lower-dimensional feature space is necessary for generalization with limited data.
In such cases, combining smaller models learned from subsets of data is empirically observed to be more efficient than training a single large model covering a wide range of flow processes \cite{zhou2025ensemble}. Boosting, stacking, weighting, and voting approaches are examples of well-established ensemble approaches in modern machine learning.
Model generalization to flows involving diverse physical processes is an active field of research, even for low-speed flows \citep{fang2023toward,rincon2023progressive,cherroud2025space,oulghelou2024machine}, and is warranted more efforts in the future.
%Furthermore, quantification of uncertainties associated with the turbulence model is crucial for the robust and safe design of TPS systems \cite{boseuse}.

Finally, machine learning methods often lack of interpretability. This is not only the case for black-box methods based on neural-network or random-forest approximations, but also for some symbolic models characterized by very complex expressions. Using sparsity-promoting techniques\citep{schmelzer2020discovery} can help obtaining simpler models that are easier to interpret. Nonlinear sensitivity analysis tools, such as Shapley factors are another avenue for better understanding the learned corrections\citep{he2022explainability}.
More in general, explainable artificial intelligence (XAI) is a promising field with high potential impact in fluid flow studies\cite{cremades2024identifying}.

These insights suggest that successful data-driven turbulence modeling requires a delicate balance between physical understanding, mathematical rigor, and practical implementation considerations. The key to reliable data-driven models appears to lie in thoughtful feature design that captures essential physics while maintaining mathematical tractability complemented by a coordinated data generation process to adequately populate feature space.

%%%%%%%%%%%%%%%%%%%%%%%%%%%%%%%%%%%%%%%%%%%%%%%%%%%%%%%%%%%%%%%%

%%%%%%%%%%%%%%%%%%%%%%%%%%%%%%%%%%%%%%%%%%%%%%%%%%%%%%%%%%%%%%%%

\section{Concluding Remarks and Perspectives}\label{sec8}

It is realistic to expect that RANS methods will continue to be popular for computing practical industrial flows in the foreseeable future. In an industrial setting, it can  be expected that  further increases in computing power will be used  to utilize advanced RANS models to shorten the design cycle rather than to yield the way to LES techniques. Turbulence modeling shortcomings remain a major limitation to the accuracy of hypersonic simulations, and this includes deficiencies in calculating the details of ZPG TBLs and SBLIs. Though there have been improvements, the concluding remarks of Marvin and Coakley~\cite{marvin1989turbulence} stating that ``the status of turbulence modeling for hypersonic flows is still far from complete'' remain mostly valid even after 35 years, partly because of a lack of concerted effort in developing Hypersonics-relevant models. The acquisition of reliable experimental data for hypersonic TBL flows remains challenging, especially at high enthalpy conditions. In the absence of experimental data at high-enthalpy conditions in the hypersonic regime, DNS and LES can be used to gain insight into flow physics and develop phenomenological models and novel turbulence modeling methodologies for chemically reacting boundary layers including radiation physics. We summarize key findings and discuss avenues for further improvements.

\subsection{Summary of Key Findings}

The turbulence models routinely applied for hypersonic flow predictions are direct extensions of those developed originally for incompressible flows by invoking Morkovin's hypothesis and using density-scaled variables. In simple applications, especially involving simple attached boundary-layer flows at near adiabatic smooth wall conditions, these incompressible model forms give satisfactory results with or without some adjustments to the model coefficients. However, the performance of standard forms of these turbulence models becomes unsatisfactory even for the TBL at equilibrium conditions when the wall temperature differs from the adiabatic conditions, especially at the realistically expected cold wall conditions. This is true even when the complexity of the model form is increased from linear eddy viscosity models to NLEVMs/EARSMs and RSTMs. In fact, although RSTMs have the natural potential to represent the dynamics of intercomponent energy transfer, additional modeling requireements coupled with more advanced numerical treatments in complex flow situations is a major disadvantage of these models.  Also, as discussed in Section \ref{sec:rstm}, thus far, RSTMs have not been shown to offer consistent predictive accuracy compared to  EVMs for hypersonic SBLI predictions. On the other hand, there is evidence that the NLEVMs and EARSMs improve model predictions for hypersonic SBLIs compared to the corresponding LEVMs, however, validation studies for these models are limited compared to the LEVMs and extensive validation is needed against the available hypersonic test cases. 

The choice and the form of the length scale determining equation for the turbulence models including LEVMs, NLEVMs, EARSMs, and RSTMs remains a major shortfall and investigations indicate $\omega$ is a better choice than $\epsilon$ for hypersonic flow predictions. Interestingly, modifications to the mixing-length hypothesis-based zero-equation models like Cebeci-Smith and Baldwin-Lomax can result in better predictions than the two-equation EVMs for equilibrium hypersonic TBLs at cold wall conditions and in some cases even for hypersonic SBLIs. We can consider current turbulence model choices as case-dependent and influenced by computational convenience. Overall, the performance of the standard turbulence models becomes increasingly unsatisfactory as the flow complexity and/or Mach number increase and the wall temperature ratio decreases.

Existing compressibility corrections generally do not  satisfactorily overcome  weaknesses seen in  standard models when predicting flow separation or wall heat flux. Moreover, compressibility correction investigations are mainly confined within the context of LEVMs. Extending and testing these model corrections in the framework of more sophisticated NLEVMs and EARSMs remains unexplored.

The development of more sophisticated turbulence models that can accurately capture the effects of surface roughness in hypersonic conditions is an ongoing area of research. Velocity shift in the log layer plays a key role in wall roughness modeling. The standard equivalent sand-grain approach is quite sensitive to the equivalent sand-grain height $k_s$ estimate. A constant value of $Pr_t = 0.9$ gives inaccurate predictions of wall heat flux over rough surfaces even with  dynamic corrections to correctly predict log-layer behavior. The existing thermal correction models aiming to accommodate $Pr_t$ variation within the roughness sublayer are generally sensitive to the parameter $S_{cor}$ for heat transfer predictions. In addition, there is a need for more experimental data on rough wall hypersonic flows to develop, validate, and improve the computational models.

Comparisons of turbulence model predictions with compressible engineering correlations (CECs) for ZPG TBLs require careful consideration, particularly when evaluating skin friction and wall heat transfer. The accuracy of these correlations diminishes significantly under cold wall conditions, necessitating additional validation against higher-fidelity data sources. The two most commonly used CECs of Van Driest II and Spalding–Chi for estimating the skin-friction give different results and studies reveal that their accuracy depends on the Mach number, wall temperature ratio, and Reynolds number along with the incompressible friction correlations used. At highly cooled wall conditions, these two theories result in inaccurate predictions. Moreover, heat transfer predictions using these CECs depend on the choice of the value of the Reynolds analogy factor $R_{af}$. At low enthalpy conditions with $ Ma \gtrsim 5$, different studies draw different conclusions and it is unclear which theory should be favored, especially for cold-wall cases. For high-enthalpy hypersonic flows at cold-wall conditions, the study of Goyne \textit{et al.}~\cite{goyne2003skin} found Spalding-Chi method to give better predictions than other correlations.

There is a great need to expand  experimental databases and use rigorous validation methodologies for turbulence models in the hypersonic regime. Almost all of the experiments employ low enthalpy wind tunnels while none of the existing SBLI experiments match true flight conditions. Moreover, little attention has been directed toward DNS/LES studies considering real gas effects relevant to hypersonic conditions. Consequently, the majority of the RANS-based studies consider calorically perfect gas models in the absence of validation data at high enthalpy conditions. Sensitivity studies of real gas effects under potential flight conditions reveal that SBLI predictions obtained using the popular $k$-$\omega$ family of turbulence models (Menter SST and Wilcox $k$-$\omega$) are sensitive to the gas models. Real gas simulations for these models show a reduction in separation extent, an increase in peak pressure, and an increase in peak heating post-reattachment compared to the ideal gas scenario~\cite{brown2013hypersonic}. Moreover, the severity of the simulated real gas effect can be expected to depend on turbulence modeling in the vicinity of separation and reattachment, which are regions where the existing turbulence models have been demonstrated to exhibit the greatest deficiency in their performance.

Machine learning techniques appear poised to have a positive impact on model development. To develop robust and generalizable turbulence models, however, a synergistic ecosystem needs to be created that brings together modelers, experimentalists and  high-fidelity simulation experts. While existing datasets such as those described in this paper serve as a reasonable starting point, the development of generalizable models requires the data generation process to be consistent with the specific requirements of the underlying model and features. Further, measured quantities should be informative to the underlying model discrepancy, which can be assessed using inverse problems.

%-------------------------------------------------------------------------------------

%%%%%%%%%%%%%%%%%%%%%%%%%%%%%%%%%%%%%%%%%%%%%%%%%%%%%%%%%%%%%%%%%%%%%%%%%%%%%%%%%%%%%%%%%%%%%%%%%%%%%%

\begin{acknowledgments}
E. Parish was supported by Sandia LDRD 226031, ``Data-driven closure modeling for hypersonic turbulent flows". K. Duraisamy was supported by OUSD(RE) Grant \# N00014-21-1-295 titled ``Physics-Aware Reduced Order Modeling for Nonequilibrium Plasma Flows: Implications in the Field of Hypersonic Aerothermodynamics.”
\end{acknowledgments}

%%%%%%%%%%%%%%%%%%%%%%%%%%%%%%%%%%%%%%%%%%%%%%%%%%%%%%%%%%%%%%%%%%%%

\section*{Declaration of interests.}

The authors report no conflict of interest.

%%%%%%%%%%%%%%%%%%%%%%%%%%%%%%%%%%%%%%%%%%%%%%%%%%%%%%%%%%%%%%%%%%%%

%%%%%%%%%%%%%%%%%%%%%%%%%%%%%%%%%%%%%%%%%%%%%%%%%%%%%%%%%%%%%%%%%%%%%%%%%%%%%%%%%%%%%%%

\appendix

%%%%%%%%%%%%%%%%%%%%%%%%%%%%%%%%%%%%%%%%%%%%%%%%%%%%%%%%%%%%%%%%%%%%%%%%%%%%%%%%%%%%%%%%
\section{}\label{appendixA}

Several correlations for compressible skin friction on a flat plate exist in the literature. The most commonly used models to estimate skin friction associated with a compressible boundary layer in terms of an equivalent incompressible boundary layer are van Driest II~\cite{van1956problem}, White–Christoph~\cite{white1972simple}, and Spalding-Chi~\cite{spalding1964drag}. These correlation theories use a compressibility transformation idea:
    \begin{equation}
        C_f = \dfrac{1}{F_c} C_{f,in},
    \end{equation}
    \begin{equation}
    \begin{split}
        Re_x = \dfrac{1}{F_x} Re_{x,in}, \\
        Re_\theta = \dfrac{1}{F_\theta} Re_{\theta,in}, 
    \end{split}
    \end{equation}
where $C_{f,in}$, $Re_{x,in}$, and $Re_{\theta,in}$ are the equivalent `incompressible' skin friction coefficient and surface distance Reynolds number, and momentum thickness Reynolds number, respectively. On the other hand, $F_c$ is the skin friction transformation function whereas $F_x$ and $F_\theta$ are surface distance and momentum thickness based Reynolds number transformation functions, respectively. The commonly used incompressible skin friction correlations are the $C_{f,in}$ vs $Re_{\theta,in}$ correlation by: \\
Kármán–Schoenherr~\cite{hopkins1971evaluation}:
    \begin{equation}
       C_{f,in} = \dfrac{1}{\log_{10} (2 Re_{\theta,in})[17.075 \log_{10} (2 Re_{\theta,in})+14.832]},
    \end{equation}
Smits \textit{et al.}~\cite{smits1983low} :
    \begin{equation}
       C_{f,in} = 0.024 Re_{\theta,in}^{-1/4},
    \end{equation}
Coles–Fernholz~\cite{nagib2007approach} :
    \begin{equation}
       C_{f,in} = 2 \left[ 2.604 \log Re_{\theta,in} + 4.127 \right]^{-2}, 
    \end{equation}
and $C_{f,in}$ vs $Re_{x,in}$ correlation by: \\
White \cite{white2006viscous} :
    \begin{equation}
        C_{f,in} = \dfrac{0.455}{F_c \ln^{2} (0.06 Re_x F_x)}; F_x = \dfrac{F_{\theta}}{F_c}.
    \end{equation}
The transformation factor $F_c$ for the van Driest II, Spalding-Chi, and White-Christoph theories can be written as~~\cite{rumsey2010compressibility}
\begin{equation}
     F_c = \dfrac{T_{aw}/T_e - 1}{(\sin^{-1}A + \sin^{-1}B)^2},
 \end{equation}
 where $T_e$ refers to free stream or boundary layer edge temperature, and 
 \begin{equation}
     T_{aw} = T_e \left( 1 + r \dfrac{\gamma-1}{2} M_{e}^2 \right). 
 \end{equation}
 The recovery factor $r$ is taken to be $0.9$. The $A$ and $B$ are given by 
 \begin{equation}
     A = \dfrac{2 a^2 - b }{(b^2 + 4 a^2)^{1/2}},
 \end{equation}
 \begin{equation}
     B = \dfrac{b }{(b^2 + 4 a^2)^{1/2}},
 \end{equation}
 where 
  \begin{equation}
     a = \left( r \dfrac{\gamma-1}{2} M_{e}^2 \dfrac{T_e}{T_{aw}} \right),
 \end{equation}
 \begin{equation}
     b = \dfrac{T_{aw}}{T_w} -1 = \dfrac{T_e}{T_{aw}}  \left( 1 + r \dfrac{\gamma-1}{2} M_{e}^2 \right) - 1.
 \end{equation}
The transformation factor $F_\theta$ is different between the three theories and is given by \\
van Driest II : 
 \begin{equation}
     F_{\theta} = \dfrac{\mu_e}{\mu_w},
 \end{equation}
Spalding-Chi :
 \begin{equation}
      F_{\theta} = \left( \dfrac{T_{w}}{T_e} \right)^{-0.702}  \left( \dfrac{T_{aw}}{T_w} \right)^{0.772}, 
 \end{equation}
 White-Christoph :
 \begin{equation}
      F_{\theta} = \sqrt{F_c} \dfrac{\mu_e}{\mu_w} \left( \dfrac{T_{e}}{T_w} \right)^{1/2}. 
 \end{equation}

%\nocite{*}
\bibliography{Manuscript_Final_Version_R1_PoF}% Produces the bibliography via BibTeX.

\begin{thebibliography}{346}%
\makeatletter
\providecommand \@ifxundefined [1]{%
 \@ifx{#1\undefined}
}%
\providecommand \@ifnum [1]{%
 \ifnum #1\expandafter \@firstoftwo
 \else \expandafter \@secondoftwo
 \fi
}%
\providecommand \@ifx [1]{%
 \ifx #1\expandafter \@firstoftwo
 \else \expandafter \@secondoftwo
 \fi
}%
\providecommand \natexlab [1]{#1}%
\providecommand \enquote  [1]{``#1''}%
\providecommand \bibnamefont  [1]{#1}%
\providecommand \bibfnamefont [1]{#1}%
\providecommand \citenamefont [1]{#1}%
\providecommand \href@noop [0]{\@secondoftwo}%
\providecommand \href [0]{\begingroup \@sanitize@url \@href}%
\providecommand \@href[1]{\@@startlink{#1}\@@href}%
\providecommand \@@href[1]{\endgroup#1\@@endlink}%
\providecommand \@sanitize@url [0]{\catcode `\\12\catcode `\$12\catcode `\&12\catcode `\#12\catcode `\^12\catcode `\_12\catcode `\%12\relax}%
\providecommand \@@startlink[1]{}%
\providecommand \@@endlink[0]{}%
\providecommand \url  [0]{\begingroup\@sanitize@url \@url }%
\providecommand \@url [1]{\endgroup\@href {#1}{\urlprefix }}%
\providecommand \urlprefix  [0]{URL }%
\providecommand \Eprint [0]{\href }%
\providecommand \doibase [0]{http://dx.doi.org/}%
\providecommand \selectlanguage [0]{\@gobble}%
\providecommand \bibinfo  [0]{\@secondoftwo}%
\providecommand \bibfield  [0]{\@secondoftwo}%
\providecommand \translation [1]{[#1]}%
\providecommand \BibitemOpen [0]{}%
\providecommand \bibitemStop [0]{}%
\providecommand \bibitemNoStop [0]{.\EOS\space}%
\providecommand \EOS [0]{\spacefactor3000\relax}%
\providecommand \BibitemShut  [1]{\csname bibitem#1\endcsname}%
\let\auto@bib@innerbib\@empty
%</preamble>
\bibitem [{\citenamefont {Anderson}(1989)}]{anderson1989hypersonic}%
  \BibitemOpen
  \bibfield  {author} {\bibinfo {author} {\bibfnamefont {J.~D.}\ \bibnamefont {Anderson}},\ }\href@noop {} {\emph {\bibinfo {title} {Hypersonic and high temperature gas dynamics}}}\ (\bibinfo  {publisher} {Aiaa},\ \bibinfo {year} {1989})\BibitemShut {NoStop}%
\bibitem [{\citenamefont {Urzay}(2020)}]{urzay2020physical}%
  \BibitemOpen
  \bibfield  {author} {\bibinfo {author} {\bibfnamefont {J.}~\bibnamefont {Urzay}},\ }\bibfield  {title} {\enquote {\bibinfo {title} {The physical characteristics of hypersonic flows},}\ }\href@noop {} {\bibfield  {journal} {\bibinfo  {journal} {In ME356 Hypersonic Aerothermodynamics class notes. Center for Turbulence Research, Stanford University}\ } (\bibinfo {year} {2020})}\BibitemShut {NoStop}%
\bibitem [{\citenamefont {Moin}\ and\ \citenamefont {Mahesh}(1998)}]{moin1998direct}%
  \BibitemOpen
  \bibfield  {author} {\bibinfo {author} {\bibfnamefont {P.}~\bibnamefont {Moin}}\ and\ \bibinfo {author} {\bibfnamefont {K.}~\bibnamefont {Mahesh}},\ }\bibfield  {title} {\enquote {\bibinfo {title} {Direct numerical simulation: a tool in turbulence research},}\ }\href@noop {} {\bibfield  {journal} {\bibinfo  {journal} {Annual review of fluid mechanics}\ }\textbf {\bibinfo {volume} {30}},\ \bibinfo {pages} {539--578} (\bibinfo {year} {1998})}\BibitemShut {NoStop}%
\bibitem [{\citenamefont {Zhiyin}(2015)}]{zhiyin2015large}%
  \BibitemOpen
  \bibfield  {author} {\bibinfo {author} {\bibfnamefont {Y.}~\bibnamefont {Zhiyin}},\ }\bibfield  {title} {\enquote {\bibinfo {title} {Large-eddy simulation: Past, present and the future},}\ }\href@noop {} {\bibfield  {journal} {\bibinfo  {journal} {Chinese journal of Aeronautics}\ }\textbf {\bibinfo {volume} {28}},\ \bibinfo {pages} {11--24} (\bibinfo {year} {2015})}\BibitemShut {NoStop}%
\bibitem [{\citenamefont {Choi}\ and\ \citenamefont {Moin}(2012)}]{choi2012grid}%
  \BibitemOpen
  \bibfield  {author} {\bibinfo {author} {\bibfnamefont {H.}~\bibnamefont {Choi}}\ and\ \bibinfo {author} {\bibfnamefont {P.}~\bibnamefont {Moin}},\ }\bibfield  {title} {\enquote {\bibinfo {title} {Grid-point requirements for large eddy simulation: Chapman’s estimates revisited},}\ }\href@noop {} {\bibfield  {journal} {\bibinfo  {journal} {Physics of fluids}\ }\textbf {\bibinfo {volume} {24}} (\bibinfo {year} {2012})}\BibitemShut {NoStop}%
\bibitem [{\citenamefont {Yang}\ and\ \citenamefont {Griffin}(2021)}]{yang2021grid}%
  \BibitemOpen
  \bibfield  {author} {\bibinfo {author} {\bibfnamefont {X.~I.}\ \bibnamefont {Yang}}\ and\ \bibinfo {author} {\bibfnamefont {K.~P.}\ \bibnamefont {Griffin}},\ }\bibfield  {title} {\enquote {\bibinfo {title} {Grid-point and time-step requirements for direct numerical simulation and large-eddy simulation},}\ }\href@noop {} {\bibfield  {journal} {\bibinfo  {journal} {Physics of Fluids}\ }\textbf {\bibinfo {volume} {33}} (\bibinfo {year} {2021})}\BibitemShut {NoStop}%
\bibitem [{\citenamefont {Bose}\ and\ \citenamefont {Park}(2018)}]{bose2018wall}%
  \BibitemOpen
  \bibfield  {author} {\bibinfo {author} {\bibfnamefont {S.~T.}\ \bibnamefont {Bose}}\ and\ \bibinfo {author} {\bibfnamefont {G.~I.}\ \bibnamefont {Park}},\ }\bibfield  {title} {\enquote {\bibinfo {title} {Wall-modeled large-eddy simulation for complex turbulent flows},}\ }\href@noop {} {\bibfield  {journal} {\bibinfo  {journal} {Annual review of fluid mechanics}\ }\textbf {\bibinfo {volume} {50}},\ \bibinfo {pages} {535--561} (\bibinfo {year} {2018})}\BibitemShut {NoStop}%
\bibitem [{\citenamefont {Spalart}(2009)}]{spalart2009detached}%
  \BibitemOpen
  \bibfield  {author} {\bibinfo {author} {\bibfnamefont {P.~R.}\ \bibnamefont {Spalart}},\ }\bibfield  {title} {\enquote {\bibinfo {title} {Detached-eddy simulation},}\ }\href@noop {} {\bibfield  {journal} {\bibinfo  {journal} {Annual review of fluid mechanics}\ }\textbf {\bibinfo {volume} {41}},\ \bibinfo {pages} {181--202} (\bibinfo {year} {2009})}\BibitemShut {NoStop}%
\bibitem [{\citenamefont {Sciacovelli}\ \emph {et~al.}(2023)\citenamefont {Sciacovelli}, \citenamefont {Cannici}, \citenamefont {Passiatore},\ and\ \citenamefont {Cinnella}}]{sciacovelli2023priori}%
  \BibitemOpen
  \bibfield  {author} {\bibinfo {author} {\bibfnamefont {L.}~\bibnamefont {Sciacovelli}}, \bibinfo {author} {\bibfnamefont {A.}~\bibnamefont {Cannici}}, \bibinfo {author} {\bibfnamefont {D.}~\bibnamefont {Passiatore}}, \ and\ \bibinfo {author} {\bibfnamefont {P.}~\bibnamefont {Cinnella}},\ }\bibfield  {title} {\enquote {\bibinfo {title} {A priori tests of turbulence models for compressible flows},}\ }\href@noop {} {\bibfield  {journal} {\bibinfo  {journal} {International Journal of Numerical Methods for Heat \& Fluid Flow}\ } (\bibinfo {year} {2023})}\BibitemShut {NoStop}%
\bibitem [{\citenamefont {Marchenay}, \citenamefont {Olazabal~Loum{\'e}},\ and\ \citenamefont {Chedevergne}(2022)}]{marchenay2022hypersonic}%
  \BibitemOpen
  \bibfield  {author} {\bibinfo {author} {\bibfnamefont {Y.}~\bibnamefont {Marchenay}}, \bibinfo {author} {\bibfnamefont {M.}~\bibnamefont {Olazabal~Loum{\'e}}}, \ and\ \bibinfo {author} {\bibfnamefont {F.}~\bibnamefont {Chedevergne}},\ }\bibfield  {title} {\enquote {\bibinfo {title} {Hypersonic turbulent flow reynolds-averaged navier--stokes simulations with roughness and blowing effects},}\ }\href@noop {} {\bibfield  {journal} {\bibinfo  {journal} {Journal of Spacecraft and Rockets}\ }\textbf {\bibinfo {volume} {59}},\ \bibinfo {pages} {1686--1696} (\bibinfo {year} {2022})}\BibitemShut {NoStop}%
\bibitem [{\citenamefont {Marvin}(1983)}]{marvin1983turbulence}%
  \BibitemOpen
  \bibfield  {author} {\bibinfo {author} {\bibfnamefont {J.~G.}\ \bibnamefont {Marvin}},\ }\bibfield  {title} {\enquote {\bibinfo {title} {Turbulence modeling for computational aerodynamics},}\ }\href@noop {} {\bibfield  {journal} {\bibinfo  {journal} {AIAA Journal}\ }\textbf {\bibinfo {volume} {21}},\ \bibinfo {pages} {941--955} (\bibinfo {year} {1983})}\BibitemShut {NoStop}%
\bibitem [{\citenamefont {Marvin}\ and\ \citenamefont {Coakley}(1989)}]{marvin1989turbulence}%
  \BibitemOpen
  \bibfield  {author} {\bibinfo {author} {\bibfnamefont {J.~G.}\ \bibnamefont {Marvin}}\ and\ \bibinfo {author} {\bibfnamefont {T.~J.}\ \bibnamefont {Coakley}},\ }\href@noop {} {\enquote {\bibinfo {title} {Turbulence modeling for hypersonic flows},}\ }\bibinfo {type} {Tech. Rep.}\ (\bibinfo {year} {1989})\BibitemShut {NoStop}%
\bibitem [{\citenamefont {Roy}\ and\ \citenamefont {Blottner}(2006)}]{roy2006review}%
  \BibitemOpen
  \bibfield  {author} {\bibinfo {author} {\bibfnamefont {C.~J.}\ \bibnamefont {Roy}}\ and\ \bibinfo {author} {\bibfnamefont {F.~G.}\ \bibnamefont {Blottner}},\ }\bibfield  {title} {\enquote {\bibinfo {title} {Review and assessment of turbulence models for hypersonic flows},}\ }\href@noop {} {\bibfield  {journal} {\bibinfo  {journal} {Progress in aerospace sciences}\ }\textbf {\bibinfo {volume} {42}},\ \bibinfo {pages} {469--530} (\bibinfo {year} {2006})}\BibitemShut {NoStop}%
\bibitem [{\citenamefont {Spalart}\ and\ \citenamefont {Allmaras}(1992)}]{spalart1992one}%
  \BibitemOpen
  \bibfield  {author} {\bibinfo {author} {\bibfnamefont {P.}~\bibnamefont {Spalart}}\ and\ \bibinfo {author} {\bibfnamefont {S.}~\bibnamefont {Allmaras}},\ }\bibfield  {title} {\enquote {\bibinfo {title} {A one-equation turbulence model for aerodynamic flows},}\ }in\ \href@noop {} {\emph {\bibinfo {booktitle} {30th aerospace sciences meeting and exhibit}}}\ (\bibinfo {year} {1992})\ p.\ \bibinfo {pages} {439}\BibitemShut {NoStop}%
\bibitem [{\citenamefont {Settles}\ and\ \citenamefont {Dodson}(1991)}]{settles1991}%
  \BibitemOpen
  \bibfield  {author} {\bibinfo {author} {\bibfnamefont {G.}~\bibnamefont {Settles}}\ and\ \bibinfo {author} {\bibfnamefont {L.}~\bibnamefont {Dodson}},\ }\href@noop {} {\enquote {\bibinfo {title} {Hypersonic shock/boundary-layer interaction database},}\ }\bibinfo {type} {Tech. Rep.}\ (\bibinfo  {institution} {NASA CR 177577},\ \bibinfo {year} {1991})\BibitemShut {NoStop}%
\bibitem [{\citenamefont {Marvin}, \citenamefont {Brown},\ and\ \citenamefont {Gnoffo}(2013)}]{marvin2013experimental}%
  \BibitemOpen
  \bibfield  {author} {\bibinfo {author} {\bibfnamefont {J.~G.}\ \bibnamefont {Marvin}}, \bibinfo {author} {\bibfnamefont {J.~L.}\ \bibnamefont {Brown}}, \ and\ \bibinfo {author} {\bibfnamefont {P.~A.}\ \bibnamefont {Gnoffo}},\ }\href@noop {} {\enquote {\bibinfo {title} {Experimental database with baseline cfd solutions: 2-d and axisymmetric hypersonic shock-wave/turbulent-boundary-layer interactions},}\ }\bibinfo {type} {Tech. Rep.}\ (\bibinfo {year} {2013})\BibitemShut {NoStop}%
\bibitem [{\citenamefont {Bardina}, \citenamefont {Coakley},\ and\ \citenamefont {Marvin}(1992)}]{bardina1992two}%
  \BibitemOpen
  \bibfield  {author} {\bibinfo {author} {\bibfnamefont {J.}~\bibnamefont {Bardina}}, \bibinfo {author} {\bibfnamefont {T.}~\bibnamefont {Coakley}}, \ and\ \bibinfo {author} {\bibfnamefont {J.}~\bibnamefont {Marvin}},\ }\bibfield  {title} {\enquote {\bibinfo {title} {Two-equation turbulence modeling for 3-d hypersonic flows},}\ }in\ \href@noop {} {\emph {\bibinfo {booktitle} {AlAA 4th International Aerospace Planes Conference}}}\ (\bibinfo {year} {1992})\ p.\ \bibinfo {pages} {5064}\BibitemShut {NoStop}%
\bibitem [{\citenamefont {Bardina}(1994)}]{bardina1994three}%
  \BibitemOpen
  \bibfield  {author} {\bibinfo {author} {\bibfnamefont {J.}~\bibnamefont {Bardina}},\ }\bibfield  {title} {\enquote {\bibinfo {title} {Three-dimensional navier-stokes method with two-equation turbulence models for efficient numerical simulation of hypersonic flows},}\ }in\ \href@noop {} {\emph {\bibinfo {booktitle} {30th Joint Propulsion Conference and Exhibit}}}\ (\bibinfo {year} {1994})\ p.\ \bibinfo {pages} {2950}\BibitemShut {NoStop}%
\bibitem [{\citenamefont {Smits}, \citenamefont {Martin},\ and\ \citenamefont {Girimaji}(2009)}]{smits2009current}%
  \BibitemOpen
  \bibfield  {author} {\bibinfo {author} {\bibfnamefont {L.}~\bibnamefont {Smits}}, \bibinfo {author} {\bibfnamefont {P.}~\bibnamefont {Martin}}, \ and\ \bibinfo {author} {\bibfnamefont {S.}~\bibnamefont {Girimaji}},\ }\bibfield  {title} {\enquote {\bibinfo {title} {Current status of basic research in hypersonic turbulence},}\ }in\ \href@noop {} {\emph {\bibinfo {booktitle} {47th AIAA Aerospace Sciences Meeting including The New Horizons Forum and Aerospace Exposition}}}\ (\bibinfo {year} {2009})\ p.\ \bibinfo {pages} {151}\BibitemShut {NoStop}%
\bibitem [{\citenamefont {Georgiadis}\ \emph {et~al.}(2014)\citenamefont {Georgiadis}, \citenamefont {Yoder}, \citenamefont {Vyas},\ and\ \citenamefont {Engblom}}]{georgiadis2014status}%
  \BibitemOpen
  \bibfield  {author} {\bibinfo {author} {\bibfnamefont {N.~J.}\ \bibnamefont {Georgiadis}}, \bibinfo {author} {\bibfnamefont {D.~A.}\ \bibnamefont {Yoder}}, \bibinfo {author} {\bibfnamefont {M.~A.}\ \bibnamefont {Vyas}}, \ and\ \bibinfo {author} {\bibfnamefont {W.~A.}\ \bibnamefont {Engblom}},\ }\bibfield  {title} {\enquote {\bibinfo {title} {Status of turbulence modeling for hypersonic propulsion flowpaths},}\ }\href@noop {} {\bibfield  {journal} {\bibinfo  {journal} {Theoretical and Computational Fluid Dynamics}\ }\textbf {\bibinfo {volume} {28}},\ \bibinfo {pages} {295--318} (\bibinfo {year} {2014})}\BibitemShut {NoStop}%
\bibitem [{\citenamefont {Gnoffo}, \citenamefont {Berry},\ and\ \citenamefont {Van~Norman}(2013)}]{gnoffo2013uncertainty}%
  \BibitemOpen
  \bibfield  {author} {\bibinfo {author} {\bibfnamefont {P.~A.}\ \bibnamefont {Gnoffo}}, \bibinfo {author} {\bibfnamefont {S.~A.}\ \bibnamefont {Berry}}, \ and\ \bibinfo {author} {\bibfnamefont {J.~W.}\ \bibnamefont {Van~Norman}},\ }\bibfield  {title} {\enquote {\bibinfo {title} {Uncertainty assessments of hypersonic shock wave-turbulent boundary-layer interactions at compression corners},}\ }\href@noop {} {\bibfield  {journal} {\bibinfo  {journal} {Journal of Spacecraft and Rockets}\ }\textbf {\bibinfo {volume} {50}},\ \bibinfo {pages} {69--95} (\bibinfo {year} {2013})}\BibitemShut {NoStop}%
\bibitem [{\citenamefont {Brown}(2002)}]{brown2002turbulence}%
  \BibitemOpen
  \bibfield  {author} {\bibinfo {author} {\bibfnamefont {J.}~\bibnamefont {Brown}},\ }\bibfield  {title} {\enquote {\bibinfo {title} {Turbulence model validation for hypersonic flows},}\ }in\ \href@noop {} {\emph {\bibinfo {booktitle} {8th AIAA/ASME Joint Thermophysics and Heat Transfer Conference}}}\ (\bibinfo {year} {2002})\ p.\ \bibinfo {pages} {3308}\BibitemShut {NoStop}%
\bibitem [{\citenamefont {Aiken}\ \emph {et~al.}(2022)\citenamefont {Aiken}, \citenamefont {Boyd}, \citenamefont {Duan},\ and\ \citenamefont {Huang}}]{aiken2022assessment}%
  \BibitemOpen
  \bibfield  {author} {\bibinfo {author} {\bibfnamefont {T.~T.}\ \bibnamefont {Aiken}}, \bibinfo {author} {\bibfnamefont {I.~D.}\ \bibnamefont {Boyd}}, \bibinfo {author} {\bibfnamefont {L.}~\bibnamefont {Duan}}, \ and\ \bibinfo {author} {\bibfnamefont {J.}~\bibnamefont {Huang}},\ }\bibfield  {title} {\enquote {\bibinfo {title} {Assessment of reynolds averaged navier-stokes models for a hypersonic cold-wall turbulent boundary layer},}\ }in\ \href@noop {} {\emph {\bibinfo {booktitle} {AIAA SciTech 2022 Forum}}}\ (\bibinfo {year} {2022})\ p.\ \bibinfo {pages} {0586}\BibitemShut {NoStop}%
\bibitem [{\citenamefont {Huang}, \citenamefont {Bretzke},\ and\ \citenamefont {Duan}(2019)}]{huang2019assessment}%
  \BibitemOpen
  \bibfield  {author} {\bibinfo {author} {\bibfnamefont {J.}~\bibnamefont {Huang}}, \bibinfo {author} {\bibfnamefont {J.-V.}\ \bibnamefont {Bretzke}}, \ and\ \bibinfo {author} {\bibfnamefont {L.}~\bibnamefont {Duan}},\ }\bibfield  {title} {\enquote {\bibinfo {title} {Assessment of turbulence models in a hypersonic cold-wall turbulent boundary layer},}\ }\href@noop {} {\bibfield  {journal} {\bibinfo  {journal} {Fluids}\ }\textbf {\bibinfo {volume} {4}},\ \bibinfo {pages} {37} (\bibinfo {year} {2019})}\BibitemShut {NoStop}%
\bibitem [{\citenamefont {Rumsey}(2010)}]{rumsey2010compressibility}%
  \BibitemOpen
  \bibfield  {author} {\bibinfo {author} {\bibfnamefont {C.~L.}\ \bibnamefont {Rumsey}},\ }\bibfield  {title} {\enquote {\bibinfo {title} {Compressibility considerations for kw turbulence models in hypersonic boundary-layer applications},}\ }\href@noop {} {\bibfield  {journal} {\bibinfo  {journal} {Journal of Spacecraft and Rockets}\ }\textbf {\bibinfo {volume} {47}},\ \bibinfo {pages} {11--20} (\bibinfo {year} {2010})}\BibitemShut {NoStop}%
\bibitem [{\citenamefont {Smith}\ and\ \citenamefont {Cebeci}(1967)}]{smith1967numerical}%
  \BibitemOpen
  \bibfield  {author} {\bibinfo {author} {\bibfnamefont {A.}~\bibnamefont {Smith}}\ and\ \bibinfo {author} {\bibfnamefont {T.}~\bibnamefont {Cebeci}},\ }\href@noop {} {\emph {\bibinfo {title} {Numerical solution of the turbulent-boundary-layer equations}}}\ (\bibinfo  {publisher} {Douglas Aircraft Company, Douglas Aircraft Division},\ \bibinfo {year} {1967})\BibitemShut {NoStop}%
\bibitem [{\citenamefont {Baldwin}\ and\ \citenamefont {Lomax}(1978)}]{baldwin1978thin}%
  \BibitemOpen
  \bibfield  {author} {\bibinfo {author} {\bibfnamefont {B.}~\bibnamefont {Baldwin}}\ and\ \bibinfo {author} {\bibfnamefont {H.}~\bibnamefont {Lomax}},\ }\bibfield  {title} {\enquote {\bibinfo {title} {Thin-layer approximation and algebraic model for separated turbulentflows},}\ }in\ \href@noop {} {\emph {\bibinfo {booktitle} {16th aerospace sciences meeting}}}\ (\bibinfo {year} {1978})\ p.\ \bibinfo {pages} {257}\BibitemShut {NoStop}%
\bibitem [{\citenamefont {Menter}(1994)}]{menter1994two}%
  \BibitemOpen
  \bibfield  {author} {\bibinfo {author} {\bibfnamefont {F.~R.}\ \bibnamefont {Menter}},\ }\bibfield  {title} {\enquote {\bibinfo {title} {Two-equation eddy-viscosity turbulence models for engineering applications},}\ }\href@noop {} {\bibfield  {journal} {\bibinfo  {journal} {AIAA journal}\ }\textbf {\bibinfo {volume} {32}},\ \bibinfo {pages} {1598--1605} (\bibinfo {year} {1994})}\BibitemShut {NoStop}%
\bibitem [{\citenamefont {Passiatore}\ \emph {et~al.}(2021)\citenamefont {Passiatore}, \citenamefont {Sciacovelli}, \citenamefont {Cinnella},\ and\ \citenamefont {Pascazio}}]{passiatore2021finite}%
  \BibitemOpen
  \bibfield  {author} {\bibinfo {author} {\bibfnamefont {D.}~\bibnamefont {Passiatore}}, \bibinfo {author} {\bibfnamefont {L.}~\bibnamefont {Sciacovelli}}, \bibinfo {author} {\bibfnamefont {P.}~\bibnamefont {Cinnella}}, \ and\ \bibinfo {author} {\bibfnamefont {G.}~\bibnamefont {Pascazio}},\ }\bibfield  {title} {\enquote {\bibinfo {title} {Finite-rate chemistry effects in turbulent hypersonic boundary layers: A direct numerical simulation study},}\ }\href@noop {} {\bibfield  {journal} {\bibinfo  {journal} {Physical Review Fluids}\ }\textbf {\bibinfo {volume} {6}},\ \bibinfo {pages} {054604} (\bibinfo {year} {2021})}\BibitemShut {NoStop}%
\bibitem [{\citenamefont {Passiatore}\ \emph {et~al.}(2022)\citenamefont {Passiatore}, \citenamefont {Sciacovelli}, \citenamefont {Cinnella},\ and\ \citenamefont {Pascazio}}]{passiatore2022thermochemical}%
  \BibitemOpen
  \bibfield  {author} {\bibinfo {author} {\bibfnamefont {D.}~\bibnamefont {Passiatore}}, \bibinfo {author} {\bibfnamefont {L.}~\bibnamefont {Sciacovelli}}, \bibinfo {author} {\bibfnamefont {P.}~\bibnamefont {Cinnella}}, \ and\ \bibinfo {author} {\bibfnamefont {G.}~\bibnamefont {Pascazio}},\ }\bibfield  {title} {\enquote {\bibinfo {title} {Thermochemical non-equilibrium effects in turbulent hypersonic boundary layers},}\ }\href@noop {} {\bibfield  {journal} {\bibinfo  {journal} {Journal of Fluid Mechanics}\ }\textbf {\bibinfo {volume} {941}},\ \bibinfo {pages} {A21} (\bibinfo {year} {2022})}\BibitemShut {NoStop}%
\bibitem [{\citenamefont {Candler}(2019)}]{candler2019rate}%
  \BibitemOpen
  \bibfield  {author} {\bibinfo {author} {\bibfnamefont {G.~V.}\ \bibnamefont {Candler}},\ }\bibfield  {title} {\enquote {\bibinfo {title} {Rate effects in hypersonic flows},}\ }\href@noop {} {\bibfield  {journal} {\bibinfo  {journal} {Annual Review of Fluid Mechanics}\ }\textbf {\bibinfo {volume} {51}},\ \bibinfo {pages} {379--402} (\bibinfo {year} {2019})}\BibitemShut {NoStop}%
\bibitem [{\citenamefont {Park}(1989)}]{park1989nonequilibrium}%
  \BibitemOpen
  \bibfield  {author} {\bibinfo {author} {\bibfnamefont {C.}~\bibnamefont {Park}},\ }\bibfield  {title} {\enquote {\bibinfo {title} {Nonequilibrium hypersonic aerothermodynamics},}\ }\href@noop {} {\  (\bibinfo {year} {1989})}\BibitemShut {NoStop}%
\bibitem [{\citenamefont {Georgiadis}, \citenamefont {Mankbadi},\ and\ \citenamefont {Vyas}(2014)}]{georgiadis2014turbulence}%
  \BibitemOpen
  \bibfield  {author} {\bibinfo {author} {\bibfnamefont {N.~J.}\ \bibnamefont {Georgiadis}}, \bibinfo {author} {\bibfnamefont {M.~R.}\ \bibnamefont {Mankbadi}}, \ and\ \bibinfo {author} {\bibfnamefont {M.~A.}\ \bibnamefont {Vyas}},\ }\bibfield  {title} {\enquote {\bibinfo {title} {Turbulence model effects on rans simulations of the hifire flight 2 ground test configurations},}\ }in\ \href@noop {} {\emph {\bibinfo {booktitle} {52nd Aerospace Sciences Meeting}}}\ (\bibinfo {year} {2014})\ p.\ \bibinfo {pages} {0624}\BibitemShut {NoStop}%
\bibitem [{\citenamefont {Bowersox}(2009)}]{bowersox2009extension}%
  \BibitemOpen
  \bibfield  {author} {\bibinfo {author} {\bibfnamefont {R.~D.}\ \bibnamefont {Bowersox}},\ }\bibfield  {title} {\enquote {\bibinfo {title} {Extension of equilibrium turbulent heat flux models to high-speed shear flows},}\ }\href@noop {} {\bibfield  {journal} {\bibinfo  {journal} {Journal of fluid mechanics}\ }\textbf {\bibinfo {volume} {633}},\ \bibinfo {pages} {61--70} (\bibinfo {year} {2009})}\BibitemShut {NoStop}%
\bibitem [{\citenamefont {Bowersox}\ and\ \citenamefont {North}(2010)}]{bowersox2010algebraic}%
  \BibitemOpen
  \bibfield  {author} {\bibinfo {author} {\bibfnamefont {R.~D.}\ \bibnamefont {Bowersox}}\ and\ \bibinfo {author} {\bibfnamefont {S.~W.}\ \bibnamefont {North}},\ }\bibfield  {title} {\enquote {\bibinfo {title} {Algebraic turbulent energy flux models for hypersonic shear flows},}\ }\href@noop {} {\bibfield  {journal} {\bibinfo  {journal} {Progress in Aerospace Sciences}\ }\textbf {\bibinfo {volume} {46}},\ \bibinfo {pages} {49--61} (\bibinfo {year} {2010})}\BibitemShut {NoStop}%
\bibitem [{\citenamefont {Urzay}\ and\ \citenamefont {Di~Renzo}(2021)}]{urzay2021engineering}%
  \BibitemOpen
  \bibfield  {author} {\bibinfo {author} {\bibfnamefont {J.}~\bibnamefont {Urzay}}\ and\ \bibinfo {author} {\bibfnamefont {M.}~\bibnamefont {Di~Renzo}},\ }\bibfield  {title} {\enquote {\bibinfo {title} {Engineering aspects of hypersonic turbulent flows at suborbital enthalpies},}\ }\href@noop {} {\bibfield  {journal} {\bibinfo  {journal} {Annual Research Briefs, Center for Turbulence Research}\ ,\ \bibinfo {pages} {7--32}} (\bibinfo {year} {2021})}\BibitemShut {NoStop}%
\bibitem [{\citenamefont {Di~Renzo}\ and\ \citenamefont {Urzay}(2021)}]{direnzo2021direct}%
  \BibitemOpen
  \bibfield  {author} {\bibinfo {author} {\bibfnamefont {M.}~\bibnamefont {Di~Renzo}}\ and\ \bibinfo {author} {\bibfnamefont {J.}~\bibnamefont {Urzay}},\ }\bibfield  {title} {\enquote {\bibinfo {title} {Direct numerical simulation of a hypersonic transitional boundary layer at suborbital enthalpies},}\ }\href@noop {} {\bibfield  {journal} {\bibinfo  {journal} {Journal of Fluid Mechanics}\ }\textbf {\bibinfo {volume} {912}} (\bibinfo {year} {2021})}\BibitemShut {NoStop}%
\bibitem [{\citenamefont {Volpiani}(2021)}]{volpiani2021numerical}%
  \BibitemOpen
  \bibfield  {author} {\bibinfo {author} {\bibfnamefont {P.~S.}\ \bibnamefont {Volpiani}},\ }\bibfield  {title} {\enquote {\bibinfo {title} {Numerical strategy to perform direct numerical simulations of hypersonic shock/boundary-layer interaction in chemical nonequilibrium},}\ }\href@noop {} {\bibfield  {journal} {\bibinfo  {journal} {Shock Waves}\ }\textbf {\bibinfo {volume} {31}},\ \bibinfo {pages} {361--378} (\bibinfo {year} {2021})}\BibitemShut {NoStop}%
\bibitem [{\citenamefont {Passiatore}\ \emph {et~al.}(2023)\citenamefont {Passiatore}, \citenamefont {Sciacovelli}, \citenamefont {Cinnella},\ and\ \citenamefont {Pascazio}}]{passiatore2023shock}%
  \BibitemOpen
  \bibfield  {author} {\bibinfo {author} {\bibfnamefont {D.}~\bibnamefont {Passiatore}}, \bibinfo {author} {\bibfnamefont {L.}~\bibnamefont {Sciacovelli}}, \bibinfo {author} {\bibfnamefont {P.}~\bibnamefont {Cinnella}}, \ and\ \bibinfo {author} {\bibfnamefont {G.}~\bibnamefont {Pascazio}},\ }\bibfield  {title} {\enquote {\bibinfo {title} {Shock impingement on a transitional hypersonic high-enthalpy boundary layer},}\ }\href@noop {} {\bibfield  {journal} {\bibinfo  {journal} {Physical Review Fluids}\ }\textbf {\bibinfo {volume} {8}},\ \bibinfo {pages} {044601} (\bibinfo {year} {2023})}\BibitemShut {NoStop}%
\bibitem [{\citenamefont {Di~Renzo}, \citenamefont {Williams},\ and\ \citenamefont {Pirozzoli}(2024)}]{di2024stagnation}%
  \BibitemOpen
  \bibfield  {author} {\bibinfo {author} {\bibfnamefont {M.}~\bibnamefont {Di~Renzo}}, \bibinfo {author} {\bibfnamefont {C.}~\bibnamefont {Williams}}, \ and\ \bibinfo {author} {\bibfnamefont {S.}~\bibnamefont {Pirozzoli}},\ }\bibfield  {title} {\enquote {\bibinfo {title} {Stagnation enthalpy effects on hypersonic turbulent compression corner flow at moderate reynolds numbers},}\ }\href@noop {} {\bibfield  {journal} {\bibinfo  {journal} {Physical Review Fluids}\ }\textbf {\bibinfo {volume} {9}},\ \bibinfo {pages} {033401} (\bibinfo {year} {2024})}\BibitemShut {NoStop}%
\bibitem [{\citenamefont {Duraisamy}, \citenamefont {Iaccarino},\ and\ \citenamefont {Xiao}(2019)}]{duraisamy2019turbulence}%
  \BibitemOpen
  \bibfield  {author} {\bibinfo {author} {\bibfnamefont {K.}~\bibnamefont {Duraisamy}}, \bibinfo {author} {\bibfnamefont {G.}~\bibnamefont {Iaccarino}}, \ and\ \bibinfo {author} {\bibfnamefont {H.}~\bibnamefont {Xiao}},\ }\bibfield  {title} {\enquote {\bibinfo {title} {Turbulence modeling in the age of data},}\ }\href@noop {} {\bibfield  {journal} {\bibinfo  {journal} {Annual review of fluid mechanics}\ }\textbf {\bibinfo {volume} {51}},\ \bibinfo {pages} {357--377} (\bibinfo {year} {2019})}\BibitemShut {NoStop}%
\bibitem [{\citenamefont {Duan}\ and\ \citenamefont {Mart{\'\i}n}(2011)}]{duan2011assessment}%
  \BibitemOpen
  \bibfield  {author} {\bibinfo {author} {\bibfnamefont {L.}~\bibnamefont {Duan}}\ and\ \bibinfo {author} {\bibfnamefont {M.~P.}\ \bibnamefont {Mart{\'\i}n}},\ }\bibfield  {title} {\enquote {\bibinfo {title} {Assessment of turbulence-chemistry interaction in hypersonic turbulent boundary layers},}\ }\href@noop {} {\bibfield  {journal} {\bibinfo  {journal} {AIAA journal}\ }\textbf {\bibinfo {volume} {49}},\ \bibinfo {pages} {172--184} (\bibinfo {year} {2011})}\BibitemShut {NoStop}%
\bibitem [{\citenamefont {Huang}\ \emph {et~al.}(2020)\citenamefont {Huang}, \citenamefont {Nicholson}, \citenamefont {Duan}, \citenamefont {Choudhari},\ and\ \citenamefont {Bowersox}}]{huang2020simulation}%
  \BibitemOpen
  \bibfield  {author} {\bibinfo {author} {\bibfnamefont {J.}~\bibnamefont {Huang}}, \bibinfo {author} {\bibfnamefont {G.~L.}\ \bibnamefont {Nicholson}}, \bibinfo {author} {\bibfnamefont {L.}~\bibnamefont {Duan}}, \bibinfo {author} {\bibfnamefont {M.~M.}\ \bibnamefont {Choudhari}}, \ and\ \bibinfo {author} {\bibfnamefont {R.~D.}\ \bibnamefont {Bowersox}},\ }\bibfield  {title} {\enquote {\bibinfo {title} {Simulation and modeling of cold-wall hypersonic turbulent boundary layers on flat plate},}\ }in\ \href@noop {} {\emph {\bibinfo {booktitle} {AIAA Scitech 2020 Forum}}}\ (\bibinfo {year} {2020})\ p.\ \bibinfo {pages} {0571}\BibitemShut {NoStop}%
\bibitem [{\citenamefont {Horstman}\ and\ \citenamefont {Owen}(1972)}]{horstman1972turbulent}%
  \BibitemOpen
  \bibfield  {author} {\bibinfo {author} {\bibfnamefont {C.}~\bibnamefont {Horstman}}\ and\ \bibinfo {author} {\bibfnamefont {F.}~\bibnamefont {Owen}},\ }\bibfield  {title} {\enquote {\bibinfo {title} {Turbulent properties of a compressible boundary layer.}}\ }\href@noop {} {\bibfield  {journal} {\bibinfo  {journal} {AIAA Journal}\ }\textbf {\bibinfo {volume} {10}},\ \bibinfo {pages} {1418--1424} (\bibinfo {year} {1972})}\BibitemShut {NoStop}%
\bibitem [{\citenamefont {Owen}\ and\ \citenamefont {Horstman}(1972)}]{owen1972structure}%
  \BibitemOpen
  \bibfield  {author} {\bibinfo {author} {\bibfnamefont {F.}~\bibnamefont {Owen}}\ and\ \bibinfo {author} {\bibfnamefont {C.}~\bibnamefont {Horstman}},\ }\bibfield  {title} {\enquote {\bibinfo {title} {On the structure of hypersonic turbulent boundary layers},}\ }\href@noop {} {\bibfield  {journal} {\bibinfo  {journal} {Journal of Fluid Mechanics}\ }\textbf {\bibinfo {volume} {53}},\ \bibinfo {pages} {611--636} (\bibinfo {year} {1972})}\BibitemShut {NoStop}%
\bibitem [{\citenamefont {Owen}, \citenamefont {Horstman},\ and\ \citenamefont {Kussoy}(1975)}]{owen1975mean}%
  \BibitemOpen
  \bibfield  {author} {\bibinfo {author} {\bibfnamefont {F.}~\bibnamefont {Owen}}, \bibinfo {author} {\bibfnamefont {C.}~\bibnamefont {Horstman}}, \ and\ \bibinfo {author} {\bibfnamefont {M.}~\bibnamefont {Kussoy}},\ }\bibfield  {title} {\enquote {\bibinfo {title} {Mean and fluctuating flow measurements of a fully-developed, non-adiabatic, hypersonic boundary layer},}\ }\href@noop {} {\bibfield  {journal} {\bibinfo  {journal} {Journal of Fluid Mechanics}\ }\textbf {\bibinfo {volume} {70}},\ \bibinfo {pages} {393--413} (\bibinfo {year} {1975})}\BibitemShut {NoStop}%
\bibitem [{\citenamefont {Mikulla}\ and\ \citenamefont {Horstman}(1976)}]{mikulla1976turbulence}%
  \BibitemOpen
  \bibfield  {author} {\bibinfo {author} {\bibfnamefont {V.}~\bibnamefont {Mikulla}}\ and\ \bibinfo {author} {\bibfnamefont {C.~C.}\ \bibnamefont {Horstman}},\ }\bibfield  {title} {\enquote {\bibinfo {title} {Turbulence measurements in hypersonic shock-wave boundary-layer interaction flows},}\ }\href@noop {} {\bibfield  {journal} {\bibinfo  {journal} {AIAA Journal}\ }\textbf {\bibinfo {volume} {14}},\ \bibinfo {pages} {568--575} (\bibinfo {year} {1976})}\BibitemShut {NoStop}%
\bibitem [{\citenamefont {Sahoo}, \citenamefont {Schultze},\ and\ \citenamefont {Smits}(2009)}]{sahoo2009effects}%
  \BibitemOpen
  \bibfield  {author} {\bibinfo {author} {\bibfnamefont {D.}~\bibnamefont {Sahoo}}, \bibinfo {author} {\bibfnamefont {M.}~\bibnamefont {Schultze}}, \ and\ \bibinfo {author} {\bibfnamefont {A.}~\bibnamefont {Smits}},\ }\bibfield  {title} {\enquote {\bibinfo {title} {Effects of roughness on a turbulent bloundary layer in hypersonic flow},}\ }in\ \href@noop {} {\emph {\bibinfo {booktitle} {39th AIAA Fluid Dynamics Conference}}}\ (\bibinfo {year} {2009})\ p.\ \bibinfo {pages} {3678}\BibitemShut {NoStop}%
\bibitem [{\citenamefont {Baumgartner}(1997)}]{baumgartner1997turbulence}%
  \BibitemOpen
  \bibfield  {author} {\bibinfo {author} {\bibfnamefont {M.~L.}\ \bibnamefont {Baumgartner}},\ }\href@noop {} {\emph {\bibinfo {title} {Turbulence structure in a hypersonic boundary layer}}}\ (\bibinfo  {publisher} {Princeton University},\ \bibinfo {year} {1997})\BibitemShut {NoStop}%
\bibitem [{\citenamefont {McGinley}, \citenamefont {Spina},\ and\ \citenamefont {Sheplak}(1994)}]{mcginley1994turbulence}%
  \BibitemOpen
  \bibfield  {author} {\bibinfo {author} {\bibfnamefont {C.}~\bibnamefont {McGinley}}, \bibinfo {author} {\bibfnamefont {E.}~\bibnamefont {Spina}}, \ and\ \bibinfo {author} {\bibfnamefont {M.}~\bibnamefont {Sheplak}},\ }\bibfield  {title} {\enquote {\bibinfo {title} {Turbulence measurements in a mach 11 helium boundary layer},}\ }in\ \href@noop {} {\emph {\bibinfo {booktitle} {Fluid Dynamics Conference}}}\ (\bibinfo {year} {1994})\ p.\ \bibinfo {pages} {2364}\BibitemShut {NoStop}%
\bibitem [{\citenamefont {So}, \citenamefont {Gatski},\ and\ \citenamefont {Sommer}(1998)}]{so1998morkovin}%
  \BibitemOpen
  \bibfield  {author} {\bibinfo {author} {\bibfnamefont {R.}~\bibnamefont {So}}, \bibinfo {author} {\bibfnamefont {T.}~\bibnamefont {Gatski}}, \ and\ \bibinfo {author} {\bibfnamefont {T.}~\bibnamefont {Sommer}},\ }\bibfield  {title} {\enquote {\bibinfo {title} {Morkovin hypothesis and the modeling of wall-bounded compressible turbulent flows},}\ }\href@noop {} {\bibfield  {journal} {\bibinfo  {journal} {AIAA journal}\ }\textbf {\bibinfo {volume} {36}},\ \bibinfo {pages} {1583--1592} (\bibinfo {year} {1998})}\BibitemShut {NoStop}%
\bibitem [{\citenamefont {Pope}(1975)}]{pope1975more}%
  \BibitemOpen
  \bibfield  {author} {\bibinfo {author} {\bibfnamefont {S.~B.}\ \bibnamefont {Pope}},\ }\bibfield  {title} {\enquote {\bibinfo {title} {A more general effective-viscosity hypothesis},}\ }\href@noop {} {\bibfield  {journal} {\bibinfo  {journal} {Journal of Fluid Mechanics}\ }\textbf {\bibinfo {volume} {72}},\ \bibinfo {pages} {331--340} (\bibinfo {year} {1975})}\BibitemShut {NoStop}%
\bibitem [{\citenamefont {Boussinesq}(1877)}]{boussinesq1877essai}%
  \BibitemOpen
  \bibfield  {author} {\bibinfo {author} {\bibfnamefont {J.}~\bibnamefont {Boussinesq}},\ }\href@noop {} {\emph {\bibinfo {title} {Essai sur la th{\'e}orie des eaux courantes}}}\ (\bibinfo  {publisher} {Imprimerie nationale},\ \bibinfo {year} {1877})\BibitemShut {NoStop}%
\bibitem [{\citenamefont {Chen}, \citenamefont {Gan},\ and\ \citenamefont {Fu}(2024)}]{chen2024improved}%
  \BibitemOpen
  \bibfield  {author} {\bibinfo {author} {\bibfnamefont {X.}~\bibnamefont {Chen}}, \bibinfo {author} {\bibfnamefont {J.}~\bibnamefont {Gan}}, \ and\ \bibinfo {author} {\bibfnamefont {L.}~\bibnamefont {Fu}},\ }\bibfield  {title} {\enquote {\bibinfo {title} {An improved baldwin--lomax algebraic wall model for high-speed canonical turbulent boundary layers using established scalings},}\ }\href@noop {} {\bibfield  {journal} {\bibinfo  {journal} {Journal of Fluid Mechanics}\ }\textbf {\bibinfo {volume} {987}},\ \bibinfo {pages} {A7} (\bibinfo {year} {2024})}\BibitemShut {NoStop}%
\bibitem [{\citenamefont {Dilley}\ and\ \citenamefont {McClinton}(2001)}]{dilley2001evaluation}%
  \BibitemOpen
  \bibfield  {author} {\bibinfo {author} {\bibfnamefont {A.~D.}\ \bibnamefont {Dilley}}\ and\ \bibinfo {author} {\bibfnamefont {C.~R.}\ \bibnamefont {McClinton}},\ }\href@noop {} {\enquote {\bibinfo {title} {Evaluation of cfd turbulent heating prediction techniques and comparison with hypersonic experimental data},}\ }\bibinfo {type} {Tech. Rep.}\ (\bibinfo {year} {2001})\BibitemShut {NoStop}%
\bibitem [{\citenamefont {Hejranfar}, \citenamefont {Esfahanian},\ and\ \citenamefont {Kamali-Moghadam}(2012)}]{hejranfar2012dual}%
  \BibitemOpen
  \bibfield  {author} {\bibinfo {author} {\bibfnamefont {K.}~\bibnamefont {Hejranfar}}, \bibinfo {author} {\bibfnamefont {V.}~\bibnamefont {Esfahanian}}, \ and\ \bibinfo {author} {\bibfnamefont {R.}~\bibnamefont {Kamali-Moghadam}},\ }\bibfield  {title} {\enquote {\bibinfo {title} {Dual-code solution procedure for efficient computing equilibrium hypersonic axisymmetric transitional/turbulent flows},}\ }\href@noop {} {\bibfield  {journal} {\bibinfo  {journal} {Aerospace Science and Technology}\ }\textbf {\bibinfo {volume} {21}},\ \bibinfo {pages} {64--74} (\bibinfo {year} {2012})}\BibitemShut {NoStop}%
\bibitem [{\citenamefont {Horstman}(1987)}]{horstman1987prediction}%
  \BibitemOpen
  \bibfield  {author} {\bibinfo {author} {\bibfnamefont {C.}~\bibnamefont {Horstman}},\ }\bibfield  {title} {\enquote {\bibinfo {title} {Prediction of hypersonic shock-wave/turbulent-boundary-layer interaction flows},}\ }in\ \href@noop {} {\emph {\bibinfo {booktitle} {19th AIAA, Fluid Dynamics, Plasma Dynamics, and Lasers Conference}}}\ (\bibinfo {year} {1987})\ p.\ \bibinfo {pages} {1367}\BibitemShut {NoStop}%
\bibitem [{\citenamefont {Narayanswami}, \citenamefont {HORSTMAN},\ and\ \citenamefont {KNIGHT}(1993)}]{narayanswami1993numerical}%
  \BibitemOpen
  \bibfield  {author} {\bibinfo {author} {\bibfnamefont {N.}~\bibnamefont {Narayanswami}}, \bibinfo {author} {\bibfnamefont {C.}~\bibnamefont {HORSTMAN}}, \ and\ \bibinfo {author} {\bibfnamefont {D.}~\bibnamefont {KNIGHT}},\ }\bibfield  {title} {\enquote {\bibinfo {title} {Numerical simulation of crossing/turbulent boundary layer interaction at mach 8.3 comparison of zero and two-equation turbulence models},}\ }in\ \href@noop {} {\emph {\bibinfo {booktitle} {31st Aerospace Sciences Meeting}}}\ (\bibinfo {year} {1993})\ p.\ \bibinfo {pages} {779}\BibitemShut {NoStop}%
\bibitem [{\citenamefont {Panaras}(2015)}]{panaras2015turbulence}%
  \BibitemOpen
  \bibfield  {author} {\bibinfo {author} {\bibfnamefont {A.~G.}\ \bibnamefont {Panaras}},\ }\bibfield  {title} {\enquote {\bibinfo {title} {Turbulence modeling of flows with extensive crossflow separation},}\ }\href@noop {} {\bibfield  {journal} {\bibinfo  {journal} {Aerospace}\ }\textbf {\bibinfo {volume} {2}},\ \bibinfo {pages} {461--481} (\bibinfo {year} {2015})}\BibitemShut {NoStop}%
\bibitem [{\citenamefont {Hu}\ and\ \citenamefont {Rizzi}(1995)}]{hu1995turbulent}%
  \BibitemOpen
  \bibfield  {author} {\bibinfo {author} {\bibfnamefont {J.}~\bibnamefont {Hu}}\ and\ \bibinfo {author} {\bibfnamefont {A.}~\bibnamefont {Rizzi}},\ }\bibfield  {title} {\enquote {\bibinfo {title} {Turbulent flow in supersonic and hypersonic nozzles},}\ }\href@noop {} {\bibfield  {journal} {\bibinfo  {journal} {AIAA journal}\ }\textbf {\bibinfo {volume} {33}},\ \bibinfo {pages} {1634--1640} (\bibinfo {year} {1995})}\BibitemShut {NoStop}%
\bibitem [{\citenamefont {KIM}\ and\ \citenamefont {HARLOFF}(1988)}]{kim1988hypersonic}%
  \BibitemOpen
  \bibfield  {author} {\bibinfo {author} {\bibfnamefont {S.}~\bibnamefont {KIM}}\ and\ \bibinfo {author} {\bibfnamefont {G.}~\bibnamefont {HARLOFF}},\ }\bibfield  {title} {\enquote {\bibinfo {title} {Hypersonic turbulent wall boundary layer computations},}\ }in\ \href@noop {} {\emph {\bibinfo {booktitle} {24th Joint Propulsion Conference}}}\ (\bibinfo {year} {1988})\ p.\ \bibinfo {pages} {2829}\BibitemShut {NoStop}%
\bibitem [{\citenamefont {McDonald}\ and\ \citenamefont {Camarata}(1968)}]{mcdonald1968extended}%
  \BibitemOpen
  \bibfield  {author} {\bibinfo {author} {\bibfnamefont {H.}~\bibnamefont {McDonald}}\ and\ \bibinfo {author} {\bibfnamefont {F.}~\bibnamefont {Camarata}},\ }\bibfield  {title} {\enquote {\bibinfo {title} {An extended mixing length approach for computing the turbulent boundary layer development},}\ }in\ \href@noop {} {\emph {\bibinfo {booktitle} {Proceedings, Stanford Conference on Computation of Turbulent Boundary Layers}}},\ Vol.~\bibinfo {volume} {1}\ (\bibinfo {year} {1968})\ pp.\ \bibinfo {pages} {83--98}\BibitemShut {NoStop}%
\bibitem [{\citenamefont {Ng}, \citenamefont {Ajmani},\ and\ \citenamefont {Taylor~III}(1989)}]{ng1989turbulence}%
  \BibitemOpen
  \bibfield  {author} {\bibinfo {author} {\bibfnamefont {W.-F.}\ \bibnamefont {Ng}}, \bibinfo {author} {\bibfnamefont {K.}~\bibnamefont {Ajmani}}, \ and\ \bibinfo {author} {\bibfnamefont {A.}~\bibnamefont {Taylor~III}},\ }\bibfield  {title} {\enquote {\bibinfo {title} {Turbulence modeling in a hypersonic inlet},}\ }\href@noop {} {\bibfield  {journal} {\bibinfo  {journal} {AIAA journal}\ }\textbf {\bibinfo {volume} {27}},\ \bibinfo {pages} {1354--1360} (\bibinfo {year} {1989})}\BibitemShut {NoStop}%
\bibitem [{\citenamefont {Goldberg}\ \emph {et~al.}(2000)\citenamefont {Goldberg}, \citenamefont {Batten}, \citenamefont {Palaniswamy}, \citenamefont {Chakravarthy},\ and\ \citenamefont {Peroomian}}]{goldberg2000hypersonic}%
  \BibitemOpen
  \bibfield  {author} {\bibinfo {author} {\bibfnamefont {U.}~\bibnamefont {Goldberg}}, \bibinfo {author} {\bibfnamefont {P.}~\bibnamefont {Batten}}, \bibinfo {author} {\bibfnamefont {S.}~\bibnamefont {Palaniswamy}}, \bibinfo {author} {\bibfnamefont {S.}~\bibnamefont {Chakravarthy}}, \ and\ \bibinfo {author} {\bibfnamefont {O.}~\bibnamefont {Peroomian}},\ }\bibfield  {title} {\enquote {\bibinfo {title} {Hypersonic flow predictions using linear and nonlinear turbulence closures},}\ }\href@noop {} {\bibfield  {journal} {\bibinfo  {journal} {Journal of Aircraft}\ }\textbf {\bibinfo {volume} {37}},\ \bibinfo {pages} {671--675} (\bibinfo {year} {2000})}\BibitemShut {NoStop}%
\bibitem [{\citenamefont {Menter}(1997)}]{menter1997eddy}%
  \BibitemOpen
  \bibfield  {author} {\bibinfo {author} {\bibfnamefont {F.}~\bibnamefont {Menter}},\ }\bibfield  {title} {\enquote {\bibinfo {title} {Eddy viscosity transport equations and their relation to the k-$\varepsilon$ model},}\ }\href@noop {} {\bibfield  {journal} {\bibinfo  {journal} {Journal of Fluids Engineering}\ }\textbf {\bibinfo {volume} {119}},\ \bibinfo {pages} {876--884} (\bibinfo {year} {1997})}\BibitemShut {NoStop}%
\bibitem [{\citenamefont {Paciorri}\ \emph {et~al.}(1997)\citenamefont {Paciorri}, \citenamefont {Dieudonne}, \citenamefont {Degrez}, \citenamefont {Charbonnier}, \citenamefont {Deconinck}, \citenamefont {Paciorri}, \citenamefont {Dieudonne}, \citenamefont {Degrez}, \citenamefont {Charbonnier},\ and\ \citenamefont {Deconinck}}]{paciorri1997validation}%
  \BibitemOpen
  \bibfield  {author} {\bibinfo {author} {\bibfnamefont {R.}~\bibnamefont {Paciorri}}, \bibinfo {author} {\bibfnamefont {W.}~\bibnamefont {Dieudonne}}, \bibinfo {author} {\bibfnamefont {G.}~\bibnamefont {Degrez}}, \bibinfo {author} {\bibfnamefont {J.-M.}\ \bibnamefont {Charbonnier}}, \bibinfo {author} {\bibfnamefont {H.}~\bibnamefont {Deconinck}}, \bibinfo {author} {\bibfnamefont {R.}~\bibnamefont {Paciorri}}, \bibinfo {author} {\bibfnamefont {W.}~\bibnamefont {Dieudonne}}, \bibinfo {author} {\bibfnamefont {G.}~\bibnamefont {Degrez}}, \bibinfo {author} {\bibfnamefont {J.-M.}\ \bibnamefont {Charbonnier}}, \ and\ \bibinfo {author} {\bibfnamefont {H.}~\bibnamefont {Deconinck}},\ }\bibfield  {title} {\enquote {\bibinfo {title} {Validation of the spalart-allmaras turbulence model for application in hypersonic flows},}\ }in\ \href@noop {} {\emph {\bibinfo {booktitle} {28th Fluid Dynamics Conference}}}\ (\bibinfo {year} {1997})\ p.\ \bibinfo {pages} {2023}\BibitemShut {NoStop}%
\bibitem [{\citenamefont {Nance}\ and\ \citenamefont {Hassan}(1999)}]{nance1999turbulence}%
  \BibitemOpen
  \bibfield  {author} {\bibinfo {author} {\bibfnamefont {R.}~\bibnamefont {Nance}}\ and\ \bibinfo {author} {\bibfnamefont {H.}~\bibnamefont {Hassan}},\ }\bibfield  {title} {\enquote {\bibinfo {title} {Turbulence modeling of shock-dominated flows with a k-zeta formulation},}\ }in\ \href@noop {} {\emph {\bibinfo {booktitle} {37th Aerospace Sciences Meeting and Exhibit}}}\ (\bibinfo {year} {1999})\ p.\ \bibinfo {pages} {153}\BibitemShut {NoStop}%
\bibitem [{\citenamefont {Xiao}, \citenamefont {Hassan},\ and\ \citenamefont {Baurle}(2007)}]{xiao2007modeling}%
  \BibitemOpen
  \bibfield  {author} {\bibinfo {author} {\bibfnamefont {X.}~\bibnamefont {Xiao}}, \bibinfo {author} {\bibfnamefont {H.~A.}\ \bibnamefont {Hassan}}, \ and\ \bibinfo {author} {\bibfnamefont {R.~A.}\ \bibnamefont {Baurle}},\ }\bibfield  {title} {\enquote {\bibinfo {title} {Modeling scramjet flows with variable turbulent prandtl and schmidt numbers},}\ }\href@noop {} {\bibfield  {journal} {\bibinfo  {journal} {AIAA journal}\ }\textbf {\bibinfo {volume} {45}},\ \bibinfo {pages} {1415--1423} (\bibinfo {year} {2007})}\BibitemShut {NoStop}%
\bibitem [{\citenamefont {Xiao}\ \emph {et~al.}(2007)\citenamefont {Xiao}, \citenamefont {Hassan}, \citenamefont {Edwards},\ and\ \citenamefont {Gaffney~Jr}}]{xiao2007role}%
  \BibitemOpen
  \bibfield  {author} {\bibinfo {author} {\bibfnamefont {X.}~\bibnamefont {Xiao}}, \bibinfo {author} {\bibfnamefont {H.}~\bibnamefont {Hassan}}, \bibinfo {author} {\bibfnamefont {J.}~\bibnamefont {Edwards}}, \ and\ \bibinfo {author} {\bibfnamefont {R.}~\bibnamefont {Gaffney~Jr}},\ }\bibfield  {title} {\enquote {\bibinfo {title} {Role of turbulent prandtl numbers on heat flux at hypersonic mach numbers},}\ }\href@noop {} {\bibfield  {journal} {\bibinfo  {journal} {AIAA journal}\ }\textbf {\bibinfo {volume} {45}},\ \bibinfo {pages} {806--813} (\bibinfo {year} {2007})}\BibitemShut {NoStop}%
\bibitem [{\citenamefont {Robinson}\ and\ \citenamefont {Hassan}(1998)}]{robinson1998further}%
  \BibitemOpen
  \bibfield  {author} {\bibinfo {author} {\bibfnamefont {D.}~\bibnamefont {Robinson}}\ and\ \bibinfo {author} {\bibfnamefont {H.}~\bibnamefont {Hassan}},\ }\bibfield  {title} {\enquote {\bibinfo {title} {Further development of the k-$\zeta$(enstrophy) turbulence closure model},}\ }\href@noop {} {\bibfield  {journal} {\bibinfo  {journal} {AIAA journal}\ }\textbf {\bibinfo {volume} {36}},\ \bibinfo {pages} {1825--1833} (\bibinfo {year} {1998})}\BibitemShut {NoStop}%
\bibitem [{\citenamefont {Coakley}(1983)}]{coakley1983turbulence}%
  \BibitemOpen
  \bibfield  {author} {\bibinfo {author} {\bibfnamefont {T.}~\bibnamefont {Coakley}},\ }\bibfield  {title} {\enquote {\bibinfo {title} {Turbulence modeling methods for the compressible navier-stokes equations},}\ }in\ \href@noop {} {\emph {\bibinfo {booktitle} {16th fluid and plasmadynamics conference}}}\ (\bibinfo {year} {1983})\ p.\ \bibinfo {pages} {1693}\BibitemShut {NoStop}%
\bibitem [{\citenamefont {Smith}(1995)}]{smith1995prediction}%
  \BibitemOpen
  \bibfield  {author} {\bibinfo {author} {\bibfnamefont {B.}~\bibnamefont {Smith}},\ }\bibfield  {title} {\enquote {\bibinfo {title} {Prediction of hypersonic shock wave turbulent boundary layer interactions with the kl two equation turbulence model},}\ }in\ \href@noop {} {\emph {\bibinfo {booktitle} {33rd Aerospace Sciences Meeting and Exhibit}}}\ (\bibinfo {year} {1995})\ p.\ \bibinfo {pages} {232}\BibitemShut {NoStop}%
\bibitem [{\citenamefont {Leschziner}(2006)}]{leschziner2006modelling}%
  \BibitemOpen
  \bibfield  {author} {\bibinfo {author} {\bibfnamefont {M.}~\bibnamefont {Leschziner}},\ }\bibfield  {title} {\enquote {\bibinfo {title} {Modelling turbulent separated flow in the context of aerodynamic applications},}\ }\href@noop {} {\bibfield  {journal} {\bibinfo  {journal} {Fluid dynamics research}\ }\textbf {\bibinfo {volume} {38}},\ \bibinfo {pages} {174--210} (\bibinfo {year} {2006})}\BibitemShut {NoStop}%
\bibitem [{\citenamefont {SHIH}(1993)}]{shih1993realisable}%
  \BibitemOpen
  \bibfield  {author} {\bibinfo {author} {\bibfnamefont {T.}~\bibnamefont {SHIH}},\ }\bibfield  {title} {\enquote {\bibinfo {title} {A realisable reynolds stress algebraic equation model},}\ }\href@noop {} {\bibfield  {journal} {\bibinfo  {journal} {NASA TM-105993}\ } (\bibinfo {year} {1993})}\BibitemShut {NoStop}%
\bibitem [{\citenamefont {Lien}(1996)}]{lien1996low}%
  \BibitemOpen
  \bibfield  {author} {\bibinfo {author} {\bibfnamefont {F.-S.}\ \bibnamefont {Lien}},\ }\bibfield  {title} {\enquote {\bibinfo {title} {Low-reynolds-number eddy-viscosity modelling based on non-linear stress-strain/vorticity relations},}\ }in\ \href@noop {} {\emph {\bibinfo {booktitle} {Proc. 3rd Symposium On Engineering Turbulence Modelling and Measurements}}}\ (\bibinfo {year} {1996})\ pp.\ \bibinfo {pages} {1--10}\BibitemShut {NoStop}%
\bibitem [{\citenamefont {Holden}(1992)}]{holden1992turbulent}%
  \BibitemOpen
  \bibfield  {author} {\bibinfo {author} {\bibfnamefont {M.}~\bibnamefont {Holden}},\ }\bibfield  {title} {\enquote {\bibinfo {title} {Turbulent boundary layer development on curved compression surfaces},}\ }\href@noop {} {\bibfield  {journal} {\bibinfo  {journal} {Calspan Report No}\ ,\ \bibinfo {pages} {7724--1}} (\bibinfo {year} {1992})}\BibitemShut {NoStop}%
\bibitem [{\citenamefont {Coleman}\ and\ \citenamefont {Stollery}(1972)}]{coleman1972heat}%
  \BibitemOpen
  \bibfield  {author} {\bibinfo {author} {\bibfnamefont {G.}~\bibnamefont {Coleman}}\ and\ \bibinfo {author} {\bibfnamefont {J.}~\bibnamefont {Stollery}},\ }\bibfield  {title} {\enquote {\bibinfo {title} {Heat transfer from hypersonic turbulent flow at a wedge compression corner},}\ }\href@noop {} {\bibfield  {journal} {\bibinfo  {journal} {Journal of Fluid Mechanics}\ }\textbf {\bibinfo {volume} {56}},\ \bibinfo {pages} {741--752} (\bibinfo {year} {1972})}\BibitemShut {NoStop}%
\bibitem [{\citenamefont {Zhang}, \citenamefont {Craft},\ and\ \citenamefont {Iacovides}(2022)}]{zhang2022application}%
  \BibitemOpen
  \bibfield  {author} {\bibinfo {author} {\bibfnamefont {H.}~\bibnamefont {Zhang}}, \bibinfo {author} {\bibfnamefont {T.}~\bibnamefont {Craft}}, \ and\ \bibinfo {author} {\bibfnamefont {H.}~\bibnamefont {Iacovides}},\ }\bibfield  {title} {\enquote {\bibinfo {title} {Application of linear and nonlinear two-equation turbulence models in hypersonic flows},}\ }\href@noop {} {\bibfield  {journal} {\bibinfo  {journal} {AIAA Journal}\ }\textbf {\bibinfo {volume} {60}},\ \bibinfo {pages} {3472--3486} (\bibinfo {year} {2022})}\BibitemShut {NoStop}%
\bibitem [{\citenamefont {Craft}, \citenamefont {Iacovides},\ and\ \citenamefont {Yoon}(2000)}]{craft2000progress}%
  \BibitemOpen
  \bibfield  {author} {\bibinfo {author} {\bibfnamefont {T.}~\bibnamefont {Craft}}, \bibinfo {author} {\bibfnamefont {H.}~\bibnamefont {Iacovides}}, \ and\ \bibinfo {author} {\bibfnamefont {J.}~\bibnamefont {Yoon}},\ }\bibfield  {title} {\enquote {\bibinfo {title} {Progress in the use of non-linear two-equation models in the computation of convective heat-transfer in impinging and separated flows},}\ }\href@noop {} {\bibfield  {journal} {\bibinfo  {journal} {Flow, Turbulence and Combustion}\ }\textbf {\bibinfo {volume} {63}},\ \bibinfo {pages} {59--80} (\bibinfo {year} {2000})}\BibitemShut {NoStop}%
\bibitem [{\citenamefont {Craft}, \citenamefont {Launder},\ and\ \citenamefont {Suga}(1996)}]{craft1996development}%
  \BibitemOpen
  \bibfield  {author} {\bibinfo {author} {\bibfnamefont {T.}~\bibnamefont {Craft}}, \bibinfo {author} {\bibfnamefont {B.}~\bibnamefont {Launder}}, \ and\ \bibinfo {author} {\bibfnamefont {K.}~\bibnamefont {Suga}},\ }\bibfield  {title} {\enquote {\bibinfo {title} {Development and application of a cubic eddy-viscosity model of turbulence},}\ }\href@noop {} {\bibfield  {journal} {\bibinfo  {journal} {International Journal of Heat and Fluid Flow}\ }\textbf {\bibinfo {volume} {17}},\ \bibinfo {pages} {108--115} (\bibinfo {year} {1996})}\BibitemShut {NoStop}%
\bibitem [{\citenamefont {Wallin}\ and\ \citenamefont {Johansson}(2000)}]{wallin2000explicit}%
  \BibitemOpen
  \bibfield  {author} {\bibinfo {author} {\bibfnamefont {S.}~\bibnamefont {Wallin}}\ and\ \bibinfo {author} {\bibfnamefont {A.~V.}\ \bibnamefont {Johansson}},\ }\bibfield  {title} {\enquote {\bibinfo {title} {An explicit algebraic reynolds stress model for incompressible and compressible turbulent flows},}\ }\href@noop {} {\bibfield  {journal} {\bibinfo  {journal} {Journal of fluid mechanics}\ }\textbf {\bibinfo {volume} {403}},\ \bibinfo {pages} {89--132} (\bibinfo {year} {2000})}\BibitemShut {NoStop}%
\bibitem [{\citenamefont {Sch{\"u}lein}(2006)}]{schulein2006skin}%
  \BibitemOpen
  \bibfield  {author} {\bibinfo {author} {\bibfnamefont {E.}~\bibnamefont {Sch{\"u}lein}},\ }\bibfield  {title} {\enquote {\bibinfo {title} {Skin friction and heat flux measurements in shock/boundary layer interaction flows},}\ }\href@noop {} {\bibfield  {journal} {\bibinfo  {journal} {AIAA journal}\ }\textbf {\bibinfo {volume} {44}},\ \bibinfo {pages} {1732--1741} (\bibinfo {year} {2006})}\BibitemShut {NoStop}%
\bibitem [{\citenamefont {Lindblad}\ \emph {et~al.}(1998)\citenamefont {Lindblad}, \citenamefont {Wallin}, \citenamefont {Johansson}, \citenamefont {Friedrich}, \citenamefont {Lechner}, \citenamefont {Krogmann}, \citenamefont {Schuelein}, \citenamefont {Courty}, \citenamefont {Ravachol},\ and\ \citenamefont {Giordano}}]{lindblad1998prediction}%
  \BibitemOpen
  \bibfield  {author} {\bibinfo {author} {\bibfnamefont {I.}~\bibnamefont {Lindblad}}, \bibinfo {author} {\bibfnamefont {S.}~\bibnamefont {Wallin}}, \bibinfo {author} {\bibfnamefont {A.}~\bibnamefont {Johansson}}, \bibinfo {author} {\bibfnamefont {R.}~\bibnamefont {Friedrich}}, \bibinfo {author} {\bibfnamefont {R.}~\bibnamefont {Lechner}}, \bibinfo {author} {\bibfnamefont {P.}~\bibnamefont {Krogmann}}, \bibinfo {author} {\bibfnamefont {E.}~\bibnamefont {Schuelein}}, \bibinfo {author} {\bibfnamefont {J.-C.}\ \bibnamefont {Courty}}, \bibinfo {author} {\bibfnamefont {M.}~\bibnamefont {Ravachol}}, \ and\ \bibinfo {author} {\bibfnamefont {D.}~\bibnamefont {Giordano}},\ }\bibfield  {title} {\enquote {\bibinfo {title} {A prediction method for high speed turbulent separated flows with experimental verification},}\ }in\ \href@noop {} {\emph {\bibinfo {booktitle} {29th AIAA, Fluid Dynamics Conference}}}\ (\bibinfo {year} {1998})\ p.\ \bibinfo {pages} {2547}\BibitemShut {NoStop}%
\bibitem [{\citenamefont {Vemula}\ and\ \citenamefont {Sinha}(2020)}]{vemula2020explicit}%
  \BibitemOpen
  \bibfield  {author} {\bibinfo {author} {\bibfnamefont {J.~B.}\ \bibnamefont {Vemula}}\ and\ \bibinfo {author} {\bibfnamefont {K.}~\bibnamefont {Sinha}},\ }\bibfield  {title} {\enquote {\bibinfo {title} {Explicit algebraic reynolds stress model for shock-dominated flows},}\ }\href@noop {} {\bibfield  {journal} {\bibinfo  {journal} {International Journal of Heat and Fluid Flow}\ }\textbf {\bibinfo {volume} {85}},\ \bibinfo {pages} {108680} (\bibinfo {year} {2020})}\BibitemShut {NoStop}%
\bibitem [{\citenamefont {Sinha}, \citenamefont {Mahesh},\ and\ \citenamefont {Candler}(2003{\natexlab{a}})}]{sinha2003modeling}%
  \BibitemOpen
  \bibfield  {author} {\bibinfo {author} {\bibfnamefont {K.}~\bibnamefont {Sinha}}, \bibinfo {author} {\bibfnamefont {K.}~\bibnamefont {Mahesh}}, \ and\ \bibinfo {author} {\bibfnamefont {G.~V.}\ \bibnamefont {Candler}},\ }\bibfield  {title} {\enquote {\bibinfo {title} {Modeling shock unsteadiness in shock/turbulence interaction},}\ }\href@noop {} {\bibfield  {journal} {\bibinfo  {journal} {Physics of fluids}\ }\textbf {\bibinfo {volume} {15}},\ \bibinfo {pages} {2290--2297} (\bibinfo {year} {2003}{\natexlab{a}})}\BibitemShut {NoStop}%
\bibitem [{\citenamefont {Raje}\ and\ \citenamefont {Sinha}(2021)}]{raje2021anisotropic}%
  \BibitemOpen
  \bibfield  {author} {\bibinfo {author} {\bibfnamefont {P.}~\bibnamefont {Raje}}\ and\ \bibinfo {author} {\bibfnamefont {K.}~\bibnamefont {Sinha}},\ }\bibfield  {title} {\enquote {\bibinfo {title} {Anisotropic sst turbulence model for shock-boundary layer interaction},}\ }\href@noop {} {\bibfield  {journal} {\bibinfo  {journal} {Computers \& Fluids}\ }\textbf {\bibinfo {volume} {228}},\ \bibinfo {pages} {105072} (\bibinfo {year} {2021})}\BibitemShut {NoStop}%
\bibitem [{\citenamefont {Rung}\ \emph {et~al.}(1999)\citenamefont {Rung}, \citenamefont {L{\"u}bcke}, \citenamefont {Franke}, \citenamefont {Xue}, \citenamefont {Thiele},\ and\ \citenamefont {Fu}}]{rung1999assessment}%
  \BibitemOpen
  \bibfield  {author} {\bibinfo {author} {\bibfnamefont {T.}~\bibnamefont {Rung}}, \bibinfo {author} {\bibfnamefont {H.}~\bibnamefont {L{\"u}bcke}}, \bibinfo {author} {\bibfnamefont {M.}~\bibnamefont {Franke}}, \bibinfo {author} {\bibfnamefont {L.}~\bibnamefont {Xue}}, \bibinfo {author} {\bibfnamefont {F.}~\bibnamefont {Thiele}}, \ and\ \bibinfo {author} {\bibfnamefont {S.}~\bibnamefont {Fu}},\ }\bibfield  {title} {\enquote {\bibinfo {title} {Assessment of explicit algebraic stress models in transonic flows},}\ }in\ \href@noop {} {\emph {\bibinfo {booktitle} {Engineering Turbulence Modelling and Experiments 4}}}\ (\bibinfo  {publisher} {Elsevier},\ \bibinfo {year} {1999})\ pp.\ \bibinfo {pages} {659--668}\BibitemShut {NoStop}%
\bibitem [{\citenamefont {Gatski}\ and\ \citenamefont {Speziale}(1993)}]{gatski1993explicit}%
  \BibitemOpen
  \bibfield  {author} {\bibinfo {author} {\bibfnamefont {T.~B.}\ \bibnamefont {Gatski}}\ and\ \bibinfo {author} {\bibfnamefont {C.~G.}\ \bibnamefont {Speziale}},\ }\bibfield  {title} {\enquote {\bibinfo {title} {On explicit algebraic stress models for complex turbulent flows},}\ }\href@noop {} {\bibfield  {journal} {\bibinfo  {journal} {Journal of fluid Mechanics}\ }\textbf {\bibinfo {volume} {254}},\ \bibinfo {pages} {59--78} (\bibinfo {year} {1993})}\BibitemShut {NoStop}%
\bibitem [{\citenamefont {Speziale}, \citenamefont {Sarkar},\ and\ \citenamefont {Gatski}(1991)}]{speziale1991modelling}%
  \BibitemOpen
  \bibfield  {author} {\bibinfo {author} {\bibfnamefont {C.~G.}\ \bibnamefont {Speziale}}, \bibinfo {author} {\bibfnamefont {S.}~\bibnamefont {Sarkar}}, \ and\ \bibinfo {author} {\bibfnamefont {T.~B.}\ \bibnamefont {Gatski}},\ }\bibfield  {title} {\enquote {\bibinfo {title} {Modelling the pressure--strain correlation of turbulence: an invariant dynamical systems approach},}\ }\href@noop {} {\bibfield  {journal} {\bibinfo  {journal} {Journal of fluid mechanics}\ }\textbf {\bibinfo {volume} {227}},\ \bibinfo {pages} {245--272} (\bibinfo {year} {1991})}\BibitemShut {NoStop}%
\bibitem [{\citenamefont {Wilcox}\ \emph {et~al.}(1998)\citenamefont {Wilcox} \emph {et~al.}}]{wilcox1998turbulence}%
  \BibitemOpen
  \bibfield  {author} {\bibinfo {author} {\bibfnamefont {D.~C.}\ \bibnamefont {Wilcox}} \emph {et~al.},\ }\href@noop {} {\emph {\bibinfo {title} {Turbulence modeling for CFD}}},\ Vol.~\bibinfo {volume} {2}\ (\bibinfo  {publisher} {DCW industries La Canada, CA},\ \bibinfo {year} {1998})\BibitemShut {NoStop}%
\bibitem [{\citenamefont {Bosco}\ \emph {et~al.}(2011)\citenamefont {Bosco}, \citenamefont {Reinartz}, \citenamefont {Brown},\ and\ \citenamefont {Boyce}}]{bosco2011investigation}%
  \BibitemOpen
  \bibfield  {author} {\bibinfo {author} {\bibfnamefont {A.}~\bibnamefont {Bosco}}, \bibinfo {author} {\bibfnamefont {B.}~\bibnamefont {Reinartz}}, \bibinfo {author} {\bibfnamefont {L.}~\bibnamefont {Brown}}, \ and\ \bibinfo {author} {\bibfnamefont {R.}~\bibnamefont {Boyce}},\ }\bibfield  {title} {\enquote {\bibinfo {title} {Investigation of a compression corner at hypersonic conditions using a reynolds stress model},}\ }in\ \href@noop {} {\emph {\bibinfo {booktitle} {17th AIAA International Space Planes and Hypersonic Systems and Technologies Conference}}}\ (\bibinfo {year} {2011})\ p.\ \bibinfo {pages} {2217}\BibitemShut {NoStop}%
\bibitem [{\citenamefont {Frauholz}\ \emph {et~al.}(2014)\citenamefont {Frauholz}, \citenamefont {Bosco}, \citenamefont {Reinartz}, \citenamefont {M{\"u}ller},\ and\ \citenamefont {Behr}}]{frauholz2014investigation}%
  \BibitemOpen
  \bibfield  {author} {\bibinfo {author} {\bibfnamefont {S.}~\bibnamefont {Frauholz}}, \bibinfo {author} {\bibfnamefont {A.}~\bibnamefont {Bosco}}, \bibinfo {author} {\bibfnamefont {B.~U.}\ \bibnamefont {Reinartz}}, \bibinfo {author} {\bibfnamefont {S.}~\bibnamefont {M{\"u}ller}}, \ and\ \bibinfo {author} {\bibfnamefont {M.}~\bibnamefont {Behr}},\ }\bibfield  {title} {\enquote {\bibinfo {title} {Investigation of hypersonic intakes using reynolds stress modeling and wavelet-based adaptation},}\ }\href@noop {} {\bibfield  {journal} {\bibinfo  {journal} {AIAA journal}\ }\textbf {\bibinfo {volume} {52}},\ \bibinfo {pages} {2765--2781} (\bibinfo {year} {2014})}\BibitemShut {NoStop}%
\bibitem [{\citenamefont {Eisfeld}\ and\ \citenamefont {Brodersen}(2005)}]{eisfeld2005advanced}%
  \BibitemOpen
  \bibfield  {author} {\bibinfo {author} {\bibfnamefont {B.}~\bibnamefont {Eisfeld}}\ and\ \bibinfo {author} {\bibfnamefont {O.}~\bibnamefont {Brodersen}},\ }\bibfield  {title} {\enquote {\bibinfo {title} {Advanced turbulence modelling and stress analysis for the dlr-f6 configuration},}\ }in\ \href@noop {} {\emph {\bibinfo {booktitle} {23rd AIAA Applied Aerodynamics Conference}}}\ (\bibinfo {year} {2005})\ p.\ \bibinfo {pages} {4727}\BibitemShut {NoStop}%
\bibitem [{\citenamefont {Gerolymos}\ \emph {et~al.}(2012)\citenamefont {Gerolymos}, \citenamefont {Lo}, \citenamefont {Vallet},\ and\ \citenamefont {Younis}}]{gerolymos2012term}%
  \BibitemOpen
  \bibfield  {author} {\bibinfo {author} {\bibfnamefont {G.}~\bibnamefont {Gerolymos}}, \bibinfo {author} {\bibfnamefont {C.}~\bibnamefont {Lo}}, \bibinfo {author} {\bibfnamefont {I.}~\bibnamefont {Vallet}}, \ and\ \bibinfo {author} {\bibfnamefont {B.}~\bibnamefont {Younis}},\ }\bibfield  {title} {\enquote {\bibinfo {title} {Term-by-term analysis of near-wall second-moment closures},}\ }\href@noop {} {\bibfield  {journal} {\bibinfo  {journal} {AIAA journal}\ }\textbf {\bibinfo {volume} {50}},\ \bibinfo {pages} {2848--2864} (\bibinfo {year} {2012})}\BibitemShut {NoStop}%
\bibitem [{\citenamefont {Gerolymos}\ and\ \citenamefont {Vallet}(2002)}]{gerolymos2002wall}%
  \BibitemOpen
  \bibfield  {author} {\bibinfo {author} {\bibfnamefont {G.}~\bibnamefont {Gerolymos}}\ and\ \bibinfo {author} {\bibfnamefont {I.}~\bibnamefont {Vallet}},\ }\bibfield  {title} {\enquote {\bibinfo {title} {Wall-normal-free near-wall reynolds-stress model for 3-d turbomachinery flows},}\ }\href@noop {} {\bibfield  {journal} {\bibinfo  {journal} {AIAA Journal}\ }\textbf {\bibinfo {volume} {40}},\ \bibinfo {pages} {199208} (\bibinfo {year} {2002})}\BibitemShut {NoStop}%
\bibitem [{\citenamefont {Gerolymos}, \citenamefont {Sauret},\ and\ \citenamefont {Vallet}(2004)}]{gerolymos2004contribution}%
  \BibitemOpen
  \bibfield  {author} {\bibinfo {author} {\bibfnamefont {G.~A.}\ \bibnamefont {Gerolymos}}, \bibinfo {author} {\bibfnamefont {E.}~\bibnamefont {Sauret}}, \ and\ \bibinfo {author} {\bibfnamefont {I.}~\bibnamefont {Vallet}},\ }\bibfield  {title} {\enquote {\bibinfo {title} {Contribution to single-point closure reynolds-stress modelling of inhomogeneous flow},}\ }\href@noop {} {\bibfield  {journal} {\bibinfo  {journal} {Theoretical and Computational Fluid Dynamics}\ }\textbf {\bibinfo {volume} {17}},\ \bibinfo {pages} {407--431} (\bibinfo {year} {2004})}\BibitemShut {NoStop}%
\bibitem [{\citenamefont {Duan}(2024{\natexlab{a}})}]{M5DNS_SBLI}%
  \BibitemOpen
  \bibfield  {author} {\bibinfo {author} {\bibfnamefont {L.}~\bibnamefont {Duan}},\ }\href@noop {} {\enquote {\bibinfo {title} {Nasa turbulence modeling resource},}\ }\bibinfo {howpublished} {\url{https://turbmodels.larc.nasa.gov/Other_DNS_Data/highspeed_curvedwalls.html}} (\bibinfo {year} {2024}{\natexlab{a}}),\ \bibinfo {note} {last accessed on December 18, 2024}\BibitemShut {NoStop}%
\bibitem [{\citenamefont {Nicholson}, \citenamefont {Duan},\ and\ \citenamefont {Bisek}(2024)}]{nicholson2024direct}%
  \BibitemOpen
  \bibfield  {author} {\bibinfo {author} {\bibfnamefont {G.~L.}\ \bibnamefont {Nicholson}}, \bibinfo {author} {\bibfnamefont {L.}~\bibnamefont {Duan}}, \ and\ \bibinfo {author} {\bibfnamefont {N.~J.}\ \bibnamefont {Bisek}},\ }\bibfield  {title} {\enquote {\bibinfo {title} {Direct numerical simulation database of high-speed flow over parameterized curved walls},}\ }\href@noop {} {\bibfield  {journal} {\bibinfo  {journal} {AIAA Journal}\ }\textbf {\bibinfo {volume} {62}},\ \bibinfo {pages} {2095--2118} (\bibinfo {year} {2024})}\BibitemShut {NoStop}%
\bibitem [{\citenamefont {Colonna}, \citenamefont {Tuttafesta},\ and\ \citenamefont {Giordano}(2001)}]{colonna2001numerical}%
  \BibitemOpen
  \bibfield  {author} {\bibinfo {author} {\bibfnamefont {G.}~\bibnamefont {Colonna}}, \bibinfo {author} {\bibfnamefont {M.}~\bibnamefont {Tuttafesta}}, \ and\ \bibinfo {author} {\bibfnamefont {D.}~\bibnamefont {Giordano}},\ }\bibfield  {title} {\enquote {\bibinfo {title} {Numerical methods to solve euler equations in one-dimensional steady nozzle flow},}\ }\href@noop {} {\bibfield  {journal} {\bibinfo  {journal} {Computer physics communications}\ }\textbf {\bibinfo {volume} {138}},\ \bibinfo {pages} {213--221} (\bibinfo {year} {2001})}\BibitemShut {NoStop}%
\bibitem [{\citenamefont {Park}, \citenamefont {Jaffe},\ and\ \citenamefont {Partridge}(2001)}]{park2001chemical}%
  \BibitemOpen
  \bibfield  {author} {\bibinfo {author} {\bibfnamefont {C.}~\bibnamefont {Park}}, \bibinfo {author} {\bibfnamefont {R.~L.}\ \bibnamefont {Jaffe}}, \ and\ \bibinfo {author} {\bibfnamefont {H.}~\bibnamefont {Partridge}},\ }\bibfield  {title} {\enquote {\bibinfo {title} {Chemical-kinetic parameters of hyperbolic earth entry},}\ }\href@noop {} {\bibfield  {journal} {\bibinfo  {journal} {Journal of Thermophysics and Heat transfer}\ }\textbf {\bibinfo {volume} {15}},\ \bibinfo {pages} {76--90} (\bibinfo {year} {2001})}\BibitemShut {NoStop}%
\bibitem [{\citenamefont {Park}(1988)}]{park1988two}%
  \BibitemOpen
  \bibfield  {author} {\bibinfo {author} {\bibfnamefont {C.}~\bibnamefont {Park}},\ }\bibfield  {title} {\enquote {\bibinfo {title} {{Two-temperature interpretation of dissociation rate data for N2 and O2}},}\ }in\ \href@noop {} {\emph {\bibinfo {booktitle} {26th Aerspace Sciences Meeting}}}\ (\bibinfo {year} {1988})\ p.\ \bibinfo {pages} {458}\BibitemShut {NoStop}%
\bibitem [{\citenamefont {Sarkar}\ \emph {et~al.}(1991)\citenamefont {Sarkar}, \citenamefont {Erlebacher}, \citenamefont {Hussaini},\ and\ \citenamefont {Kreiss}}]{sarkar1991analysis}%
  \BibitemOpen
  \bibfield  {author} {\bibinfo {author} {\bibfnamefont {S.}~\bibnamefont {Sarkar}}, \bibinfo {author} {\bibfnamefont {G.}~\bibnamefont {Erlebacher}}, \bibinfo {author} {\bibfnamefont {M.~Y.}\ \bibnamefont {Hussaini}}, \ and\ \bibinfo {author} {\bibfnamefont {H.~O.}\ \bibnamefont {Kreiss}},\ }\bibfield  {title} {\enquote {\bibinfo {title} {The analysis and modelling of dilatational terms in compressible turbulence},}\ }\href@noop {} {\bibfield  {journal} {\bibinfo  {journal} {Journal of Fluid Mechanics}\ }\textbf {\bibinfo {volume} {227}},\ \bibinfo {pages} {473--493} (\bibinfo {year} {1991})}\BibitemShut {NoStop}%
\bibitem [{\citenamefont {Zeman}(1990)}]{zeman1990dilatation}%
  \BibitemOpen
  \bibfield  {author} {\bibinfo {author} {\bibfnamefont {O.}~\bibnamefont {Zeman}},\ }\bibfield  {title} {\enquote {\bibinfo {title} {Dilatation dissipation: the concept and application in modeling compressible mixing layers},}\ }\href@noop {} {\bibfield  {journal} {\bibinfo  {journal} {Physics of Fluids A: Fluid Dynamics}\ }\textbf {\bibinfo {volume} {2}},\ \bibinfo {pages} {178--188} (\bibinfo {year} {1990})}\BibitemShut {NoStop}%
\bibitem [{\citenamefont {Wilcox}(1992)}]{wilcox1992dilatation}%
  \BibitemOpen
  \bibfield  {author} {\bibinfo {author} {\bibfnamefont {D.~C.}\ \bibnamefont {Wilcox}},\ }\bibfield  {title} {\enquote {\bibinfo {title} {Dilatation-dissipation corrections for advanced turbulence models},}\ }\href@noop {} {\bibfield  {journal} {\bibinfo  {journal} {AIAA journal}\ }\textbf {\bibinfo {volume} {30}},\ \bibinfo {pages} {2639--2646} (\bibinfo {year} {1992})}\BibitemShut {NoStop}%
\bibitem [{\citenamefont {Zhang}, \citenamefont {Duan},\ and\ \citenamefont {Choudhari}(2018)}]{zhang2018direct}%
  \BibitemOpen
  \bibfield  {author} {\bibinfo {author} {\bibfnamefont {C.}~\bibnamefont {Zhang}}, \bibinfo {author} {\bibfnamefont {L.}~\bibnamefont {Duan}}, \ and\ \bibinfo {author} {\bibfnamefont {M.~M.}\ \bibnamefont {Choudhari}},\ }\bibfield  {title} {\enquote {\bibinfo {title} {Direct numerical simulation database for supersonic and hypersonic turbulent boundary layers},}\ }\href@noop {} {\bibfield  {journal} {\bibinfo  {journal} {AIAA journal}\ }\textbf {\bibinfo {volume} {56}},\ \bibinfo {pages} {4297--4311} (\bibinfo {year} {2018})}\BibitemShut {NoStop}%
\bibitem [{\citenamefont {Grasso}\ and\ \citenamefont {Falconi}(1993)}]{grasso1993high}%
  \BibitemOpen
  \bibfield  {author} {\bibinfo {author} {\bibfnamefont {F.}~\bibnamefont {Grasso}}\ and\ \bibinfo {author} {\bibfnamefont {D.}~\bibnamefont {Falconi}},\ }\bibfield  {title} {\enquote {\bibinfo {title} {High-speed turbulence modeling of shock-wave/boundary-layer interaction},}\ }\href@noop {} {\bibfield  {journal} {\bibinfo  {journal} {AIAA journal}\ }\textbf {\bibinfo {volume} {31}},\ \bibinfo {pages} {1199--1206} (\bibinfo {year} {1993})}\BibitemShut {NoStop}%
\bibitem [{\citenamefont {Speziale}\ and\ \citenamefont {Sarkar}(1989)}]{speziale1989preliminary}%
  \BibitemOpen
  \bibfield  {author} {\bibinfo {author} {\bibfnamefont {C.~G.}\ \bibnamefont {Speziale}}\ and\ \bibinfo {author} {\bibfnamefont {S.}~\bibnamefont {Sarkar}},\ }\href@noop {} {\enquote {\bibinfo {title} {A preliminary compressible second-order closure model for high speed flows},}\ }\bibinfo {type} {Tech. Rep.}\ (\bibinfo {year} {1989})\BibitemShut {NoStop}%
\bibitem [{\citenamefont {Sarkar}(1992)}]{sarkar1992pressure}%
  \BibitemOpen
  \bibfield  {author} {\bibinfo {author} {\bibfnamefont {S.}~\bibnamefont {Sarkar}},\ }\bibfield  {title} {\enquote {\bibinfo {title} {The pressure--dilatation correlation in compressible flows},}\ }\href@noop {} {\bibfield  {journal} {\bibinfo  {journal} {Physics of Fluids A: Fluid Dynamics}\ }\textbf {\bibinfo {volume} {4}},\ \bibinfo {pages} {2674--2682} (\bibinfo {year} {1992})}\BibitemShut {NoStop}%
\bibitem [{\citenamefont {Vuong}\ and\ \citenamefont {Coakley}(1987)}]{vuong1987modeling}%
  \BibitemOpen
  \bibfield  {author} {\bibinfo {author} {\bibfnamefont {S.}~\bibnamefont {Vuong}}\ and\ \bibinfo {author} {\bibfnamefont {T.}~\bibnamefont {Coakley}},\ }\bibfield  {title} {\enquote {\bibinfo {title} {Modeling of turbulence for hypersonic flows with and without separation},}\ }in\ \href@noop {} {\emph {\bibinfo {booktitle} {25th AIAA Aerospace Sciences Meeting}}}\ (\bibinfo {year} {1987})\ p.\ \bibinfo {pages} {286}\BibitemShut {NoStop}%
\bibitem [{\citenamefont {Huang}\ and\ \citenamefont {Coakley}(1993{\natexlab{a}})}]{huang1993calculations}%
  \BibitemOpen
  \bibfield  {author} {\bibinfo {author} {\bibfnamefont {P.}~\bibnamefont {Huang}}\ and\ \bibinfo {author} {\bibfnamefont {T.}~\bibnamefont {Coakley}},\ }\bibfield  {title} {\enquote {\bibinfo {title} {Calculations of supersonic and hypersonic flows using compressible wall functions},}\ }in\ \href@noop {} {\emph {\bibinfo {booktitle} {Engineering Turbulence Modelling and Experiments}}}\ (\bibinfo  {publisher} {Elsevier},\ \bibinfo {year} {1993})\ pp.\ \bibinfo {pages} {731--739}\BibitemShut {NoStop}%
\bibitem [{\citenamefont {Veera}\ and\ \citenamefont {Sinha}(2009)}]{veera2009modeling}%
  \BibitemOpen
  \bibfield  {author} {\bibinfo {author} {\bibfnamefont {V.~K.}\ \bibnamefont {Veera}}\ and\ \bibinfo {author} {\bibfnamefont {K.}~\bibnamefont {Sinha}},\ }\bibfield  {title} {\enquote {\bibinfo {title} {Modeling the effect of upstream temperature fluctuations on shock/homogeneous turbulence interaction},}\ }\href@noop {} {\bibfield  {journal} {\bibinfo  {journal} {Physics of fluids}\ }\textbf {\bibinfo {volume} {21}} (\bibinfo {year} {2009})}\BibitemShut {NoStop}%
\bibitem [{\citenamefont {Rathi}\ and\ \citenamefont {Sinha}(2024)}]{rathi2024simulation}%
  \BibitemOpen
  \bibfield  {author} {\bibinfo {author} {\bibfnamefont {H.}~\bibnamefont {Rathi}}\ and\ \bibinfo {author} {\bibfnamefont {K.}~\bibnamefont {Sinha}},\ }\bibfield  {title} {\enquote {\bibinfo {title} {Simulation of hypersonic shock-boundary layer interaction using shock-strength dependent turbulence model},}\ }\href@noop {} {\bibfield  {journal} {\bibinfo  {journal} {AIAA Journal}\ ,\ \bibinfo {pages} {1--14}} (\bibinfo {year} {2024})}\BibitemShut {NoStop}%
\bibitem [{\citenamefont {Catris}\ and\ \citenamefont {Aupoix}(2000)}]{catris2000density}%
  \BibitemOpen
  \bibfield  {author} {\bibinfo {author} {\bibfnamefont {S.}~\bibnamefont {Catris}}\ and\ \bibinfo {author} {\bibfnamefont {B.}~\bibnamefont {Aupoix}},\ }\bibfield  {title} {\enquote {\bibinfo {title} {Density corrections for turbulence models},}\ }\href@noop {} {\bibfield  {journal} {\bibinfo  {journal} {Aerospace Science and Technology}\ }\textbf {\bibinfo {volume} {4}},\ \bibinfo {pages} {1--11} (\bibinfo {year} {2000})}\BibitemShut {NoStop}%
\bibitem [{\citenamefont {Pecnik}\ and\ \citenamefont {Patel}(2017)}]{Pecnik2017scaling}%
  \BibitemOpen
  \bibfield  {author} {\bibinfo {author} {\bibfnamefont {R.}~\bibnamefont {Pecnik}}\ and\ \bibinfo {author} {\bibfnamefont {A.}~\bibnamefont {Patel}},\ }\bibfield  {title} {\enquote {\bibinfo {title} {Scaling and modelling of turbulence in variable property channel flows},}\ }\href@noop {} {\bibfield  {journal} {\bibinfo  {journal} {Journal of Fluid Mechanics}\ }\textbf {\bibinfo {volume} {823}},\ \bibinfo {pages} {R1--1--11} (\bibinfo {year} {2017})}\BibitemShut {NoStop}%
\bibitem [{\citenamefont {Otero}\ \emph {et~al.}(2018)\citenamefont {Otero}, \citenamefont {Patel}, \citenamefont {Diez},\ and\ \citenamefont {Pecnik}}]{otero2018turbulence}%
  \BibitemOpen
  \bibfield  {author} {\bibinfo {author} {\bibfnamefont {G.}~\bibnamefont {Otero}}, \bibinfo {author} {\bibfnamefont {A.}~\bibnamefont {Patel}}, \bibinfo {author} {\bibfnamefont {R.}~\bibnamefont {Diez}}, \ and\ \bibinfo {author} {\bibfnamefont {R.}~\bibnamefont {Pecnik}},\ }\bibfield  {title} {\enquote {\bibinfo {title} {Turbulence modelling for flows with strong variations in thermo-physical properties},}\ }\href@noop {} {\bibfield  {journal} {\bibinfo  {journal} {Journal of Heat and Fluid Flow}\ }\textbf {\bibinfo {volume} {823}},\ \bibinfo {pages} {R1} (\bibinfo {year} {2018})}\BibitemShut {NoStop}%
\bibitem [{\citenamefont {Hasan}\ \emph {et~al.}(2023)\citenamefont {Hasan}, \citenamefont {Larsson}, \citenamefont {Pirozzoli},\ and\ \citenamefont {Pecnik}}]{hasan2023incorporating}%
  \BibitemOpen
  \bibfield  {author} {\bibinfo {author} {\bibfnamefont {A.~M.}\ \bibnamefont {Hasan}}, \bibinfo {author} {\bibfnamefont {J.}~\bibnamefont {Larsson}}, \bibinfo {author} {\bibfnamefont {S.}~\bibnamefont {Pirozzoli}}, \ and\ \bibinfo {author} {\bibfnamefont {R.}~\bibnamefont {Pecnik}},\ }\bibfield  {title} {\enquote {\bibinfo {title} {Incorporating intrinsic compressibility effects in velocity transformations for wall-bounded turbulent flows},}\ }\href@noop {} {\bibfield  {journal} {\bibinfo  {journal} {Physical Review Fluids}\ }\textbf {\bibinfo {volume} {8}},\ \bibinfo {pages} {L112601} (\bibinfo {year} {2023})}\BibitemShut {NoStop}%
\bibitem [{\citenamefont {Danis}\ and\ \citenamefont {Durbin}(2022)}]{danis2022compressibility}%
  \BibitemOpen
  \bibfield  {author} {\bibinfo {author} {\bibfnamefont {M.~E.}\ \bibnamefont {Danis}}\ and\ \bibinfo {author} {\bibfnamefont {P.}~\bibnamefont {Durbin}},\ }\bibfield  {title} {\enquote {\bibinfo {title} {Compressibility correction to k- $\omega$ models for hypersonic turbulent boundary layers},}\ }\href@noop {} {\bibfield  {journal} {\bibinfo  {journal} {AIAA Journal}\ }\textbf {\bibinfo {volume} {60}},\ \bibinfo {pages} {6225--6234} (\bibinfo {year} {2022})}\BibitemShut {NoStop}%
\bibitem [{\citenamefont {Nicholson}\ \emph {et~al.}(2021{\natexlab{a}})\citenamefont {Nicholson}, \citenamefont {Huang}, \citenamefont {Duan},\ and\ \citenamefont {Choudhari}}]{nicholson2021simulation}%
  \BibitemOpen
  \bibfield  {author} {\bibinfo {author} {\bibfnamefont {G.}~\bibnamefont {Nicholson}}, \bibinfo {author} {\bibfnamefont {J.}~\bibnamefont {Huang}}, \bibinfo {author} {\bibfnamefont {L.}~\bibnamefont {Duan}}, \ and\ \bibinfo {author} {\bibfnamefont {M.~M.}\ \bibnamefont {Choudhari}},\ }\bibfield  {title} {\enquote {\bibinfo {title} {Simulation and modeling of hypersonic turbulent boundary layers subject to adverse pressure gradients due to streamline curvature},}\ }in\ \href@noop {} {\emph {\bibinfo {booktitle} {AIAA Aviation 2021 Forum}}}\ (\bibinfo {year} {2021})\ p.\ \bibinfo {pages} {2891}\BibitemShut {NoStop}%
\bibitem [{\citenamefont {Nicholson}\ \emph {et~al.}(2021{\natexlab{b}})\citenamefont {Nicholson}, \citenamefont {Huang}, \citenamefont {Duan}, \citenamefont {Choudhari},\ and\ \citenamefont {Bowersox}}]{nicholson2021fav}%
  \BibitemOpen
  \bibfield  {author} {\bibinfo {author} {\bibfnamefont {G.}~\bibnamefont {Nicholson}}, \bibinfo {author} {\bibfnamefont {J.}~\bibnamefont {Huang}}, \bibinfo {author} {\bibfnamefont {L.}~\bibnamefont {Duan}}, \bibinfo {author} {\bibfnamefont {M.~M.}\ \bibnamefont {Choudhari}}, \ and\ \bibinfo {author} {\bibfnamefont {R.~D.}\ \bibnamefont {Bowersox}},\ }\bibfield  {title} {\enquote {\bibinfo {title} {Simulation and modeling of hypersonic turbulent boundary layers subject to favorable pressure gradients due to streamline curvature},}\ }in\ \href@noop {} {\emph {\bibinfo {booktitle} {AIAA Scitech 2021 Forum}}}\ (\bibinfo {year} {2021})\ p.\ \bibinfo {pages} {1672}\BibitemShut {NoStop}%
\bibitem [{\citenamefont {Huang}, \citenamefont {Duan},\ and\ \citenamefont {Choudhari}(2022)}]{huang2022direct}%
  \BibitemOpen
  \bibfield  {author} {\bibinfo {author} {\bibfnamefont {J.}~\bibnamefont {Huang}}, \bibinfo {author} {\bibfnamefont {L.}~\bibnamefont {Duan}}, \ and\ \bibinfo {author} {\bibfnamefont {M.~M.}\ \bibnamefont {Choudhari}},\ }\bibfield  {title} {\enquote {\bibinfo {title} {Direct numerical simulation of hypersonic turbulent boundary layers: effect of spatial evolution and reynolds number},}\ }\href@noop {} {\bibfield  {journal} {\bibinfo  {journal} {Journal of Fluid Mechanics}\ }\textbf {\bibinfo {volume} {937}},\ \bibinfo {pages} {A3} (\bibinfo {year} {2022})}\BibitemShut {NoStop}%
\bibitem [{\citenamefont {Gnoffo}, \citenamefont {Berry},\ and\ \citenamefont {Van~Norman}(2011)}]{gnoffo2011uncertainty}%
  \BibitemOpen
  \bibfield  {author} {\bibinfo {author} {\bibfnamefont {P.}~\bibnamefont {Gnoffo}}, \bibinfo {author} {\bibfnamefont {S.}~\bibnamefont {Berry}}, \ and\ \bibinfo {author} {\bibfnamefont {J.}~\bibnamefont {Van~Norman}},\ }\bibfield  {title} {\enquote {\bibinfo {title} {Uncertainty assessments of 2d and axisymmetric hypersonic shock wave-turbulent boundary layer interaction simulations at compression corners},}\ }in\ \href@noop {} {\emph {\bibinfo {booktitle} {42nd AIAA thermophysics conference}}}\ (\bibinfo {year} {2011})\ p.\ \bibinfo {pages} {3142}\BibitemShut {NoStop}%
\bibitem [{\citenamefont {Maeder}, \citenamefont {Adams},\ and\ \citenamefont {Kleiser}(2001)}]{maeder2001direct}%
  \BibitemOpen
  \bibfield  {author} {\bibinfo {author} {\bibfnamefont {T.}~\bibnamefont {Maeder}}, \bibinfo {author} {\bibfnamefont {N.~A.}\ \bibnamefont {Adams}}, \ and\ \bibinfo {author} {\bibfnamefont {L.}~\bibnamefont {Kleiser}},\ }\bibfield  {title} {\enquote {\bibinfo {title} {Direct simulation of turbulent supersonic boundary layers by an extended temporal approach},}\ }\href@noop {} {\bibfield  {journal} {\bibinfo  {journal} {Journal of Fluid Mechanics}\ }\textbf {\bibinfo {volume} {429}},\ \bibinfo {pages} {187--216} (\bibinfo {year} {2001})}\BibitemShut {NoStop}%
\bibitem [{\citenamefont {Duan}, \citenamefont {Beekman},\ and\ \citenamefont {Martin}(2010)}]{duan2010direct}%
  \BibitemOpen
  \bibfield  {author} {\bibinfo {author} {\bibfnamefont {L.}~\bibnamefont {Duan}}, \bibinfo {author} {\bibfnamefont {I.}~\bibnamefont {Beekman}}, \ and\ \bibinfo {author} {\bibfnamefont {M.}~\bibnamefont {Martin}},\ }\bibfield  {title} {\enquote {\bibinfo {title} {Direct numerical simulation of hypersonic turbulent boundary layers. part 2. effect of wall temperature},}\ }\href@noop {} {\bibfield  {journal} {\bibinfo  {journal} {Journal of Fluid Mechanics}\ }\textbf {\bibinfo {volume} {655}},\ \bibinfo {pages} {419--445} (\bibinfo {year} {2010})}\BibitemShut {NoStop}%
\bibitem [{\citenamefont {Duan}, \citenamefont {Beekman},\ and\ \citenamefont {Martin}(2011)}]{duan2011direct}%
  \BibitemOpen
  \bibfield  {author} {\bibinfo {author} {\bibfnamefont {L.}~\bibnamefont {Duan}}, \bibinfo {author} {\bibfnamefont {I.}~\bibnamefont {Beekman}}, \ and\ \bibinfo {author} {\bibfnamefont {M.}~\bibnamefont {Martin}},\ }\bibfield  {title} {\enquote {\bibinfo {title} {Direct numerical simulation of hypersonic turbulent boundary layers. part 3. effect of mach number},}\ }\href@noop {} {\bibfield  {journal} {\bibinfo  {journal} {Journal of Fluid Mechanics}\ }\textbf {\bibinfo {volume} {672}},\ \bibinfo {pages} {245--267} (\bibinfo {year} {2011})}\BibitemShut {NoStop}%
\bibitem [{\citenamefont {Liang}(2012)}]{liang2012dns}%
  \BibitemOpen
  \bibfield  {author} {\bibinfo {author} {\bibfnamefont {X.}~\bibnamefont {Liang}},\ }\bibfield  {title} {\enquote {\bibinfo {title} {Dns and analysis of a spatially evolving hypersonic turbulent boundary layer over a flat plate at mach 8},}\ }\href@noop {} {\bibfield  {journal} {\bibinfo  {journal} {SCIENTIA SINICA Physica, Mechanica \& Astronomica}\ }\textbf {\bibinfo {volume} {42}},\ \bibinfo {pages} {282} (\bibinfo {year} {2012})}\BibitemShut {NoStop}%
\bibitem [{\citenamefont {Liang}\ and\ \citenamefont {Li}(2013)}]{liang2013dns}%
  \BibitemOpen
  \bibfield  {author} {\bibinfo {author} {\bibfnamefont {X.}~\bibnamefont {Liang}}\ and\ \bibinfo {author} {\bibfnamefont {X.}~\bibnamefont {Li}},\ }\bibfield  {title} {\enquote {\bibinfo {title} {Dns of a spatially evolving hypersonic turbulent boundary layer at mach 8},}\ }\href@noop {} {\bibfield  {journal} {\bibinfo  {journal} {Science China Physics, Mechanics and Astronomy}\ }\textbf {\bibinfo {volume} {56}},\ \bibinfo {pages} {1408--1418} (\bibinfo {year} {2013})}\BibitemShut {NoStop}%
\bibitem [{\citenamefont {Nicholson}(2024)}]{Ni24}%
  \BibitemOpen
  \bibfield  {author} {\bibinfo {author} {\bibfnamefont {G.}~\bibnamefont {Nicholson}},\ }\emph {\bibinfo {title} {{Simulation and Modeling of Hypersonic Turbulent Boundary Layers Subject to Favorable and Adverse Pressure Gradients Due to Streamline Curvature }}},\ \href@noop {} {Ph.D. thesis},\ \bibinfo  {school} {Ohio State University} (\bibinfo {year} {2024})\BibitemShut {NoStop}%
\bibitem [{\citenamefont {Roy}\ and\ \citenamefont {Sinha}(2019{\natexlab{a}})}]{roy2019turbulent}%
  \BibitemOpen
  \bibfield  {author} {\bibinfo {author} {\bibfnamefont {S.}~\bibnamefont {Roy}}\ and\ \bibinfo {author} {\bibfnamefont {K.}~\bibnamefont {Sinha}},\ }\bibfield  {title} {\enquote {\bibinfo {title} {Turbulent heat flux model for hypersonic shock--boundary layer interaction},}\ }\href@noop {} {\bibfield  {journal} {\bibinfo  {journal} {AIAA Journal}\ }\textbf {\bibinfo {volume} {57}},\ \bibinfo {pages} {3624--3629} (\bibinfo {year} {2019}{\natexlab{a}})}\BibitemShut {NoStop}%
\bibitem [{\citenamefont {Roy}, \citenamefont {Pathak},\ and\ \citenamefont {Sinha}(2018)}]{roy2018variable}%
  \BibitemOpen
  \bibfield  {author} {\bibinfo {author} {\bibfnamefont {S.}~\bibnamefont {Roy}}, \bibinfo {author} {\bibfnamefont {U.}~\bibnamefont {Pathak}}, \ and\ \bibinfo {author} {\bibfnamefont {K.}~\bibnamefont {Sinha}},\ }\bibfield  {title} {\enquote {\bibinfo {title} {Variable turbulent prandtl number model for shock/boundary-layer interaction},}\ }\href@noop {} {\bibfield  {journal} {\bibinfo  {journal} {AIAA Journal}\ }\textbf {\bibinfo {volume} {56}},\ \bibinfo {pages} {342--355} (\bibinfo {year} {2018})}\BibitemShut {NoStop}%
\bibitem [{\citenamefont {Barone}, \citenamefont {Parish},\ and\ \citenamefont {Jordan}(2024{\natexlab{a}})}]{BaPaJo24}%
  \BibitemOpen
  \bibfield  {author} {\bibinfo {author} {\bibfnamefont {M.}~\bibnamefont {Barone}}, \bibinfo {author} {\bibfnamefont {E.}~\bibnamefont {Parish}}, \ and\ \bibinfo {author} {\bibfnamefont {C.}~\bibnamefont {Jordan}},\ }\bibfield  {title} {\enquote {\bibinfo {title} {{Data-Driven Modifications to the Spalart--Allmaras Turbulence Model for Supersonic and Hypersonic Boundary Layers}},}\ }in\ \href@noop {} {\emph {\bibinfo {booktitle} {AIAA SciTech}}}\ (\bibinfo {year} {2024})\BibitemShut {NoStop}%
\bibitem [{\citenamefont {Ribner}(1954)}]{Ribner_1954}%
  \BibitemOpen
  \bibfield  {author} {\bibinfo {author} {\bibfnamefont {H.~S.}\ \bibnamefont {Ribner}},\ }\href@noop {} {\emph {\bibinfo {title} {Convection of a pattern of vorticity through a shock wave}}},\ Vol.\ \bibinfo {volume} {1164}\ (\bibinfo  {publisher} {NACA},\ \bibinfo {year} {1954})\BibitemShut {NoStop}%
\bibitem [{\citenamefont {Ryu}\ and\ \citenamefont {Livescu}(2014)}]{Ryu_Livescu_2014}%
  \BibitemOpen
  \bibfield  {author} {\bibinfo {author} {\bibfnamefont {J.}~\bibnamefont {Ryu}}\ and\ \bibinfo {author} {\bibfnamefont {D.}~\bibnamefont {Livescu}},\ }\bibfield  {title} {\enquote {\bibinfo {title} {Turbulence structure behind the shock in canonical shock--vortical turbulence interaction},}\ }\href@noop {} {\bibfield  {journal} {\bibinfo  {journal} {Journal of Fluid Mechanics}\ }\textbf {\bibinfo {volume} {756}},\ \bibinfo {pages} {R1} (\bibinfo {year} {2014})}\BibitemShut {NoStop}%
\bibitem [{\citenamefont {Lele}(1992)}]{Lele_1992}%
  \BibitemOpen
  \bibfield  {author} {\bibinfo {author} {\bibfnamefont {S.~K.}\ \bibnamefont {Lele}},\ }\bibfield  {title} {\enquote {\bibinfo {title} {Shock-jump relations in a turbulent flow},}\ }\href@noop {} {\bibfield  {journal} {\bibinfo  {journal} {Physics of Fluids A: Fluid Dynamics}\ }\textbf {\bibinfo {volume} {4}},\ \bibinfo {pages} {2900--2905} (\bibinfo {year} {1992})}\BibitemShut {NoStop}%
\bibitem [{\citenamefont {Zank}\ \emph {et~al.}(2002)\citenamefont {Zank}, \citenamefont {Zhou}, \citenamefont {Matthaeus},\ and\ \citenamefont {Rice}}]{Zank_Zhou_Matthaeus_Rice_2002}%
  \BibitemOpen
  \bibfield  {author} {\bibinfo {author} {\bibfnamefont {G.~P.}\ \bibnamefont {Zank}}, \bibinfo {author} {\bibfnamefont {Y.}~\bibnamefont {Zhou}}, \bibinfo {author} {\bibfnamefont {W.~H.}\ \bibnamefont {Matthaeus}}, \ and\ \bibinfo {author} {\bibfnamefont {W.}~\bibnamefont {Rice}},\ }\bibfield  {title} {\enquote {\bibinfo {title} {The interaction of turbulence with shock waves: A basic model},}\ }\href@noop {} {\bibfield  {journal} {\bibinfo  {journal} {Physics of Fluids}\ }\textbf {\bibinfo {volume} {14}},\ \bibinfo {pages} {3766--3774} (\bibinfo {year} {2002})}\BibitemShut {NoStop}%
\bibitem [{\citenamefont {Jacquin}, \citenamefont {Cambon},\ and\ \citenamefont {Blin}(1993)}]{Jacquin_Cambon_Blin_1993}%
  \BibitemOpen
  \bibfield  {author} {\bibinfo {author} {\bibfnamefont {L.}~\bibnamefont {Jacquin}}, \bibinfo {author} {\bibfnamefont {C.}~\bibnamefont {Cambon}}, \ and\ \bibinfo {author} {\bibfnamefont {E.}~\bibnamefont {Blin}},\ }\bibfield  {title} {\enquote {\bibinfo {title} {Turbulence amplification by a shock wave and rapid distortion theory},}\ }\href@noop {} {\bibfield  {journal} {\bibinfo  {journal} {Physics of Fluids A: Fluid Dynamics}\ }\textbf {\bibinfo {volume} {5}},\ \bibinfo {pages} {2539--2550} (\bibinfo {year} {1993})}\BibitemShut {NoStop}%
\bibitem [{\citenamefont {Griffond}\ and\ \citenamefont {Soulard}(2012)}]{Griffond_Soulard_2012}%
  \BibitemOpen
  \bibfield  {author} {\bibinfo {author} {\bibfnamefont {J.}~\bibnamefont {Griffond}}\ and\ \bibinfo {author} {\bibfnamefont {O.}~\bibnamefont {Soulard}},\ }\bibfield  {title} {\enquote {\bibinfo {title} {Evolution of axisymmetric weakly turbulent mixtures interacting with shock or rarefaction waves},}\ }\href@noop {} {\bibfield  {journal} {\bibinfo  {journal} {Physics of Fluids}\ }\textbf {\bibinfo {volume} {24}} (\bibinfo {year} {2012})}\BibitemShut {NoStop}%
\bibitem [{\citenamefont {Howarth}(1953)}]{Howarth1953}%
  \BibitemOpen
  \bibfield  {author} {\bibinfo {author} {\bibfnamefont {L.}~\bibnamefont {Howarth}},\ }\href@noop {} {\emph {\bibinfo {title} {Modern developments in fluid dynamics : high speed flow}}},\ The Oxford engineering science series\ (\bibinfo  {publisher} {Oxford at the Clarendon Press},\ \bibinfo {year} {1953})\BibitemShut {NoStop}%
\bibitem [{\citenamefont {Schwarzkopf}\ \emph {et~al.}(2016)\citenamefont {Schwarzkopf}, \citenamefont {Livescu}, \citenamefont {Baltzer}, \citenamefont {Gore},\ and\ \citenamefont {Ristorcelli}}]{Schwarzkopf_Livescu_Baltzer_Gore_Ristorcelli_2016}%
  \BibitemOpen
  \bibfield  {author} {\bibinfo {author} {\bibfnamefont {J.~D.}\ \bibnamefont {Schwarzkopf}}, \bibinfo {author} {\bibfnamefont {D.}~\bibnamefont {Livescu}}, \bibinfo {author} {\bibfnamefont {J.~R.}\ \bibnamefont {Baltzer}}, \bibinfo {author} {\bibfnamefont {R.~A.}\ \bibnamefont {Gore}}, \ and\ \bibinfo {author} {\bibfnamefont {J.}~\bibnamefont {Ristorcelli}},\ }\bibfield  {title} {\enquote {\bibinfo {title} {A two-length scale turbulence model for single-phase multi-fluid mixing},}\ }\href@noop {} {\bibfield  {journal} {\bibinfo  {journal} {Flow, Turbulence and Combustion}\ }\textbf {\bibinfo {volume} {96}},\ \bibinfo {pages} {1--43} (\bibinfo {year} {2016})}\BibitemShut {NoStop}%
\bibitem [{\citenamefont {Chen}\ and\ \citenamefont {Donzis}(2019)}]{Chen_Donzis_2019}%
  \BibitemOpen
  \bibfield  {author} {\bibinfo {author} {\bibfnamefont {C.~H.}\ \bibnamefont {Chen}}\ and\ \bibinfo {author} {\bibfnamefont {D.~A.}\ \bibnamefont {Donzis}},\ }\bibfield  {title} {\enquote {\bibinfo {title} {Shock--turbulence interactions at high turbulence intensities},}\ }\href@noop {} {\bibfield  {journal} {\bibinfo  {journal} {Journal of Fluid Mechanics}\ }\textbf {\bibinfo {volume} {870}},\ \bibinfo {pages} {813--847} (\bibinfo {year} {2019})}\BibitemShut {NoStop}%
\bibitem [{\citenamefont {Lacombe}\ \emph {et~al.}(2021)\citenamefont {Lacombe}, \citenamefont {Roy}, \citenamefont {Sinha}, \citenamefont {Karl},\ and\ \citenamefont {Hickey}}]{Lacombe_Roy_Sinha_Karl_Hickey_2021}%
  \BibitemOpen
  \bibfield  {author} {\bibinfo {author} {\bibfnamefont {F.}~\bibnamefont {Lacombe}}, \bibinfo {author} {\bibfnamefont {S.}~\bibnamefont {Roy}}, \bibinfo {author} {\bibfnamefont {K.}~\bibnamefont {Sinha}}, \bibinfo {author} {\bibfnamefont {S.}~\bibnamefont {Karl}}, \ and\ \bibinfo {author} {\bibfnamefont {J.-P.}\ \bibnamefont {Hickey}},\ }\bibfield  {title} {\enquote {\bibinfo {title} {Characteristic scales in shock--turbulence interaction},}\ }\href@noop {} {\bibfield  {journal} {\bibinfo  {journal} {AIAA Journal}\ }\textbf {\bibinfo {volume} {59}},\ \bibinfo {pages} {526--532} (\bibinfo {year} {2021})}\BibitemShut {NoStop}%
\bibitem [{\citenamefont {Braun}\ and\ \citenamefont {Gore}(2018)}]{Braun_Gore_2018}%
  \BibitemOpen
  \bibfield  {author} {\bibinfo {author} {\bibfnamefont {N.}~\bibnamefont {Braun}}\ and\ \bibinfo {author} {\bibfnamefont {R.}~\bibnamefont {Gore}},\ }\bibfield  {title} {\enquote {\bibinfo {title} {On primitive variable behaviour near shocks in ensemble-averaged methods},}\ }\href@noop {} {\bibfield  {journal} {\bibinfo  {journal} {Journal of Turbulence}\ }\textbf {\bibinfo {volume} {19}},\ \bibinfo {pages} {868--888} (\bibinfo {year} {2018})}\BibitemShut {NoStop}%
\bibitem [{\citenamefont {Sinha}, \citenamefont {Mahesh},\ and\ \citenamefont {Candler}(2003{\natexlab{b}})}]{Sinha_Mahesh_Candler_2003}%
  \BibitemOpen
  \bibfield  {author} {\bibinfo {author} {\bibfnamefont {K.}~\bibnamefont {Sinha}}, \bibinfo {author} {\bibfnamefont {K.}~\bibnamefont {Mahesh}}, \ and\ \bibinfo {author} {\bibfnamefont {G.~V.}\ \bibnamefont {Candler}},\ }\bibfield  {title} {\enquote {\bibinfo {title} {Modeling shock unsteadiness in shock/turbulence interaction},}\ }\href@noop {} {\bibfield  {journal} {\bibinfo  {journal} {Physics of fluids}\ }\textbf {\bibinfo {volume} {15}},\ \bibinfo {pages} {2290--2297} (\bibinfo {year} {2003}{\natexlab{b}})}\BibitemShut {NoStop}%
\bibitem [{\citenamefont {Zhang}\ \emph {et~al.}(2017{\natexlab{a}})\citenamefont {Zhang}, \citenamefont {Gao}, \citenamefont {Jiang},\ and\ \citenamefont {Lee}}]{Zhang_2017}%
  \BibitemOpen
  \bibfield  {author} {\bibinfo {author} {\bibfnamefont {Z.}~\bibnamefont {Zhang}}, \bibinfo {author} {\bibfnamefont {Z.}~\bibnamefont {Gao}}, \bibinfo {author} {\bibfnamefont {C.}~\bibnamefont {Jiang}}, \ and\ \bibinfo {author} {\bibfnamefont {C.-H.}\ \bibnamefont {Lee}},\ }\bibfield  {title} {\enquote {\bibinfo {title} {A rans model correction on unphysical over-prediction of turbulent quantities across shock wave},}\ }\href@noop {} {\bibfield  {journal} {\bibinfo  {journal} {International Journal of Heat and Mass Transfer}\ }\textbf {\bibinfo {volume} {106}},\ \bibinfo {pages} {1107--1119} (\bibinfo {year} {2017}{\natexlab{a}})}\BibitemShut {NoStop}%
\bibitem [{\citenamefont {Prasad}\ and\ \citenamefont {Gaitonde}(2023)}]{prasad2023turbulence}%
  \BibitemOpen
  \bibfield  {author} {\bibinfo {author} {\bibfnamefont {C.}~\bibnamefont {Prasad}}\ and\ \bibinfo {author} {\bibfnamefont {D.~V.}\ \bibnamefont {Gaitonde}},\ }\bibfield  {title} {\enquote {\bibinfo {title} {Turbulence modeling of 3d high-speed flows with upstream-informed corrections},}\ }\href@noop {} {\bibfield  {journal} {\bibinfo  {journal} {Shock Waves}\ }\textbf {\bibinfo {volume} {33}},\ \bibinfo {pages} {99--115} (\bibinfo {year} {2023})}\BibitemShut {NoStop}%
\bibitem [{\citenamefont {Zhang}\ \emph {et~al.}(2017{\natexlab{b}})\citenamefont {Zhang}, \citenamefont {Gao}, \citenamefont {Jiang},\ and\ \citenamefont {Lee}}]{Zhang_Gao_Jiang_Lee_2017}%
  \BibitemOpen
  \bibfield  {author} {\bibinfo {author} {\bibfnamefont {Z.}~\bibnamefont {Zhang}}, \bibinfo {author} {\bibfnamefont {Z.}~\bibnamefont {Gao}}, \bibinfo {author} {\bibfnamefont {C.}~\bibnamefont {Jiang}}, \ and\ \bibinfo {author} {\bibfnamefont {C.-H.}\ \bibnamefont {Lee}},\ }\bibfield  {title} {\enquote {\bibinfo {title} {A rans model correction on unphysical over-prediction of turbulent quantities across shock wave},}\ }\href@noop {} {\bibfield  {journal} {\bibinfo  {journal} {International Journal of Heat and Mass Transfer}\ }\textbf {\bibinfo {volume} {106}},\ \bibinfo {pages} {1107--1119} (\bibinfo {year} {2017}{\natexlab{b}})}\BibitemShut {NoStop}%
\bibitem [{\citenamefont {Tian}\ \emph {et~al.}(2023)\citenamefont {Tian}, \citenamefont {Gao}, \citenamefont {Jiang},\ and\ \citenamefont {Lee}}]{Tian_Gao_Jiang_Lee_2023}%
  \BibitemOpen
  \bibfield  {author} {\bibinfo {author} {\bibfnamefont {Y.}~\bibnamefont {Tian}}, \bibinfo {author} {\bibfnamefont {Z.}~\bibnamefont {Gao}}, \bibinfo {author} {\bibfnamefont {C.}~\bibnamefont {Jiang}}, \ and\ \bibinfo {author} {\bibfnamefont {C.-H.}\ \bibnamefont {Lee}},\ }\bibfield  {title} {\enquote {\bibinfo {title} {A correction for reynolds-averaged-navier--stokes turbulence model under the effect of shock waves in hypersonic flows},}\ }\href@noop {} {\bibfield  {journal} {\bibinfo  {journal} {International Journal for Numerical Methods in Fluids}\ }\textbf {\bibinfo {volume} {95}},\ \bibinfo {pages} {313--333} (\bibinfo {year} {2023})}\BibitemShut {NoStop}%
\bibitem [{\citenamefont {Vemula}\ and\ \citenamefont {Sinha}(2017)}]{Vemula_Sinha_2017}%
  \BibitemOpen
  \bibfield  {author} {\bibinfo {author} {\bibfnamefont {J.~B.}\ \bibnamefont {Vemula}}\ and\ \bibinfo {author} {\bibfnamefont {K.}~\bibnamefont {Sinha}},\ }\bibfield  {title} {\enquote {\bibinfo {title} {Reynolds stress models applied to canonical shock-turbulence interaction},}\ }\href@noop {} {\bibfield  {journal} {\bibinfo  {journal} {Journal of Turbulence}\ }\textbf {\bibinfo {volume} {18}},\ \bibinfo {pages} {653--687} (\bibinfo {year} {2017})}\BibitemShut {NoStop}%
\bibitem [{\citenamefont {Karl}, \citenamefont {Hickey},\ and\ \citenamefont {Lacombe}(2019)}]{Karl_Hickey_Lacombe_2019}%
  \BibitemOpen
  \bibfield  {author} {\bibinfo {author} {\bibfnamefont {S.}~\bibnamefont {Karl}}, \bibinfo {author} {\bibfnamefont {J.-P.}\ \bibnamefont {Hickey}}, \ and\ \bibinfo {author} {\bibfnamefont {F.}~\bibnamefont {Lacombe}},\ }\bibfield  {title} {\enquote {\bibinfo {title} {Reynolds stress models for shock-turbulence interaction},}\ }in\ \href@noop {} {\emph {\bibinfo {booktitle} {31st International Symposium on Shock Waves 1: Fundamentals 31}}}\ (\bibinfo {organization} {Springer},\ \bibinfo {year} {2019})\ pp.\ \bibinfo {pages} {511--517}\BibitemShut {NoStop}%
\bibitem [{\citenamefont {Holden}\ \emph {et~al.}(2010)\citenamefont {Holden}, \citenamefont {Wadhams}, \citenamefont {MacLean},\ and\ \citenamefont {Mundy}}]{holden2010experimental}%
  \BibitemOpen
  \bibfield  {author} {\bibinfo {author} {\bibfnamefont {M.}~\bibnamefont {Holden}}, \bibinfo {author} {\bibfnamefont {T.}~\bibnamefont {Wadhams}}, \bibinfo {author} {\bibfnamefont {M.}~\bibnamefont {MacLean}}, \ and\ \bibinfo {author} {\bibfnamefont {E.}~\bibnamefont {Mundy}},\ }\bibfield  {title} {\enquote {\bibinfo {title} {Experimental studies of shock wave/turbulent boundary layer interaction in high reynolds number supersonic and hypersonic flows to evaluate the performance of cfd codes},}\ }in\ \href@noop {} {\emph {\bibinfo {booktitle} {40th fluid dynamics conference and exhibit}}}\ (\bibinfo {year} {2010})\ p.\ \bibinfo {pages} {4468}\BibitemShut {NoStop}%
\bibitem [{\citenamefont {Kandula}\ and\ \citenamefont {Wilcox}(1995)}]{KaWi95}%
  \BibitemOpen
  \bibfield  {author} {\bibinfo {author} {\bibfnamefont {M.}~\bibnamefont {Kandula}}\ and\ \bibinfo {author} {\bibfnamefont {D.}~\bibnamefont {Wilcox}},\ }\bibfield  {title} {\enquote {\bibinfo {title} {An examination of k-omega turbulence model for boundary layers, free shear layers and separated flows},}\ }in\ \href@noop {} {\emph {\bibinfo {booktitle} {Fluid dynamics conference}}}\ (\bibinfo {year} {1995})\ p.\ \bibinfo {pages} {2317}\BibitemShut {NoStop}%
\bibitem [{\citenamefont {Parish}\ \emph {et~al.}(2024{\natexlab{a}})\citenamefont {Parish}, \citenamefont {Barone}, \citenamefont {Ching}, \citenamefont {Jordan}, \citenamefont {Miller}, \citenamefont {Nicholson}, \citenamefont {Gitushi}, \citenamefont {Beresh}, \citenamefont {Gupta},\ and\ \citenamefont {Duraisamy}}]{parish2024report}%
  \BibitemOpen
  \bibfield  {author} {\bibinfo {author} {\bibfnamefont {E.}~\bibnamefont {Parish}}, \bibinfo {author} {\bibfnamefont {M.}~\bibnamefont {Barone}}, \bibinfo {author} {\bibfnamefont {D.}~\bibnamefont {Ching}}, \bibinfo {author} {\bibfnamefont {C.}~\bibnamefont {Jordan}}, \bibinfo {author} {\bibfnamefont {N.}~\bibnamefont {Miller}}, \bibinfo {author} {\bibfnamefont {G.}~\bibnamefont {Nicholson}}, \bibinfo {author} {\bibfnamefont {K.}~\bibnamefont {Gitushi}}, \bibinfo {author} {\bibfnamefont {S.}~\bibnamefont {Beresh}}, \bibinfo {author} {\bibfnamefont {N.}~\bibnamefont {Gupta}}, \ and\ \bibinfo {author} {\bibfnamefont {K.}~\bibnamefont {Duraisamy}},\ }\href@noop {} {\enquote {\bibinfo {title} {{Data-driven closure modeling for hypersonic turbulent flows}},}\ }\bibinfo {type} {Tech. Rep.}\ (\bibinfo  {institution} {Sandia National Laboratories},\ \bibinfo {year} {2024})\BibitemShut {NoStop}%
\bibitem [{\citenamefont {Huang}, \citenamefont {Bradshaw},\ and\ \citenamefont {Coakley}(1994)}]{huang1994turbulence}%
  \BibitemOpen
  \bibfield  {author} {\bibinfo {author} {\bibfnamefont {P.}~\bibnamefont {Huang}}, \bibinfo {author} {\bibfnamefont {P.}~\bibnamefont {Bradshaw}}, \ and\ \bibinfo {author} {\bibfnamefont {T.}~\bibnamefont {Coakley}},\ }\bibfield  {title} {\enquote {\bibinfo {title} {Turbulence models for compressible boundary layers},}\ }\href@noop {} {\bibfield  {journal} {\bibinfo  {journal} {AIAA journal}\ }\textbf {\bibinfo {volume} {32}},\ \bibinfo {pages} {735--740} (\bibinfo {year} {1994})}\BibitemShut {NoStop}%
\bibitem [{\citenamefont {Hoste}\ \emph {et~al.}(2024)\citenamefont {Hoste}, \citenamefont {Gibbons}, \citenamefont {Ecker}, \citenamefont {Amato}, \citenamefont {Knight}, \citenamefont {Sattarov}, \citenamefont {Thiry}, \citenamefont {Hickey}, \citenamefont {Hizir}, \citenamefont {K\"{o}kt\"{u}rk}, \citenamefont {Castelino}, \citenamefont {Viti}, \citenamefont {Roldan}, \citenamefont {Qiang}, \citenamefont {Coder}, \citenamefont {Baurle},\ and\ \citenamefont {White}}]{Hoste2024}%
  \BibitemOpen
  \bibfield  {author} {\bibinfo {author} {\bibfnamefont {J.-J.}\ \bibnamefont {Hoste}}, \bibinfo {author} {\bibfnamefont {N.}~\bibnamefont {Gibbons}}, \bibinfo {author} {\bibfnamefont {T.}~\bibnamefont {Ecker}}, \bibinfo {author} {\bibfnamefont {C.}~\bibnamefont {Amato}}, \bibinfo {author} {\bibfnamefont {D.}~\bibnamefont {Knight}}, \bibinfo {author} {\bibfnamefont {A.}~\bibnamefont {Sattarov}}, \bibinfo {author} {\bibfnamefont {O.}~\bibnamefont {Thiry}}, \bibinfo {author} {\bibfnamefont {J.-P.}\ \bibnamefont {Hickey}}, \bibinfo {author} {\bibfnamefont {F.}~\bibnamefont {Hizir}}, \bibinfo {author} {\bibfnamefont {T.}~\bibnamefont {K\"{o}kt\"{u}rk}}, \bibinfo {author} {\bibfnamefont {N.}~\bibnamefont {Castelino}}, \bibinfo {author} {\bibfnamefont {V.}~\bibnamefont {Viti}}, \bibinfo {author} {\bibfnamefont {M.}~\bibnamefont {Roldan}}, \bibinfo {author} {\bibfnamefont {S.}~\bibnamefont {Qiang}}, \bibinfo {author} {\bibfnamefont {J.}~\bibnamefont {Coder}}, \bibinfo {author} {\bibfnamefont {R.}~\bibnamefont
  {Baurle}}, \ and\ \bibinfo {author} {\bibfnamefont {J.}~\bibnamefont {White}},\ }\bibfield  {title} {\enquote {\bibinfo {title} {A review of {RANS} modeling for hypersonic large cone-flares},}\ }\href@noop {} {\bibfield  {journal} {\bibinfo  {journal} {Physics of Fluids}\ }\textbf {\bibinfo {volume} {(under review)}} (\bibinfo {year} {2024})}\BibitemShut {NoStop}%
\bibitem [{\citenamefont {Coakley}\ \emph {et~al.}(1994)\citenamefont {Coakley}, \citenamefont {Horstman}, \citenamefont {Marvin}, \citenamefont {Viegas}, \citenamefont {Bardina}, \citenamefont {Huang},\ and\ \citenamefont {Kussoy}}]{coakley1994turbulence}%
  \BibitemOpen
  \bibfield  {author} {\bibinfo {author} {\bibfnamefont {T.}~\bibnamefont {Coakley}}, \bibinfo {author} {\bibfnamefont {C.}~\bibnamefont {Horstman}}, \bibinfo {author} {\bibfnamefont {J.}~\bibnamefont {Marvin}}, \bibinfo {author} {\bibfnamefont {J.}~\bibnamefont {Viegas}}, \bibinfo {author} {\bibfnamefont {J.}~\bibnamefont {Bardina}}, \bibinfo {author} {\bibfnamefont {P.}~\bibnamefont {Huang}}, \ and\ \bibinfo {author} {\bibfnamefont {M.}~\bibnamefont {Kussoy}},\ }\href@noop {} {\enquote {\bibinfo {title} {Turbulence compressibility corrections},}\ }\bibinfo {type} {Tech. Rep.}\ (\bibinfo {year} {1994})\BibitemShut {NoStop}%
\bibitem [{\citenamefont {Huang}\ and\ \citenamefont {Coakley}(1993{\natexlab{b}})}]{huang1993turbulence}%
  \BibitemOpen
  \bibfield  {author} {\bibinfo {author} {\bibfnamefont {P.}~\bibnamefont {Huang}}\ and\ \bibinfo {author} {\bibfnamefont {T.}~\bibnamefont {Coakley}},\ }\bibfield  {title} {\enquote {\bibinfo {title} {Turbulence modeling for complex hypersonic flows},}\ }in\ \href@noop {} {\emph {\bibinfo {booktitle} {31st Aerospace Sciences Meeting}}}\ (\bibinfo {year} {1993})\ p.\ \bibinfo {pages} {200}\BibitemShut {NoStop}%
\bibitem [{\citenamefont {Roy}\ and\ \citenamefont {Sinha}(2019{\natexlab{b}})}]{SuSi19}%
  \BibitemOpen
  \bibfield  {author} {\bibinfo {author} {\bibfnamefont {S.}~\bibnamefont {Roy}}\ and\ \bibinfo {author} {\bibfnamefont {K.}~\bibnamefont {Sinha}},\ }\bibfield  {title} {\enquote {\bibinfo {title} {Turbulent heat flux model for hypersonic shock--boundary layer interaction},}\ }\href@noop {} {\bibfield  {journal} {\bibinfo  {journal} {AIAA Journal}\ }\textbf {\bibinfo {volume} {57}},\ \bibinfo {pages} {3624--3629} (\bibinfo {year} {2019}{\natexlab{b}})}\BibitemShut {NoStop}%
\bibitem [{\citenamefont {Prince}, \citenamefont {Vannahme},\ and\ \citenamefont {Stollery}(2005)}]{prince2005experiments}%
  \BibitemOpen
  \bibfield  {author} {\bibinfo {author} {\bibfnamefont {S.}~\bibnamefont {Prince}}, \bibinfo {author} {\bibfnamefont {M.}~\bibnamefont {Vannahme}}, \ and\ \bibinfo {author} {\bibfnamefont {J.}~\bibnamefont {Stollery}},\ }\bibfield  {title} {\enquote {\bibinfo {title} {Experiments on the hypersonic turbulent shock-wave/boundary-layer interaction and the effects of surface roughness},}\ }\href@noop {} {\bibfield  {journal} {\bibinfo  {journal} {The Aeronautical Journal}\ }\textbf {\bibinfo {volume} {109}},\ \bibinfo {pages} {177--184} (\bibinfo {year} {2005})}\BibitemShut {NoStop}%
\bibitem [{\citenamefont {Holden}, \citenamefont {Wadhams},\ and\ \citenamefont {Mundy}(2008)}]{holden2008review}%
  \BibitemOpen
  \bibfield  {author} {\bibinfo {author} {\bibfnamefont {M.}~\bibnamefont {Holden}}, \bibinfo {author} {\bibfnamefont {T.}~\bibnamefont {Wadhams}}, \ and\ \bibinfo {author} {\bibfnamefont {E.}~\bibnamefont {Mundy}},\ }\bibfield  {title} {\enquote {\bibinfo {title} {A review of experimental studies of surface roughness and blowing on the heat transfer and skin friction to nosetips and slender cones in high mach numbers flows},}\ }in\ \href@noop {} {\emph {\bibinfo {booktitle} {40th Thermophysics Conference}}}\ (\bibinfo {year} {2008})\ p.\ \bibinfo {pages} {3907}\BibitemShut {NoStop}%
\bibitem [{\citenamefont {Peltier}, \citenamefont {Humble},\ and\ \citenamefont {Bowersox}(2016)}]{peltier2016crosshatch}%
  \BibitemOpen
  \bibfield  {author} {\bibinfo {author} {\bibfnamefont {S.}~\bibnamefont {Peltier}}, \bibinfo {author} {\bibfnamefont {R.}~\bibnamefont {Humble}}, \ and\ \bibinfo {author} {\bibfnamefont {R.}~\bibnamefont {Bowersox}},\ }\bibfield  {title} {\enquote {\bibinfo {title} {Crosshatch roughness distortions on a hypersonic turbulent boundary layer},}\ }\href@noop {} {\bibfield  {journal} {\bibinfo  {journal} {Physics of Fluids}\ }\textbf {\bibinfo {volume} {28}} (\bibinfo {year} {2016})}\BibitemShut {NoStop}%
\bibitem [{\citenamefont {Berg}(1977)}]{berg1977surface}%
  \BibitemOpen
  \bibfield  {author} {\bibinfo {author} {\bibfnamefont {D.~E.}\ \bibnamefont {Berg}},\ }\href@noop {} {\emph {\bibinfo {title} {Surface roughness effects on the hypersonic turbulent boundary layer.}}}\ (\bibinfo  {publisher} {California Institute of Technology},\ \bibinfo {year} {1977})\BibitemShut {NoStop}%
\bibitem [{\citenamefont {Finson}\ and\ \citenamefont {Clarke}(1980)}]{finson1980effect}%
  \BibitemOpen
  \bibfield  {author} {\bibinfo {author} {\bibfnamefont {M.}~\bibnamefont {Finson}}\ and\ \bibinfo {author} {\bibfnamefont {A.}~\bibnamefont {Clarke}},\ }\bibfield  {title} {\enquote {\bibinfo {title} {The effect of surface roughness character on turbulent re-entry heating},}\ }in\ \href@noop {} {\emph {\bibinfo {booktitle} {15th Thermophysics Conference}}}\ (\bibinfo {year} {1980})\ p.\ \bibinfo {pages} {1459}\BibitemShut {NoStop}%
\bibitem [{\citenamefont {Aupoix}(2015{\natexlab{a}})}]{aupoix2015roughness}%
  \BibitemOpen
  \bibfield  {author} {\bibinfo {author} {\bibfnamefont {B.}~\bibnamefont {Aupoix}},\ }\bibfield  {title} {\enquote {\bibinfo {title} {Roughness corrections for the k--$\omega$ shear stress transport model: Status and proposals},}\ }\href@noop {} {\bibfield  {journal} {\bibinfo  {journal} {Journal of Fluids Engineering}\ }\textbf {\bibinfo {volume} {137}},\ \bibinfo {pages} {021202} (\bibinfo {year} {2015}{\natexlab{a}})}\BibitemShut {NoStop}%
\bibitem [{\citenamefont {Goddard~Jr}(1959)}]{goddard1959effect}%
  \BibitemOpen
  \bibfield  {author} {\bibinfo {author} {\bibfnamefont {F.~E.}\ \bibnamefont {Goddard~Jr}},\ }\bibfield  {title} {\enquote {\bibinfo {title} {Effect of uniformly distributed roughness on turbulent skin-friction drag at supersonic speeds},}\ }\href@noop {} {\bibfield  {journal} {\bibinfo  {journal} {Journal of the Aerospace Sciences}\ }\textbf {\bibinfo {volume} {26}},\ \bibinfo {pages} {1--15} (\bibinfo {year} {1959})}\BibitemShut {NoStop}%
\bibitem [{\citenamefont {Lin}\ and\ \citenamefont {Bywater}(1982)}]{lin1982turbulence}%
  \BibitemOpen
  \bibfield  {author} {\bibinfo {author} {\bibfnamefont {T.}~\bibnamefont {Lin}}\ and\ \bibinfo {author} {\bibfnamefont {R.}~\bibnamefont {Bywater}},\ }\bibfield  {title} {\enquote {\bibinfo {title} {Turbulence models for high-speed, rough-wall boundary layers},}\ }\href@noop {} {\bibfield  {journal} {\bibinfo  {journal} {AIAA Journal}\ }\textbf {\bibinfo {volume} {20}},\ \bibinfo {pages} {325--333} (\bibinfo {year} {1982})}\BibitemShut {NoStop}%
\bibitem [{\citenamefont {Aupoix}(2015{\natexlab{b}})}]{aupoix2015improved}%
  \BibitemOpen
  \bibfield  {author} {\bibinfo {author} {\bibfnamefont {B.}~\bibnamefont {Aupoix}},\ }\bibfield  {title} {\enquote {\bibinfo {title} {Improved heat transfer predictions on rough surfaces},}\ }\href@noop {} {\bibfield  {journal} {\bibinfo  {journal} {International Journal of Heat and Fluid Flow}\ }\textbf {\bibinfo {volume} {56}},\ \bibinfo {pages} {160--171} (\bibinfo {year} {2015}{\natexlab{b}})}\BibitemShut {NoStop}%
\bibitem [{\citenamefont {Olazabal-Loum{\'e}}\ \emph {et~al.}(2017)\citenamefont {Olazabal-Loum{\'e}}, \citenamefont {Danvin}, \citenamefont {Mathiaud}, \citenamefont {Aupoix},\ and\ \citenamefont {Toulouse}}]{olazabal2017study}%
  \BibitemOpen
  \bibfield  {author} {\bibinfo {author} {\bibfnamefont {M.}~\bibnamefont {Olazabal-Loum{\'e}}}, \bibinfo {author} {\bibfnamefont {F.}~\bibnamefont {Danvin}}, \bibinfo {author} {\bibfnamefont {J.}~\bibnamefont {Mathiaud}}, \bibinfo {author} {\bibfnamefont {B.}~\bibnamefont {Aupoix}}, \ and\ \bibinfo {author} {\bibfnamefont {O.}~\bibnamefont {Toulouse}},\ }\bibfield  {title} {\enquote {\bibinfo {title} {Study on k-$\omega$ shear stress transport model corrections applied to rough wall turbulent hypersonic boundary layers},}\ }in\ \href@noop {} {\emph {\bibinfo {booktitle} {Seventh European Conference for Aeronautics and Space Sciences}}}\ (\bibinfo {year} {2017})\BibitemShut {NoStop}%
\bibitem [{\citenamefont {Olazabal-Loum{\'e}}\ \emph {et~al.}(2019)\citenamefont {Olazabal-Loum{\'e}}, \citenamefont {Chedevergne}, \citenamefont {Danvin},\ and\ \citenamefont {Mathiaud}}]{olazabal2019roughness}%
  \BibitemOpen
  \bibfield  {author} {\bibinfo {author} {\bibfnamefont {M.}~\bibnamefont {Olazabal-Loum{\'e}}}, \bibinfo {author} {\bibfnamefont {F.}~\bibnamefont {Chedevergne}}, \bibinfo {author} {\bibfnamefont {F.}~\bibnamefont {Danvin}}, \ and\ \bibinfo {author} {\bibfnamefont {J.}~\bibnamefont {Mathiaud}},\ }\bibfield  {title} {\enquote {\bibinfo {title} {Roughness corrections applied to the simulation of turbulent hypersonic flows},}\ }in\ \href@noop {} {\emph {\bibinfo {booktitle} {EUCASS 2019}}}\ (\bibinfo {year} {2019})\BibitemShut {NoStop}%
\bibitem [{\citenamefont {Bukva}\ \emph {et~al.}(2021)\citenamefont {Bukva}, \citenamefont {Zhang}, \citenamefont {Christopher},\ and\ \citenamefont {Hickey}}]{Bukva2021}%
  \BibitemOpen
  \bibfield  {author} {\bibinfo {author} {\bibfnamefont {A.}~\bibnamefont {Bukva}}, \bibinfo {author} {\bibfnamefont {K.}~\bibnamefont {Zhang}}, \bibinfo {author} {\bibfnamefont {N.}~\bibnamefont {Christopher}}, \ and\ \bibinfo {author} {\bibfnamefont {J.-P.}\ \bibnamefont {Hickey}},\ }\bibfield  {title} {\enquote {\bibinfo {title} {Assessment of turbulence modeling for massively-cooled turbulent boundary layer flows with transpiration cooling},}\ }\href@noop {} {\bibfield  {journal} {\bibinfo  {journal} {Physics of Fluids}\ }\textbf {\bibinfo {volume} {33}} (\bibinfo {year} {2021})}\BibitemShut {NoStop}%
\bibitem [{\citenamefont {Wilcox}(1988)}]{wilcox1988reassessment}%
  \BibitemOpen
  \bibfield  {author} {\bibinfo {author} {\bibfnamefont {D.~C.}\ \bibnamefont {Wilcox}},\ }\bibfield  {title} {\enquote {\bibinfo {title} {Reassessment of the scale-determining equation for advanced turbulence models},}\ }\href@noop {} {\bibfield  {journal} {\bibinfo  {journal} {AIAA journal}\ }\textbf {\bibinfo {volume} {26}},\ \bibinfo {pages} {1299--1310} (\bibinfo {year} {1988})}\BibitemShut {NoStop}%
\bibitem [{\citenamefont {Marchenay}, \citenamefont {Chedevergne},\ and\ \citenamefont {Olazabal~Loum{\'e}}(2021)}]{marchenay2021modeling}%
  \BibitemOpen
  \bibfield  {author} {\bibinfo {author} {\bibfnamefont {Y.}~\bibnamefont {Marchenay}}, \bibinfo {author} {\bibfnamefont {F.}~\bibnamefont {Chedevergne}}, \ and\ \bibinfo {author} {\bibfnamefont {M.}~\bibnamefont {Olazabal~Loum{\'e}}},\ }\bibfield  {title} {\enquote {\bibinfo {title} {Modeling of combined effects of surface roughness and blowing for reynolds-averaged navier--stokes turbulence models},}\ }\href@noop {} {\bibfield  {journal} {\bibinfo  {journal} {Physics of Fluids}\ }\textbf {\bibinfo {volume} {33}} (\bibinfo {year} {2021})}\BibitemShut {NoStop}%
\bibitem [{\citenamefont {Spalding}(1971)}]{spalding1971concentration}%
  \BibitemOpen
  \bibfield  {author} {\bibinfo {author} {\bibfnamefont {D.}~\bibnamefont {Spalding}},\ }\bibfield  {title} {\enquote {\bibinfo {title} {Concentration fluctuations in a round turbulent free jet},}\ }\href@noop {} {\bibfield  {journal} {\bibinfo  {journal} {Chemical Engineering Science}\ }\textbf {\bibinfo {volume} {26}},\ \bibinfo {pages} {95--107} (\bibinfo {year} {1971})}\BibitemShut {NoStop}%
\bibitem [{\citenamefont {Launder}(2005)}]{launder2005heat}%
  \BibitemOpen
  \bibfield  {author} {\bibinfo {author} {\bibfnamefont {B.~E.}\ \bibnamefont {Launder}},\ }\bibfield  {title} {\enquote {\bibinfo {title} {Heat and mass transport},}\ }\href@noop {} {\bibfield  {journal} {\bibinfo  {journal} {Turbulence}\ ,\ \bibinfo {pages} {231--287}} (\bibinfo {year} {2005})}\BibitemShut {NoStop}%
\bibitem [{\citenamefont {Tominaga}\ and\ \citenamefont {Stathopoulos}(2007)}]{tominaga2007turbulent}%
  \BibitemOpen
  \bibfield  {author} {\bibinfo {author} {\bibfnamefont {Y.}~\bibnamefont {Tominaga}}\ and\ \bibinfo {author} {\bibfnamefont {T.}~\bibnamefont {Stathopoulos}},\ }\bibfield  {title} {\enquote {\bibinfo {title} {Turbulent schmidt numbers for cfd analysis with various types of flowfield},}\ }\href@noop {} {\bibfield  {journal} {\bibinfo  {journal} {Atmospheric Environment}\ }\textbf {\bibinfo {volume} {41}},\ \bibinfo {pages} {8091--8099} (\bibinfo {year} {2007})}\BibitemShut {NoStop}%
\bibitem [{\citenamefont {Marquardt}, \citenamefont {Klaas},\ and\ \citenamefont {Schr{\"o}der}(2020)}]{marquardt2020experimental}%
  \BibitemOpen
  \bibfield  {author} {\bibinfo {author} {\bibfnamefont {P.}~\bibnamefont {Marquardt}}, \bibinfo {author} {\bibfnamefont {M.}~\bibnamefont {Klaas}}, \ and\ \bibinfo {author} {\bibfnamefont {W.}~\bibnamefont {Schr{\"o}der}},\ }\bibfield  {title} {\enquote {\bibinfo {title} {Experimental investigation of the turbulent schmidt number in supersonic film cooling with shock interaction},}\ }\href@noop {} {\bibfield  {journal} {\bibinfo  {journal} {Experiments in Fluids}\ }\textbf {\bibinfo {volume} {61}},\ \bibinfo {pages} {160} (\bibinfo {year} {2020})}\BibitemShut {NoStop}%
\bibitem [{\citenamefont {Xiang}\ \emph {et~al.}(2020)\citenamefont {Xiang}, \citenamefont {Yang}, \citenamefont {Xie}, \citenamefont {Li},\ and\ \citenamefont {Ren}}]{xiang2020turbulence}%
  \BibitemOpen
  \bibfield  {author} {\bibinfo {author} {\bibfnamefont {Z.}~\bibnamefont {Xiang}}, \bibinfo {author} {\bibfnamefont {S.}~\bibnamefont {Yang}}, \bibinfo {author} {\bibfnamefont {S.}~\bibnamefont {Xie}}, \bibinfo {author} {\bibfnamefont {J.}~\bibnamefont {Li}}, \ and\ \bibinfo {author} {\bibfnamefont {H.}~\bibnamefont {Ren}},\ }\bibfield  {title} {\enquote {\bibinfo {title} {Turbulence--chemistry interaction models with finite-rate chemistry and compressibility correction for simulation of supersonic turbulent combustion},}\ }\href@noop {} {\bibfield  {journal} {\bibinfo  {journal} {Engineering Applications of Computational Fluid Mechanics}\ }\textbf {\bibinfo {volume} {14}},\ \bibinfo {pages} {1546--1561} (\bibinfo {year} {2020})}\BibitemShut {NoStop}%
\bibitem [{\citenamefont {Bray}\ \emph {et~al.}(2006)\citenamefont {Bray}, \citenamefont {Champion}, \citenamefont {Libby},\ and\ \citenamefont {Swaminathan}}]{bray2006finite}%
  \BibitemOpen
  \bibfield  {author} {\bibinfo {author} {\bibfnamefont {K.}~\bibnamefont {Bray}}, \bibinfo {author} {\bibfnamefont {M.}~\bibnamefont {Champion}}, \bibinfo {author} {\bibfnamefont {P.}~\bibnamefont {Libby}}, \ and\ \bibinfo {author} {\bibfnamefont {N.}~\bibnamefont {Swaminathan}},\ }\bibfield  {title} {\enquote {\bibinfo {title} {Finite rate chemistry and presumed pdf models for premixed turbulent combustion},}\ }\href@noop {} {\bibfield  {journal} {\bibinfo  {journal} {Combustion and Flame}\ }\textbf {\bibinfo {volume} {146}},\ \bibinfo {pages} {665--673} (\bibinfo {year} {2006})}\BibitemShut {NoStop}%
\bibitem [{\citenamefont {Baranwal}, \citenamefont {Donzis},\ and\ \citenamefont {Bowersox}(2020)}]{baranwal2020vibrational}%
  \BibitemOpen
  \bibfield  {author} {\bibinfo {author} {\bibfnamefont {A.}~\bibnamefont {Baranwal}}, \bibinfo {author} {\bibfnamefont {D.~A.}\ \bibnamefont {Donzis}}, \ and\ \bibinfo {author} {\bibfnamefont {R.~D.}\ \bibnamefont {Bowersox}},\ }\bibfield  {title} {\enquote {\bibinfo {title} {Vibrational turbulent prandtl number in flows with thermal non-equilibrium},}\ }in\ \href@noop {} {\emph {\bibinfo {booktitle} {AIAA Scitech 2020 Forum}}}\ (\bibinfo {year} {2020})\ p.\ \bibinfo {pages} {2052}\BibitemShut {NoStop}%
\bibitem [{\citenamefont {Longo}(2004)}]{longo2004modelling}%
  \BibitemOpen
  \bibfield  {author} {\bibinfo {author} {\bibfnamefont {J.}~\bibnamefont {Longo}},\ }\bibfield  {title} {\enquote {\bibinfo {title} {Modelling of hypersonic flow phenomena},}\ }\href@noop {} {\bibfield  {journal} {\bibinfo  {journal} {Critical Technologies for Hypersonic Vehicle Development Technology--RTO/AVT/VKI Lecture Series}\ ,\ \bibinfo {pages} {10--14}} (\bibinfo {year} {2004})}\BibitemShut {NoStop}%
\bibitem [{\citenamefont {Mansour}\ \emph {et~al.}(2024)\citenamefont {Mansour}, \citenamefont {Panerai}, \citenamefont {Lachaud},\ and\ \citenamefont {Magin}}]{mansour2024flow}%
  \BibitemOpen
  \bibfield  {author} {\bibinfo {author} {\bibfnamefont {N.~N.}\ \bibnamefont {Mansour}}, \bibinfo {author} {\bibfnamefont {F.}~\bibnamefont {Panerai}}, \bibinfo {author} {\bibfnamefont {J.}~\bibnamefont {Lachaud}}, \ and\ \bibinfo {author} {\bibfnamefont {T.}~\bibnamefont {Magin}},\ }\bibfield  {title} {\enquote {\bibinfo {title} {Flow mechanics in ablative thermal protection systems},}\ }\href@noop {} {\bibfield  {journal} {\bibinfo  {journal} {Annual Review of Fluid Mechanics}\ }\textbf {\bibinfo {volume} {56}},\ \bibinfo {pages} {549--575} (\bibinfo {year} {2024})}\BibitemShut {NoStop}%
\bibitem [{\citenamefont {Schneider}(2010)}]{schneider2010hypersonic}%
  \BibitemOpen
  \bibfield  {author} {\bibinfo {author} {\bibfnamefont {S.~P.}\ \bibnamefont {Schneider}},\ }\bibfield  {title} {\enquote {\bibinfo {title} {Hypersonic boundary-layer transition with ablation and blowing},}\ }\href@noop {} {\bibfield  {journal} {\bibinfo  {journal} {Journal of Spacecraft and Rockets}\ }\textbf {\bibinfo {volume} {47}},\ \bibinfo {pages} {225--237} (\bibinfo {year} {2010})}\BibitemShut {NoStop}%
\bibitem [{\citenamefont {Yang}, \citenamefont {Wang},\ and\ \citenamefont {Gao}(2022)}]{Yang_Wang_Gao_2022}%
  \BibitemOpen
  \bibfield  {author} {\bibinfo {author} {\bibfnamefont {Z.}~\bibnamefont {Yang}}, \bibinfo {author} {\bibfnamefont {S.}~\bibnamefont {Wang}}, \ and\ \bibinfo {author} {\bibfnamefont {Z.}~\bibnamefont {Gao}},\ }\bibfield  {title} {\enquote {\bibinfo {title} {Studies on effects of wall temperature variation on heat transfer in hypersonic laminar boundary layer},}\ }\href@noop {} {\bibfield  {journal} {\bibinfo  {journal} {International Journal of Heat and Mass Transfer}\ }\textbf {\bibinfo {volume} {190}},\ \bibinfo {pages} {122790} (\bibinfo {year} {2022})}\BibitemShut {NoStop}%
\bibitem [{\citenamefont {Cheng}\ and\ \citenamefont {Fu}(2024)}]{Cheng_Fu_2024}%
  \BibitemOpen
  \bibfield  {author} {\bibinfo {author} {\bibfnamefont {C.}~\bibnamefont {Cheng}}\ and\ \bibinfo {author} {\bibfnamefont {L.}~\bibnamefont {Fu}},\ }\bibfield  {title} {\enquote {\bibinfo {title} {Mean temperature scalings in compressible wall turbulence},}\ }\href@noop {} {\bibfield  {journal} {\bibinfo  {journal} {Physical Review Fluids}\ }\textbf {\bibinfo {volume} {9}},\ \bibinfo {pages} {054610} (\bibinfo {year} {2024})}\BibitemShut {NoStop}%
\bibitem [{\citenamefont {Lewis}\ and\ \citenamefont {Hickey}(2023)}]{Lewis_Hickey_2023}%
  \BibitemOpen
  \bibfield  {author} {\bibinfo {author} {\bibfnamefont {M.~T.}\ \bibnamefont {Lewis}}\ and\ \bibinfo {author} {\bibfnamefont {J.-P.}\ \bibnamefont {Hickey}},\ }\bibfield  {title} {\enquote {\bibinfo {title} {Conjugate heat transfer in high-speed external flows: A review},}\ }\href@noop {} {\bibfield  {journal} {\bibinfo  {journal} {Journal of Thermophysics and Heat Transfer}\ }\textbf {\bibinfo {volume} {37}},\ \bibinfo {pages} {697--712} (\bibinfo {year} {2023})}\BibitemShut {NoStop}%
\bibitem [{\citenamefont {Yoder}(2016)}]{Yoder_2016}%
  \BibitemOpen
  \bibfield  {author} {\bibinfo {author} {\bibfnamefont {D.~A.}\ \bibnamefont {Yoder}},\ }\bibfield  {title} {\enquote {\bibinfo {title} {Comparison of turbulent thermal diffusivity and scalar variance models},}\ }in\ \href@noop {} {\emph {\bibinfo {booktitle} {54th AIAA aerospace sciences meeting}}}\ (\bibinfo {year} {2016})\ p.\ \bibinfo {pages} {1561}\BibitemShut {NoStop}%
\bibitem [{\citenamefont {Yang}\ \emph {et~al.}(2021)\citenamefont {Yang}, \citenamefont {Iacovides}, \citenamefont {Craft},\ and\ \citenamefont {Apsley}}]{Yang_Iacovides_Craft_Apsley_2021}%
  \BibitemOpen
  \bibfield  {author} {\bibinfo {author} {\bibfnamefont {G.}~\bibnamefont {Yang}}, \bibinfo {author} {\bibfnamefont {H.}~\bibnamefont {Iacovides}}, \bibinfo {author} {\bibfnamefont {T.}~\bibnamefont {Craft}}, \ and\ \bibinfo {author} {\bibfnamefont {D.}~\bibnamefont {Apsley}},\ }\bibfield  {title} {\enquote {\bibinfo {title} {Rans model development on temperature variance in conjugate heat transfer},}\ }\href@noop {} {\bibfield  {journal} {\bibinfo  {journal} {Journal of Turbulence}\ }\textbf {\bibinfo {volume} {22}},\ \bibinfo {pages} {180--207} (\bibinfo {year} {2021})}\BibitemShut {NoStop}%
\bibitem [{\citenamefont {Maheu}, \citenamefont {Moureau},\ and\ \citenamefont {Domingo}(2012)}]{Maheu_Moureau_Domingo_2012}%
  \BibitemOpen
  \bibfield  {author} {\bibinfo {author} {\bibfnamefont {N.}~\bibnamefont {Maheu}}, \bibinfo {author} {\bibfnamefont {V.}~\bibnamefont {Moureau}}, \ and\ \bibinfo {author} {\bibfnamefont {P.}~\bibnamefont {Domingo}},\ }\bibfield  {title} {\enquote {\bibinfo {title} {Large-eddy simulations of flow and heat transfer around a low-mach number turbine blade},}\ }in\ \href@noop {} {\emph {\bibinfo {booktitle} {THMT-12. Proceedings of the Seventh International Symposium On Turbulence Heat and Mass Transfer}}}\ (\bibinfo {organization} {Begel House Inc.},\ \bibinfo {year} {2012})\BibitemShut {NoStop}%
\bibitem [{\citenamefont {Gelain}(2021)}]{Gelain2021}%
  \BibitemOpen
  \bibfield  {author} {\bibinfo {author} {\bibfnamefont {M.}~\bibnamefont {Gelain}},\ }\emph {\bibinfo {title} {Aerothermal characterisation of a surface heat exchanger implemented in a turbofan by-pass duct}},\ \href@noop {} {Ph.D. thesis},\ \bibinfo  {school} {Universit{\'e} Paris-Saclay} (\bibinfo {year} {2021})\BibitemShut {NoStop}%
\bibitem [{\citenamefont {Muller}\ \emph {et~al.}(2024{\natexlab{a}})\citenamefont {Muller}, \citenamefont {Dutta}, \citenamefont {Boisvert},\ and\ \citenamefont {Oefelein}}]{Muller_Dutta_Boisvert_Oefelein_2024}%
  \BibitemOpen
  \bibfield  {author} {\bibinfo {author} {\bibfnamefont {J.~A.}\ \bibnamefont {Muller}}, \bibinfo {author} {\bibfnamefont {M.}~\bibnamefont {Dutta}}, \bibinfo {author} {\bibfnamefont {J.}~\bibnamefont {Boisvert}}, \ and\ \bibinfo {author} {\bibfnamefont {J.~C.}\ \bibnamefont {Oefelein}},\ }\bibfield  {title} {\enquote {\bibinfo {title} {Investigation of conjugate heat transfer in wall-modeled large eddy simulation of high-speed compressible wall-bounded flows},}\ }in\ \href@noop {} {\emph {\bibinfo {booktitle} {Turbo Expo: Power for Land, Sea, and Air}}},\ Vol.\ \bibinfo {volume} {88094}\ (\bibinfo {organization} {American Society of Mechanical Engineers},\ \bibinfo {year} {2024})\ p.\ \bibinfo {pages} {V013T13A034}\BibitemShut {NoStop}%
\bibitem [{\citenamefont {Muller}\ \emph {et~al.}(2024{\natexlab{b}})\citenamefont {Muller}, \citenamefont {Boisvert}, \citenamefont {Dutta},\ and\ \citenamefont {Oefelein}}]{muller2024investigation}%
  \BibitemOpen
  \bibfield  {author} {\bibinfo {author} {\bibfnamefont {J.}~\bibnamefont {Muller}}, \bibinfo {author} {\bibfnamefont {J.}~\bibnamefont {Boisvert}}, \bibinfo {author} {\bibfnamefont {M.}~\bibnamefont {Dutta}}, \ and\ \bibinfo {author} {\bibfnamefont {J.}~\bibnamefont {Oefelein}},\ }\bibfield  {title} {\enquote {\bibinfo {title} {Investigation of loosely-coupled conjugate heat transfer with wall-modeled large eddy simulation in a mach 2.5 flow},}\ }in\ \href@noop {} {\emph {\bibinfo {booktitle} {AIAA AVIATION FORUM AND ASCEND 2024}}}\ (\bibinfo {year} {2024})\ p.\ \bibinfo {pages} {4175}\BibitemShut {NoStop}%
\bibitem [{\citenamefont {Marvin}(2024)}]{Hypersonic_SBLI_Expt}%
  \BibitemOpen
  \bibfield  {author} {\bibinfo {author} {\bibfnamefont {J.}~\bibnamefont {Marvin}},\ }\href@noop {} {\enquote {\bibinfo {title} {Nasa turbulence modeling resource},}\ }\bibinfo {howpublished} {\url{https://turbmodels.larc.nasa.gov/Other_exp_Data/sbli_various_marvin_exp.html}} (\bibinfo {year} {2024}),\ \bibinfo {note} {last accessed on December 18, 2024}\BibitemShut {NoStop}%
\bibitem [{\citenamefont {Winkler}\ and\ \citenamefont {Cha}(1959)}]{winkler1959investigation}%
  \BibitemOpen
  \bibfield  {author} {\bibinfo {author} {\bibfnamefont {A.~M.}\ \bibnamefont {Winkler}}\ and\ \bibinfo {author} {\bibfnamefont {M.~H.}\ \bibnamefont {Cha}},\ }\href@noop {} {\emph {\bibinfo {title} {Investigation of flat plate hypersonic turbulent boundary layers with heat transfer at a Mach number of 5.2}}}\ (\bibinfo  {publisher} {NAVORD},\ \bibinfo {year} {1959})\BibitemShut {NoStop}%
\bibitem [{\citenamefont {Winkler}(1961)}]{winkler1961investigation}%
  \BibitemOpen
  \bibfield  {author} {\bibinfo {author} {\bibfnamefont {E.~M.}\ \bibnamefont {Winkler}},\ }\bibfield  {title} {\enquote {\bibinfo {title} {Investigation of flat-plate hypersonic, turbulent boundary layers with heat transfer},}\ }\href@noop {} {\bibfield  {journal} {\bibinfo  {journal} {Journal of Applied Mechanics}\ }\textbf {\bibinfo {volume} {28}},\ \bibinfo {pages} {323} (\bibinfo {year} {1961})}\BibitemShut {NoStop}%
\bibitem [{\citenamefont {Danberg}(1964)}]{danberg1964characteristics}%
  \BibitemOpen
  \bibfield  {author} {\bibinfo {author} {\bibfnamefont {J.~E.}\ \bibnamefont {Danberg}},\ }\bibfield  {title} {\enquote {\bibinfo {title} {Characteristics of the turbulent boundary layer with heat and mass transfer at m= 6.7},}\ }\href@noop {} {\  (\bibinfo {year} {1964})}\BibitemShut {NoStop}%
\bibitem [{\citenamefont {Danberg}(1967)}]{danberg1967characteristics}%
  \BibitemOpen
  \bibfield  {author} {\bibinfo {author} {\bibfnamefont {J.~E.}\ \bibnamefont {Danberg}},\ }\bibfield  {title} {\enquote {\bibinfo {title} {Characteristics of the turbulent boundary layer with heat and mass transfer: data tabulation},}\ }\href@noop {} {\bibfield  {journal} {\bibinfo  {journal} {NOLTR 675-6}\ } (\bibinfo {year} {1967})}\BibitemShut {NoStop}%
\bibitem [{\citenamefont {Young}(1965)}]{young1965experimental}%
  \BibitemOpen
  \bibfield  {author} {\bibinfo {author} {\bibfnamefont {F.~L.}\ \bibnamefont {Young}},\ }\href@noop {} {\enquote {\bibinfo {title} {Experimental investigation of the effects of surface roughness on compressible turbulent boundary layer skin friction and heat transfer},}\ }\bibinfo {type} {Tech. Rep.}\ (\bibinfo  {institution} {Technical Report DLR-532, CR- 21 Defense Research Laboratory, University of Texas, Austin},\ \bibinfo {year} {1965})\BibitemShut {NoStop}%
\bibitem [{\citenamefont {Neal~Jr}(1966)}]{neal1966study}%
  \BibitemOpen
  \bibfield  {author} {\bibinfo {author} {\bibfnamefont {L.}~\bibnamefont {Neal~Jr}},\ }\href@noop {} {\enquote {\bibinfo {title} {A study of the pressure, heat transfer, and skin friction on sharp and blunt flat plates at mach 6.8},}\ }\bibinfo {type} {Tech. Rep.}\ (\bibinfo {year} {1966})\BibitemShut {NoStop}%
\bibitem [{\citenamefont {Wallace}(1967)}]{wallace1967hypersonic}%
  \BibitemOpen
  \bibfield  {author} {\bibinfo {author} {\bibfnamefont {J.}~\bibnamefont {Wallace}},\ }\bibfield  {title} {\enquote {\bibinfo {title} {Hypersonic turbulent boundary layer studies at cold wall temperatures},}\ }\href@noop {} {\bibfield  {journal} {\bibinfo  {journal} {Proceedings of the 1967 Heat Transfer and Fluid Mechanics Institute held at the University of California, San Diego, La Jolla, California}\ } (\bibinfo {year} {1967})}\BibitemShut {NoStop}%
\bibitem [{\citenamefont {Hopkins}(1969)}]{hopkins1969summary}%
  \BibitemOpen
  \bibfield  {author} {\bibinfo {author} {\bibfnamefont {E.~J.}\ \bibnamefont {Hopkins}},\ }\href@noop {} {\emph {\bibinfo {title} {Summary and correlation of skin-friction and heat-transfer data for a hypersonic turbulent boundary layer on simple shapes}}},\ Vol.\ \bibinfo {volume} {5089}\ (\bibinfo  {publisher} {National Aeronautics and Space Administration},\ \bibinfo {year} {1969})\BibitemShut {NoStop}%
\bibitem [{\citenamefont {Voisinet}\ and\ \citenamefont {Lee}(1972)}]{voisinet1972measurements}%
  \BibitemOpen
  \bibfield  {author} {\bibinfo {author} {\bibfnamefont {R.}~\bibnamefont {Voisinet}}\ and\ \bibinfo {author} {\bibfnamefont {R.~E.}\ \bibnamefont {Lee}},\ }\href@noop {} {\enquote {\bibinfo {title} {Measurements of a mach 4.9 zero-pressure-gradient turbulent boundary layer with heat transfer. part 1: Data compilation},}\ }\bibinfo {type} {Tech. Rep.}\ (\bibinfo  {institution} {Technical Report NOLTR 72-232 (White Oak, Silver Spring, MD: United States Naval Ordnance Laboratory)},\ \bibinfo {year} {1972})\BibitemShut {NoStop}%
\bibitem [{\citenamefont {HOLDEN}(1972)}]{holden1972shock}%
  \BibitemOpen
  \bibfield  {author} {\bibinfo {author} {\bibfnamefont {M.}~\bibnamefont {HOLDEN}},\ }\bibfield  {title} {\enquote {\bibinfo {title} {Shock wave-turbulent boundary layer interaction in hypersonic flow},}\ }in\ \href@noop {} {\emph {\bibinfo {booktitle} {10th Aerospace Sciences Meeting}}}\ (\bibinfo {organization} {American Institute of Aeronautics and Astronautics},\ \bibinfo {year} {1972})\BibitemShut {NoStop}%
\bibitem [{\citenamefont {Watson}, \citenamefont {Harris},\ and\ \citenamefont {ANDERS}(1973)}]{watson1973measurements}%
  \BibitemOpen
  \bibfield  {author} {\bibinfo {author} {\bibfnamefont {R.}~\bibnamefont {Watson}}, \bibinfo {author} {\bibfnamefont {J.}~\bibnamefont {Harris}}, \ and\ \bibinfo {author} {\bibfnamefont {J.}~\bibnamefont {ANDERS}, \bibfnamefont {JR}},\ }\bibfield  {title} {\enquote {\bibinfo {title} {Measurements in a transitional/turbulent mach 10 boundary layer at high-reynolds numbers},}\ }in\ \href@noop {} {\emph {\bibinfo {booktitle} {11th Aerospace Sciences Meeting}}}\ (\bibinfo {year} {1973})\ p.\ \bibinfo {pages} {165}\BibitemShut {NoStop}%
\bibitem [{\citenamefont {Goyne}, \citenamefont {Stalker},\ and\ \citenamefont {Paull}(2003)}]{goyne2003skin}%
  \BibitemOpen
  \bibfield  {author} {\bibinfo {author} {\bibfnamefont {C.}~\bibnamefont {Goyne}}, \bibinfo {author} {\bibfnamefont {R.}~\bibnamefont {Stalker}}, \ and\ \bibinfo {author} {\bibfnamefont {A.}~\bibnamefont {Paull}},\ }\bibfield  {title} {\enquote {\bibinfo {title} {Skin-friction measurements in high-enthalpy hypersonic boundary layers},}\ }\href@noop {} {\bibfield  {journal} {\bibinfo  {journal} {Journal of Fluid Mechanics}\ }\textbf {\bibinfo {volume} {485}},\ \bibinfo {pages} {1--32} (\bibinfo {year} {2003})}\BibitemShut {NoStop}%
\bibitem [{\citenamefont {Hill}(1959)}]{hill1959turbulent}%
  \BibitemOpen
  \bibfield  {author} {\bibinfo {author} {\bibfnamefont {F.}~\bibnamefont {Hill}},\ }\bibfield  {title} {\enquote {\bibinfo {title} {Turbulent boundary layer measurements at mach numbers from 8 to 10},}\ }\href@noop {} {\bibfield  {journal} {\bibinfo  {journal} {The Physics of Fluids}\ }\textbf {\bibinfo {volume} {2}},\ \bibinfo {pages} {668--680} (\bibinfo {year} {1959})}\BibitemShut {NoStop}%
\bibitem [{\citenamefont {Lee}, \citenamefont {Yanta},\ and\ \citenamefont {Leonas}(1969)}]{lee1969velocity}%
  \BibitemOpen
  \bibfield  {author} {\bibinfo {author} {\bibfnamefont {R.~E.}\ \bibnamefont {Lee}}, \bibinfo {author} {\bibfnamefont {W.~J.}\ \bibnamefont {Yanta}}, \ and\ \bibinfo {author} {\bibfnamefont {A.~C.}\ \bibnamefont {Leonas}},\ }\href@noop {} {\emph {\bibinfo {title} {Velocity Profile, Skin-friction Balance and Heat-transfer Measurements of the Turbulent Boundary Layers at Mach 5 and Zero-pressure Gradient}}},\ Vol.~\bibinfo {volume} {69}\ (\bibinfo  {publisher} {United States Naval Ordnance Laboratory},\ \bibinfo {year} {1969})\BibitemShut {NoStop}%
\bibitem [{\citenamefont {Backx}(1973)}]{backx1973measurements}%
  \BibitemOpen
  \bibfield  {author} {\bibinfo {author} {\bibfnamefont {E.}~\bibnamefont {Backx}},\ }\bibfield  {title} {\enquote {\bibinfo {title} {Measurements in the mach 15 turbulent boundary layer on the wall of the longshot conical nozzle},}\ }in\ \href@noop {} {\emph {\bibinfo {booktitle} {EUROMECH, Colloquium on Heat Transfer in Turbulent Boundary Layers with Variable Fluid Properties, Goettingen, West Germany}}}\ (\bibinfo {year} {1973})\BibitemShut {NoStop}%
\bibitem [{\citenamefont {Backx}(1974)}]{backx1974experimental}%
  \BibitemOpen
  \bibfield  {author} {\bibinfo {author} {\bibfnamefont {E.}~\bibnamefont {Backx}},\ }\bibfield  {title} {\enquote {\bibinfo {title} {Experimental study of the turbulent boundary layer at mach 15 and 19. 8 in a conical nozzle},}\ }\href@noop {} {\  (\bibinfo {year} {1974})}\BibitemShut {NoStop}%
\bibitem [{\citenamefont {Backx}\ and\ \citenamefont {Richards}(1976)}]{backx1976high}%
  \BibitemOpen
  \bibfield  {author} {\bibinfo {author} {\bibfnamefont {E.}~\bibnamefont {Backx}}\ and\ \bibinfo {author} {\bibfnamefont {B.}~\bibnamefont {Richards}},\ }\bibfield  {title} {\enquote {\bibinfo {title} {A high mach number turbulent boundary-layer study},}\ }\href@noop {} {\bibfield  {journal} {\bibinfo  {journal} {AIAA Journal}\ }\textbf {\bibinfo {volume} {14}},\ \bibinfo {pages} {1159--1160} (\bibinfo {year} {1976})}\BibitemShut {NoStop}%
\bibitem [{\citenamefont {Samuels}, \citenamefont {Peterson},\ and\ \citenamefont {Adcock}(1967)}]{samuels1967experimental}%
  \BibitemOpen
  \bibfield  {author} {\bibinfo {author} {\bibfnamefont {R.~D.}\ \bibnamefont {Samuels}}, \bibinfo {author} {\bibfnamefont {J.~B.}\ \bibnamefont {Peterson}}, \ and\ \bibinfo {author} {\bibfnamefont {J.~B.}\ \bibnamefont {Adcock}},\ }\href@noop {} {\emph {\bibinfo {title} {Experimental investigation of the turbulent boundary layer at a Mach number of 6 with heat transfer at high Reynolds numbers}}},\ Vol.\ \bibinfo {volume} {3858}\ (\bibinfo  {publisher} {National Aeronautics and Space Administration},\ \bibinfo {year} {1967})\BibitemShut {NoStop}%
\bibitem [{\citenamefont {Murray}, \citenamefont {Hillier},\ and\ \citenamefont {Williams}(2013)}]{murray2013experimental}%
  \BibitemOpen
  \bibfield  {author} {\bibinfo {author} {\bibfnamefont {N.}~\bibnamefont {Murray}}, \bibinfo {author} {\bibfnamefont {R.}~\bibnamefont {Hillier}}, \ and\ \bibinfo {author} {\bibfnamefont {S.}~\bibnamefont {Williams}},\ }\bibfield  {title} {\enquote {\bibinfo {title} {Experimental investigation of axisymmetric hypersonic shock-wave/turbulent-boundary-layer interactions},}\ }\href@noop {} {\bibfield  {journal} {\bibinfo  {journal} {Journal of Fluid Mechanics}\ }\textbf {\bibinfo {volume} {714}},\ \bibinfo {pages} {152--189} (\bibinfo {year} {2013})}\BibitemShut {NoStop}%
\bibitem [{\citenamefont {Kimmel}(1993)}]{kimmel1993experimental}%
  \BibitemOpen
  \bibfield  {author} {\bibinfo {author} {\bibfnamefont {R.}~\bibnamefont {Kimmel}},\ }\bibfield  {title} {\enquote {\bibinfo {title} {Experimental transition zone lengths in pressure gradient in hypersonic flow},}\ }\href@noop {} {\bibfield  {journal} {\bibinfo  {journal} {ASME-PUBLICATIONS-FED}\ }\textbf {\bibinfo {volume} {151}},\ \bibinfo {pages} {117--117} (\bibinfo {year} {1993})}\BibitemShut {NoStop}%
\bibitem [{\citenamefont {KIMMEL}(1997)}]{kimmel1997effect}%
  \BibitemOpen
  \bibfield  {author} {\bibinfo {author} {\bibfnamefont {R.}~\bibnamefont {KIMMEL}},\ }\bibfield  {title} {\enquote {\bibinfo {title} {The effect of pressure gradients on transition zone length in hypersonic boundary layers},}\ }\href@noop {} {\bibfield  {journal} {\bibinfo  {journal} {Journal of fluids engineering}\ }\textbf {\bibinfo {volume} {119}},\ \bibinfo {pages} {36--41} (\bibinfo {year} {1997})}\BibitemShut {NoStop}%
\bibitem [{\citenamefont {Duan}(2024{\natexlab{b}})}]{Hypersonic_FP_DNS}%
  \BibitemOpen
  \bibfield  {author} {\bibinfo {author} {\bibfnamefont {L.}~\bibnamefont {Duan}},\ }\href@noop {} {\enquote {\bibinfo {title} {Nasa turbulence modeling resource},}\ }\bibinfo {howpublished} {\url{https://turbmodels.larc.nasa.gov/Other_DNS_Data/supersonic_hypersonic_flatplate.html}} (\bibinfo {year} {2024}{\natexlab{b}}),\ \bibinfo {note} {last accessed on December 18, 2024}\BibitemShut {NoStop}%
\bibitem [{\citenamefont {Xin-Liang}, \citenamefont {De-Xun},\ and\ \citenamefont {Yan-Wen}(2006)}]{xin2006direct}%
  \BibitemOpen
  \bibfield  {author} {\bibinfo {author} {\bibfnamefont {L.}~\bibnamefont {Xin-Liang}}, \bibinfo {author} {\bibfnamefont {F.}~\bibnamefont {De-Xun}}, \ and\ \bibinfo {author} {\bibfnamefont {M.}~\bibnamefont {Yan-Wen}},\ }\bibfield  {title} {\enquote {\bibinfo {title} {Direct numerical simulation of a spatially evolving supersonic turbulent boundary layer at ma= 6},}\ }\href@noop {} {\bibfield  {journal} {\bibinfo  {journal} {Chinese Physics Letters}\ }\textbf {\bibinfo {volume} {23}},\ \bibinfo {pages} {1519} (\bibinfo {year} {2006})}\BibitemShut {NoStop}%
\bibitem [{\citenamefont {Martin}(2007)}]{martin2007direct}%
  \BibitemOpen
  \bibfield  {author} {\bibinfo {author} {\bibfnamefont {M.~P.}\ \bibnamefont {Martin}},\ }\bibfield  {title} {\enquote {\bibinfo {title} {Direct numerical simulation of hypersonic turbulent boundary layers. part 1. initialization and comparison with experiments},}\ }\href@noop {} {\bibfield  {journal} {\bibinfo  {journal} {Journal of Fluid Mechanics}\ }\textbf {\bibinfo {volume} {570}},\ \bibinfo {pages} {347--364} (\bibinfo {year} {2007})}\BibitemShut {NoStop}%
\bibitem [{\citenamefont {Chu}, \citenamefont {Zhuang},\ and\ \citenamefont {Lu}(2013)}]{chu2013effect}%
  \BibitemOpen
  \bibfield  {author} {\bibinfo {author} {\bibfnamefont {Y.-B.}\ \bibnamefont {Chu}}, \bibinfo {author} {\bibfnamefont {Y.-Q.}\ \bibnamefont {Zhuang}}, \ and\ \bibinfo {author} {\bibfnamefont {X.-Y.}\ \bibnamefont {Lu}},\ }\bibfield  {title} {\enquote {\bibinfo {title} {Effect of wall temperature on hypersonic turbulent boundary layer},}\ }\href@noop {} {\bibfield  {journal} {\bibinfo  {journal} {Journal of Turbulence}\ }\textbf {\bibinfo {volume} {14}},\ \bibinfo {pages} {37--57} (\bibinfo {year} {2013})}\BibitemShut {NoStop}%
\bibitem [{\citenamefont {Duan}, \citenamefont {Choudhari},\ and\ \citenamefont {Zhang}(2016)}]{duan2016pressure}%
  \BibitemOpen
  \bibfield  {author} {\bibinfo {author} {\bibfnamefont {L.}~\bibnamefont {Duan}}, \bibinfo {author} {\bibfnamefont {M.~M.}\ \bibnamefont {Choudhari}}, \ and\ \bibinfo {author} {\bibfnamefont {C.}~\bibnamefont {Zhang}},\ }\bibfield  {title} {\enquote {\bibinfo {title} {Pressure fluctuations induced by a hypersonic turbulent boundary layer},}\ }\href@noop {} {\bibfield  {journal} {\bibinfo  {journal} {Journal of Fluid Mechanics}\ }\textbf {\bibinfo {volume} {804}},\ \bibinfo {pages} {578--607} (\bibinfo {year} {2016})}\BibitemShut {NoStop}%
\bibitem [{\citenamefont {Zhang}, \citenamefont {Duan},\ and\ \citenamefont {Choudhari}(2017)}]{zhang2017effect}%
  \BibitemOpen
  \bibfield  {author} {\bibinfo {author} {\bibfnamefont {C.}~\bibnamefont {Zhang}}, \bibinfo {author} {\bibfnamefont {L.}~\bibnamefont {Duan}}, \ and\ \bibinfo {author} {\bibfnamefont {M.~M.}\ \bibnamefont {Choudhari}},\ }\bibfield  {title} {\enquote {\bibinfo {title} {Effect of wall cooling on boundary-layer-induced pressure fluctuations at mach 6},}\ }\href@noop {} {\bibfield  {journal} {\bibinfo  {journal} {Journal of Fluid Mechanics}\ }\textbf {\bibinfo {volume} {822}},\ \bibinfo {pages} {5--30} (\bibinfo {year} {2017})}\BibitemShut {NoStop}%
\bibitem [{\citenamefont {Sciacovelli}\ \emph {et~al.}(2020)\citenamefont {Sciacovelli}, \citenamefont {Gloerfelt}, \citenamefont {Passiatore}, \citenamefont {Cinnella},\ and\ \citenamefont {Grasso}}]{sciacovelli2020numerical}%
  \BibitemOpen
  \bibfield  {author} {\bibinfo {author} {\bibfnamefont {L.}~\bibnamefont {Sciacovelli}}, \bibinfo {author} {\bibfnamefont {X.}~\bibnamefont {Gloerfelt}}, \bibinfo {author} {\bibfnamefont {D.}~\bibnamefont {Passiatore}}, \bibinfo {author} {\bibfnamefont {P.}~\bibnamefont {Cinnella}}, \ and\ \bibinfo {author} {\bibfnamefont {F.}~\bibnamefont {Grasso}},\ }\bibfield  {title} {\enquote {\bibinfo {title} {Numerical investigation of high-speed turbulent boundary layers of dense gases},}\ }\href@noop {} {\bibfield  {journal} {\bibinfo  {journal} {Flow, Turbulence and Combustion}\ }\textbf {\bibinfo {volume} {105}},\ \bibinfo {pages} {555--579} (\bibinfo {year} {2020})}\BibitemShut {NoStop}%
\bibitem [{\citenamefont {Dang}\ \emph {et~al.}(2022)\citenamefont {Dang}, \citenamefont {Liu}, \citenamefont {Guo}, \citenamefont {Duan},\ and\ \citenamefont {Li}}]{dang2022direct}%
  \BibitemOpen
  \bibfield  {author} {\bibinfo {author} {\bibfnamefont {G.}~\bibnamefont {Dang}}, \bibinfo {author} {\bibfnamefont {S.}~\bibnamefont {Liu}}, \bibinfo {author} {\bibfnamefont {T.}~\bibnamefont {Guo}}, \bibinfo {author} {\bibfnamefont {J.}~\bibnamefont {Duan}}, \ and\ \bibinfo {author} {\bibfnamefont {X.}~\bibnamefont {Li}},\ }\bibfield  {title} {\enquote {\bibinfo {title} {Direct numerical simulation of compressible turbulence accelerated by graphics processing unit: An open-source high accuracy accelerated computational fluid dynamic software},}\ }\href@noop {} {\bibfield  {journal} {\bibinfo  {journal} {Physics of Fluids}\ }\textbf {\bibinfo {volume} {34}} (\bibinfo {year} {2022})}\BibitemShut {NoStop}%
\bibitem [{\citenamefont {Aultman}, \citenamefont {Roy},\ and\ \citenamefont {Duan}(2024)}]{aultman2024asymptotic}%
  \BibitemOpen
  \bibfield  {author} {\bibinfo {author} {\bibfnamefont {M.~T.}\ \bibnamefont {Aultman}}, \bibinfo {author} {\bibfnamefont {D.}~\bibnamefont {Roy}}, \ and\ \bibinfo {author} {\bibfnamefont {L.}~\bibnamefont {Duan}},\ }\bibfield  {title} {\enquote {\bibinfo {title} {Asymptotic near-wall behavior of a mach 6 cold-wall turbulent boundary layer},}\ }in\ \href@noop {} {\emph {\bibinfo {booktitle} {AIAA SCITECH 2024 Forum}}}\ (\bibinfo {year} {2024})\ p.\ \bibinfo {pages} {2733}\BibitemShut {NoStop}%
\bibitem [{\citenamefont {Duan}\ and\ \citenamefont {Martin}(2011)}]{duan2011direct4}%
  \BibitemOpen
  \bibfield  {author} {\bibinfo {author} {\bibfnamefont {L.}~\bibnamefont {Duan}}\ and\ \bibinfo {author} {\bibfnamefont {M.}~\bibnamefont {Martin}},\ }\bibfield  {title} {\enquote {\bibinfo {title} {Direct numerical simulation of hypersonic turbulent boundary layers. part 4. effect of high enthalpy},}\ }\href@noop {} {\bibfield  {journal} {\bibinfo  {journal} {Journal of Fluid Mechanics}\ }\textbf {\bibinfo {volume} {684}},\ \bibinfo {pages} {25--59} (\bibinfo {year} {2011})}\BibitemShut {NoStop}%
\bibitem [{\citenamefont {Appels}(1973)}]{appels1973turbulent}%
  \BibitemOpen
  \bibfield  {author} {\bibinfo {author} {\bibfnamefont {C.}~\bibnamefont {Appels}},\ }\bibfield  {title} {\enquote {\bibinfo {title} {Turbulent boundary layer separation at mach 12},}\ }\href@noop {} {\bibfield  {journal} {\bibinfo  {journal} {VKI TN-90}\ } (\bibinfo {year} {1973})}\BibitemShut {NoStop}%
\bibitem [{\citenamefont {COET}(1993)}]{coet1993experiments}%
  \BibitemOpen
  \bibfield  {author} {\bibinfo {author} {\bibfnamefont {M.}~\bibnamefont {COET}},\ }\bibfield  {title} {\enquote {\bibinfo {title} {Experiments on shock wave/boundary layer interaction in hypersonic flow},}\ }\href@noop {} {\bibfield  {journal} {\bibinfo  {journal} {La Recherche Aerospatiale(English Edition)}\ ,\ \bibinfo {pages} {61--74}} (\bibinfo {year} {1993})}\BibitemShut {NoStop}%
\bibitem [{\citenamefont {Kussoy}\ and\ \citenamefont {Horstman}(1991)}]{kussoy1991documentation}%
  \BibitemOpen
  \bibfield  {author} {\bibinfo {author} {\bibfnamefont {M.~I.}\ \bibnamefont {Kussoy}}\ and\ \bibinfo {author} {\bibfnamefont {K.}~\bibnamefont {Horstman}},\ }\bibfield  {title} {\enquote {\bibinfo {title} {Documentation of two-and three-dimensional shock-wave/turbulent-boundary-layer interaction flows at mach 8.2},}\ }\href@noop {} {\bibfield  {journal} {\bibinfo  {journal} {NASA Ames Research Center Technical Report}\ } (\bibinfo {year} {1991})}\BibitemShut {NoStop}%
\bibitem [{\citenamefont {Wadhams}\ \emph {et~al.}(2008)\citenamefont {Wadhams}, \citenamefont {Mundy}, \citenamefont {MacLean},\ and\ \citenamefont {Holden}}]{WaMaHo08}%
  \BibitemOpen
  \bibfield  {author} {\bibinfo {author} {\bibfnamefont {T.}~\bibnamefont {Wadhams}}, \bibinfo {author} {\bibfnamefont {E.}~\bibnamefont {Mundy}}, \bibinfo {author} {\bibfnamefont {M.}~\bibnamefont {MacLean}}, \ and\ \bibinfo {author} {\bibfnamefont {M.}~\bibnamefont {Holden}},\ }\bibfield  {title} {\enquote {\bibinfo {title} {Ground test studies of the hifire-1 transition experiment part 1: experimental results},}\ }\href@noop {} {\bibfield  {journal} {\bibinfo  {journal} {Journal of Spacecraft and Rockets}\ }\textbf {\bibinfo {volume} {45}},\ \bibinfo {pages} {1134--1148} (\bibinfo {year} {2008})}\BibitemShut {NoStop}%
\bibitem [{\citenamefont {Coleman}(1973{\natexlab{a}})}]{coleman1973study}%
  \BibitemOpen
  \bibfield  {author} {\bibinfo {author} {\bibfnamefont {G.}~\bibnamefont {Coleman}},\ }\bibfield  {title} {\enquote {\bibinfo {title} {A study of hypersonic boundary layers over a family of axisymmetric bodies at zero incidence: preliminary report and data tabulation},}\ }\href@noop {} {\bibfield  {journal} {\bibinfo  {journal} {Imperial College Aero Report 73-06 Imperial College of Science and Technology, London, UK}\ } (\bibinfo {year} {1973}{\natexlab{a}})}\BibitemShut {NoStop}%
\bibitem [{\citenamefont {Coleman}(1973{\natexlab{b}})}]{coleman1973hypersonic}%
  \BibitemOpen
  \bibfield  {author} {\bibinfo {author} {\bibfnamefont {G.~T.}\ \bibnamefont {Coleman}},\ }\emph {\bibinfo {title} {Hypersonic turbulent boundary layer studies}},\ \href@noop {} {Ph.D. thesis},\ \bibinfo  {school} {Imperial College London} (\bibinfo {year} {1973}{\natexlab{b}})\BibitemShut {NoStop}%
\bibitem [{\citenamefont {Coleman}\ and\ \citenamefont {Stollery}(1974)}]{coleman1974incipient}%
  \BibitemOpen
  \bibfield  {author} {\bibinfo {author} {\bibfnamefont {G.}~\bibnamefont {Coleman}}\ and\ \bibinfo {author} {\bibfnamefont {J.}~\bibnamefont {Stollery}},\ }\bibfield  {title} {\enquote {\bibinfo {title} {Incipient separation of axially symmetric hypersonic turbulent boundary layers},}\ }\href@noop {} {\bibfield  {journal} {\bibinfo  {journal} {AIAA Journal}\ }\textbf {\bibinfo {volume} {12}},\ \bibinfo {pages} {119--120} (\bibinfo {year} {1974})}\BibitemShut {NoStop}%
\bibitem [{\citenamefont {Holden}, \citenamefont {Wadhams},\ and\ \citenamefont {MacLean}(2014)}]{holden2014measurements}%
  \BibitemOpen
  \bibfield  {author} {\bibinfo {author} {\bibfnamefont {M.}~\bibnamefont {Holden}}, \bibinfo {author} {\bibfnamefont {T.}~\bibnamefont {Wadhams}}, \ and\ \bibinfo {author} {\bibfnamefont {M.}~\bibnamefont {MacLean}},\ }\bibfield  {title} {\enquote {\bibinfo {title} {Measurements in regions of shock wave/turbulent boundary layer interaction on double cone and hollow cylinder/flare configurations for open and},}\ }\href@noop {} {\bibfield  {journal} {\bibinfo  {journal} {Blind” Code Evaluation/Validation,” Tech. rep., American Institute of Aeronautics and Astronautics}\ } (\bibinfo {year} {2014})}\BibitemShut {NoStop}%
\bibitem [{\citenamefont {Holden}(2018)}]{holden2018measurements}%
  \BibitemOpen
  \bibfield  {author} {\bibinfo {author} {\bibfnamefont {M.}~\bibnamefont {Holden}},\ }\bibfield  {title} {\enquote {\bibinfo {title} {Measurements in regions of shock wave/turbulent boundary layer interaction from mach 4 to 7 at flight duplicated velocities to evaluate and improve the models of turbulence in cfd codes},}\ }in\ \href@noop {} {\emph {\bibinfo {booktitle} {2018 Fluid Dynamics Conference}}}\ (\bibinfo {year} {2018})\ p.\ \bibinfo {pages} {3706}\BibitemShut {NoStop}%
\bibitem [{\citenamefont {Holden}(1991)}]{holden1991studies}%
  \BibitemOpen
  \bibfield  {author} {\bibinfo {author} {\bibfnamefont {M.}~\bibnamefont {Holden}},\ }\bibfield  {title} {\enquote {\bibinfo {title} {Studies of the mean and unsteady structure of turbulent boundary layer separation in hypersonic flow},}\ }in\ \href@noop {} {\emph {\bibinfo {booktitle} {22nd Fluid Dynamics, Plasma Dynamics and Lasers Conference}}}\ (\bibinfo {year} {1991})\ p.\ \bibinfo {pages} {1778}\BibitemShut {NoStop}%
\bibitem [{\citenamefont {Running}\ \emph {et~al.}(2019)\citenamefont {Running}, \citenamefont {Juliano}, \citenamefont {Jewell}, \citenamefont {Borg},\ and\ \citenamefont {Kimmel}}]{running2019hypersonic}%
  \BibitemOpen
  \bibfield  {author} {\bibinfo {author} {\bibfnamefont {C.~L.}\ \bibnamefont {Running}}, \bibinfo {author} {\bibfnamefont {T.~J.}\ \bibnamefont {Juliano}}, \bibinfo {author} {\bibfnamefont {J.~S.}\ \bibnamefont {Jewell}}, \bibinfo {author} {\bibfnamefont {M.~P.}\ \bibnamefont {Borg}}, \ and\ \bibinfo {author} {\bibfnamefont {R.~L.}\ \bibnamefont {Kimmel}},\ }\bibfield  {title} {\enquote {\bibinfo {title} {Hypersonic shock-wave/boundary-layer interactions on a cone/flare},}\ }\href@noop {} {\bibfield  {journal} {\bibinfo  {journal} {Experimental Thermal and Fluid Science}\ }\textbf {\bibinfo {volume} {109}},\ \bibinfo {pages} {109911} (\bibinfo {year} {2019})}\BibitemShut {NoStop}%
\bibitem [{\citenamefont {Kussoy}\ and\ \citenamefont {Horstmann}(1975)}]{kussoy1975experimental}%
  \BibitemOpen
  \bibfield  {author} {\bibinfo {author} {\bibfnamefont {M.}~\bibnamefont {Kussoy}}\ and\ \bibinfo {author} {\bibfnamefont {C.}~\bibnamefont {Horstmann}},\ }\href@noop {} {\enquote {\bibinfo {title} {An experimental documentation of a hypersonic shock-wave turbulent boundary layer interaction flow: With and without separation},}\ }\bibinfo {type} {Tech. Rep.}\ (\bibinfo {year} {1975})\BibitemShut {NoStop}%
\bibitem [{\citenamefont {Kussoy}(1989)}]{kussoy1989documentation}%
  \BibitemOpen
  \bibfield  {author} {\bibinfo {author} {\bibfnamefont {M.~I.}\ \bibnamefont {Kussoy}},\ }\href@noop {} {\emph {\bibinfo {title} {Documentation of two-and three-dimensional hypersonic shock wave/turbulent boundary layer interaction flows}}},\ Vol.\ \bibinfo {volume} {101075}\ (\bibinfo  {publisher} {National Aeronautics and Space Administration, Ames Research Center},\ \bibinfo {year} {1989})\BibitemShut {NoStop}%
\bibitem [{\citenamefont {Rodi}, \citenamefont {Dolling},\ and\ \citenamefont {Knight}(1991)}]{rodi1991experimental}%
  \BibitemOpen
  \bibfield  {author} {\bibinfo {author} {\bibfnamefont {P.}~\bibnamefont {Rodi}}, \bibinfo {author} {\bibfnamefont {D.}~\bibnamefont {Dolling}}, \ and\ \bibinfo {author} {\bibfnamefont {D.}~\bibnamefont {Knight}},\ }\bibfield  {title} {\enquote {\bibinfo {title} {An experimental/computational study of heat transfer in sharp fin induced turbulent interactions at mach 5},}\ }in\ \href@noop {} {\emph {\bibinfo {booktitle} {22nd Fluid Dynamics, Plasma Dynamics and Lasers Conference}}}\ (\bibinfo {year} {1991})\ p.\ \bibinfo {pages} {1764}\BibitemShut {NoStop}%
\bibitem [{\citenamefont {Rodi}\ and\ \citenamefont {Dolling}(1992)}]{rodi1992experimental}%
  \BibitemOpen
  \bibfield  {author} {\bibinfo {author} {\bibfnamefont {P.}~\bibnamefont {Rodi}}\ and\ \bibinfo {author} {\bibfnamefont {D.}~\bibnamefont {Dolling}},\ }\bibfield  {title} {\enquote {\bibinfo {title} {An experimental/computational study of sharp fin induced shock wave/turbulent boundary layer interactions at mach 5-experimental results},}\ }in\ \href@noop {} {\emph {\bibinfo {booktitle} {30th Aerospace Sciences Meeting and Exhibit}}}\ (\bibinfo {year} {1992})\ p.\ \bibinfo {pages} {749}\BibitemShut {NoStop}%
\bibitem [{\citenamefont {Rodi}\ and\ \citenamefont {Dolling}(1995)}]{rodi1995behavior}%
  \BibitemOpen
  \bibfield  {author} {\bibinfo {author} {\bibfnamefont {P.}~\bibnamefont {Rodi}}\ and\ \bibinfo {author} {\bibfnamefont {D.}~\bibnamefont {Dolling}},\ }\bibfield  {title} {\enquote {\bibinfo {title} {Behavior of pressure and heat transfer in sharp fin-induced turbulent interactions},}\ }\href@noop {} {\bibfield  {journal} {\bibinfo  {journal} {AIAA journal}\ }\textbf {\bibinfo {volume} {33}},\ \bibinfo {pages} {2013--2019} (\bibinfo {year} {1995})}\BibitemShut {NoStop}%
\bibitem [{\citenamefont {Borovoy}\ \emph {et~al.}(2009)\citenamefont {Borovoy}, \citenamefont {Mosharov}, \citenamefont {Noev},\ and\ \citenamefont {Radchenko}}]{borovoy2009laminar}%
  \BibitemOpen
  \bibfield  {author} {\bibinfo {author} {\bibfnamefont {V.~Y.}\ \bibnamefont {Borovoy}}, \bibinfo {author} {\bibfnamefont {V.}~\bibnamefont {Mosharov}}, \bibinfo {author} {\bibfnamefont {A.~Y.}\ \bibnamefont {Noev}}, \ and\ \bibinfo {author} {\bibfnamefont {V.}~\bibnamefont {Radchenko}},\ }\bibfield  {title} {\enquote {\bibinfo {title} {Laminar-turbulent flow over wedges mounted on sharp and blunt plates},}\ }\href@noop {} {\bibfield  {journal} {\bibinfo  {journal} {Fluid Dynamics}\ }\textbf {\bibinfo {volume} {44}},\ \bibinfo {pages} {382--396} (\bibinfo {year} {2009})}\BibitemShut {NoStop}%
\bibitem [{\citenamefont {Borovoy}\ \emph {et~al.}(2012)\citenamefont {Borovoy}, \citenamefont {Egorov}, \citenamefont {Noev}, \citenamefont {Radchenko}, \citenamefont {Skuratov},\ and\ \citenamefont {Struminskaya}}]{borovoy20123d}%
  \BibitemOpen
  \bibfield  {author} {\bibinfo {author} {\bibfnamefont {V.~Y.}\ \bibnamefont {Borovoy}}, \bibinfo {author} {\bibfnamefont {I.~V.}\ \bibnamefont {Egorov}}, \bibinfo {author} {\bibfnamefont {A.~Y.}\ \bibnamefont {Noev}}, \bibinfo {author} {\bibfnamefont {V.~N.}\ \bibnamefont {Radchenko}}, \bibinfo {author} {\bibfnamefont {A.~S.}\ \bibnamefont {Skuratov}}, \ and\ \bibinfo {author} {\bibfnamefont {I.~V.}\ \bibnamefont {Struminskaya}},\ }\bibfield  {title} {\enquote {\bibinfo {title} {3d shock/turbulent boundary layer interaction on the plate near a wedge in presence of an entropy layer},}\ }\href@noop {} {\bibfield  {journal} {\bibinfo  {journal} {TsAGI Science Journal}\ }\textbf {\bibinfo {volume} {43}} (\bibinfo {year} {2012})}\BibitemShut {NoStop}%
\bibitem [{\citenamefont {Borovoy}\ \emph {et~al.}(2016)\citenamefont {Borovoy}, \citenamefont {Egorov}, \citenamefont {Mosharov}, \citenamefont {Radchenko}, \citenamefont {Skuratov},\ and\ \citenamefont {Struminskaya}}]{borovoy2016entropy}%
  \BibitemOpen
  \bibfield  {author} {\bibinfo {author} {\bibfnamefont {V.}~\bibnamefont {Borovoy}}, \bibinfo {author} {\bibfnamefont {I.}~\bibnamefont {Egorov}}, \bibinfo {author} {\bibfnamefont {V.}~\bibnamefont {Mosharov}}, \bibinfo {author} {\bibfnamefont {V.}~\bibnamefont {Radchenko}}, \bibinfo {author} {\bibfnamefont {A.}~\bibnamefont {Skuratov}}, \ and\ \bibinfo {author} {\bibfnamefont {I.}~\bibnamefont {Struminskaya}},\ }\bibfield  {title} {\enquote {\bibinfo {title} {Entropy-layer influence on single-fin and double-fin/boundary-layer interactions},}\ }\href@noop {} {\bibfield  {journal} {\bibinfo  {journal} {AIAA Journal}\ }\textbf {\bibinfo {volume} {54}},\ \bibinfo {pages} {443--457} (\bibinfo {year} {2016})}\BibitemShut {NoStop}%
\bibitem [{\citenamefont {Law}(1975)}]{law1975three}%
  \BibitemOpen
  \bibfield  {author} {\bibinfo {author} {\bibfnamefont {C.~H.}\ \bibnamefont {Law}},\ }\href@noop {} {\emph {\bibinfo {title} {Three-dimensional shock wave-turbulent boundary layer interactions at Mach 6}}},\ Vol.~\bibinfo {volume} {75}\ (\bibinfo  {publisher} {Aerospace Research Laboratories, Air Force Systems Command, United States~…},\ \bibinfo {year} {1975})\BibitemShut {NoStop}%
\bibitem [{\citenamefont {Holden}(1984)}]{holden1984experimental}%
  \BibitemOpen
  \bibfield  {author} {\bibinfo {author} {\bibfnamefont {M.}~\bibnamefont {Holden}},\ }\bibfield  {title} {\enquote {\bibinfo {title} {Experimental studies of quasi-two-dimensional and three-dimensional viscous interaction regions induced by skewed-shock and swept-shock boundary layer interaction},}\ }in\ \href@noop {} {\emph {\bibinfo {booktitle} {17th Fluid Dynamics, Plasma Dynamics, and Lasers Conference}}}\ (\bibinfo {year} {1984})\ p.\ \bibinfo {pages} {1677}\BibitemShut {NoStop}%
\bibitem [{\citenamefont {Sch{\"u}lein}\ and\ \citenamefont {Zheltovodov}(2001)}]{schulein2001documentation}%
  \BibitemOpen
  \bibfield  {author} {\bibinfo {author} {\bibfnamefont {E.}~\bibnamefont {Sch{\"u}lein}}\ and\ \bibinfo {author} {\bibfnamefont {A.}~\bibnamefont {Zheltovodov}},\ }\bibfield  {title} {\enquote {\bibinfo {title} {Documentation of experimental data for hypersonic 3-d shock waves},}\ }\href@noop {} {\bibfield  {journal} {\bibinfo  {journal} {Report IB 223-99 A 26 Deutsches Zentrum für Luft- und Raumfahrt e.V. (DLR), Institut für Strömungsmechanik, Göttingen, Germany}\ } (\bibinfo {year} {2001})}\BibitemShut {NoStop}%
\bibitem [{\citenamefont {Kussoy}, \citenamefont {Horstoman},\ and\ \citenamefont {Horstman}(1993)}]{kussoy1993hypersonic}%
  \BibitemOpen
  \bibfield  {author} {\bibinfo {author} {\bibfnamefont {M.}~\bibnamefont {Kussoy}}, \bibinfo {author} {\bibfnamefont {K.}~\bibnamefont {Horstoman}}, \ and\ \bibinfo {author} {\bibfnamefont {C.}~\bibnamefont {Horstman}},\ }\bibfield  {title} {\enquote {\bibinfo {title} {Hypersonic crossing shock-wave/turbulent-boundary-layer interactions},}\ }\href@noop {} {\bibfield  {journal} {\bibinfo  {journal} {AIAA journal}\ }\textbf {\bibinfo {volume} {31}},\ \bibinfo {pages} {2197--2203} (\bibinfo {year} {1993})}\BibitemShut {NoStop}%
\bibitem [{\citenamefont {Volpiani}, \citenamefont {Bernardini},\ and\ \citenamefont {Larsson}(2020)}]{volpiani2020effects}%
  \BibitemOpen
  \bibfield  {author} {\bibinfo {author} {\bibfnamefont {P.~S.}\ \bibnamefont {Volpiani}}, \bibinfo {author} {\bibfnamefont {M.}~\bibnamefont {Bernardini}}, \ and\ \bibinfo {author} {\bibfnamefont {J.}~\bibnamefont {Larsson}},\ }\bibfield  {title} {\enquote {\bibinfo {title} {Effects of a nonadiabatic wall on hypersonic shock/boundary-layer interactions},}\ }\href@noop {} {\bibfield  {journal} {\bibinfo  {journal} {Physical Review Fluids}\ }\textbf {\bibinfo {volume} {5}},\ \bibinfo {pages} {014602} (\bibinfo {year} {2020})}\BibitemShut {NoStop}%
\bibitem [{\citenamefont {Priebe}\ and\ \citenamefont {Mart{\'\i}n}(2021)}]{priebe2021turbulence}%
  \BibitemOpen
  \bibfield  {author} {\bibinfo {author} {\bibfnamefont {S.}~\bibnamefont {Priebe}}\ and\ \bibinfo {author} {\bibfnamefont {M.~P.}\ \bibnamefont {Mart{\'\i}n}},\ }\bibfield  {title} {\enquote {\bibinfo {title} {Turbulence in a hypersonic compression ramp flow},}\ }\href@noop {} {\bibfield  {journal} {\bibinfo  {journal} {Physical Review Fluids}\ }\textbf {\bibinfo {volume} {6}},\ \bibinfo {pages} {034601} (\bibinfo {year} {2021})}\BibitemShut {NoStop}%
\bibitem [{\citenamefont {Bookey}\ \emph {et~al.}(2005)\citenamefont {Bookey}, \citenamefont {Wyckham}, \citenamefont {Smits},\ and\ \citenamefont {Martin}}]{bookey2005new}%
  \BibitemOpen
  \bibfield  {author} {\bibinfo {author} {\bibfnamefont {P.}~\bibnamefont {Bookey}}, \bibinfo {author} {\bibfnamefont {C.}~\bibnamefont {Wyckham}}, \bibinfo {author} {\bibfnamefont {A.}~\bibnamefont {Smits}}, \ and\ \bibinfo {author} {\bibfnamefont {P.}~\bibnamefont {Martin}},\ }\bibfield  {title} {\enquote {\bibinfo {title} {New experimental data of stbli at dns/les accessible reynolds numbers},}\ }in\ \href@noop {} {\emph {\bibinfo {booktitle} {43rd AIAA Aerospace Sciences Meeting and Exhibit}}}\ (\bibinfo {year} {2005})\ p.\ \bibinfo {pages} {309}\BibitemShut {NoStop}%
\bibitem [{\citenamefont {Zhang}\ \emph {et~al.}(2022)\citenamefont {Zhang}, \citenamefont {Guo}, \citenamefont {Dang},\ and\ \citenamefont {Li}}]{zhang2022direct}%
  \BibitemOpen
  \bibfield  {author} {\bibinfo {author} {\bibfnamefont {J.}~\bibnamefont {Zhang}}, \bibinfo {author} {\bibfnamefont {T.}~\bibnamefont {Guo}}, \bibinfo {author} {\bibfnamefont {G.}~\bibnamefont {Dang}}, \ and\ \bibinfo {author} {\bibfnamefont {X.}~\bibnamefont {Li}},\ }\bibfield  {title} {\enquote {\bibinfo {title} {Direct numerical simulation of shock wave/turbulent boundary layer interaction in a swept compression ramp at mach 6},}\ }\href@noop {} {\bibfield  {journal} {\bibinfo  {journal} {Physics of Fluids}\ }\textbf {\bibinfo {volume} {34}} (\bibinfo {year} {2022})}\BibitemShut {NoStop}%
\bibitem [{\citenamefont {Guo}\ \emph {et~al.}(2023)\citenamefont {Guo}, \citenamefont {Zhang}, \citenamefont {Tong},\ and\ \citenamefont {Li}}]{guo2023amplification}%
  \BibitemOpen
  \bibfield  {author} {\bibinfo {author} {\bibfnamefont {T.}~\bibnamefont {Guo}}, \bibinfo {author} {\bibfnamefont {J.}~\bibnamefont {Zhang}}, \bibinfo {author} {\bibfnamefont {F.}~\bibnamefont {Tong}}, \ and\ \bibinfo {author} {\bibfnamefont {X.}~\bibnamefont {Li}},\ }\bibfield  {title} {\enquote {\bibinfo {title} {Amplification of turbulent kinetic energy and temperature fluctuation in a hypersonic turbulent boundary layer over a compression ramp},}\ }\href@noop {} {\bibfield  {journal} {\bibinfo  {journal} {Physics of Fluids}\ }\textbf {\bibinfo {volume} {35}} (\bibinfo {year} {2023})}\BibitemShut {NoStop}%
\bibitem [{\citenamefont {Fulin}\ \emph {et~al.}(2023)\citenamefont {Fulin}, \citenamefont {Junyi}, \citenamefont {Jiang}, \citenamefont {Dong},\ and\ \citenamefont {Xianxu}}]{fulin2023hypersonic}%
  \BibitemOpen
  \bibfield  {author} {\bibinfo {author} {\bibfnamefont {T.}~\bibnamefont {Fulin}}, \bibinfo {author} {\bibfnamefont {D.}~\bibnamefont {Junyi}}, \bibinfo {author} {\bibfnamefont {L.}~\bibnamefont {Jiang}}, \bibinfo {author} {\bibfnamefont {S.}~\bibnamefont {Dong}}, \ and\ \bibinfo {author} {\bibfnamefont {Y.}~\bibnamefont {Xianxu}},\ }\bibfield  {title} {\enquote {\bibinfo {title} {Hypersonic shock wave and turbulent boundary layer interaction in a sharp cone/flare model},}\ }\href@noop {} {\bibfield  {journal} {\bibinfo  {journal} {Chinese Journal of Aeronautics}\ }\textbf {\bibinfo {volume} {36}},\ \bibinfo {pages} {80--95} (\bibinfo {year} {2023})}\BibitemShut {NoStop}%
\bibitem [{\citenamefont {Holden}(1982)}]{holden1982experimental}%
  \BibitemOpen
  \bibfield  {author} {\bibinfo {author} {\bibfnamefont {M.}~\bibnamefont {Holden}},\ }\bibfield  {title} {\enquote {\bibinfo {title} {Experimental studies of surface roughness, entropy swallowing and boundary layer transition effects on the skin friction and heat transfer distribution in high speed flows},}\ }in\ \href@noop {} {\emph {\bibinfo {booktitle} {20th Aerospace Sciences Meeting}}}\ (\bibinfo {year} {1982})\ p.~\bibinfo {pages} {34}\BibitemShut {NoStop}%
\bibitem [{\citenamefont {Holden}(1989)}]{holden1989studies}%
  \BibitemOpen
  \bibfield  {author} {\bibinfo {author} {\bibfnamefont {M.}~\bibnamefont {Holden}},\ }\bibfield  {title} {\enquote {\bibinfo {title} {Studies of surface roughness and blowing effects on hypersonic turbulent boundary layers over slender cones},}\ }in\ \href@noop {} {\emph {\bibinfo {booktitle} {27th Aerospace Sciences Meeting}}}\ (\bibinfo {year} {1989})\ p.\ \bibinfo {pages} {458}\BibitemShut {NoStop}%
\bibitem [{\citenamefont {Babinsky}\ and\ \citenamefont {Edwards}(1997)}]{babinsky1997large}%
  \BibitemOpen
  \bibfield  {author} {\bibinfo {author} {\bibfnamefont {H.}~\bibnamefont {Babinsky}}\ and\ \bibinfo {author} {\bibfnamefont {J.}~\bibnamefont {Edwards}},\ }\bibfield  {title} {\enquote {\bibinfo {title} {Large-scale roughness influence on turbulent hypersonic boundary layers approaching compression corners},}\ }\href@noop {} {\bibfield  {journal} {\bibinfo  {journal} {Journal of spacecraft and rockets}\ }\textbf {\bibinfo {volume} {34}},\ \bibinfo {pages} {70--75} (\bibinfo {year} {1997})}\BibitemShut {NoStop}%
\bibitem [{\citenamefont {Hollis}(2014)}]{hollis2014distributed}%
  \BibitemOpen
  \bibfield  {author} {\bibinfo {author} {\bibfnamefont {B.~R.}\ \bibnamefont {Hollis}},\ }\bibfield  {title} {\enquote {\bibinfo {title} {Distributed roughness effects on blunt-body transition and turbulent heating},}\ }in\ \href@noop {} {\emph {\bibinfo {booktitle} {52nd Aerospace Sciences Meeting}}}\ (\bibinfo {year} {2014})\ p.\ \bibinfo {pages} {0238}\BibitemShut {NoStop}%
\bibitem [{\citenamefont {Wilder}\ and\ \citenamefont {Prabhu}(2019)}]{wilder2019rough}%
  \BibitemOpen
  \bibfield  {author} {\bibinfo {author} {\bibfnamefont {M.~C.}\ \bibnamefont {Wilder}}\ and\ \bibinfo {author} {\bibfnamefont {D.~K.}\ \bibnamefont {Prabhu}},\ }\bibfield  {title} {\enquote {\bibinfo {title} {Rough-wall turbulent heat transfer experiments in hypersonic free flight},}\ }in\ \href@noop {} {\emph {\bibinfo {booktitle} {AIAA Aviation 2019 Forum}}}\ (\bibinfo {year} {2019})\ p.\ \bibinfo {pages} {3009}\BibitemShut {NoStop}%
\bibitem [{\citenamefont {Forsyth}, \citenamefont {Hambidge},\ and\ \citenamefont {McGilvray}(2024)}]{forsyth2024experimental}%
  \BibitemOpen
  \bibfield  {author} {\bibinfo {author} {\bibfnamefont {P.~R.}\ \bibnamefont {Forsyth}}, \bibinfo {author} {\bibfnamefont {C.}~\bibnamefont {Hambidge}}, \ and\ \bibinfo {author} {\bibfnamefont {M.}~\bibnamefont {McGilvray}},\ }\bibfield  {title} {\enquote {\bibinfo {title} {Experimental assessment of hypersonic convective heat transfer augmentation due to surface roughness},}\ }\href@noop {} {\bibfield  {journal} {\bibinfo  {journal} {Journal of Thermophysics and Heat Transfer}\ ,\ \bibinfo {pages} {1--10}} (\bibinfo {year} {2024})}\BibitemShut {NoStop}%
\bibitem [{\citenamefont {Collen}\ \emph {et~al.}(2021)\citenamefont {Collen}, \citenamefont {Doherty}, \citenamefont {Subiah}, \citenamefont {Sopek}, \citenamefont {Jahn}, \citenamefont {Gildfind}, \citenamefont {Penty~Geraets}, \citenamefont {Gollan}, \citenamefont {Hambidge}, \citenamefont {Morgan} \emph {et~al.}}]{collen2021development}%
  \BibitemOpen
  \bibfield  {author} {\bibinfo {author} {\bibfnamefont {P.}~\bibnamefont {Collen}}, \bibinfo {author} {\bibfnamefont {L.~J.}\ \bibnamefont {Doherty}}, \bibinfo {author} {\bibfnamefont {S.~D.}\ \bibnamefont {Subiah}}, \bibinfo {author} {\bibfnamefont {T.}~\bibnamefont {Sopek}}, \bibinfo {author} {\bibfnamefont {I.}~\bibnamefont {Jahn}}, \bibinfo {author} {\bibfnamefont {D.}~\bibnamefont {Gildfind}}, \bibinfo {author} {\bibfnamefont {R.}~\bibnamefont {Penty~Geraets}}, \bibinfo {author} {\bibfnamefont {R.}~\bibnamefont {Gollan}}, \bibinfo {author} {\bibfnamefont {C.}~\bibnamefont {Hambidge}}, \bibinfo {author} {\bibfnamefont {R.}~\bibnamefont {Morgan}},  \emph {et~al.},\ }\bibfield  {title} {\enquote {\bibinfo {title} {Development and commissioning of the t6 stalker tunnel},}\ }\href@noop {} {\bibfield  {journal} {\bibinfo  {journal} {Experiments in Fluids}\ }\textbf {\bibinfo {volume} {62}},\ \bibinfo {pages} {1--24} (\bibinfo {year} {2021})}\BibitemShut {NoStop}%
\bibitem [{\citenamefont {Duan}\ \emph {et~al.}(2012)\citenamefont {Duan}, \citenamefont {Martin}, \citenamefont {Feldick}, \citenamefont {Modest},\ and\ \citenamefont {Levin}}]{duan2012study}%
  \BibitemOpen
  \bibfield  {author} {\bibinfo {author} {\bibfnamefont {L.}~\bibnamefont {Duan}}, \bibinfo {author} {\bibfnamefont {M.}~\bibnamefont {Martin}}, \bibinfo {author} {\bibfnamefont {A.}~\bibnamefont {Feldick}}, \bibinfo {author} {\bibfnamefont {M.}~\bibnamefont {Modest}}, \ and\ \bibinfo {author} {\bibfnamefont {D.}~\bibnamefont {Levin}},\ }\bibfield  {title} {\enquote {\bibinfo {title} {Study of turbulence-radiation interaction in hypersonic turbulent boundary layers},}\ }\href@noop {} {\bibfield  {journal} {\bibinfo  {journal} {AIAA journal}\ }\textbf {\bibinfo {volume} {50}},\ \bibinfo {pages} {447--453} (\bibinfo {year} {2012})}\BibitemShut {NoStop}%
\bibitem [{\citenamefont {VAN~DRIEST}(1956)}]{van1956problem}%
  \BibitemOpen
  \bibfield  {author} {\bibinfo {author} {\bibfnamefont {E.}~\bibnamefont {VAN~DRIEST}},\ }\bibfield  {title} {\enquote {\bibinfo {title} {The problem of aerodynamic heating},}\ }\href@noop {} {\bibfield  {journal} {\bibinfo  {journal} {Aeronaut. Eng. Rev.}\ }\textbf {\bibinfo {volume} {15}},\ \bibinfo {pages} {26--41} (\bibinfo {year} {1956})}\BibitemShut {NoStop}%
\bibitem [{\citenamefont {White}\ and\ \citenamefont {Christoph}(1972)}]{white1972simple}%
  \BibitemOpen
  \bibfield  {author} {\bibinfo {author} {\bibfnamefont {F.}~\bibnamefont {White}}\ and\ \bibinfo {author} {\bibfnamefont {G.}~\bibnamefont {Christoph}},\ }\bibfield  {title} {\enquote {\bibinfo {title} {A simple theory for the two-dimensional compressible turbulent boundary layer},}\ }\href@noop {} {\bibfield  {journal} {\bibinfo  {journal} {Journal of Fluids Engineering, Transactions of the ASME}\ }\textbf {\bibinfo {volume} {94}},\ \bibinfo {pages} {636} (\bibinfo {year} {1972})}\BibitemShut {NoStop}%
\bibitem [{\citenamefont {Spalding}\ and\ \citenamefont {Chi}(1964)}]{spalding1964drag}%
  \BibitemOpen
  \bibfield  {author} {\bibinfo {author} {\bibfnamefont {D.}~\bibnamefont {Spalding}}\ and\ \bibinfo {author} {\bibfnamefont {S.}~\bibnamefont {Chi}},\ }\bibfield  {title} {\enquote {\bibinfo {title} {The drag of a compressible turbulent boundary layer on a smooth flat plate with and without heat transfer},}\ }\href@noop {} {\bibfield  {journal} {\bibinfo  {journal} {Journal of Fluid Mechanics}\ }\textbf {\bibinfo {volume} {18}},\ \bibinfo {pages} {117--143} (\bibinfo {year} {1964})}\BibitemShut {NoStop}%
\bibitem [{\citenamefont {Hopkins}\ and\ \citenamefont {Inouye}(1971)}]{hopkins1971evaluation}%
  \BibitemOpen
  \bibfield  {author} {\bibinfo {author} {\bibfnamefont {E.~J.}\ \bibnamefont {Hopkins}}\ and\ \bibinfo {author} {\bibfnamefont {M.}~\bibnamefont {Inouye}},\ }\bibfield  {title} {\enquote {\bibinfo {title} {An evaluation of theories for predicting turbulent skin friction andheat transfer on flat plates at supersonic and hypersonic mach numbers},}\ }\href@noop {} {\bibfield  {journal} {\bibinfo  {journal} {AIAA Journal}\ }\textbf {\bibinfo {volume} {9}},\ \bibinfo {pages} {993--1003} (\bibinfo {year} {1971})}\BibitemShut {NoStop}%
\bibitem [{\citenamefont {Bhagwandin}\ and\ \citenamefont {Martin}(2023)}]{bhagwandin2023wall}%
  \BibitemOpen
  \bibfield  {author} {\bibinfo {author} {\bibfnamefont {V.~A.}\ \bibnamefont {Bhagwandin}}\ and\ \bibinfo {author} {\bibfnamefont {P.}~\bibnamefont {Martin}},\ }\bibfield  {title} {\enquote {\bibinfo {title} {Wall-resolved les of mach 6 bolt-2 hypersonic vehicle},}\ }in\ \href@noop {} {\emph {\bibinfo {booktitle} {AIAA AVIATION 2023 Forum}}}\ (\bibinfo {year} {2023})\ p.\ \bibinfo {pages} {3848}\BibitemShut {NoStop}%
\bibitem [{\citenamefont {Helm}\ and\ \citenamefont {Mart{\'\i}n}(2022)}]{helm2022large}%
  \BibitemOpen
  \bibfield  {author} {\bibinfo {author} {\bibfnamefont {C.~M.}\ \bibnamefont {Helm}}\ and\ \bibinfo {author} {\bibfnamefont {M.}~\bibnamefont {Mart{\'\i}n}},\ }\bibfield  {title} {\enquote {\bibinfo {title} {Large eddy simulation of two separated hypersonic shock/turbulent boundary layer interactions},}\ }\href@noop {} {\bibfield  {journal} {\bibinfo  {journal} {Physical Review Fluids}\ }\textbf {\bibinfo {volume} {7}},\ \bibinfo {pages} {074601} (\bibinfo {year} {2022})}\BibitemShut {NoStop}%
\bibitem [{\citenamefont {Bhagwandin}\ and\ \citenamefont {Martin}(2021)}]{bhagwandin2021shock}%
  \BibitemOpen
  \bibfield  {author} {\bibinfo {author} {\bibfnamefont {V.~A.}\ \bibnamefont {Bhagwandin}}\ and\ \bibinfo {author} {\bibfnamefont {P.}~\bibnamefont {Martin}},\ }\bibfield  {title} {\enquote {\bibinfo {title} {Les of shock-turbulent boundary layer interaction over a mach 10 hollow cylinder with flare.}}\ }in\ \href@noop {} {\emph {\bibinfo {booktitle} {AIAA AVIATION 2021 FORUM}}}\ (\bibinfo {year} {2021})\ p.\ \bibinfo {pages} {2820}\BibitemShut {NoStop}%
\bibitem [{\citenamefont {Kianvashrad}\ and\ \citenamefont {Knight}(2021)}]{kianvashrad2021large}%
  \BibitemOpen
  \bibfield  {author} {\bibinfo {author} {\bibfnamefont {N.}~\bibnamefont {Kianvashrad}}\ and\ \bibinfo {author} {\bibfnamefont {D.~D.}\ \bibnamefont {Knight}},\ }\bibfield  {title} {\enquote {\bibinfo {title} {Large eddy simulation of hypersonic cold wall flat plate},}\ }in\ \href@noop {} {\emph {\bibinfo {booktitle} {AIAA AVIATION 2021 FORUM}}}\ (\bibinfo {year} {2021})\ p.\ \bibinfo {pages} {2882}\BibitemShut {NoStop}%
\bibitem [{\citenamefont {Bhagwandin}, \citenamefont {Helm},\ and\ \citenamefont {Martin}(2019)}]{bhagwandin2019shock}%
  \BibitemOpen
  \bibfield  {author} {\bibinfo {author} {\bibfnamefont {V.~A.}\ \bibnamefont {Bhagwandin}}, \bibinfo {author} {\bibfnamefont {C.~M.}\ \bibnamefont {Helm}}, \ and\ \bibinfo {author} {\bibfnamefont {P.}~\bibnamefont {Martin}},\ }\bibfield  {title} {\enquote {\bibinfo {title} {Shock-turbulent boundary layer interactions in separated compression corners at mach 10.}}\ }in\ \href@noop {} {\emph {\bibinfo {booktitle} {AIAA Scitech 2019 Forum}}}\ (\bibinfo {year} {2019})\ p.\ \bibinfo {pages} {1129}\BibitemShut {NoStop}%
\bibitem [{\citenamefont {Fang}\ \emph {et~al.}(2015)\citenamefont {Fang}, \citenamefont {Yao}, \citenamefont {Zheltovodov},\ and\ \citenamefont {Lu}}]{fang2015large}%
  \BibitemOpen
  \bibfield  {author} {\bibinfo {author} {\bibfnamefont {J.}~\bibnamefont {Fang}}, \bibinfo {author} {\bibfnamefont {Y.}~\bibnamefont {Yao}}, \bibinfo {author} {\bibfnamefont {A.}~\bibnamefont {Zheltovodov}}, \ and\ \bibinfo {author} {\bibfnamefont {L.}~\bibnamefont {Lu}},\ }\bibfield  {title} {\enquote {\bibinfo {title} {Large-eddy simulation of a three-dimensional hypersonic shock wave turbulent boundary layer interaction of a single-fin},}\ }in\ \href@noop {} {\emph {\bibinfo {booktitle} {53rd AIAA Aerospace Sciences Meeting}}}\ (\bibinfo {year} {2015})\ p.\ \bibinfo {pages} {1062}\BibitemShut {NoStop}%
\bibitem [{\citenamefont {Fang}\ \emph {et~al.}(2017)\citenamefont {Fang}, \citenamefont {Yao}, \citenamefont {Zheltovodov},\ and\ \citenamefont {Lu}}]{fang2017investigation}%
  \BibitemOpen
  \bibfield  {author} {\bibinfo {author} {\bibfnamefont {J.}~\bibnamefont {Fang}}, \bibinfo {author} {\bibfnamefont {Y.}~\bibnamefont {Yao}}, \bibinfo {author} {\bibfnamefont {A.~A.}\ \bibnamefont {Zheltovodov}}, \ and\ \bibinfo {author} {\bibfnamefont {L.}~\bibnamefont {Lu}},\ }\bibfield  {title} {\enquote {\bibinfo {title} {Investigation of three-dimensional shock wave/turbulent-boundary-layer interaction initiated by a single fin},}\ }\href@noop {} {\bibfield  {journal} {\bibinfo  {journal} {AIAA Journal}\ }\textbf {\bibinfo {volume} {55}},\ \bibinfo {pages} {509--523} (\bibinfo {year} {2017})}\BibitemShut {NoStop}%
\bibitem [{\citenamefont {Neeb}, \citenamefont {Saile},\ and\ \citenamefont {G{\"u}lhan}(2015)}]{neeb2015experimental}%
  \BibitemOpen
  \bibfield  {author} {\bibinfo {author} {\bibfnamefont {D.}~\bibnamefont {Neeb}}, \bibinfo {author} {\bibfnamefont {D.}~\bibnamefont {Saile}}, \ and\ \bibinfo {author} {\bibfnamefont {A.}~\bibnamefont {G{\"u}lhan}},\ }\bibfield  {title} {\enquote {\bibinfo {title} {Experimental flow characterization and heat flux augmentation analysis of a hypersonic turbulent boundary layer along a rough surface},}\ }in\ \href@noop {} {\emph {\bibinfo {booktitle} {Proceedings of the 8th European Symposium on Aerothermodynamics for Space Vehicles}}},\ \bibinfo {series and number} {\bibinfo {number} {89873}}\ (\bibinfo {year} {2015})\ pp.\ \bibinfo {pages} {1--15}\BibitemShut {NoStop}%
\bibitem [{\citenamefont {Menter}\ \emph {et~al.}(2003)\citenamefont {Menter}, \citenamefont {Kuntz}, \citenamefont {Langtry} \emph {et~al.}}]{menter2003ten}%
  \BibitemOpen
  \bibfield  {author} {\bibinfo {author} {\bibfnamefont {F.~R.}\ \bibnamefont {Menter}}, \bibinfo {author} {\bibfnamefont {M.}~\bibnamefont {Kuntz}}, \bibinfo {author} {\bibfnamefont {R.}~\bibnamefont {Langtry}},  \emph {et~al.},\ }\bibfield  {title} {\enquote {\bibinfo {title} {Ten years of industrial experience with the sst turbulence model},}\ }\href@noop {} {\bibfield  {journal} {\bibinfo  {journal} {Turbulence, heat and mass transfer}\ }\textbf {\bibinfo {volume} {4}},\ \bibinfo {pages} {625--632} (\bibinfo {year} {2003})}\BibitemShut {NoStop}%
\bibitem [{\citenamefont {Georgiadis}\ and\ \citenamefont {Yoder}(2013)}]{georgiadis2013recalibration}%
  \BibitemOpen
  \bibfield  {author} {\bibinfo {author} {\bibfnamefont {N.}~\bibnamefont {Georgiadis}}\ and\ \bibinfo {author} {\bibfnamefont {D.}~\bibnamefont {Yoder}},\ }\bibfield  {title} {\enquote {\bibinfo {title} {Recalibration of the shear stress transport model to improve calculation of shock separated flows},}\ }in\ \href@noop {} {\emph {\bibinfo {booktitle} {51st AIAA Aerospace Sciences Meeting Including the New Horizons Forum and Aerospace Exposition}}}\ (\bibinfo {year} {2013})\ p.\ \bibinfo {pages} {685}\BibitemShut {NoStop}%
\bibitem [{\citenamefont {Kamenetskiy}\ \emph {et~al.}(2014)\citenamefont {Kamenetskiy}, \citenamefont {Bussoletti}, \citenamefont {Hilmes}, \citenamefont {Venkatakrishnan}, \citenamefont {Wigton},\ and\ \citenamefont {Johnson}}]{kamenetskiy2014numerical}%
  \BibitemOpen
  \bibfield  {author} {\bibinfo {author} {\bibfnamefont {D.~S.}\ \bibnamefont {Kamenetskiy}}, \bibinfo {author} {\bibfnamefont {J.~E.}\ \bibnamefont {Bussoletti}}, \bibinfo {author} {\bibfnamefont {C.~L.}\ \bibnamefont {Hilmes}}, \bibinfo {author} {\bibfnamefont {V.}~\bibnamefont {Venkatakrishnan}}, \bibinfo {author} {\bibfnamefont {L.~B.}\ \bibnamefont {Wigton}}, \ and\ \bibinfo {author} {\bibfnamefont {F.~T.}\ \bibnamefont {Johnson}},\ }\bibfield  {title} {\enquote {\bibinfo {title} {Numerical evidence of multiple solutions for the reynolds-averaged navier--stokes equations},}\ }\href@noop {} {\bibfield  {journal} {\bibinfo  {journal} {AIAA journal}\ }\textbf {\bibinfo {volume} {52}},\ \bibinfo {pages} {1686--1698} (\bibinfo {year} {2014})}\BibitemShut {NoStop}%
\bibitem [{\citenamefont {Zhu}\ \emph {et~al.}(2020)\citenamefont {Zhu}, \citenamefont {Zhang}, \citenamefont {Wang},\ and\ \citenamefont {Zhang}}]{zhu2020analysis}%
  \BibitemOpen
  \bibfield  {author} {\bibinfo {author} {\bibfnamefont {Z.}~\bibnamefont {Zhu}}, \bibinfo {author} {\bibfnamefont {X.}~\bibnamefont {Zhang}}, \bibinfo {author} {\bibfnamefont {X.}~\bibnamefont {Wang}}, \ and\ \bibinfo {author} {\bibfnamefont {L.}~\bibnamefont {Zhang}},\ }\bibfield  {title} {\enquote {\bibinfo {title} {Analysis of compressibility corrections for turbulence models in hypersonic boundary-layer applications},}\ }\href@noop {} {\bibfield  {journal} {\bibinfo  {journal} {Journal of Spacecraft and Rockets}\ }\textbf {\bibinfo {volume} {57}},\ \bibinfo {pages} {364--372} (\bibinfo {year} {2020})}\BibitemShut {NoStop}%
\bibitem [{\citenamefont {Guohua}, \citenamefont {Xiaogang},\ and\ \citenamefont {Meiliang}(2012)}]{guohua2012assessment}%
  \BibitemOpen
  \bibfield  {author} {\bibinfo {author} {\bibfnamefont {T.}~\bibnamefont {Guohua}}, \bibinfo {author} {\bibfnamefont {D.}~\bibnamefont {Xiaogang}}, \ and\ \bibinfo {author} {\bibfnamefont {M.}~\bibnamefont {Meiliang}},\ }\bibfield  {title} {\enquote {\bibinfo {title} {Assessment of two turbulence models and some compressibility corrections for hypersonic compression corners by high-order difference schemes},}\ }\href@noop {} {\bibfield  {journal} {\bibinfo  {journal} {Chinese Journal of Aeronautics}\ }\textbf {\bibinfo {volume} {25}},\ \bibinfo {pages} {25--32} (\bibinfo {year} {2012})}\BibitemShut {NoStop}%
\bibitem [{\citenamefont {Jones}\ and\ \citenamefont {Launder}(1972)}]{jones1972prediction}%
  \BibitemOpen
  \bibfield  {author} {\bibinfo {author} {\bibfnamefont {W.~P.}\ \bibnamefont {Jones}}\ and\ \bibinfo {author} {\bibfnamefont {B.~E.}\ \bibnamefont {Launder}},\ }\bibfield  {title} {\enquote {\bibinfo {title} {The prediction of laminarization with a two-equation model of turbulence},}\ }\href@noop {} {\bibfield  {journal} {\bibinfo  {journal} {International journal of heat and mass transfer}\ }\textbf {\bibinfo {volume} {15}},\ \bibinfo {pages} {301--314} (\bibinfo {year} {1972})}\BibitemShut {NoStop}%
\bibitem [{\citenamefont {Launder}\ and\ \citenamefont {Sharma}(1974)}]{launder1974application}%
  \BibitemOpen
  \bibfield  {author} {\bibinfo {author} {\bibfnamefont {B.~E.}\ \bibnamefont {Launder}}\ and\ \bibinfo {author} {\bibfnamefont {B.~I.}\ \bibnamefont {Sharma}},\ }\bibfield  {title} {\enquote {\bibinfo {title} {Application of the energy-dissipation model of turbulence to the calculation of flow near a spinning disc},}\ }\href@noop {} {\bibfield  {journal} {\bibinfo  {journal} {Letters in heat and mass transfer}\ }\textbf {\bibinfo {volume} {1}},\ \bibinfo {pages} {131--137} (\bibinfo {year} {1974})}\BibitemShut {NoStop}%
\bibitem [{\citenamefont {Rodi}(1991)}]{rodi1991experience}%
  \BibitemOpen
  \bibfield  {author} {\bibinfo {author} {\bibfnamefont {W.}~\bibnamefont {Rodi}},\ }\bibfield  {title} {\enquote {\bibinfo {title} {Experience with two-layer models combining the k-epsilon model with a one-equation model near the wall},}\ }in\ \href@noop {} {\emph {\bibinfo {booktitle} {29th Aerospace sciences meeting}}}\ (\bibinfo {year} {1991})\ p.\ \bibinfo {pages} {216}\BibitemShut {NoStop}%
\bibitem [{\citenamefont {Wilcox}(2008)}]{wilcox2008formulation}%
  \BibitemOpen
  \bibfield  {author} {\bibinfo {author} {\bibfnamefont {D.~C.}\ \bibnamefont {Wilcox}},\ }\bibfield  {title} {\enquote {\bibinfo {title} {Formulation of the kw turbulence model revisited},}\ }\href@noop {} {\bibfield  {journal} {\bibinfo  {journal} {AIAA journal}\ }\textbf {\bibinfo {volume} {46}},\ \bibinfo {pages} {2823--2838} (\bibinfo {year} {2008})}\BibitemShut {NoStop}%
\bibitem [{\citenamefont {Murray}(2007)}]{murray2007three}%
  \BibitemOpen
  \bibfield  {author} {\bibinfo {author} {\bibfnamefont {N.~P.}\ \bibnamefont {Murray}},\ }\emph {\bibinfo {title} {Three-dimensional turbulent shock-wave: boundary-layer interactions in hypersonic flows}},\ \href@noop {} {Ph.D. thesis},\ \bibinfo  {school} {Imperial College London (University of London)} (\bibinfo {year} {2007})\BibitemShut {NoStop}%
\bibitem [{\citenamefont {Paciorri}\ \emph {et~al.}(1998)\citenamefont {Paciorri}, \citenamefont {Dieudonn{\'e}}, \citenamefont {Degrez}, \citenamefont {Charbonnier},\ and\ \citenamefont {Deconinck}}]{paciorri1998exploring}%
  \BibitemOpen
  \bibfield  {author} {\bibinfo {author} {\bibfnamefont {R.}~\bibnamefont {Paciorri}}, \bibinfo {author} {\bibfnamefont {W.}~\bibnamefont {Dieudonn{\'e}}}, \bibinfo {author} {\bibfnamefont {G.}~\bibnamefont {Degrez}}, \bibinfo {author} {\bibfnamefont {J.-M.}\ \bibnamefont {Charbonnier}}, \ and\ \bibinfo {author} {\bibfnamefont {H.}~\bibnamefont {Deconinck}},\ }\bibfield  {title} {\enquote {\bibinfo {title} {Exploring the validity of the spalart-allmaras turbulence model for hypersonic flows},}\ }\href@noop {} {\bibfield  {journal} {\bibinfo  {journal} {Journal of Spacecraft and Rockets}\ }\textbf {\bibinfo {volume} {35}},\ \bibinfo {pages} {121--126} (\bibinfo {year} {1998})}\BibitemShut {NoStop}%
\bibitem [{\citenamefont {Delery}\ and\ \citenamefont {Panaras}(1996)}]{delery1996shock}%
  \BibitemOpen
  \bibfield  {author} {\bibinfo {author} {\bibfnamefont {J.~M.}\ \bibnamefont {Delery}}\ and\ \bibinfo {author} {\bibfnamefont {A.}~\bibnamefont {Panaras}},\ }\bibfield  {title} {\enquote {\bibinfo {title} {Shock-wave/boundary-layer interactions in high-mach-number flows},}\ }\href@noop {} {\bibfield  {journal} {\bibinfo  {journal} {AGARD ADVISORY REPORT AGARD AR}\ ,\ \bibinfo {pages} {2--1}} (\bibinfo {year} {1996})}\BibitemShut {NoStop}%
\bibitem [{\citenamefont {Zhang}, \citenamefont {Wu},\ and\ \citenamefont {Wang}(2010)}]{zhang2010turbulence}%
  \BibitemOpen
  \bibfield  {author} {\bibinfo {author} {\bibfnamefont {X.-H.}\ \bibnamefont {Zhang}}, \bibinfo {author} {\bibfnamefont {Y.-Z.}\ \bibnamefont {Wu}}, \ and\ \bibinfo {author} {\bibfnamefont {J.-F.}\ \bibnamefont {Wang}},\ }\bibfield  {title} {\enquote {\bibinfo {title} {Turbulence models for accurate aerothermal prediction in hypersonic flows},}\ }\href@noop {} {\bibfield  {journal} {\bibinfo  {journal} {Modern Physics Letters B}\ }\textbf {\bibinfo {volume} {24}},\ \bibinfo {pages} {1345--1348} (\bibinfo {year} {2010})}\BibitemShut {NoStop}%
\bibitem [{\citenamefont {Brown}(2011)}]{brown2011shock}%
  \BibitemOpen
  \bibfield  {author} {\bibinfo {author} {\bibfnamefont {J.}~\bibnamefont {Brown}},\ }\bibfield  {title} {\enquote {\bibinfo {title} {Shock wave impingement on boundary layers at hypersonic speeds: computational analysis and uncertainty},}\ }in\ \href@noop {} {\emph {\bibinfo {booktitle} {42nd AIAA Thermophysics Conference}}}\ (\bibinfo {year} {2011})\ p.\ \bibinfo {pages} {3143}\BibitemShut {NoStop}%
\bibitem [{\citenamefont {Pasha}\ and\ \citenamefont {Sinha}(2012)}]{pasha2012simulation}%
  \BibitemOpen
  \bibfield  {author} {\bibinfo {author} {\bibfnamefont {A.~A.}\ \bibnamefont {Pasha}}\ and\ \bibinfo {author} {\bibfnamefont {K.}~\bibnamefont {Sinha}},\ }\bibfield  {title} {\enquote {\bibinfo {title} {Simulation of hypersonic shock/turbulent boundary-layer interactions using shock-unsteadiness model},}\ }\href@noop {} {\bibfield  {journal} {\bibinfo  {journal} {Journal of Propulsion and Power}\ }\textbf {\bibinfo {volume} {28}},\ \bibinfo {pages} {46--60} (\bibinfo {year} {2012})}\BibitemShut {NoStop}%
\bibitem [{\citenamefont {Brown}(2013)}]{brown2013hypersonic}%
  \BibitemOpen
  \bibfield  {author} {\bibinfo {author} {\bibfnamefont {J.~L.}\ \bibnamefont {Brown}},\ }\bibfield  {title} {\enquote {\bibinfo {title} {Hypersonic shock wave impingement on turbulent boundary layers: computational analysis and uncertainty},}\ }\href@noop {} {\bibfield  {journal} {\bibinfo  {journal} {Journal of Spacecraft and Rockets}\ }\textbf {\bibinfo {volume} {50}},\ \bibinfo {pages} {96--123} (\bibinfo {year} {2013})}\BibitemShut {NoStop}%
\bibitem [{\citenamefont {Elfstrom}(1972)}]{elfstrom1972turbulent}%
  \BibitemOpen
  \bibfield  {author} {\bibinfo {author} {\bibfnamefont {G.}~\bibnamefont {Elfstrom}},\ }\bibfield  {title} {\enquote {\bibinfo {title} {Turbulent hypersonic flow at a wedge-compression corner},}\ }\href@noop {} {\bibfield  {journal} {\bibinfo  {journal} {Journal of fluid Mechanics}\ }\textbf {\bibinfo {volume} {53}},\ \bibinfo {pages} {113--127} (\bibinfo {year} {1972})}\BibitemShut {NoStop}%
\bibitem [{\citenamefont {Coakley}\ and\ \citenamefont {Huang}(1992)}]{coakley1992turbulence}%
  \BibitemOpen
  \bibfield  {author} {\bibinfo {author} {\bibfnamefont {T.}~\bibnamefont {Coakley}}\ and\ \bibinfo {author} {\bibfnamefont {P.}~\bibnamefont {Huang}},\ }\bibfield  {title} {\enquote {\bibinfo {title} {Turbulence modeling for high speed flows},}\ }in\ \href@noop {} {\emph {\bibinfo {booktitle} {30th Aerospace Sciences Meeting and Exhibit}}}\ (\bibinfo {year} {1992})\ p.\ \bibinfo {pages} {436}\BibitemShut {NoStop}%
\bibitem [{\citenamefont {Horstman}(1992)}]{horstman1992hypersonic}%
  \BibitemOpen
  \bibfield  {author} {\bibinfo {author} {\bibfnamefont {C.}~\bibnamefont {Horstman}},\ }\bibfield  {title} {\enquote {\bibinfo {title} {Hypersonic shock-wave turbulent-boundary-layer interaction flows},}\ }\href@noop {} {\bibfield  {journal} {\bibinfo  {journal} {AIAA journal}\ }\textbf {\bibinfo {volume} {30}},\ \bibinfo {pages} {1480--1481} (\bibinfo {year} {1992})}\BibitemShut {NoStop}%
\bibitem [{\citenamefont {Raje}(2021)}]{raje2021thesis}%
  \BibitemOpen
  \bibfield  {author} {\bibinfo {author} {\bibfnamefont {P.}~\bibnamefont {Raje}},\ }\href@noop {} {\emph {\bibinfo {title} {Advanced two-equation turbulence models for computing shock-dominated flows in aerospace applications}}}\ (\bibinfo  {publisher} {Indian Institute of Technology Bombay},\ \bibinfo {year} {2021})\BibitemShut {NoStop}%
\bibitem [{\citenamefont {Jie}\ and\ \citenamefont {Jie}(2011)}]{jie2011stress}%
  \BibitemOpen
  \bibfield  {author} {\bibinfo {author} {\bibfnamefont {T.}~\bibnamefont {Jie}}\ and\ \bibinfo {author} {\bibfnamefont {J.}~\bibnamefont {Jie}},\ }\bibfield  {title} {\enquote {\bibinfo {title} {Stress limiter consideration for k-omega turbulence models in shock-wave/turbulent boundary-layer interactions in supersonic and hypersonic flows},}\ }in\ \href@noop {} {\emph {\bibinfo {booktitle} {20th AIAA Computational Fluid Dynamics Conference}}}\ (\bibinfo {year} {2011})\ p.\ \bibinfo {pages} {3980}\BibitemShut {NoStop}%
\bibitem [{\citenamefont {Li}\ \emph {et~al.}(2021)\citenamefont {Li}, \citenamefont {Zeng}, \citenamefont {Chen}, \citenamefont {Zhang},\ and\ \citenamefont {Yan}}]{li2021bayesian}%
  \BibitemOpen
  \bibfield  {author} {\bibinfo {author} {\bibfnamefont {J.-p.}\ \bibnamefont {Li}}, \bibinfo {author} {\bibfnamefont {F.-z.}\ \bibnamefont {Zeng}}, \bibinfo {author} {\bibfnamefont {S.-s.}\ \bibnamefont {Chen}}, \bibinfo {author} {\bibfnamefont {K.-l.}\ \bibnamefont {Zhang}}, \ and\ \bibinfo {author} {\bibfnamefont {C.}~\bibnamefont {Yan}},\ }\bibfield  {title} {\enquote {\bibinfo {title} {Bayesian model evaluation of three k--$\omega$ turbulence models for hypersonic shock wave--boundary layer interaction flows},}\ }\href@noop {} {\bibfield  {journal} {\bibinfo  {journal} {Acta Astronautica}\ }\textbf {\bibinfo {volume} {189}},\ \bibinfo {pages} {143--157} (\bibinfo {year} {2021})}\BibitemShut {NoStop}%
\bibitem [{\citenamefont {Steelant}(2002)}]{steelant2002effect}%
  \BibitemOpen
  \bibfield  {author} {\bibinfo {author} {\bibfnamefont {J.}~\bibnamefont {Steelant}},\ }\bibfield  {title} {\enquote {\bibinfo {title} {Effect of a compressibility correction on different turbulence models},}\ }in\ \href@noop {} {\emph {\bibinfo {booktitle} {Engineering Turbulence Modelling and Experiments 5}}}\ (\bibinfo  {publisher} {Elsevier},\ \bibinfo {year} {2002})\ pp.\ \bibinfo {pages} {207--216}\BibitemShut {NoStop}%
\bibitem [{\citenamefont {Coratekin}, \citenamefont {Van~Keuk},\ and\ \citenamefont {Ballmann}(2004)}]{coratekin2004performance}%
  \BibitemOpen
  \bibfield  {author} {\bibinfo {author} {\bibfnamefont {T.}~\bibnamefont {Coratekin}}, \bibinfo {author} {\bibfnamefont {J.}~\bibnamefont {Van~Keuk}}, \ and\ \bibinfo {author} {\bibfnamefont {J.}~\bibnamefont {Ballmann}},\ }\bibfield  {title} {\enquote {\bibinfo {title} {Performance of upwind schemes and turbulence models in hypersonic flows},}\ }\href@noop {} {\bibfield  {journal} {\bibinfo  {journal} {AIAA journal}\ }\textbf {\bibinfo {volume} {42}},\ \bibinfo {pages} {945--957} (\bibinfo {year} {2004})}\BibitemShut {NoStop}%
\bibitem [{\citenamefont {Olsen}, \citenamefont {Coakley},\ and\ \citenamefont {Lillard}(2005)}]{olsen2005lag}%
  \BibitemOpen
  \bibfield  {author} {\bibinfo {author} {\bibfnamefont {M.}~\bibnamefont {Olsen}}, \bibinfo {author} {\bibfnamefont {T.}~\bibnamefont {Coakley}}, \ and\ \bibinfo {author} {\bibfnamefont {R.}~\bibnamefont {Lillard}},\ }\bibfield  {title} {\enquote {\bibinfo {title} {The lag model applied to high speed flows},}\ }in\ \href@noop {} {\emph {\bibinfo {booktitle} {43rd AIAA Aerospace Sciences Meeting and Exhibit}}}\ (\bibinfo {year} {2005})\ p.\ \bibinfo {pages} {101}\BibitemShut {NoStop}%
\bibitem [{\citenamefont {Zhang}\ \emph {et~al.}(2015)\citenamefont {Zhang}, \citenamefont {Gao}, \citenamefont {Jiang},\ and\ \citenamefont {Lee}}]{zhang2015turbulence}%
  \BibitemOpen
  \bibfield  {author} {\bibinfo {author} {\bibfnamefont {Z.}~\bibnamefont {Zhang}}, \bibinfo {author} {\bibfnamefont {Z.}~\bibnamefont {Gao}}, \bibinfo {author} {\bibfnamefont {C.}~\bibnamefont {Jiang}}, \ and\ \bibinfo {author} {\bibfnamefont {C.}~\bibnamefont {Lee}},\ }\bibfield  {title} {\enquote {\bibinfo {title} {Turbulence model modification for fake amplification of turbulence kinetic energy by shock-turbulence interation},}\ }in\ \href@noop {} {\emph {\bibinfo {booktitle} {51st AIAA/SAE/ASEE Joint Propulsion Conference}}}\ (\bibinfo {year} {2015})\ p.\ \bibinfo {pages} {4205}\BibitemShut {NoStop}%
\bibitem [{\citenamefont {Zhang}(2021)}]{zhang2021turbulence}%
  \BibitemOpen
  \bibfield  {author} {\bibinfo {author} {\bibfnamefont {H.}~\bibnamefont {Zhang}},\ }\href@noop {} {\emph {\bibinfo {title} {Turbulence Modelling for Aerodynamic Heating in Hypersonic Flows}}}\ (\bibinfo  {publisher} {The University of Manchester (United Kingdom)},\ \bibinfo {year} {2021})\BibitemShut {NoStop}%
\bibitem [{\citenamefont {Holden}(1978)}]{holden1978study}%
  \BibitemOpen
  \bibfield  {author} {\bibinfo {author} {\bibfnamefont {M.}~\bibnamefont {Holden}},\ }\bibfield  {title} {\enquote {\bibinfo {title} {A study of flow separation in regions of shock wave-boundary layer interaction in hypersonic flow},}\ }in\ \href@noop {} {\emph {\bibinfo {booktitle} {11th Fluid and PlasmaDynamics Conference}}}\ (\bibinfo {year} {1978})\ p.\ \bibinfo {pages} {1169}\BibitemShut {NoStop}%
\bibitem [{\citenamefont {Barone}, \citenamefont {Nicholson},\ and\ \citenamefont {Duan}(2022)}]{barone2022internal}%
  \BibitemOpen
  \bibfield  {author} {\bibinfo {author} {\bibfnamefont {M.}~\bibnamefont {Barone}}, \bibinfo {author} {\bibfnamefont {G.~L.}\ \bibnamefont {Nicholson}}, \ and\ \bibinfo {author} {\bibfnamefont {L.}~\bibnamefont {Duan}},\ }\bibfield  {title} {\enquote {\bibinfo {title} {Internal energy balance and aerodynamic heating predictions for hypersonic turbulent boundary layers},}\ }\href@noop {} {\bibfield  {journal} {\bibinfo  {journal} {Physical Review Fluids}\ }\textbf {\bibinfo {volume} {7}},\ \bibinfo {pages} {084604} (\bibinfo {year} {2022})}\BibitemShut {NoStop}%
\bibitem [{\citenamefont {Smith}(1996)}]{smith1996prediction}%
  \BibitemOpen
  \bibfield  {author} {\bibinfo {author} {\bibfnamefont {B.~R.}\ \bibnamefont {Smith}},\ }\bibfield  {title} {\enquote {\bibinfo {title} {Prediction of hypersonic shock-wave/turbulent boundary-layer interactions},}\ }\href@noop {} {\bibfield  {journal} {\bibinfo  {journal} {Journal of spacecraft and rockets}\ }\textbf {\bibinfo {volume} {33}},\ \bibinfo {pages} {614--619} (\bibinfo {year} {1996})}\BibitemShut {NoStop}%
\bibitem [{\citenamefont {Barone}, \citenamefont {Parish},\ and\ \citenamefont {Jordan}(2024{\natexlab{b}})}]{barone2024data}%
  \BibitemOpen
  \bibfield  {author} {\bibinfo {author} {\bibfnamefont {M.~F.}\ \bibnamefont {Barone}}, \bibinfo {author} {\bibfnamefont {E.}~\bibnamefont {Parish}}, \ and\ \bibinfo {author} {\bibfnamefont {C.}~\bibnamefont {Jordan}},\ }\bibfield  {title} {\enquote {\bibinfo {title} {Data-driven modifications to the spalart-allmaras turbulence model for supersonic and hypersonic boundary layers},}\ }in\ \href@noop {} {\emph {\bibinfo {booktitle} {AIAA SCITECH 2024 Forum}}}\ (\bibinfo {year} {2024})\ p.\ \bibinfo {pages} {0071}\BibitemShut {NoStop}%
\bibitem [{\citenamefont {Holden}, \citenamefont {Wadhams},\ and\ \citenamefont {MacLean}(2013)}]{holden2013measurements}%
  \BibitemOpen
  \bibfield  {author} {\bibinfo {author} {\bibfnamefont {M.}~\bibnamefont {Holden}}, \bibinfo {author} {\bibfnamefont {T.}~\bibnamefont {Wadhams}}, \ and\ \bibinfo {author} {\bibfnamefont {M.}~\bibnamefont {MacLean}},\ }\bibfield  {title} {\enquote {\bibinfo {title} {Measurements in regions of shock wave/turbulent boundary layer interaction from mach 4 to 10 for open and “blind” code evaluation/validation},}\ }in\ \href@noop {} {\emph {\bibinfo {booktitle} {21st AIAA Computational Fluid Dynamics Conference}}}\ (\bibinfo {year} {2013})\ p.\ \bibinfo {pages} {2836}\BibitemShut {NoStop}%
\bibitem [{\citenamefont {Pecnik}\ \emph {et~al.}(2012)\citenamefont {Pecnik}, \citenamefont {Terrapon}, \citenamefont {Ham}, \citenamefont {Iaccarino},\ and\ \citenamefont {Pitsch}}]{pecnik2012reynolds}%
  \BibitemOpen
  \bibfield  {author} {\bibinfo {author} {\bibfnamefont {R.}~\bibnamefont {Pecnik}}, \bibinfo {author} {\bibfnamefont {V.~E.}\ \bibnamefont {Terrapon}}, \bibinfo {author} {\bibfnamefont {F.}~\bibnamefont {Ham}}, \bibinfo {author} {\bibfnamefont {G.}~\bibnamefont {Iaccarino}}, \ and\ \bibinfo {author} {\bibfnamefont {H.}~\bibnamefont {Pitsch}},\ }\bibfield  {title} {\enquote {\bibinfo {title} {Reynolds-averaged navier-stokes simulations of the hyshot ii scramjet},}\ }\href@noop {} {\bibfield  {journal} {\bibinfo  {journal} {AIAA journal}\ }\textbf {\bibinfo {volume} {50}},\ \bibinfo {pages} {1717--1732} (\bibinfo {year} {2012})}\BibitemShut {NoStop}%
\bibitem [{\citenamefont {Barth}, \citenamefont {Wheatley},\ and\ \citenamefont {Smart}(2015)}]{barth2015effects}%
  \BibitemOpen
  \bibfield  {author} {\bibinfo {author} {\bibfnamefont {J.~E.}\ \bibnamefont {Barth}}, \bibinfo {author} {\bibfnamefont {V.}~\bibnamefont {Wheatley}}, \ and\ \bibinfo {author} {\bibfnamefont {M.~K.}\ \bibnamefont {Smart}},\ }\bibfield  {title} {\enquote {\bibinfo {title} {Effects of hydrogen fuel injection in a mach 12 scramjet inlet},}\ }\href@noop {} {\bibfield  {journal} {\bibinfo  {journal} {AIAA Journal}\ }\textbf {\bibinfo {volume} {53}},\ \bibinfo {pages} {2907--2919} (\bibinfo {year} {2015})}\BibitemShut {NoStop}%
\bibitem [{\citenamefont {Dai}\ \emph {et~al.}(2022)\citenamefont {Dai}, \citenamefont {Sun}, \citenamefont {Zhuo}, \citenamefont {Zhou}, \citenamefont {Zhou},\ and\ \citenamefont {Yue}}]{dai2022numerical}%
  \BibitemOpen
  \bibfield  {author} {\bibinfo {author} {\bibfnamefont {C.}~\bibnamefont {Dai}}, \bibinfo {author} {\bibfnamefont {B.}~\bibnamefont {Sun}}, \bibinfo {author} {\bibfnamefont {C.}~\bibnamefont {Zhuo}}, \bibinfo {author} {\bibfnamefont {S.}~\bibnamefont {Zhou}}, \bibinfo {author} {\bibfnamefont {C.}~\bibnamefont {Zhou}}, \ and\ \bibinfo {author} {\bibfnamefont {L.}~\bibnamefont {Yue}},\ }\bibfield  {title} {\enquote {\bibinfo {title} {Numerical study of high temperature non-equilibrium effects of double-wedge in hypervelocity flow},}\ }\href@noop {} {\bibfield  {journal} {\bibinfo  {journal} {Aerospace Science and Technology}\ }\textbf {\bibinfo {volume} {124}},\ \bibinfo {pages} {107526} (\bibinfo {year} {2022})}\BibitemShut {NoStop}%
\bibitem [{\citenamefont {Reddy}\ and\ \citenamefont {Sinha}(2009)}]{reddy2009hypersonic}%
  \BibitemOpen
  \bibfield  {author} {\bibinfo {author} {\bibfnamefont {D.~S.~K.}\ \bibnamefont {Reddy}}\ and\ \bibinfo {author} {\bibfnamefont {K.}~\bibnamefont {Sinha}},\ }\bibfield  {title} {\enquote {\bibinfo {title} {Hypersonic turbulent flow simulation of fire ii reentry vehicle afterbody},}\ }\href@noop {} {\bibfield  {journal} {\bibinfo  {journal} {Journal of Spacecraft and Rockets}\ }\textbf {\bibinfo {volume} {46}},\ \bibinfo {pages} {745--757} (\bibinfo {year} {2009})}\BibitemShut {NoStop}%
\bibitem [{\citenamefont {Bosco}(2011)}]{bosco2011reynolds}%
  \BibitemOpen
  \bibfield  {author} {\bibinfo {author} {\bibfnamefont {A.}~\bibnamefont {Bosco}},\ }\emph {\bibinfo {title} {Reynolds stress model for hypersonic flows}},\ \href@noop {} {Ph.D. thesis},\ \bibinfo  {school} {Aachen, Techn. Hochsch., Diss., 2011} (\bibinfo {year} {2011})\BibitemShut {NoStop}%
\bibitem [{\citenamefont {Daly}\ and\ \citenamefont {Harlow}(1970)}]{daly1970transport}%
  \BibitemOpen
  \bibfield  {author} {\bibinfo {author} {\bibfnamefont {B.~J.}\ \bibnamefont {Daly}}\ and\ \bibinfo {author} {\bibfnamefont {F.~H.}\ \bibnamefont {Harlow}},\ }\bibfield  {title} {\enquote {\bibinfo {title} {Transport equations in turbulence},}\ }\href@noop {} {\bibfield  {journal} {\bibinfo  {journal} {Physics of fluids}\ }\textbf {\bibinfo {volume} {13}},\ \bibinfo {pages} {2634--2649} (\bibinfo {year} {1970})}\BibitemShut {NoStop}%
\bibitem [{\citenamefont {Abe}\ and\ \citenamefont {Suga}(2001)}]{abe2001towards}%
  \BibitemOpen
  \bibfield  {author} {\bibinfo {author} {\bibfnamefont {K.}~\bibnamefont {Abe}}\ and\ \bibinfo {author} {\bibfnamefont {K.}~\bibnamefont {Suga}},\ }\bibfield  {title} {\enquote {\bibinfo {title} {Towards the development of a reynolds-averaged algebraic turbulent scalar-flux model},}\ }\href@noop {} {\bibfield  {journal} {\bibinfo  {journal} {International Journal of Heat and Fluid Flow}\ }\textbf {\bibinfo {volume} {22}},\ \bibinfo {pages} {19--29} (\bibinfo {year} {2001})}\BibitemShut {NoStop}%
\bibitem [{\citenamefont {Shin}\ \emph {et~al.}(2008)\citenamefont {Shin}, \citenamefont {An}, \citenamefont {Choi}, \citenamefont {Kim},\ and\ \citenamefont {Kim}}]{shin2008elliptic}%
  \BibitemOpen
  \bibfield  {author} {\bibinfo {author} {\bibfnamefont {J.~K.}\ \bibnamefont {Shin}}, \bibinfo {author} {\bibfnamefont {J.~S.}\ \bibnamefont {An}}, \bibinfo {author} {\bibfnamefont {Y.~D.}\ \bibnamefont {Choi}}, \bibinfo {author} {\bibfnamefont {Y.~C.}\ \bibnamefont {Kim}}, \ and\ \bibinfo {author} {\bibfnamefont {M.~S.}\ \bibnamefont {Kim}},\ }\bibfield  {title} {\enquote {\bibinfo {title} {Elliptic relaxation second moment closure for the turbulent heat fluxes},}\ }\href@noop {} {\bibfield  {journal} {\bibinfo  {journal} {Journal of Turbulence}\ ,\ \bibinfo {pages} {N3}} (\bibinfo {year} {2008})}\BibitemShut {NoStop}%
\bibitem [{\citenamefont {Weatheritt}\ \emph {et~al.}(2020)\citenamefont {Weatheritt}, \citenamefont {Zhao}, \citenamefont {Sandberg}, \citenamefont {Mizukami},\ and\ \citenamefont {Tanimoto}}]{weatheritt2020data}%
  \BibitemOpen
  \bibfield  {author} {\bibinfo {author} {\bibfnamefont {J.}~\bibnamefont {Weatheritt}}, \bibinfo {author} {\bibfnamefont {Y.}~\bibnamefont {Zhao}}, \bibinfo {author} {\bibfnamefont {R.~D.}\ \bibnamefont {Sandberg}}, \bibinfo {author} {\bibfnamefont {S.}~\bibnamefont {Mizukami}}, \ and\ \bibinfo {author} {\bibfnamefont {K.}~\bibnamefont {Tanimoto}},\ }\bibfield  {title} {\enquote {\bibinfo {title} {Data-driven scalar-flux model development with application to jet in cross flow},}\ }\href@noop {} {\bibfield  {journal} {\bibinfo  {journal} {International Journal of Heat and Mass Transfer}\ }\textbf {\bibinfo {volume} {147}},\ \bibinfo {pages} {118931} (\bibinfo {year} {2020})}\BibitemShut {NoStop}%
\bibitem [{\citenamefont {Zhao}\ \emph {et~al.}(2020)\citenamefont {Zhao}, \citenamefont {Akolekar}, \citenamefont {Weatheritt}, \citenamefont {Michelassi},\ and\ \citenamefont {Sandberg}}]{zhao2020rans}%
  \BibitemOpen
  \bibfield  {author} {\bibinfo {author} {\bibfnamefont {Y.}~\bibnamefont {Zhao}}, \bibinfo {author} {\bibfnamefont {H.~D.}\ \bibnamefont {Akolekar}}, \bibinfo {author} {\bibfnamefont {J.}~\bibnamefont {Weatheritt}}, \bibinfo {author} {\bibfnamefont {V.}~\bibnamefont {Michelassi}}, \ and\ \bibinfo {author} {\bibfnamefont {R.~D.}\ \bibnamefont {Sandberg}},\ }\bibfield  {title} {\enquote {\bibinfo {title} {Rans turbulence model development using cfd-driven machine learning},}\ }\href@noop {} {\bibfield  {journal} {\bibinfo  {journal} {Journal of Computational Physics}\ }\textbf {\bibinfo {volume} {411}},\ \bibinfo {pages} {109413} (\bibinfo {year} {2020})}\BibitemShut {NoStop}%
\bibitem [{\citenamefont {Duraisamy}(2021)}]{duraisamy2021perspectives}%
  \BibitemOpen
  \bibfield  {author} {\bibinfo {author} {\bibfnamefont {K.}~\bibnamefont {Duraisamy}},\ }\bibfield  {title} {\enquote {\bibinfo {title} {Perspectives on machine learning-augmented reynolds-averaged and large eddy simulation models of turbulence},}\ }\href@noop {} {\bibfield  {journal} {\bibinfo  {journal} {Physical Review Fluids}\ }\textbf {\bibinfo {volume} {6}},\ \bibinfo {pages} {050504} (\bibinfo {year} {2021})}\BibitemShut {NoStop}%
\bibitem [{\citenamefont {Parish}\ and\ \citenamefont {Duraisamy}(2016)}]{parish2016paradigm}%
  \BibitemOpen
  \bibfield  {author} {\bibinfo {author} {\bibfnamefont {E.~J.}\ \bibnamefont {Parish}}\ and\ \bibinfo {author} {\bibfnamefont {K.}~\bibnamefont {Duraisamy}},\ }\bibfield  {title} {\enquote {\bibinfo {title} {A paradigm for data-driven predictive modeling using field inversion and machine learning},}\ }\href@noop {} {\bibfield  {journal} {\bibinfo  {journal} {Journal of computational physics}\ }\textbf {\bibinfo {volume} {305}},\ \bibinfo {pages} {758--774} (\bibinfo {year} {2016})}\BibitemShut {NoStop}%
\bibitem [{\citenamefont {Ling}, \citenamefont {Kurzawski},\ and\ \citenamefont {Templeton}(2016)}]{ling2016reynolds}%
  \BibitemOpen
  \bibfield  {author} {\bibinfo {author} {\bibfnamefont {J.}~\bibnamefont {Ling}}, \bibinfo {author} {\bibfnamefont {A.}~\bibnamefont {Kurzawski}}, \ and\ \bibinfo {author} {\bibfnamefont {J.}~\bibnamefont {Templeton}},\ }\bibfield  {title} {\enquote {\bibinfo {title} {Reynolds averaged turbulence modelling using deep neural networks with embedded invariance},}\ }\href@noop {} {\bibfield  {journal} {\bibinfo  {journal} {Journal of Fluid Mechanics}\ }\textbf {\bibinfo {volume} {807}},\ \bibinfo {pages} {155--166} (\bibinfo {year} {2016})}\BibitemShut {NoStop}%
\bibitem [{\citenamefont {Weatheritt}\ and\ \citenamefont {Sandberg}(2016)}]{weatheritt2016novel}%
  \BibitemOpen
  \bibfield  {author} {\bibinfo {author} {\bibfnamefont {J.}~\bibnamefont {Weatheritt}}\ and\ \bibinfo {author} {\bibfnamefont {R.}~\bibnamefont {Sandberg}},\ }\bibfield  {title} {\enquote {\bibinfo {title} {A novel evolutionary algorithm applied to algebraic modifications of the {RANS} stress--strain relationship},}\ }\href@noop {} {\bibfield  {journal} {\bibinfo  {journal} {Journal of Computational Physics}\ }\textbf {\bibinfo {volume} {325}},\ \bibinfo {pages} {22--37} (\bibinfo {year} {2016})}\BibitemShut {NoStop}%
\bibitem [{\citenamefont {Schmelzer}, \citenamefont {Dwight},\ and\ \citenamefont {Cinnella}(2020)}]{schmelzer2020discovery}%
  \BibitemOpen
  \bibfield  {author} {\bibinfo {author} {\bibfnamefont {M.}~\bibnamefont {Schmelzer}}, \bibinfo {author} {\bibfnamefont {R.~P.}\ \bibnamefont {Dwight}}, \ and\ \bibinfo {author} {\bibfnamefont {P.}~\bibnamefont {Cinnella}},\ }\bibfield  {title} {\enquote {\bibinfo {title} {Discovery of algebraic \text{Reynolds-stress} models using sparse symbolic regression},}\ }\href@noop {} {\bibfield  {journal} {\bibinfo  {journal} {Flow, Turbulence and Combustion}\ }\textbf {\bibinfo {volume} {104}},\ \bibinfo {pages} {579--603} (\bibinfo {year} {2020})}\BibitemShut {NoStop}%
\bibitem [{\citenamefont {Cherroud}\ \emph {et~al.}(2022)\citenamefont {Cherroud}, \citenamefont {Merle}, \citenamefont {Cinnella},\ and\ \citenamefont {Gloerfelt}}]{cherroud2022sparse}%
  \BibitemOpen
  \bibfield  {author} {\bibinfo {author} {\bibfnamefont {S.}~\bibnamefont {Cherroud}}, \bibinfo {author} {\bibfnamefont {X.}~\bibnamefont {Merle}}, \bibinfo {author} {\bibfnamefont {P.}~\bibnamefont {Cinnella}}, \ and\ \bibinfo {author} {\bibfnamefont {X.}~\bibnamefont {Gloerfelt}},\ }\bibfield  {title} {\enquote {\bibinfo {title} {Sparse bayesian learning of explicit algebraic reynolds-stress models for turbulent separated flows},}\ }\href@noop {} {\bibfield  {journal} {\bibinfo  {journal} {International Journal of Heat and Fluid Flow}\ }\textbf {\bibinfo {volume} {98}},\ \bibinfo {pages} {109047} (\bibinfo {year} {2022})}\BibitemShut {NoStop}%
\bibitem [{\citenamefont {Sa{\"\i}di}\ \emph {et~al.}(2022)\citenamefont {Sa{\"\i}di}, \citenamefont {Schmelzer}, \citenamefont {Cinnella},\ and\ \citenamefont {Grasso}}]{saidi2022}%
  \BibitemOpen
  \bibfield  {author} {\bibinfo {author} {\bibfnamefont {I.~B.~H.}\ \bibnamefont {Sa{\"\i}di}}, \bibinfo {author} {\bibfnamefont {M.}~\bibnamefont {Schmelzer}}, \bibinfo {author} {\bibfnamefont {P.}~\bibnamefont {Cinnella}}, \ and\ \bibinfo {author} {\bibfnamefont {F.}~\bibnamefont {Grasso}},\ }\bibfield  {title} {\enquote {\bibinfo {title} {{CFD}-driven symbolic identification of algebraic reynolds-stress models},}\ }\href@noop {} {\bibfield  {journal} {\bibinfo  {journal} {Journal of Computational Physics}\ }\textbf {\bibinfo {volume} {457}},\ \bibinfo {pages} {111037} (\bibinfo {year} {2022})}\BibitemShut {NoStop}%
\bibitem [{\citenamefont {Kaandorp}\ and\ \citenamefont {Dwight}(2020)}]{kaandorp2020data}%
  \BibitemOpen
  \bibfield  {author} {\bibinfo {author} {\bibfnamefont {M.~L.}\ \bibnamefont {Kaandorp}}\ and\ \bibinfo {author} {\bibfnamefont {R.~P.}\ \bibnamefont {Dwight}},\ }\bibfield  {title} {\enquote {\bibinfo {title} {Data-driven modelling of the reynolds stress tensor using random forests with invariance},}\ }\href@noop {} {\bibfield  {journal} {\bibinfo  {journal} {Computers \& Fluids}\ }\textbf {\bibinfo {volume} {202}},\ \bibinfo {pages} {104497} (\bibinfo {year} {2020})}\BibitemShut {NoStop}%
\bibitem [{\citenamefont {Holland}, \citenamefont {Baeder},\ and\ \citenamefont {Duraisamy}(2019{\natexlab{a}})}]{holland2019towards}%
  \BibitemOpen
  \bibfield  {author} {\bibinfo {author} {\bibfnamefont {J.~R.}\ \bibnamefont {Holland}}, \bibinfo {author} {\bibfnamefont {J.~D.}\ \bibnamefont {Baeder}}, \ and\ \bibinfo {author} {\bibfnamefont {K.}~\bibnamefont {Duraisamy}},\ }\bibfield  {title} {\enquote {\bibinfo {title} {Towards integrated field inversion and machine learning with embedded neural networks for rans modeling},}\ }in\ \href@noop {} {\emph {\bibinfo {booktitle} {AIAA Scitech 2019 forum}}}\ (\bibinfo {year} {2019})\ p.\ \bibinfo {pages} {1884}\BibitemShut {NoStop}%
\bibitem [{\citenamefont {Holland}, \citenamefont {Baeder},\ and\ \citenamefont {Duraisamy}(2019{\natexlab{b}})}]{holland2019field}%
  \BibitemOpen
  \bibfield  {author} {\bibinfo {author} {\bibfnamefont {J.~R.}\ \bibnamefont {Holland}}, \bibinfo {author} {\bibfnamefont {J.~D.}\ \bibnamefont {Baeder}}, \ and\ \bibinfo {author} {\bibfnamefont {K.}~\bibnamefont {Duraisamy}},\ }\bibfield  {title} {\enquote {\bibinfo {title} {Field inversion and machine learning with embedded neural networks: Physics-consistent neural network training},}\ }in\ \href@noop {} {\emph {\bibinfo {booktitle} {AIAA Aviation 2019 Forum}}}\ (\bibinfo {year} {2019})\ p.\ \bibinfo {pages} {3200}\BibitemShut {NoStop}%
\bibitem [{\citenamefont {Srivastava}\ and\ \citenamefont {Duraisamy}(2021)}]{srivastava2021generalizable}%
  \BibitemOpen
  \bibfield  {author} {\bibinfo {author} {\bibfnamefont {V.}~\bibnamefont {Srivastava}}\ and\ \bibinfo {author} {\bibfnamefont {K.}~\bibnamefont {Duraisamy}},\ }\bibfield  {title} {\enquote {\bibinfo {title} {Generalizable physics-constrained modeling using learning and inference assisted by feature-space engineering},}\ }\href@noop {} {\bibfield  {journal} {\bibinfo  {journal} {Physical Review Fluids}\ }\textbf {\bibinfo {volume} {6}},\ \bibinfo {pages} {124602} (\bibinfo {year} {2021})}\BibitemShut {NoStop}%
\bibitem [{\citenamefont {Cinnella}(2024)}]{cinnella2024data}%
  \BibitemOpen
  \bibfield  {author} {\bibinfo {author} {\bibfnamefont {P.}~\bibnamefont {Cinnella}},\ }\bibfield  {title} {\enquote {\bibinfo {title} {Data-driven turbulence modeling},}\ }\href@noop {} {\bibfield  {journal} {\bibinfo  {journal} {arXiv preprint arXiv:2404.09074}\ } (\bibinfo {year} {2024})}\BibitemShut {NoStop}%
\bibitem [{\citenamefont {Duraisamy}(2025)}]{duraisamy2025book}%
  \BibitemOpen
  \bibfield  {author} {\bibinfo {author} {\bibfnamefont {K.}~\bibnamefont {Duraisamy}},\ }\href@noop {} {\emph {\bibinfo {title} {Data Driven Analysis and Modeling of Turbulent Flows}}},\ Vol.~\bibinfo {volume} {1}\ (\bibinfo  {publisher} {Elsevier},\ \bibinfo {year} {2025})\BibitemShut {NoStop}%
\bibitem [{\citenamefont {Huang}\ \emph {et~al.}(2017)\citenamefont {Huang}, \citenamefont {Duan}, \citenamefont {Wang}, \citenamefont {Sun},\ and\ \citenamefont {Xiao}}]{huang2017high}%
  \BibitemOpen
  \bibfield  {author} {\bibinfo {author} {\bibfnamefont {J.}~\bibnamefont {Huang}}, \bibinfo {author} {\bibfnamefont {L.}~\bibnamefont {Duan}}, \bibinfo {author} {\bibfnamefont {J.}~\bibnamefont {Wang}}, \bibinfo {author} {\bibfnamefont {R.}~\bibnamefont {Sun}}, \ and\ \bibinfo {author} {\bibfnamefont {H.}~\bibnamefont {Xiao}},\ }\bibfield  {title} {\enquote {\bibinfo {title} {High-mach-number turbulence modeling using machine learning and direct numerical simulation database},}\ }in\ \href@noop {} {\emph {\bibinfo {booktitle} {55th AIAA Aerospace Sciences Meeting}}}\ (\bibinfo {year} {2017})\ p.\ \bibinfo {pages} {0315}\BibitemShut {NoStop}%
\bibitem [{\citenamefont {Wang}\ \emph {et~al.}(2019)\citenamefont {Wang}, \citenamefont {Huang}, \citenamefont {Duan},\ and\ \citenamefont {Xiao}}]{wang2019prediction}%
  \BibitemOpen
  \bibfield  {author} {\bibinfo {author} {\bibfnamefont {J.-X.}\ \bibnamefont {Wang}}, \bibinfo {author} {\bibfnamefont {J.}~\bibnamefont {Huang}}, \bibinfo {author} {\bibfnamefont {L.}~\bibnamefont {Duan}}, \ and\ \bibinfo {author} {\bibfnamefont {H.}~\bibnamefont {Xiao}},\ }\bibfield  {title} {\enquote {\bibinfo {title} {{Prediction of Reynolds stresses in high-Mach-number turbulent boundary layers using physics-informed machine learning}},}\ }\href@noop {} {\bibfield  {journal} {\bibinfo  {journal} {Theoretical and Computational Fluid Dynamics}\ }\textbf {\bibinfo {volume} {33}},\ \bibinfo {pages} {1--19} (\bibinfo {year} {2019})}\BibitemShut {NoStop}%
\bibitem [{\citenamefont {Parish}\ \emph {et~al.}(2023)\citenamefont {Parish}, \citenamefont {Ching}, \citenamefont {Miller}, \citenamefont {Beresh},\ and\ \citenamefont {Barone}}]{PaChMi23}%
  \BibitemOpen
  \bibfield  {author} {\bibinfo {author} {\bibfnamefont {E.}~\bibnamefont {Parish}}, \bibinfo {author} {\bibfnamefont {D.~S.}\ \bibnamefont {Ching}}, \bibinfo {author} {\bibfnamefont {N.~E.}\ \bibnamefont {Miller}}, \bibinfo {author} {\bibfnamefont {S.~J.}\ \bibnamefont {Beresh}}, \ and\ \bibinfo {author} {\bibfnamefont {M.~F.}\ \bibnamefont {Barone}},\ }\bibfield  {title} {\enquote {\bibinfo {title} {Turbulence modeling for compressible flows using discrepancy tensor-basis neural networks and extrapolation detection},}\ }in\ \href@noop {} {\emph {\bibinfo {booktitle} {AIAA SciTech 2023 Forum}}}\ (\bibinfo {year} {2023})\ p.\ \bibinfo {pages} {2126}\BibitemShut {NoStop}%
\bibitem [{\citenamefont {Jordan}(2023)}]{Jo23}%
  \BibitemOpen
  \bibfield  {author} {\bibinfo {author} {\bibfnamefont {C.}~\bibnamefont {Jordan}},\ }\emph {\bibinfo {title} {{Turbulence Model Development for Hypersonic Shock Wave Boundary Layer Interactions }}},\ \href@noop {} {Ph.D. thesis},\ \bibinfo  {school} {North Carolina State University} (\bibinfo {year} {2023})\BibitemShut {NoStop}%
\bibitem [{\citenamefont {Parish}\ \emph {et~al.}(2024{\natexlab{b}})\citenamefont {Parish}, \citenamefont {Ching}, \citenamefont {Jordan}, \citenamefont {Nicholson}, \citenamefont {Miller}, \citenamefont {Beresh}, \citenamefont {Barone}, \citenamefont {Gupta},\ and\ \citenamefont {Duraisamy}}]{PaChJo24}%
  \BibitemOpen
  \bibfield  {author} {\bibinfo {author} {\bibfnamefont {E.}~\bibnamefont {Parish}}, \bibinfo {author} {\bibfnamefont {D.~S.}\ \bibnamefont {Ching}}, \bibinfo {author} {\bibfnamefont {C.}~\bibnamefont {Jordan}}, \bibinfo {author} {\bibfnamefont {G.}~\bibnamefont {Nicholson}}, \bibinfo {author} {\bibfnamefont {N.~E.}\ \bibnamefont {Miller}}, \bibinfo {author} {\bibfnamefont {S.}~\bibnamefont {Beresh}}, \bibinfo {author} {\bibfnamefont {M.}~\bibnamefont {Barone}}, \bibinfo {author} {\bibfnamefont {N.}~\bibnamefont {Gupta}}, \ and\ \bibinfo {author} {\bibfnamefont {K.}~\bibnamefont {Duraisamy}},\ }\bibfield  {title} {\enquote {\bibinfo {title} {Data-driven turbulent prandtl number modeling for hypersonic shock--boundary-layer interactions},}\ }\href@noop {} {\bibfield  {journal} {\bibinfo  {journal} {AIAA Journal}\ ,\ \bibinfo {pages} {1--22}} (\bibinfo {year} {2024}{\natexlab{b}})}\BibitemShut {NoStop}%
\bibitem [{\citenamefont {Singh}, \citenamefont {Medida},\ and\ \citenamefont {Duraisamy}(2017)}]{singh2017machine}%
  \BibitemOpen
  \bibfield  {author} {\bibinfo {author} {\bibfnamefont {A.~P.}\ \bibnamefont {Singh}}, \bibinfo {author} {\bibfnamefont {S.}~\bibnamefont {Medida}}, \ and\ \bibinfo {author} {\bibfnamefont {K.}~\bibnamefont {Duraisamy}},\ }\bibfield  {title} {\enquote {\bibinfo {title} {Machine-learning-augmented predictive modeling of turbulent separated flows over airfoils},}\ }\href@noop {} {\bibfield  {journal} {\bibinfo  {journal} {AIAA journal}\ }\textbf {\bibinfo {volume} {55}},\ \bibinfo {pages} {2215--2227} (\bibinfo {year} {2017})}\BibitemShut {NoStop}%
\bibitem [{\citenamefont {Chowdhary}\ \emph {et~al.}(2022)\citenamefont {Chowdhary}, \citenamefont {Hoang}, \citenamefont {Lee}, \citenamefont {Ray}, \citenamefont {Weirs},\ and\ \citenamefont {Carnes}}]{chowdhary2022calibrating}%
  \BibitemOpen
  \bibfield  {author} {\bibinfo {author} {\bibfnamefont {K.}~\bibnamefont {Chowdhary}}, \bibinfo {author} {\bibfnamefont {C.}~\bibnamefont {Hoang}}, \bibinfo {author} {\bibfnamefont {K.}~\bibnamefont {Lee}}, \bibinfo {author} {\bibfnamefont {J.}~\bibnamefont {Ray}}, \bibinfo {author} {\bibfnamefont {V.~G.}\ \bibnamefont {Weirs}}, \ and\ \bibinfo {author} {\bibfnamefont {B.}~\bibnamefont {Carnes}},\ }\bibfield  {title} {\enquote {\bibinfo {title} {Calibrating hypersonic turbulence flow models with the hifire-1 experiment using data-driven machine-learned models},}\ }\href@noop {} {\bibfield  {journal} {\bibinfo  {journal} {Computer Methods in Applied Mechanics and Engineering}\ }\textbf {\bibinfo {volume} {401}},\ \bibinfo {pages} {115396} (\bibinfo {year} {2022})}\BibitemShut {NoStop}%
\bibitem [{\citenamefont {Zafar}\ \emph {et~al.}(2024)\citenamefont {Zafar}, \citenamefont {Zhou}, \citenamefont {Roy}, \citenamefont {Stelter},\ and\ \citenamefont {Xiao}}]{zafar2024data}%
  \BibitemOpen
  \bibfield  {author} {\bibinfo {author} {\bibfnamefont {M.~I.}\ \bibnamefont {Zafar}}, \bibinfo {author} {\bibfnamefont {X.}~\bibnamefont {Zhou}}, \bibinfo {author} {\bibfnamefont {C.~J.}\ \bibnamefont {Roy}}, \bibinfo {author} {\bibfnamefont {D.}~\bibnamefont {Stelter}}, \ and\ \bibinfo {author} {\bibfnamefont {H.}~\bibnamefont {Xiao}},\ }\bibfield  {title} {\enquote {\bibinfo {title} {Data-driven turbulence modeling approach for cold-wall hypersonic boundary layers},}\ }\href@noop {} {\bibfield  {journal} {\bibinfo  {journal} {arXiv preprint arXiv:2406.17446}\ } (\bibinfo {year} {2024})}\BibitemShut {NoStop}%
\bibitem [{\citenamefont {Ling}\ and\ \citenamefont {Templeton}(2015)}]{ling2015evaluation}%
  \BibitemOpen
  \bibfield  {author} {\bibinfo {author} {\bibfnamefont {J.}~\bibnamefont {Ling}}\ and\ \bibinfo {author} {\bibfnamefont {J.}~\bibnamefont {Templeton}},\ }\bibfield  {title} {\enquote {\bibinfo {title} {Evaluation of machine learning algorithms for prediction of regions of high {Reynolds averaged Navier Stokes} uncertainty},}\ }\href@noop {} {\bibfield  {journal} {\bibinfo  {journal} {Physics of Fluids (1994-present)}\ }\textbf {\bibinfo {volume} {27}},\ \bibinfo {pages} {085103} (\bibinfo {year} {2015})}\BibitemShut {NoStop}%
\bibitem [{\citenamefont {Wu}, \citenamefont {Xiao},\ and\ \citenamefont {Paterson}(2018)}]{wu2018data-driven}%
  \BibitemOpen
  \bibfield  {author} {\bibinfo {author} {\bibfnamefont {J.-L.}\ \bibnamefont {Wu}}, \bibinfo {author} {\bibfnamefont {H.}~\bibnamefont {Xiao}}, \ and\ \bibinfo {author} {\bibfnamefont {E.}~\bibnamefont {Paterson}},\ }\bibfield  {title} {\enquote {\bibinfo {title} {Physics-informed machine learning approach for augmenting turbulence models: A comprehensive framework},}\ }\href@noop {} {\bibfield  {journal} {\bibinfo  {journal} {Physical Review Fluids}\ }\textbf {\bibinfo {volume} {3}},\ \bibinfo {pages} {074602} (\bibinfo {year} {2018})}\BibitemShut {NoStop}%
\bibitem [{\citenamefont {Beneddine}(2023)}]{beneddine2023nonlinear}%
  \BibitemOpen
  \bibfield  {author} {\bibinfo {author} {\bibfnamefont {S.}~\bibnamefont {Beneddine}},\ }\bibfield  {title} {\enquote {\bibinfo {title} {Nonlinear input feature reduction for data-based physical modeling},}\ }\href@noop {} {\bibfield  {journal} {\bibinfo  {journal} {Journal of Computational Physics}\ }\textbf {\bibinfo {volume} {474}},\ \bibinfo {pages} {111832} (\bibinfo {year} {2023})}\BibitemShut {NoStop}%
\bibitem [{\citenamefont {Zhou}(2025)}]{zhou2025ensemble}%
  \BibitemOpen
  \bibfield  {author} {\bibinfo {author} {\bibfnamefont {Z.-H.}\ \bibnamefont {Zhou}},\ }\href@noop {} {\emph {\bibinfo {title} {Ensemble methods: foundations and algorithms}}}\ (\bibinfo  {publisher} {CRC press},\ \bibinfo {year} {2025})\BibitemShut {NoStop}%
\bibitem [{\citenamefont {Fang}\ \emph {et~al.}(2023)\citenamefont {Fang}, \citenamefont {Zhao}, \citenamefont {Waschkowski}, \citenamefont {Ooi},\ and\ \citenamefont {Sandberg}}]{fang2023toward}%
  \BibitemOpen
  \bibfield  {author} {\bibinfo {author} {\bibfnamefont {Y.}~\bibnamefont {Fang}}, \bibinfo {author} {\bibfnamefont {Y.}~\bibnamefont {Zhao}}, \bibinfo {author} {\bibfnamefont {F.}~\bibnamefont {Waschkowski}}, \bibinfo {author} {\bibfnamefont {A.~S.}\ \bibnamefont {Ooi}}, \ and\ \bibinfo {author} {\bibfnamefont {R.~D.}\ \bibnamefont {Sandberg}},\ }\bibfield  {title} {\enquote {\bibinfo {title} {Toward more general turbulence models via multicase computational-fluid-dynamics-driven training},}\ }\href@noop {} {\bibfield  {journal} {\bibinfo  {journal} {AIAA Journal}\ }\textbf {\bibinfo {volume} {61}},\ \bibinfo {pages} {2100--2115} (\bibinfo {year} {2023})}\BibitemShut {NoStop}%
\bibitem [{\citenamefont {Rinc{\'o}n}\ \emph {et~al.}(2023)\citenamefont {Rinc{\'o}n}, \citenamefont {Amarloo}, \citenamefont {Reclari}, \citenamefont {Yang},\ and\ \citenamefont {Abkar}}]{rincon2023progressive}%
  \BibitemOpen
  \bibfield  {author} {\bibinfo {author} {\bibfnamefont {M.~J.}\ \bibnamefont {Rinc{\'o}n}}, \bibinfo {author} {\bibfnamefont {A.}~\bibnamefont {Amarloo}}, \bibinfo {author} {\bibfnamefont {M.}~\bibnamefont {Reclari}}, \bibinfo {author} {\bibfnamefont {X.~I.}\ \bibnamefont {Yang}}, \ and\ \bibinfo {author} {\bibfnamefont {M.}~\bibnamefont {Abkar}},\ }\bibfield  {title} {\enquote {\bibinfo {title} {Progressive augmentation of reynolds stress tensor models for secondary flow prediction by computational fluid dynamics driven surrogate optimisation},}\ }\href@noop {} {\bibfield  {journal} {\bibinfo  {journal} {International Journal of Heat and Fluid Flow}\ }\textbf {\bibinfo {volume} {104}},\ \bibinfo {pages} {109242} (\bibinfo {year} {2023})}\BibitemShut {NoStop}%
\bibitem [{\citenamefont {Cherroud}\ \emph {et~al.}(2025)\citenamefont {Cherroud}, \citenamefont {Merle}, \citenamefont {Cinnella},\ and\ \citenamefont {Gloerfelt}}]{cherroud2025space}%
  \BibitemOpen
  \bibfield  {author} {\bibinfo {author} {\bibfnamefont {S.}~\bibnamefont {Cherroud}}, \bibinfo {author} {\bibfnamefont {X.}~\bibnamefont {Merle}}, \bibinfo {author} {\bibfnamefont {P.}~\bibnamefont {Cinnella}}, \ and\ \bibinfo {author} {\bibfnamefont {X.}~\bibnamefont {Gloerfelt}},\ }\bibfield  {title} {\enquote {\bibinfo {title} {Space-dependent aggregation of stochastic data-driven turbulence models},}\ }\href@noop {} {\bibfield  {journal} {\bibinfo  {journal} {Journal of Computational Physics}\ ,\ \bibinfo {pages} {113793}} (\bibinfo {year} {2025})}\BibitemShut {NoStop}%
\bibitem [{\citenamefont {Oulghelou}\ \emph {et~al.}(2024)\citenamefont {Oulghelou}, \citenamefont {Cherroud}, \citenamefont {Merle},\ and\ \citenamefont {Cinnella}}]{oulghelou2024machine}%
  \BibitemOpen
  \bibfield  {author} {\bibinfo {author} {\bibfnamefont {M.}~\bibnamefont {Oulghelou}}, \bibinfo {author} {\bibfnamefont {S.}~\bibnamefont {Cherroud}}, \bibinfo {author} {\bibfnamefont {X.}~\bibnamefont {Merle}}, \ and\ \bibinfo {author} {\bibfnamefont {P.}~\bibnamefont {Cinnella}},\ }\bibfield  {title} {\enquote {\bibinfo {title} {Machine-learning-assisted blending of data-driven turbulence models},}\ }\href@noop {} {\bibfield  {journal} {\bibinfo  {journal} {arXiv preprint arXiv:2410.14431}\ } (\bibinfo {year} {2024})}\BibitemShut {NoStop}%
\bibitem [{\citenamefont {He}\ \emph {et~al.}(2022)\citenamefont {He}, \citenamefont {Tan}, \citenamefont {Rigas},\ and\ \citenamefont {Vahdati}}]{he2022explainability}%
  \BibitemOpen
  \bibfield  {author} {\bibinfo {author} {\bibfnamefont {X.}~\bibnamefont {He}}, \bibinfo {author} {\bibfnamefont {J.}~\bibnamefont {Tan}}, \bibinfo {author} {\bibfnamefont {G.}~\bibnamefont {Rigas}}, \ and\ \bibinfo {author} {\bibfnamefont {M.}~\bibnamefont {Vahdati}},\ }\bibfield  {title} {\enquote {\bibinfo {title} {On the explainability of machine-learning-assisted turbulence modeling for transonic flows},}\ }\href@noop {} {\bibfield  {journal} {\bibinfo  {journal} {International Journal of Heat and Fluid Flow}\ }\textbf {\bibinfo {volume} {97}},\ \bibinfo {pages} {109038} (\bibinfo {year} {2022})}\BibitemShut {NoStop}%
\bibitem [{\citenamefont {Cremades}\ \emph {et~al.}(2024)\citenamefont {Cremades}, \citenamefont {Hoyas}, \citenamefont {Deshpande}, \citenamefont {Quintero}, \citenamefont {Lellep}, \citenamefont {Lee}, \citenamefont {Monty}, \citenamefont {Hutchins}, \citenamefont {Linkmann}, \citenamefont {Marusic} \emph {et~al.}}]{cremades2024identifying}%
  \BibitemOpen
  \bibfield  {author} {\bibinfo {author} {\bibfnamefont {A.}~\bibnamefont {Cremades}}, \bibinfo {author} {\bibfnamefont {S.}~\bibnamefont {Hoyas}}, \bibinfo {author} {\bibfnamefont {R.}~\bibnamefont {Deshpande}}, \bibinfo {author} {\bibfnamefont {P.}~\bibnamefont {Quintero}}, \bibinfo {author} {\bibfnamefont {M.}~\bibnamefont {Lellep}}, \bibinfo {author} {\bibfnamefont {W.~J.}\ \bibnamefont {Lee}}, \bibinfo {author} {\bibfnamefont {J.~P.}\ \bibnamefont {Monty}}, \bibinfo {author} {\bibfnamefont {N.}~\bibnamefont {Hutchins}}, \bibinfo {author} {\bibfnamefont {M.}~\bibnamefont {Linkmann}}, \bibinfo {author} {\bibfnamefont {I.}~\bibnamefont {Marusic}},  \emph {et~al.},\ }\bibfield  {title} {\enquote {\bibinfo {title} {Identifying regions of importance in wall-bounded turbulence through explainable deep learning},}\ }\href@noop {} {\bibfield  {journal} {\bibinfo  {journal} {Nature Communications}\ }\textbf {\bibinfo {volume} {15}},\ \bibinfo {pages} {3864} (\bibinfo {year} {2024})}\BibitemShut {NoStop}%
\bibitem [{\citenamefont {Smits}, \citenamefont {Matheson},\ and\ \citenamefont {Joubert}(1983)}]{smits1983low}%
  \BibitemOpen
  \bibfield  {author} {\bibinfo {author} {\bibfnamefont {A.}~\bibnamefont {Smits}}, \bibinfo {author} {\bibfnamefont {N.}~\bibnamefont {Matheson}}, \ and\ \bibinfo {author} {\bibfnamefont {P.}~\bibnamefont {Joubert}},\ }\bibfield  {title} {\enquote {\bibinfo {title} {Low-reynolds-number turbulent boundary layers in zero and favorable pressure gradients},}\ }\href@noop {} {\bibfield  {journal} {\bibinfo  {journal} {Journal of ship research}\ }\textbf {\bibinfo {volume} {27}},\ \bibinfo {pages} {147--157} (\bibinfo {year} {1983})}\BibitemShut {NoStop}%
\bibitem [{\citenamefont {Nagib}, \citenamefont {Chauhan},\ and\ \citenamefont {Monkewitz}(2007)}]{nagib2007approach}%
  \BibitemOpen
  \bibfield  {author} {\bibinfo {author} {\bibfnamefont {H.~M.}\ \bibnamefont {Nagib}}, \bibinfo {author} {\bibfnamefont {K.~A.}\ \bibnamefont {Chauhan}}, \ and\ \bibinfo {author} {\bibfnamefont {P.~A.}\ \bibnamefont {Monkewitz}},\ }\bibfield  {title} {\enquote {\bibinfo {title} {Approach to an asymptotic state for zero pressure gradient turbulent boundary layers},}\ }\href@noop {} {\bibfield  {journal} {\bibinfo  {journal} {Philosophical Transactions of the Royal Society A: Mathematical, Physical and Engineering Sciences}\ }\textbf {\bibinfo {volume} {365}},\ \bibinfo {pages} {755--770} (\bibinfo {year} {2007})}\BibitemShut {NoStop}%
\bibitem [{\citenamefont {White}\ and\ \citenamefont {Majdalani}(2006)}]{white2006viscous}%
  \BibitemOpen
  \bibfield  {author} {\bibinfo {author} {\bibfnamefont {F.~M.}\ \bibnamefont {White}}\ and\ \bibinfo {author} {\bibfnamefont {J.}~\bibnamefont {Majdalani}},\ }\href@noop {} {\emph {\bibinfo {title} {Viscous fluid flow}}},\ Vol.~\bibinfo {volume} {3}\ (\bibinfo  {publisher} {McGraw-Hill New York},\ \bibinfo {year} {2006})\BibitemShut {NoStop}%
\end{thebibliography}%


@article{daly1970transport,
  title={Transport equations in turbulence},
  author={Daly, Bart J and Harlow, Francis H},
  journal={Physics of fluids},
  volume={13},
  number={11},
  pages={2634--2649},
  year={1970}
}

@article{zafar2024data,
  title={Data-Driven Turbulence Modeling Approach for Cold-Wall Hypersonic Boundary Layers},
  author={Zafar, Muhammad I and Zhou, Xuhui and Roy, Christopher J and Stelter, David and Xiao, Heng},
  journal={arXiv preprint arXiv:2406.17446},
  year={2024}
}

@article{zhao2020rans,
  title={RANS turbulence model development using CFD-driven machine learning},
  author={Zhao, Yaomin and Akolekar, Harshal D and Weatheritt, Jack and Michelassi, Vittorio and Sandberg, Richard D},
  journal={Journal of Computational Physics},
  volume={411},
  pages={109413},
  year={2020},
  publisher={Elsevier}
}

@article{singh2017machine,
  title={Machine-learning-augmented predictive modeling of turbulent separated flows over airfoils},
  author={Singh, Anand Pratap and Medida, Shivaji and Duraisamy, Karthik},
  journal={AIAA journal},
  volume={55},
  number={7},
  pages={2215--2227},
  year={2017},
  publisher={American Institute of Aeronautics and Astronautics}
}

@article{parish2016paradigm,
  title={A paradigm for data-driven predictive modeling using field inversion and machine learning},
  author={Parish, Eric J and Duraisamy, Karthik},
  journal={Journal of computational physics},
  volume={305},
  pages={758--774},
  year={2016},
  publisher={Elsevier}
}

@article{duraisamy2021perspectives,
  title={Perspectives on machine learning-augmented Reynolds-averaged and large eddy simulation models of turbulence},
  author={Duraisamy, Karthik},
  journal={Physical Review Fluids},
  volume={6},
  number={5},
  pages={050504},
  year={2021},
  publisher={APS}
}

@article{abe2001towards,
  title={Towards the development of a Reynolds-averaged algebraic turbulent scalar-flux model},
  author={Abe, K and Suga, K},
  journal={International Journal of Heat and Fluid Flow},
  volume={22},
  number={1},
  pages={19--29},
  year={2001},
  publisher={Elsevier}
}

@article{shin2008elliptic,
  title={Elliptic relaxation second moment closure for the turbulent heat fluxes},
  author={Shin, Jong Keun and An, Jeong Soo and Choi, Young Don and Kim, Young Chan and Kim, Min Soo},
  journal={Journal of Turbulence},
  number={9},
  pages={N3},
  year={2008},
  publisher={Taylor \& Francis}
}

@article{weatheritt2020data,
  title={Data-driven scalar-flux model development with application to jet in cross flow},
  author={Weatheritt, Jack and Zhao, Yaomin and Sandberg, Richard D and Mizukami, Satoshi and Tanimoto, Koichi},
  journal={International Journal of Heat and Mass Transfer},
  volume={147},
  pages={118931},
  year={2020},
  publisher={Elsevier}
}

@article{lozano2023machine,
  title={Machine learning building-block-flow wall model for large-eddy simulation},
  author={Lozano-Dur{\'a}n, Adri{\'a}n and Bae, H Jane},
  journal={Journal of Fluid Mechanics},
  volume={963},
  pages={A35},
  year={2023},
  publisher={Cambridge University Press}
}

@techreport{boseuse,
  title={The Use of Probabilistic Modeling for Reentry Thermal Protection System Design},
  author={Bose, Deepak and Wright, Michael J and Mansour, Nagi N},
  institution={Tech. rep., RTO-MP-AVT-147}
}


@article{longo2004modelling,
  title={Modelling of hypersonic flow phenomena},
  author={Longo, JMA},
  journal={Critical Technologies for Hypersonic Vehicle Development Technology--RTO/AVT/VKI Lecture Series},
  pages={10--14},
  year={2004},
  publisher={Citeseer}
}
@article{Hoste2024,
  title={A review of {RANS} modeling for hypersonic large cone-flares},
  author={Hoste, J.-J. and Gibbons, N. and Ecker, T. and Amato, C. and Knight, D. and Sattarov, A and Thiry, O. and Hickey, J.-P. and Hizir, F.E.  and K\"{o}kt\"{u}rk, T. and Castelino, N. and Viti, V. and Roldan, M. and Qiang, S. and Coder, J. and Baurle, R. and White, J. },
  journal={Physics of Fluids},
  volume={(under review)},
  number={},
  pages={},
  year={2024}
}

@article{CHRISTOPHER2020,
title = {DNS of turbulent flat-plate flow with transpiration cooling},
journal = {International Journal of Heat and Mass Transfer},
volume = {157},
pages = {119972},
year = {2020},
issn = {0017-9310},
doi = {https://doi.org/10.1016/j.ijheatmasstransfer.2020.119972},
url = {https://www.sciencedirect.com/science/article/pii/S0017931019355656},
author = {Nicholas Christopher and Johannes M.F. Peter and Markus J. Kloker and Jean-Pierre Hickey},
keywords = {Transpiration cooling, Compressible DNS, Turbulent boundary layer, Cooling efficiency},
abstract = {Transpiration cooling in a turbulent boundary layer on a flat plate is investigated using direct numerical simulations (DNS). The simulations are performed by solving the compressible Navier-Stokes equations at low Mach number conditions (M∞=0.3). Both the coolant and the hot gas are air, with isothermal walls and coolant at a temperature of Tw/T∞=0.5, while the blowing ratio and boundary conditions have been varied. These simulations elucidate the turbulence and heat-flux alterations due to the interaction of the coolant with the hot-gas boundary layer. By increasing the blowing ratio, the peak turbulent kinetic energy moves away from the wall to a region of shear between the low-momentum coolant and high-momentum hot gas. The reduction of the wall heat transfer is caused by the combined effects of heat advection due to the non-zero wall-normal velocity at the wall, and the reduction of the average boundary-layer temperature due to the accumulation of coolant. A new model for the latter effect is proposed which is physically realistic in the limit cases. The proposed combined model accounts for both heat advection and film accumulation and shows good agreement with the DNS data. An increase in turbulent transport of heat with increasing blowing rate is caused by the production of vortices between the coolant and hot gas. This causes a reduction in the cooling effectiveness, and can be seen near the leading edge of the transpiration region. In order to investigate wall modelling effects, simulations with uniform coolant injection have been compared to simulations with injection via many small strips. It is observed that as the strips get smaller (at fixed total mass flow rate and fixed wall porosity), the results trend towards the uniform injection case. Therefore, it is hypothesized that for small pore sizes, neglecting the effects of the individual pores in the wall boundary condition is physically justifiable.}
}

@article{Bukva2021,
    author = {Bukva, Alexander and Zhang, Kevin and Christopher, Nicholas and Hickey, Jean-Pierre},
    title = {Assessment of turbulence modeling for massively-cooled turbulent boundary layer flows with transpiration cooling},
    journal = {Physics of Fluids},
    volume = {33},
    number = {9},
    pages = {095114},
    year = {2021},
    month = {09},
    abstract = {We assess Reynolds-averaged Navier–Stokes (RANS) turbulent closures for the prediction of a turbulent boundary layer with transpiration cooling via comparison with a high-fidelity direct numerical simulation database. This study considers the canonical zero-pressure gradient, flat-plate, turbulent boundary layer over a massively cooled wall, with transpiration cooling. The simulations are conducted at a low-subsonic Mach number and we study two transpiration cooling configurations with uniform and slit injection at various blowing ratios. The DNS and RANS simulation setups are nearly identical. The RANS-based turbulence models perform well in the qualitative estimation of the velocity and thermal boundary layer evolution at low-blowing ratios (F = 0.2 and 0.6\%); more significant differences are noted at higher blowing ratios (F=2.0\%). The RANS models, especially the Spalart–Allmaras model, over-predict turbulence production near the wall which results in faster growth in the boundary thickness; this error becomes more pronounced at higher blowing ratios. Despite the greater mixing of momentum, the thermal mixing is under-predicted compared to the DNS in the uniform blowing case but over-predicted for the slit case. These results suggest that modeling errors in the temperature distribution due to turbulent thermal flux modeling can be significant even if the velocity is correctly modeled.},
    issn = {1070-6631},
    doi = {10.1063/5.0062155},
    url = {https://doi.org/10.1063/5.0062155},
    eprint = {https://pubs.aip.org/aip/pof/article-pdf/doi/10.1063/5.0062155/13883731/095114\_1\_online.pdf},
}


@InProceedings{holden2013measurements,
  author    = {Holden, MS and Wadhams, TP and MacLean, MG},
  booktitle = {21st AIAA Computational Fluid Dynamics Conference},
  title     = {Measurements in Regions of Shock Wave/Turbulent Boundary Layer Interaction from Mach 4 to 10 for Open and “Blind” Code Evaluation/Validation},
  year      = {2013},
  pages     = {2836},
}


@article{mansour2024flow,
  title={Flow Mechanics in Ablative Thermal Protection Systems},
  author={Mansour, Nagi N and Panerai, Francesco and Lachaud, Jean and Magin, Thierry},
  journal={Annual Review of Fluid Mechanics},
  volume={56},
  number={1},
  pages={549--575},
  year={2024},
  publisher={Annual Reviews}
}
@article{schneider2010hypersonic,
  title={Hypersonic boundary-layer transition with ablation and blowing},
  author={Schneider, Steven P},
  journal={Journal of Spacecraft and Rockets},
  volume={47},
  number={2},
  pages={225--237},
  year={2010}
}


@article{park1989nonequilibrium,
  title={Nonequilibrium hypersonic aerothermodynamics},
  author={Park, Chul},
  year={1989}
}

@article{direnzo2021direct,
  title={Direct numerical simulation of a hypersonic transitional boundary layer at suborbital enthalpies},
  author={Di Renzo, M. and Urzay, J.},
  journal={Journal of Fluid Mechanics},
  volume={912},
  year={2021},
  publisher={Cambridge University Press}
}
@article{di2024stagnation,
  title={Stagnation enthalpy effects on hypersonic turbulent compression corner flow at moderate Reynolds numbers},
  author={Di Renzo, M and Williams, CT and Pirozzoli, S},
  journal={Physical Review Fluids},
  volume={9},
  number={3},
  pages={033401},
  year={2024},
  publisher={APS}
}
@article{volpiani2021numerical,
  title={Numerical strategy to perform direct numerical simulations of hypersonic shock/boundary-layer interaction in chemical nonequilibrium},
  author={Volpiani, Pedro Stefanin},
  journal={Shock Waves},
  volume={31},
  number={4},
  pages={361--378},
  year={2021},
  publisher={Springer}
}


@article{colonna2001numerical,
  title={Numerical methods to solve Euler equations in one-dimensional steady nozzle flow},
  author={Colonna, G and Tuttafesta, M and Giordano, D},
  journal={Computer physics communications},
  volume={138},
  number={3},
  pages={213--221},
  year={2001},
  publisher={Elsevier}
}

@article{park2001chemical,
  title={Chemical-kinetic parameters of hyperbolic earth entry},
  author={Park, Chul and Jaffe, Richard L and Partridge, Harry},
  journal={Journal of Thermophysics and Heat transfer},
  volume={15},
  number={1},
  pages={76--90},
  year={2001}
}

@inproceedings{park1988two,
  title="{Two-temperature interpretation of dissociation rate data for N2 and O2}",
  author="Park, C.",
  booktitle="26th Aerspace Sciences Meeting",
  pages="458",
  year="1988"
}


@Article{	  pecnik2017scaling,
  title		= {Scaling and modelling of turbulence in variable property
		  channel flows},
  author	= {Pecnik, R. and Patel, A.},
  journal	= {Journal of Fluid Mechanics},
  year		= {2017},
  pages		= {R1-1--11},
  volume	= {823}
}

@Article{   otero2018turbulence,
  title   = {Turbulence modelling for flows with strong variations in thermo-physical
properties},
  author  = {Otero, G. and Patel, A. and Diez, R. and Pecnik, R.},
  journal = {Journal of Heat and Fluid Flow},
  year    = {2018},
  pages   = {R1},
  volume  = {823}
}

@article{hasan2023incorporating,
  title={Incorporating intrinsic compressibility effects in velocity transformations for wall-bounded turbulent flows},
  author={Hasan, Asif Manzoor and Larsson, Johan and Pirozzoli, Sergio and Pecnik, Rene},
  journal={Physical Review Fluids},
  volume={8},
  number={11},
  pages={L112601},
  year={2023},
  publisher={APS}
}


@article{bowersox2009extension,
  title={Extension of equilibrium turbulent heat flux models to high-speed shear flows},
  author={Bowersox, Rodney DW},
  journal={Journal of fluid mechanics},
  volume={633},
  pages={61--70},
  year={2009},
  publisher={Cambridge University Press}
}
@article{bowersox2010algebraic,
  title={Algebraic turbulent energy flux models for hypersonic shear flows},
  author={Bowersox, Rodney DW and North, Simon W},
  journal={Progress in Aerospace Sciences},
  volume={46},
  number={2-3},
  pages={49--61},
  year={2010},
  publisher={Elsevier}
}

@article{spalding1971concentration,
  title={Concentration fluctuations in a round turbulent free jet},
  author={Spalding, DB},
  journal={Chemical Engineering Science},
  volume={26},
  number={1},
  pages={95--107},
  year={1971},
  publisher={Elsevier}
}

@article{launder2005heat,
  title={Heat and mass transport},
  author={Launder, Brian E},
  journal={Turbulence},
  pages={231--287},
  year={2005},
  publisher={Springer}
}

@article{marquardt2020experimental,
  title={Experimental investigation of the turbulent Schmidt number in supersonic film cooling with shock interaction},
  author={Marquardt, Pascal and Klaas, Michael and Schr{\"o}der, Wolfgang},
  journal={Experiments in Fluids},
  volume={61},
  number={7},
  pages={160},
  year={2020},
  publisher={Springer}
}

@article{tominaga2007turbulent,
  title={Turbulent Schmidt numbers for CFD analysis with various types of flowfield},
  author={Tominaga, Yoshihide and Stathopoulos, Ted},
  journal={Atmospheric Environment},
  volume={41},
  number={37},
  pages={8091--8099},
  year={2007},
  publisher={Elsevier}
}

@article{xiang2020turbulence,
  author =        {Xiang, Z. and Yang, S. and Xie, S. and Li, J. and
                   Ren, H.},
  journal =       {Engineering Applications of Computational Fluid
                   Mechanics},
  number =        {1},
  pages =         {1546--1561},
  publisher =     {Taylor \& Francis},
  title =         {Turbulence--chemistry interaction models with
                   finite-rate chemistry and compressibility correction
                   for simulation of supersonic turbulent combustion},
  volume =        {14},
  year =          {2020},
  doi = {10.1080/19942060.2020.1842248},
}

@article{bray2006finite,
  author =        {Bray, K. N. C. and Champion, M. and Libby, P. A. and
                   Swaminathan, N.},
  journal =       {Combustion and Flame},
  number =        {4},
  pages =         {665--673},
  publisher =     {Elsevier},
  title =         {Finite rate chemistry and presumed PDF models for
                   premixed turbulent combustion},
  volume =        {146},
  year =          {2006},
  doi = {10.1016/j.combustflame.2006.07.001},
}

@inbook{baranwal2020vibrational,
  author =        {Baranwal, A. and Donzis, D. A. and Bowersox, R. D.},
  booktitle =     {AIAA Scitech 2020 Forum},
  pages =         {2052},
  title =         {Vibrational turbulent Prandtl number in flows with
                   thermal non-equilibrium},
  year =          {2020},
  doi = {10.2514/6.2020-2052},
}

@conference{BaPaJo24,
        author = {Matthew Barone and Eric Parish and Cyrus Jordan}, 
        booktitle = {AIAA SciTech},
        date-added = {2023-12-04 10:40:13 -0600},
        date-modified = {2023-12-04 10:41:01 -0600},
        title = {{Data-Driven Modifications to the Spalart--Allmaras Turbulence
Model for Supersonic and Hypersonic Boundary Layers}},
        year = {2024}}

@conference{PaChMi23,
        abstract = { View Video Presentation: https://doi.org/10.2514/6.2023-2126.vidThe Reynolds-averaged Navier--Stokes (RANS) equations remain a workhorse technology for simulating compressible fluid flows of practical interest. Due to model-form errors, however, RANS models can yield erroneous predictions that preclude their use on mission-critical problems. This work presents a data-driven turbulence modeling strategy aimed at improving RANS models for compressible fluid flows. The strategy outlined has three core aspects: (1) prediction for the discrepancy in the Reynolds stress tensor and turbulent heat flux via machine learning (ML), (2) estimating uncertainties in ML model outputs via out-of-distribution detection, and (3) multi-step training strategies to improve feature-response consistency. Results are presented across a range of cases publicly available on NASA's turbulence modeling resource involving wall-bounded flows, jet flows, and hypersonic boundary layer flows with cold walls. We find that one ML turbulence model is able to provide consistent improvements for numerous quantities-of-interest across all cases. },
        author = {Eric Parish and David S. Ching and Nathan E. Miller and Steven J. Beresh and Matthew F. Barone},
        booktitle = {AIAA SCITECH 2023 Forum},
        year = {2023},
        doi = {10.2514/6.2023-2126},
        eprint = {https://arc.aiaa.org/doi/pdf/10.2514/6.2023-2126},
        title = {{Turbulence modeling for compressible flows using discrepancy
tensor-basis neural networks and extrapolation detection}},
        url = {https://arc.aiaa.org/doi/abs/10.2514/6.2023-2126},
        bdsk-url-1 = {https://arc.aiaa.org/doi/abs/10.2514/6.2023-2126},
        bdsk-url-2 = {https://doi.org/10.2514/6.2023-2126}}
}

@phdthesis{Ni24,
        author = {Gary Nicholson}, 
        school = {Ohio State University},
        title = {{Simulation and Modeling of Hypersonic Turbulent Boundary
Layers Subject to Favorable and Adverse Pressure Gradients
Due to Streamline Curvature }},
        year = {2024}}
        

@phdthesis{Jo23,
        author = {Cyrus Jordan}, 
        date-added = {2024-06-07 16:46:01 -0500},
        date-modified = {2024-06-07 16:46:40 -0500},
        school = {North Carolina State University},
        title = {{Turbulence Model Development for Hypersonic Shock Wave
Boundary Layer Interactions }},
        year = {2023}}
        
@article{WaHuDuXi19,
        abstract = {Modeled Reynolds stress is a major source of model-form uncertainties in Reynolds-averaged Navier--Stokes (RANS) simulations. Recently, a physics-informed machine learning (PIML) approach has been proposed for reconstructing the discrepancies in RANS-modeled Reynolds stresses. The merits of the PIML framework have been demonstrated in several canonical incompressible flows. However, its performance on high-Mach-number flows is still not clear. In this work, we use the PIML approach to predict the discrepancies in RANS-modeled Reynolds stresses in high-Mach-number flat-plate turbulent boundary layers by using an existing DNS database. Specifically, the discrepancy function is first constructed using a DNS training flow and then used to correct RANS-predicted Reynolds stresses under flow conditions different from the DNS. The machine learning technique is shown to significantly improve RANS-modeled turbulent normal stresses, the turbulent kinetic energy, and the Reynolds stress anisotropy. Improvements are consistently observed when different training datasets are used. Moreover, a high-dimensional visualization technique and a distance metrics are used to provide a priori assessment of prediction confidence based only on RANS simulations. This study demonstrates that the PIML approach is a computationally affordable technique for improving the accuracy of RANS-modeled Reynolds stresses for high-Mach-number turbulent flows when there is a lack of experiments and high-fidelity simulations.},
        author = {Wang, Jian-Xun and Huang, Junji and Duan, Lian and Xiao, Heng},
        date = {2019/02/01},
        date-added = {2023-05-16 07:40:48 -0500},
        date-modified = {2023-05-16 07:40:48 -0500},
        doi = {10.1007/s00162-018-0480-2},
        id = {Wang2019},
        isbn = {1432-2250},
        journal = {Theoretical and Computational Fluid Dynamics},
        number = {1},
        pages = {1--19},
        title = {{Prediction of Reynolds stresses in high-Mach-number turbulent
boundary layers using physics-informed machine learning}},
        url = {https://doi.org/10.1007/s00162-018-0480-2},
        volume = {33},
        year = {2019},
        bdsk-url-1 = {https://doi.org/10.1007/s00162-018-0480-2}}


@article{passiatore2022thermochemical,
  title={Thermochemical non-equilibrium effects in turbulent hypersonic boundary layers},
  author={Passiatore, Donatella and Sciacovelli, Luca and Cinnella, Paola and Pascazio, G},
  journal={Journal of Fluid Mechanics},
  volume={941},
  pages={A21},
  year={2022},
  publisher={Cambridge University Press}
}

@article{passiatore2023shock,
  title={Shock impingement on a transitional hypersonic high-enthalpy boundary layer},
  author={Passiatore, Donatella and Sciacovelli, Luca and Cinnella, Paola and Pascazio, Giuseppe},
  journal={Physical Review Fluids},
  volume={8},
  number={4},
  pages={044601},
  year={2023},
  publisher={APS}
}


@book{gnoffo1989conservation,
  title={Conservation equations and physical models for hypersonic air flows in thermal and chemical nonequilibrium},
  author={Gnoffo, Peter A},
  volume={2867},
  year={1989},
  publisher={National Aeronautics and Space Administration, Office of Management~…}
}

@article{roy2006review,
  title={Review and assessment of turbulence models for hypersonic flows},
  author={Roy, Christopher J and Blottner, Frederick G},
  journal={Progress in aerospace sciences},
  volume={42},
  number={7-8},
  pages={469--530},
  year={2006},
  publisher={Elsevier}
}


@inproceedings{smits2009current,
  title={Current status of basic research in hypersonic turbulence},
  author={Smits, Lex and Martin, Pino and Girimaji, Sharath},
  booktitle={47th AIAA Aerospace Sciences Meeting including The New Horizons Forum and Aerospace Exposition},
  pages={151},
  year={2009}
}

@inproceedings{brown2002turbulence,
  title={Turbulence model validation for hypersonic flows},
  author={Brown, James},
  booktitle={8th AIAA/ASME Joint Thermophysics and Heat Transfer Conference},
  pages={3308},
  year={2002}
}

@techreport{marvin1989turbulence,
  title={Turbulence modeling for hypersonic flows},
  author={Marvin, Joseph G and Coakley, Thomas J},
  year={1989}
}

@article{marvin1992turbulence,
  title={Turbulence modeling for hypersonic flows},
  author={Marvin, JG and Coakley, TJ},
  journal={Advances in Hypersonics: Modeling Hypersonic Flows},
  pages={1--43},
  year={1992},
  publisher={Springer}
}

@article{degrez1993shock,
  title={Shock-wave/boundary-layer interaction in supersonic and hypersonic flows},
  author={Degrez, G and Simeonides, G and Delery, J and Vandromme, D and Dolling, D and Knight, D and Settles, G},
  journal={AGARD Report},
  volume={792},
  year={1993}
}

@article{bose2013uncertainty,
  title={Uncertainty assessment of hypersonic aerothermodynamics prediction capability},
  author={Bose, Deepak and Brown, James L and Prabhu, Dinesh K and Gnoffo, Peter and Johnston, Christopher O and Hollis, Brian},
  journal={Journal of Spacecraft and Rockets},
  volume={50},
  number={1},
  pages={12--18},
  year={2013}
}

@article{georgiadis2014status,
  title={Status of turbulence modeling for hypersonic propulsion flowpaths},
  author={Georgiadis, Nicholas J and Yoder, Dennis A and Vyas, Manan A and Engblom, William A},
  journal={Theoretical and Computational Fluid Dynamics},
  volume={28},
  pages={295--318},
  year={2014},
  publisher={Springer}
}

@inproceedings{goldberg2005hypersonic,
  title={Hypersonic turbulent flow predictions using CFD++},
  author={Goldberg, Uriel},
  booktitle={AIAA/CIRA 13th International Space Planes and Hypersonics Systems and Technologies Conference},
  pages={3214},
  year={2005}
}

@inproceedings{huang1993turbulence,
  title={Turbulence modeling for complex hypersonic flows},
  author={Huang, P and Coakley, T},
  booktitle={31st Aerospace Sciences Meeting},
  pages={200},
  year={1993}
}

@inproceedings{maicke2010evaluation,
  title={Evaluation of CFD codes for hypersonic flow modeling},
  author={Maicke, Brian and Barber, Timothy and Majdalani, Joe},
  booktitle={46th AIAA/ASME/SAE/ASEE joint propulsion conference \& exhibit},
  pages={7184},
  year={2010}
}

@inproceedings{marvin1992cfd,
  title={A CFD validation roadmap for hypersonic flows},
  author={Marvin, Joseph G},
  booktitle={Fluid Dynamics Panel Meeting and Symposium on Theoretical and Experimental Methods in Hypersonic Flows},
  number={NASA-TM-103935},
  year={1992}
}

@techreport{coakley1994turbulence,
  title={Turbulence compressibility corrections},
  author={Coakley, TJ and Horstman, CC and Marvin, JG and Viegas, JR and Bardina, JE and Huang, PG and Kussoy, MI},
  year={1994}
}

@inproceedings{paciorri1997validation,
  title={Validation of the Spalart-Allmaras turbulence model for application in hypersonic flows},
  author={Paciorri, Renato and Dieudonne, W and Degrez, G and Charbonnier, J-M and Deconinck, H and Paciorri, R and Dieudonne, W and Degrez, G and Charbonnier, J-M and Deconinck, H},
  booktitle={28th Fluid Dynamics Conference},
  pages={2023},
  year={1997}
}

@article{roy2003methodology,
  title={Methodology for turbulence model validation: application to hypersonic flows},
  author={Roy, Christopher J and Blottner, Frederick G},
  journal={Journal of spacecraft and rockets},
  volume={40},
  number={3},
  pages={313--325},
  year={2003}
}

@article{rumsey2010compressibility,
  title={Compressibility considerations for kw turbulence models in hypersonic boundary-layer applications},
  author={Rumsey, Christopher L},
  journal={Journal of Spacecraft and Rockets},
  volume={47},
  number={1},
  pages={11--20},
  year={2010}
}

@article{schmisseur2015hypersonics,
  title={Hypersonics into the 21st century: A perspective on AFOSR-sponsored research in aerothermodynamics},
  author={Schmisseur, John D},
  journal={Progress in Aerospace Sciences},
  volume={72},
  pages={3--16},
  year={2015},
  publisher={Elsevier}
}


@inproceedings{wilcox1991progress,
  title={Progress in hypersonic turbulence modeling},
  author={Wilcox, David},
  booktitle={22nd Fluid Dynamics, Plasma Dynamics and Lasers Conference},
  pages={1785},
  year={1991}
}


@inproceedings{gnoffo2011uncertainty,
  title={Uncertainty assessments of 2D and axisymmetric hypersonic shock wave-turbulent boundary layer interaction simulations at compression corners},
  author={Gnoffo, Peter and Berry, Scott and Van Norman, John},
  booktitle={42nd AIAA thermophysics conference},
  pages={3142},
  year={2011}
}

@inproceedings{brown2011shock,
  title={Shock wave impingement on boundary layers at hypersonic speeds: computational analysis and uncertainty},
  author={Brown, James},
  booktitle={42nd AIAA Thermophysics Conference},
  pages={3143},
  year={2011}
}

@article{gnoffo2013uncertainty,
  title={Uncertainty assessments of hypersonic shock wave-turbulent boundary-layer interactions at compression corners},
  author={Gnoffo, Peter A and Berry, Scott A and Van Norman, John W},
  journal={Journal of Spacecraft and Rockets},
  volume={50},
  number={1},
  pages={69--95},
  year={2013}
}

@article{goldberg2000hypersonic,
  title={Hypersonic flow predictions using linear and nonlinear turbulence closures},
  author={Goldberg, U and Batten, P and Palaniswamy, S and Chakravarthy, S and Peroomian, O},
  journal={Journal of Aircraft},
  volume={37},
  number={4},
  pages={671--675},
  year={2000}
}

@article{goldberg2001hypersonic,
  title={Hypersonic flow heat transfer prediction using single equation turbulence models},
  author={Goldberg, U},
  journal={J. Heat Transfer},
  volume={123},
  number={1},
  pages={65--69},
  year={2001}
}

@article{brown2013hypersonic,
  title={Hypersonic shock wave impingement on turbulent boundary layers: computational analysis and uncertainty},
  author={Brown, James L},
  journal={Journal of Spacecraft and Rockets},
  volume={50},
  number={1},
  pages={96--123},
  year={2013}
}


@techreport{marvin2013experimental,
  title={Experimental database with baseline CFD solutions: 2-D and axisymmetric hypersonic shock-wave/turbulent-boundary-layer interactions},
  author={Marvin, Joseph G and Brown, James L and Gnoffo, Peter A},
  year={2013}
}

@article{schulein2006skin,
  title={Skin friction and heat flux measurements in shock/boundary layer interaction flows},
  author={Sch{\"u}lein, Erich},
  journal={AIAA journal},
  volume={44},
  number={8},
  pages={1732--1741},
  year={2006}
}

@inproceedings{rumsey2015recent,
  title={Recent developments on the turbulence modeling resource website},
  author={Rumsey, Christopher L},
  booktitle={22nd AIAA Computational Fluid Dynamics Conference},
  pages={2927},
  year={2015}
}

@inproceedings{holden2010experimental,
  title={Experimental studies of shock wave/turbulent boundary layer interaction in high Reynolds number supersonic and hypersonic flows to evaluate the performance of CFD codes},
  author={Holden, Michael and Wadhams, Timothy and MacLean, Matthew and Mundy, Erik},
  booktitle={40th fluid dynamics conference and exhibit},
  pages={4468},
  year={2010}
}

@article{raje2021anisotropic,
  title={Anisotropic SST turbulence model for shock-boundary layer interaction},
  author={Raje, Pratikkumar and Sinha, Krishnendu},
  journal={Computers \& Fluids},
  volume={228},
  pages={105072},
  year={2021},
  publisher={Elsevier}
}

@inproceedings{holden2012review,
  title={Review of basic research and development programs conducted in the lens facilities in hypersonic flows},
  author={Holden, Michael and Wadhams, Timothy and MacLean, Matthew and Dufrene, Aaron and Mundy, Erik and Marineau, Eric},
  booktitle={50th AIAA Aerospace Sciences Meeting including the New Horizons Forum and Aerospace Exposition},
  pages={469},
  year={2012}
}

@inproceedings{waligura2022investigation,
  title={Investigation of spalart-allmaras turbulence model modifications for hypersonic flows utilizing output-based grid adaptation},
  author={Waligura, Carter J and Couchman, Benjamin L and Galbraith, Marshall C and Allmaras, Steven R and Harris, Wesley L},
  booktitle={AIAA SciTech 2022 Forum},
  pages={0587},
  year={2022}
}

@article{pasha2012simulation,
  title={Simulation of hypersonic shock/turbulent boundary-layer interactions using shock-unsteadiness model},
  author={Pasha, Amjad A and Sinha, Krishnendu},
  journal={Journal of Propulsion and Power},
  volume={28},
  number={1},
  pages={46--60},
  year={2012}
}

@article{xiao2007modeling,
  title={Modeling scramjet flows with variable turbulent Prandtl and Schmidt numbers},
  author={Xiao, Xudong and Hassan, Hassan A and Baurle, Robert A},
  journal={AIAA journal},
  volume={45},
  number={6},
  pages={1415--1423},
  year={2007}
}

@article{pecnik2012reynolds,
  title={Reynolds-averaged Navier-Stokes simulations of the HyShot II scramjet},
  author={Pecnik, Ren{\'e} and Terrapon, Vincent E and Ham, Frank and Iaccarino, Gianluca and Pitsch, Heinz},
  journal={AIAA journal},
  volume={50},
  number={8},
  pages={1717--1732},
  year={2012}
}

@inproceedings{georgiadis2013recalibration,
  title={Recalibration of the shear stress transport model to improve calculation of shock separated flows},
  author={Georgiadis, Nicholas and Yoder, Dennis},
  booktitle={51st AIAA Aerospace Sciences Meeting Including the New Horizons Forum and Aerospace Exposition},
  pages={685},
  year={2013}
}

@inproceedings{coakley1992turbulence,
  title={Turbulence modeling for high speed flows},
  author={Coakley, T and Huang, P},
  booktitle={30th Aerospace Sciences Meeting and Exhibit},
  pages={436},
  year={1992}
}

@article{zhang2022application,
  title={Application of linear and nonlinear two-equation turbulence models in hypersonic flows},
  author={Zhang, Haoyuan and Craft, Timothy and Iacovides, Hector},
  journal={AIAA Journal},
  volume={60},
  number={6},
  pages={3472--3486},
  year={2022},
  publisher={American Institute of Aeronautics and Astronautics}
}

@article{zhao2019uncertainty,
  title={Uncertainty and sensitivity analysis of SST turbulence model on hypersonic flow heat transfer},
  author={Zhao, Yatian and Yan, Chao and Wang, Xiaoyong and Liu, Hongkang and Zhang, Wei},
  journal={International Journal of Heat and Mass Transfer},
  volume={136},
  pages={808--820},
  year={2019},
  publisher={Elsevier}
}

@article{huang2019assessment,
  title={Assessment of turbulence models in a hypersonic cold-wall turbulent boundary layer},
  author={Huang, Junji and Bretzke, Jorge-Valentino and Duan, Lian},
  journal={Fluids},
  volume={4},
  number={1},
  pages={37},
  year={2019},
  publisher={MDPI}
}

@article{lagares2021turbulence,
  title={Turbulence modeling in hypersonic turbulent boundary layers subject to convex wall curvature},
  author={Lagares-Nieves, Christian J and Santiago, Jean and Araya, Guillermo},
  journal={AIAA Journal},
  volume={59},
  number={12},
  pages={4935--4954},
  year={2021},
  publisher={American Institute of Aeronautics and Astronautics}
}

@article{paciorri1998exploring,
  title={Exploring the validity of the Spalart-Allmaras turbulence model for hypersonic flows},
  author={Paciorri, Renato and Dieudonn{\'e}, Walter and Degrez, G{\'e}rard and Charbonnier, J-M and Deconinck, Herman},
  journal={Journal of Spacecraft and Rockets},
  volume={35},
  number={2},
  pages={121--126},
  year={1998}
}

@article{coratekin2004performance,
  title={Performance of upwind schemes and turbulence models in hypersonic flows},
  author={Coratekin, Turan and Van Keuk, J{\"o}rn and Ballmann, Josef},
  journal={AIAA journal},
  volume={42},
  number={5},
  pages={945--957},
  year={2004}
}

@article{chowdhary2022calibrating,
  title={Calibrating hypersonic turbulence flow models with the HIFiRE-1 experiment using data-driven machine-learned models},
  author={Chowdhary, Kenny and Hoang, Chi and Lee, Kookjin and Ray, Jaideep and Weirs, V Gregory and Carnes, Brian},
  journal={Computer Methods in Applied Mechanics and Engineering},
  volume={401},
  pages={115396},
  year={2022},
  publisher={Elsevier}
}

@inproceedings{yoon2007computational,
  title={Computational challenges in hypersonic flow simulations},
  author={Yoon, Seokkwan and Gnoffo, Peter and White, Jeffery and Thomas, James},
  booktitle={39th AIAA Thermophysics Conference},
  pages={4265},
  year={2007}
}

@article{sarma2000physico,
  title={Physico--chemical modelling in hypersonic flow simulation},
  author={Sarma, GSR},
  journal={Progress in Aerospace Sciences},
  volume={36},
  number={3-4},
  pages={281--349},
  year={2000},
  publisher={Elsevier}
}

@book{anderson1989hypersonic,
  title={Hypersonic and high temperature gas dynamics},
  author={Anderson, John David},
  year={1989},
  publisher={Aiaa}
}

@inproceedings{coakley1990assessment,
  title={An assessment and application of turbulence models for hypersonic flows},
  author={Coakley, TJ and Viegas, JR and Huang, PG and Rubesin, MW},
  booktitle={National Aero-Space Plane Technology Symposium},
  number={PAPER-106},
  year={1990}
}

@inproceedings{rubesin1989turbulence,
  title={Turbulence modeling for aerodynamic flows},
  author={RUBESIN, MORRIS},
  booktitle={27th Aerospace Sciences Meeting},
  pages={606},
  year={1989}
}

@article{candler2015advances,
  title={Advances in computational fluid dynamics methods for hypersonic flows},
  author={Candler, Graham V and Subbareddy, Pramod K and Brock, Joseph M},
  journal={Journal of Spacecraft and Rockets},
  volume={52},
  number={1},
  pages={17--28},
  year={2015},
  publisher={American Institute of Aeronautics and Astronautics}
}

@article{li2021bayesian,
  title={Bayesian model evaluation of three k--$\omega$ turbulence models for hypersonic shock wave--boundary layer interaction flows},
  author={Li, Jin-ping and Zeng, Fan-zhi and Chen, Shu-sheng and Zhang, Kai-ling and Yan, Chao},
  journal={Acta Astronautica},
  volume={189},
  pages={143--157},
  year={2021},
  publisher={Elsevier}
}

@inproceedings{candler2015next,
  title={Next-generation CFD for hypersonic and aerothermal flows},
  author={Candler, Graham V},
  booktitle={22nd AIAA Computational Fluid Dynamics Conference},
  pages={3048},
  year={2015}
}

@inproceedings{bardina1992two,
  title={Two-equation turbulence modeling for 3-d hypersonic flows},
  author={Bardina, J and Coakley, T and Marvin, J},
  booktitle={AlAA 4th International Aerospace Planes Conference},
  pages={5064},
  year={1992}
}

@article{nguyen2006modeling,
  title={Modeling and simulation of turbulent reacting flow around a hypersonic space probe},
  author={Nguyen-Bui, NT-H and Duffa, G},
  journal={Journal of spacecraft and rockets},
  volume={43},
  number={4},
  pages={919--923},
  year={2006}
}

@article{roy2019turbulent,
  title={Turbulent heat flux model for hypersonic shock--boundary layer interaction},
  author={Roy, Subhajit and Sinha, Krishnendu},
  journal={AIAA Journal},
  volume={57},
  number={8},
  pages={3624--3629},
  year={2019},
  publisher={American Institute of Aeronautics and Astronautics}
}

@inproceedings{erb2018investigation,
  title={Investigation of turbulence model uncertainty for supersonic/hypersonic shock wave-boundary layer interaction predictions},
  author={Erb, Aaron J and Hosder, Serhat},
  booktitle={22nd AIAA International Space Planes and Hypersonics Systems and Technologies Conference},
  pages={5195},
  year={2018}
}

@article{candler2019rate,
  title={Rate effects in hypersonic flows},
  author={Candler, Graham V},
  journal={Annual Review of Fluid Mechanics},
  volume={51},
  number={1},
  pages={379--402},
  year={2019},
  publisher={Annual Reviews}
}

@article{hu1995turbulent,
  title={Turbulent flow in supersonic and hypersonic nozzles},
  author={Hu, Jiasen and Rizzi, Arthur},
  journal={AIAA journal},
  volume={33},
  number={9},
  pages={1634--1640},
  year={1995}
}

@article{reddy2009hypersonic,
  title={Hypersonic turbulent flow simulation of Fire II reentry vehicle afterbody},
  author={Reddy, D Siva K and Sinha, Krishnendu},
  journal={Journal of Spacecraft and Rockets},
  volume={46},
  number={4},
  pages={745--757},
  year={2009}
}

@inproceedings{bardina1994three,
  title={Three-dimensional Navier-Stokes method with two-equation turbulence models for efficient numerical simulation of hypersonic flows},
  author={Bardina, J},
  booktitle={30th Joint Propulsion Conference and Exhibit},
  pages={2950},
  year={1994}
}

@inproceedings{nicholson2021simulation,
  title={Simulation and modeling of hypersonic turbulent boundary layers subject to adverse pressure gradients due to streamline curvature},
  author={Nicholson, Gary and Huang, Junji and Duan, Lian and Choudhari, Meelan M},
  booktitle={AIAA Aviation 2021 Forum},
  pages={2891},
  year={2021}
}

@article{yancheng2012evaluation,
  title={Evaluation of Turbulence Models in Predicting Hypersonic and Subsonic Base Flows Using Grid Adaptation Techniques},
  author={Yancheng, YOU and Buanga, Bj{\"o}rn and Hannemann, Volker and Luedeke, Heinrich},
  journal={Chinese Journal of Aeronautics},
  volume={25},
  number={3},
  pages={325--334},
  year={2012},
  publisher={Elsevier}
}


@inproceedings{huang2020simulation,
  title={Simulation and modeling of cold-wall hypersonic turbulent boundary layers on flat plate},
  author={Huang, Junji and Nicholson, Gary L and Duan, Lian and Choudhari, Meelan M and Bowersox, Rodney D},
  booktitle={AIAA Scitech 2020 Forum},
  pages={0571},
  year={2020}
}

@article{ng1989turbulence,
  title={Turbulence modeling in a hypersonic inlet},
  author={Ng, Wing-Fai and Ajmani, K and Taylor III, AC},
  journal={AIAA journal},
  volume={27},
  number={10},
  pages={1354--1360},
  year={1989}
}

@inproceedings{bardina1994three-1,
  title={Three-Dimensional Navier-Stokes Simulations with Two-Equation Turbulence Models of Intersecting Shock-Waves/Turbulent Boundary Layer at Mach 8.3},
  author={Bardina, Jorge E and Coakley, Thomas J},
  booktitle={Applied Aerodynamics Conference},
  number={NASA-CR-201979},
  year={1994}
}

@article{yentsch2014performance,
  title={Performance of turbulence modeling in simulation of the HIFiRE-1 flight test},
  author={Yentsch, Robert J and Gaitonde, Datta V and Kimmel, Roger},
  journal={Journal of Spacecraft and Rockets},
  volume={51},
  number={1},
  pages={117--127},
  year={2014},
  publisher={American Institute of Aeronautics and Astronautics}
}

@article{smith1996prediction,
  title={Prediction of hypersonic shock-wave/turbulent boundary-layer interactions},
  author={Smith, Brian R},
  journal={Journal of spacecraft and rockets},
  volume={33},
  number={5},
  pages={614--619},
  year={1996}
}

@inproceedings{jie2011stress,
  title={Stress limiter consideration for k-omega turbulence models in shock-wave/turbulent boundary-layer interactions in supersonic and hypersonic flows},
  author={Jie, Tan and Jie, Jin},
  booktitle={20th AIAA Computational Fluid Dynamics Conference},
  pages={3980},
  year={2011}
}

@inproceedings{georgiadis2014turbulence,
  title={Turbulence model effects on RANS simulations of the HIFiRE flight 2 ground test configurations},
  author={Georgiadis, Nicholas J and Mankbadi, Mina R and Vyas, Manan A},
  booktitle={52nd Aerospace Sciences Meeting},
  pages={0624},
  year={2014}
}

@article{zhang2010turbulence,
  title={Turbulence models for accurate aerothermal prediction in hypersonic flows},
  author={Zhang, Xiang-Hong and Wu, Yi-Zao and Wang, Jiang-Feng},
  journal={Modern Physics Letters B},
  volume={24},
  number={13},
  pages={1345--1348},
  year={2010},
  publisher={World Scientific}
}

@inproceedings{aiken2022assessment,
  title={Assessment of Reynolds Averaged Navier-Stokes models for a hypersonic cold-wall turbulent boundary layer},
  author={Aiken, Timothy T and Boyd, Iain D and Duan, Lian and Huang, Junji},
  booktitle={AIAA SciTech 2022 Forum},
  pages={0586},
  year={2022}
}

@article{kral1998recent,
  title={Recent experience with different turbulence models applied to the calculation of flow over aircraft components},
  author={Kral, LD},
  journal={Progress in Aerospace Sciences},
  volume={34},
  number={7-8},
  pages={481--541},
  year={1998},
  publisher={Elsevier}
}

@article{danis2022compressibility,
  title={Compressibility correction to k- $\omega$ models for hypersonic turbulent boundary layers},
  author={Danis, Mustafa E and Durbin, Paul},
  journal={AIAA Journal},
  volume={60},
  number={11},
  pages={6225--6234},
  year={2022},
  publisher={American Institute of Aeronautics and Astronautics}
}

@incollection{huang1993calculations,
  title={Calculations of supersonic and hypersonic flows using compressible wall functions},
  author={Huang, PG and Coakley, TJ},
  booktitle={Engineering Turbulence Modelling and Experiments},
  pages={731--739},
  year={1993},
  publisher={Elsevier}
}

@article{knight2017assessment,
  title={Assessment of predictive capabilities for aerodynamic heating in hypersonic flow},
  author={Knight, Doyle and Chazot, Olivier and Austin, Joanna and Badr, Mohammad Ali and Candler, Graham and Celik, Bayram and de Rosa, Donato and Donelli, Raffaele and Komives, Jeffrey and Lani, Andrea and others},
  journal={Progress in Aerospace Sciences},
  volume={90},
  pages={39--53},
  year={2017},
  publisher={Elsevier}
}

@article{xiao2007role,
  title={Role of turbulent Prandtl numbers on heat flux at hypersonic Mach numbers},
  author={Xiao, Xudong and Hassan, HA and Edwards, JR and Gaffney Jr, RL},
  journal={AIAA journal},
  volume={45},
  number={4},
  pages={806--813},
  year={2007}
}

@article{hoskin2024discontinuous,
  title={Discontinuous Galerkin methods for hypersonic flows},
  author={Hoskin, Dominique S and Van Heyningen, R Loek and Nguyen, Ngoc Cuong and Vila-P{\'e}rez, Jordi and Harris, Wesley L and Peraire, Jaime},
  journal={Progress in Aerospace Sciences},
  volume={146},
  pages={100999},
  year={2024},
  publisher={Elsevier}
}

@inproceedings{holden2006historical,
  title={Historical review of experimental studies and prediction methods to describe laminar and turbulent shock wave/boundary layer interactions in hypersonic flows},
  author={Holden, Michael},
  booktitle={44th AIAA Aerospace Sciences Meeting and Exhibit},
  pages={494},
  year={2006}
}

@book{babinsky2011shock,
  title={Shock wave-boundary-layer interactions},
  author={Babinsky, Holger and Harvey, John K},
  volume={32},
  year={2011},
  publisher={Cambridge University Press}
}

@article{gaitonde2015progress,
  title={Progress in shock wave/boundary layer interactions},
  author={Gaitonde, Datta V},
  journal={Progress in Aerospace Sciences},
  volume={72},
  pages={80--99},
  year={2015},
  publisher={Elsevier}
}

@article{gaitonde2023dynamics,
  title={Dynamics of three-dimensional shock-wave/boundary-layer interactions},
  author={Gaitonde, Datta V and Adler, Michael C},
  journal={Annual Review of Fluid Mechanics},
  volume={55},
  number={1},
  pages={291--321},
  year={2023},
  publisher={Annual Reviews}
}

@article{shyy1997compressibility,
  title={Compressibility effects in modeling complex turbulent flows},
  author={Shyy, Wei and Krishnamurty, Venkata S},
  journal={Progress in aerospace sciences},
  volume={33},
  number={9-10},
  pages={587--645},
  year={1997},
  publisher={Elsevier}
}

@book{gatski2013compressibility,
  title={Compressibility, turbulence and high speed flow},
  author={Gatski, Thomas B and Bonnet, Jean-Paul},
  year={2013},
  publisher={Academic Press}
}

@article{catris2000density,
  title={Density corrections for turbulence models},
  author={Catris, St{\'e}phane and Aupoix, Bertrand},
  journal={Aerospace Science and Technology},
  volume={4},
  number={1},
  pages={1--11},
  year={2000},
  publisher={Elsevier}
}

@article{dai2022influence,
  title={Influence of high temperature non-equilibrium effects on Mach 12 scramjet inlet},
  author={Dai, Chunliang and Sun, Bo and Zhou, Shengbing and Zhuo, Changfei and Zhou, Changsheng and Yue, Lianjie},
  journal={Acta Astronautica},
  volume={193},
  pages={237--254},
  year={2022},
  publisher={Elsevier}
}

@article{barth2015effects,
  title={Effects of hydrogen fuel injection in a Mach 12 scramjet inlet},
  author={Barth, James E and Wheatley, Vincent and Smart, Michael K},
  journal={AIAA Journal},
  volume={53},
  number={10},
  pages={2907--2919},
  year={2015},
  publisher={American Institute of Aeronautics and Astronautics}
}

@article{dai2022numerical,
  title={Numerical study of high temperature non-equilibrium effects of double-wedge in hypervelocity flow},
  author={Dai, Chunliang and Sun, Bo and Zhuo, Changfei and Zhou, Shengbing and Zhou, Changsheng and Yue, Lianjie},
  journal={Aerospace Science and Technology},
  volume={124},
  pages={107526},
  year={2022},
  publisher={Elsevier}
}

@article{sciacovelli2023priori,
  title={A priori tests of turbulence models for compressible flows},
  author={Sciacovelli, Luca and Cannici, Aron and Passiatore, Donatella and Cinnella, Paola},
  journal={International Journal of Numerical Methods for Heat \& Fluid Flow},
  year={2023},
  publisher={Emerald Publishing Limited}
}

@article{chen2024improved,
  title={An improved Baldwin--Lomax algebraic wall model for high-speed canonical turbulent boundary layers using established scalings},
  author={Chen, Xianliang and Gan, Jianping and Fu, Lin},
  journal={Journal of Fluid Mechanics},
  volume={987},
  pages={A7},
  year={2024},
  publisher={Cambridge University Press}
}

@article{wilcox1992dilatation,
  title={Dilatation-dissipation corrections for advanced turbulence models},
  author={Wilcox, David C},
  journal={AIAA journal},
  volume={30},
  number={11},
  pages={2639--2646},
  year={1992}
}

@article{urzay2018supersonic,
  title={Supersonic combustion in air-breathing propulsion systems for hypersonic flight},
  author={Urzay, Javier},
  journal={Annual Review of Fluid Mechanics},
  volume={50},
  number={1},
  pages={593--627},
  year={2018},
  publisher={Annual Reviews}
}

@article{urzay2021engineering,
  title={Engineering aspects of hypersonic turbulent flows at suborbital enthalpies},
  author={Urzay, J. and Di Renzo, M.},
  journal={Annual Research Briefs, Center for Turbulence Research},
  pages={7--32},
  year={2021}
}

@article{sarma2000physico,
  title={Physico--chemical modelling in hypersonic flow simulation},
  author={Sarma, GSR},
  journal={Progress in Aerospace Sciences},
  volume={36},
  number={3-4},
  pages={281--349},
  year={2000},
  publisher={Elsevier}
}

@inproceedings{molchanov2013three,
  title={Three-equation k-$\varepsilon$-Vn turbulence model for high-speed flows},
  author={Molchanov, Alexander M},
  booktitle={43rd AIAA Fluid Dynamics Conference},
  pages={3181},
  year={2013}
}

@article{duraisamy2019turbulence,
  title={Turbulence modeling in the age of data},
  author={Duraisamy, Karthik and Iaccarino, Gianluca and Xiao, Heng},
  journal={Annual review of fluid mechanics},
  volume={51},
  number={1},
  pages={357--377},
  year={2019},
  publisher={Annual Reviews}
}

@book{wilcox1998turbulence,
  title={Turbulence modeling for CFD},
  author={Wilcox, David C and others},
  volume={2},
  year={1998},
  publisher={DCW industries La Canada, CA}
}

@article{dolling2001fifty,
  title={Fifty years of shock-wave/boundary-layer interaction research: what next?},
  author={Dolling, David S},
  journal={AIAA journal},
  volume={39},
  number={8},
  pages={1517--1531},
  year={2001}
}

@inproceedings{zhang2015turbulence,
  title={Turbulence model modification for fake amplification of Turbulence Kinetic Energy by Shock-Turbulence Interation},
  author={Zhang, Zhichao and Gao, Zhenxun and Jiang, Chongwen and Lee, Chunhian},
  booktitle={51st AIAA/SAE/ASEE Joint Propulsion Conference},
  pages={4205},
  year={2015}
}

@article{lee1997interaction,
  title={Interaction of isotropic turbulence with shock waves: effect of shock strength},
  author={Lee, Sangsan and Lele, Sanjiva K and Moin, Parviz},
  journal={Journal of Fluid Mechanics},
  volume={340},
  pages={225--247},
  year={1997},
  publisher={Cambridge University Press}
}

@conference{KaWi95,
	author = {M Kandula and D Wilcox},
	booktitle = {Fluid Dynamics Conference},
	doi = {10.2514/6.1995-2317},
	eprint = {https://arc.aiaa.org/doi/pdf/10.2514/6.1995-2317},
	title = {An examination of k-omega turbulence model for boundary layers, free shear layers and separated flows},
    year = {1995},
	url = {https://arc.aiaa.org/doi/abs/10.2514/6.1995-2317},
	bdsk-url-1 = {https://arc.aiaa.org/doi/abs/10.2514/6.1995-2317},
	bdsk-url-2 = {https://doi.org/10.2514/6.1995-2317}}


@article{menter1994two,
  title={Two-equation eddy-viscosity turbulence models for engineering applications},
  author={Menter, Florian R},
  journal={AIAA journal},
  volume={32},
  number={8},
  pages={1598--1605},
  year={1994}
}

@incollection{steelant2002effect,
  title={Effect of a compressibility correction on different turbulence models},
  author={Steelant, Johan},
  booktitle={Engineering Turbulence Modelling and Experiments 5},
  pages={207--216},
  year={2002},
  publisher={Elsevier}
}

@article{huang1994turbulence,
  title={Turbulence models for compressible boundary layers},
  author={Huang, PG and Bradshaw, P and Coakley, TJ},
  journal={AIAA journal},
  volume={32},
  number={4},
  pages={735--740},
  year={1994}
}

@article{duan2011direct,
  title={Direct numerical simulation of hypersonic turbulent boundary layers. Part 3. Effect of Mach number},
  author={Duan, Lian and Beekman, I and Martin, MP},
  journal={Journal of Fluid Mechanics},
  volume={672},
  pages={245--267},
  year={2011},
  publisher={Cambridge University Press}
}

@article{rose1973ratio,
  title={Ratio of Reynolds shear stress to turbulence kinetic energy in a boundary layer},
  author={Rose, William C and Murphy, John D},
  journal={Physics of Fluids},
  volume={16},
  number={6},
  pages={935--937},
  year={1973}
}

@article{pope1975more,
  title={A more general effective-viscosity hypothesis},
  author={Pope, Stephen B},
  journal={Journal of Fluid Mechanics},
  volume={72},
  number={2},
  pages={331--340},
  year={1975},
  publisher={Cambridge University Press}
}

@article{raje2016physically,
  title={A physically consistent and numerically robust k-$\epsilon$ model for computing turbulent flows with shock waves},
  author={Raje, Pratikkumar and Sinha, Krishnendu},
  journal={Computers \& Fluids},
  volume={136},
  pages={35--47},
  year={2016},
  publisher={Elsevier}
}

@inproceedings{rodi1976new,
  title={A new algebraic relation for calculating the Reynolds stresses},
  author={Rodi, W},
  booktitle={Gesellschaft Angewandte Mathematik und Mechanik Workshop Paris France},
  volume={56},
  pages={219},
  year={1976}
}

@article{craft1996development,
  title={Development and application of a cubic eddy-viscosity model of turbulence},
  author={Craft, TJ and Launder, BE and Suga, K},
  journal={International Journal of Heat and Fluid Flow},
  volume={17},
  number={2},
  pages={108--115},
  year={1996},
  publisher={Elsevier}
}

@article{gatski2000nonlinear,
  title={Nonlinear eddy viscosity and algebraic stress models for solving complex turbulent flows},
  author={Gatski, TB and Jongen, T},
  journal={Progress in Aerospace Sciences},
  volume={36},
  number={8},
  pages={655--682},
  year={2000},
  publisher={Elsevier}
}

@inproceedings{lindblad1998prediction,
  title={A prediction method for high speed turbulent separated flows with experimental verification},
  author={Lindblad, I and Wallin, S and Johansson, A and Friedrich, R and Lechner, R and Krogmann, P and Schuelein, E and Courty, J-C and Ravachol, M and Giordano, D},
  booktitle={29th AIAA, Fluid Dynamics Conference},
  pages={2547},
  year={1998}
}

@article{wallin2000explicit,
  title={An explicit algebraic Reynolds stress model for incompressible and compressible turbulent flows},
  author={Wallin, Stefan and Johansson, Arne V},
  journal={Journal of fluid mechanics},
  volume={403},
  pages={89--132},
  year={2000},
  publisher={Cambridge University Press}
}

@article{vemula2020explicit,
  title={Explicit algebraic Reynolds stress model for shock-dominated flows},
  author={Vemula, Jagadish Babu and Sinha, Krishnendu},
  journal={International Journal of Heat and Fluid Flow},
  volume={85},
  pages={108680},
  year={2020},
  publisher={Elsevier}
}
 @article{Vemula_Sinha_2017, title={Reynolds stress models applied to canonical shock-turbulence interaction}, volume={18}, ISSN={1468-5248}, DOI={10.1080/14685248.2017.1317923}, number={7}, journal={Journal of Turbulence}, author={Vemula, Jagadish Babu and Sinha, Krishnendu}, year={2017}, month=jul, pages={653–687}, language={en} }


@article{gatski1993explicit,
  title={On explicit algebraic stress models for complex turbulent flows},
  author={Gatski, Thomas B and Speziale, Charles G},
  journal={Journal of fluid Mechanics},
  volume={254},
  pages={59--78},
  year={1993},
  publisher={Cambridge University Press}
}

@article{launder1975progress,
  title={Progress in the development of a Reynolds-stress turbulence closure},
  author={Launder, Brian Edward and Reece, G Jr and Rodi, W},
  journal={Journal of fluid mechanics},
  volume={68},
  number={3},
  pages={537--566},
  year={1975},
  publisher={Cambridge University Press}
}

@article{rotta1951statistische,
  title={Statistische theorie nichthomogener turbulenz},
  author={Rotta, JC},
  journal={Zeitschrift f{\"u}r Physik},
  volume={129},
  pages={547--572},
  year={1951},
  publisher={Springer}
}

@article{jongen1998general,
  title={General explicit algebraic stress relations and best approximation for three-dimensional flows},
  author={Jongen, T and Gatski, TB},
  journal={International Journal of Engineering Science},
  volume={36},
  number={7-8},
  pages={739--763},
  year={1998},
  publisher={Elsevier}
}

@book{boussinesq1877essai,
  title={Essai sur la th{\'e}orie des eaux courantes},
  author={Boussinesq, Jean},
  year={1877},
  publisher={Imprimerie nationale}
}


@article{prandtl1925,
author = {Prandtl, L.},
title = {7. Bericht über Untersuchungen zur ausgebildeten Turbulenz},
journal = {ZAMM - Journal of Applied Mathematics and Mechanics / Zeitschrift für Angewandte Mathematik und Mechanik},
volume = {5},
number = {2},
pages = {136-139},
doi = {https://doi.org/10.1002/zamm.19250050212},
url = {https://onlinelibrary.wiley.com/doi/abs/10.1002/zamm.19250050212},
eprint = {https://onlinelibrary.wiley.com/doi/pdf/10.1002/zamm.19250050212},
year = {1925}
}

@article{chou1945velocity,
  title={On velocity correlations and the solutions of the equations of turbulent fluctuation},
  author={Chou, Pei-Yuan},
  journal={Quarterly of applied mathematics},
  volume={3},
  number={1},
  pages={38--54},
  year={1945}
}

@inproceedings{kolmogorov1941equations,
  title={Equations of turbulent motion in an incompressible fluid},
  author={Kolmogorov, Andrej Nikolaevich},
  booktitle={Dokl. Akad. Nauk SSSR},
  volume={30},
  pages={299--303},
  year={1941}
}

@article{sarkar1991analysis,
  title={The analysis and modelling of dilatational terms in compressible turbulence},
  author={Sarkar, Sutanu and Erlebacher, Gordon and Hussaini, M Yousuff and Kreiss, Heinz Otto},
  journal={Journal of Fluid Mechanics},
  volume={227},
  pages={473--493},
  year={1991},
  publisher={Cambridge University Press}
}

@article{settles1994hypersonic,
  title={Hypersonic shock/boundary-layer interaction database: new and corrected data},
  author={Settles, Gary S and Dodson, Lori J},
  journal={Pennsylvania State Univ. Report},
  year={1994}
}

@article{zhang2018direct,
  title={Direct numerical simulation database for supersonic and hypersonic turbulent boundary layers},
  author={Zhang, Chao and Duan, Lian and Choudhari, Meelan M},
  journal={AIAA journal},
  volume={56},
  number={11},
  pages={4297--4311},
  year={2018},
  publisher={American Institute of Aeronautics and Astronautics}
}

@article{murray2013experimental,
  title={Experimental investigation of axisymmetric hypersonic shock-wave/turbulent-boundary-layer interactions},
  author={Murray, N and Hillier, R and Williams, S},
  journal={Journal of Fluid Mechanics},
  volume={714},
  pages={152--189},
  year={2013},
  publisher={Cambridge University Press}
}

@phdthesis{williams2005three,
  title={Three-dimensional separation of a hypersonic boundary layer},
  author={Williams, Simon},
  year={2005},
  school={Imperial College London}
}

@phdthesis{murray2007three,
  title={Three-dimensional turbulent shock-wave: boundary-layer interactions in hypersonic flows},
  author={Murray, Neil Paul},
  year={2007},
  school={Imperial College London (University of London)}
}

@phdthesis{coleman1973hypersonic,
  title={Hypersonic turbulent boundary layer studies},
  author={Coleman, Graham Trevor},
  year={1973},
  school={Imperial College London}
}

@article{coleman1972heat,
  title={Heat transfer from hypersonic turbulent flow at a wedge compression corner},
  author={Coleman, GT and Stollery, JL},
  journal={Journal of Fluid Mechanics},
  volume={56},
  number={4},
  pages={741--752},
  year={1972},
  publisher={Cambridge University Press}
}

@article{elfstrom1972turbulent,
  title={Turbulent hypersonic flow at a wedge-compression corner},
  author={Elfstrom, GM},
  journal={Journal of fluid Mechanics},
  volume={53},
  number={1},
  pages={113--127},
  year={1972},
  publisher={Cambridge University Press}
}

@article{kussoy1991documentation,
  title={Documentation of two-and three-dimensional shock-wave/turbulent-boundary-layer interaction flows at Mach 8.2},
  author={Kussoy, Marvin I and Horstman, KC},
  journal={NASA Ames Research Center Technical Report},
  year={1991}
}
@techreport{parish2024report,
  title={{Data-driven closure modeling for
hypersonic turbulent flows}},
  author={Parish, Eric and Barone, Matthew and Ching, David and Jordan, Cyrus and Miller, Nathan and Nicholson, Gary and Gitushi, Kevin and Beresh, Steven and Gupta, Niloy and Duraisamy, Karthik},
  year={2024},
  institution={Sandia National Laboratories}
}
@techreport{holden1988studies,
  title={Studies of the structure of attached and separated regions of viscous/inviscid interaction and the effects of combined surface roughness and blowing in high-Reynolds-number hypersonic flows. Final report, 1 August 1985-1 June 1988},
  author={Holden, MS and Bergman, R and Harvey, J and Duryea, G and Moselle, J},
  year={1988},
  institution={Calspan UB Research Center, Buffalo, NY (USA)}
}

@article{holden1986shock,
  title={Shock wave/turbulent boundary layer interaction in high-Reynolds number hypersonic flows},
  author={Holden, MS and Havener, AG and Lee, CH},
  journal={Calspan-University of Buffalo Research Centre, CUBRC-86681},
  year={1986}
}

@inproceedings{holden1991studies,
  title={Studies of the mean and unsteady structure of turbulent boundary layer separation in hypersonic flow},
  author={Holden, M},
  booktitle={22nd Fluid Dynamics, Plasma Dynamics and Lasers Conference},
  pages={1778},
  year={1991}
}

@book{kussoy1989documentation,
  title={Documentation of two-and three-dimensional hypersonic shock wave/turbulent boundary layer interaction flows},
  author={Kussoy, Marvin I},
  volume={101075},
  year={1989},
  publisher={National Aeronautics and Space Administration, Ames Research Center}
}

@article{volpiani2020effects,
  title={Effects of a nonadiabatic wall on hypersonic shock/boundary-layer interactions},
  author={Volpiani, Pedro S and Bernardini, Matteo and Larsson, Johan},
  journal={Physical Review Fluids},
  volume={5},
  number={1},
  pages={014602},
  year={2020},
  publisher={APS}
}

@inproceedings{volpiani2019using,
  title={Using large-eddy simulations to design a new hypersonic shock/boundary-layer interaction experiment},
  author={Volpiani, Pedro S and Wagner, Alexander and Bernardini, Matteo and Larsson, Johan},
  booktitle={AIAA Scitech 2019 Forum},
  pages={0098},
  year={2019}
}

@article{duan2011direct4,
  title={Direct numerical simulation of hypersonic turbulent boundary layers. Part 4. Effect of high enthalpy},
  author={Duan, Lian and Martin, MP},
  journal={Journal of Fluid Mechanics},
  volume={684},
  pages={25--59},
  year={2011},
  publisher={Cambridge University Press}
}


@article{morkovin1962effects,
  title={Effects of compressibility on turbulent flows},
  author={Morkovin, Mark V},
  journal={M{\'e}canique de la Turbulence},
  volume={367},
  number={380},
  pages={26},
  year={1962},
  publisher={CNRS Paris}
}

@article{passiatore2021finite,
  title={Finite-rate chemistry effects in turbulent hypersonic boundary layers: A direct numerical simulation study},
  author={Passiatore, Donatella and Sciacovelli, Luca and Cinnella, Paola and Pascazio, Giuseppe},
  journal={Physical Review Fluids},
  volume={6},
  number={5},
  pages={054604},
  year={2021},
  publisher={APS}
}


@article{zeman1990dilatation,
  title={Dilatation dissipation: the concept and application in modeling compressible mixing layers},
  author={Zeman, Otto},
  journal={Physics of Fluids A: Fluid Dynamics},
  volume={2},
  number={2},
  pages={178--188},
  year={1990},
  publisher={American Institute of Physics}
}

@phdthesis{kim2016non,
  title={Non-equilibrium effects on hypersonic turbulent boundary layers},
  author={Kim, Pilbum},
  year={2016},
  school={University of California, Los Angeles}
}

@inproceedings{baurle2004modeling,
  title={Modeling of high speed reacting flows: established practices and future challenges},
  author={Baurle, Robert},
  booktitle={42nd AIAA aerospace sciences meeting and exhibit},
  pages={267},
  year={2004}
}

@inproceedings{ebrahimi2004overview,
  title={An overview of computational fluid dynamics for application to advanced propulsion systems},
  author={Ebrahimi, Houshang},
  booktitle={37th AIAA Thermophysics Conference},
  pages={2370},
  year={2004}
}

@article{dang2022direct,
  title={Direct numerical simulation of compressible turbulence accelerated by graphics processing unit: An open-source high accuracy accelerated computational fluid dynamic software},
  author={Dang, Guanlin and Liu, Shiwei and Guo, Tongbiao and Duan, Junyi and Li, Xinliang},
  journal={Physics of Fluids},
  volume={34},
  number={12},
  year={2022},
  publisher={AIP Publishing}
}

@article{priebe2021turbulence,
  title={Turbulence in a hypersonic compression ramp flow},
  author={Priebe, Stephan and Mart{\'\i}n, M Pino},
  journal={Physical Review Fluids},
  volume={6},
  number={3},
  pages={034601},
  year={2021},
  publisher={APS}
}

@article{yu2023coherent,
  title={Coherent structures and turbulent model refinement in oblique shock/hypersonic turbulent boundary layer interactions},
  author={Yu, Ming and Sun, Dong and Zhou, QingQing and Liu, PengXin and Yuan, XianXu},
  journal={Physics of Fluids},
  volume={35},
  number={8},
  year={2023},
  publisher={AIP Publishing}
}


@article{helm2022large,
  title={Large eddy simulation of two separated hypersonic shock/turbulent boundary layer interactions},
  author={Helm, Clara M and Mart{\'\i}n, MP},
  journal={Physical Review Fluids},
  volume={7},
  number={7},
  pages={074601},
  year={2022},
  publisher={APS}
}

@article{zuo2023hypersonic,
  title={Hypersonic shock wave/turbulent boundary layer interaction over a compression ramp},
  author={Zuo, Feng-Yuan},
  journal={AIAA Journal},
  volume={61},
  number={4},
  pages={1579--1595},
  year={2023},
  publisher={American Institute of Aeronautics and Astronautics}
}

@article{zhang2022direct,
  title={Direct numerical simulation of shock wave/turbulent boundary layer interaction in a swept compression ramp at Mach 6},
  author={Zhang, Ji and Guo, Tongbiao and Dang, Guanlin and Li, Xinliang},
  journal={Physics of Fluids},
  volume={34},
  number={11},
  year={2022},
  publisher={AIP Publishing}
}

@article{guo2023amplification,
  title={Amplification of turbulent kinetic energy and temperature fluctuation in a hypersonic turbulent boundary layer over a compression ramp},
  author={Guo, Tongbiao and Zhang, Ji and Tong, Fulin and Li, Xinliang},
  journal={Physics of Fluids},
  volume={35},
  number={4},
  year={2023},
  publisher={AIP Publishing}
}

@article{chen2022wall,
  title={LES wall modeling for heat transfer at high speeds},
  author={Chen, Peng ES and Lv, Yu and Xu, Haosen HA and Shi, Yipeng and Yang, Xiang IA},
  journal={Physical Review Fluids},
  volume={7},
  number={1},
  pages={014608},
  year={2022},
  publisher={APS}
}

@article{fulin2023hypersonic,
  title={Hypersonic shock wave and turbulent boundary layer interaction in a sharp cone/flare model},
  author={Fulin, TONG and Junyi, DUAN and Jiang, LAI and Dong, SUN and Xianxu, YUAN},
  journal={Chinese Journal of Aeronautics},
  volume={36},
  number={3},
  pages={80--95},
  year={2023},
  publisher={Elsevier}
}

@inproceedings{sebastian2022influence,
  title={Influence of crossflow Mach number on spanwise-inclined jet injection},
  author={Sebastian, Robin and Schreyer, Anne-Marie},
  booktitle={AIAA Scitech 2022 Forum},
  pages={1363},
  year={2022}
}

@article{qi2023large,
  title={Large-eddy simulation of a hypersonic turbulent boundary layer over a compression corner},
  author={Qi, Han and Li, Xinliang and Ji, Xiangxin and Tong, Fulin and Yu, Changping},
  journal={AIP Advances},
  volume={13},
  number={2},
  year={2023},
  publisher={AIP Publishing}
}

@article{guo2024wall,
  title={Wall skin friction analysis in a hypersonic turbulent boundary layer over a compression ramp},
  author={Guo, Tongbiao and Zhang, Ji and Zhu, Yanhua and Li, Xinliang},
  journal={Journal of Fluid Mechanics},
  volume={988},
  pages={A23},
  year={2024},
  publisher={Cambridge University Press}
}

@article{di2024stagnation,
  title={Stagnation enthalpy effects on hypersonic turbulent compression corner flow at moderate Reynolds numbers},
  author={Di Renzo, M and Williams, CT and Pirozzoli, S},
  journal={Physical Review Fluids},
  volume={9},
  number={3},
  pages={033401},
  year={2024},
  publisher={APS}
}

@inproceedings{bhagwandin2021shock,
  title={LES of Shock-Turbulent Boundary Layer Interaction over a Mach 10 Hollow Cylinder with Flare.},
  author={Bhagwandin, Vishal A and Martin, Pino},
  booktitle={AIAA AVIATION 2021 FORUM},
  pages={2820},
  year={2021}
}

@article{ninni2023simulation,
  title={Simulation of High-Enthalpy Turbulent Shock Wave/Boundary Layer Interaction Using a RANS Approach},
  author={Ninni, Davide and Bonelli, Francesco and Pascazio, Giuseppe},
  journal={Aerotecnica Missili \& Spazio},
  volume={102},
  number={4},
  pages={323--335},
  year={2023},
  publisher={Springer}
}

@article{helm2021scaling,
  title={Scaling of hypersonic shock/turbulent boundary layer interactions},
  author={Helm, Clara M and Mart{\'\i}n, MP},
  journal={Physical Review Fluids},
  volume={6},
  number={7},
  pages={074607},
  year={2021},
  publisher={APS}
}

@article{helm2021characterization,
  title={Characterization of the shear layer in separated shock/turbulent boundary layer interactions},
  author={Helm, Clara M and Mart{\'\i}n, M Pino and Williams, Owen JH},
  journal={Journal of Fluid Mechanics},
  volume={912},
  pages={A7},
  year={2021},
  publisher={Cambridge University Press}
}

@inproceedings{muir2024assessment,
  title={Assessment of Reynolds-Averaged Navier-Stokes Simulations for Predicting Scaling Trends in Shock-Separated Boundary Layers},
  author={Muir, David P and Martin, Pino},
  booktitle={AIAA AVIATION FORUM AND ASCEND 2024},
  pages={3550},
  year={2024}
}

@inproceedings{bhagwandin2023wall,
  title={Wall-Resolved LES of Mach 6 BoLT-2 Hypersonic Vehicle},
  author={Bhagwandin, Vishal A and Martin, Pino},
  booktitle={AIAA AVIATION 2023 Forum},
  pages={3848},
  year={2023}
}

@article{grube2021reynolds,
  title={Reynolds stress anisotropy in shock/isotropic turbulence interactions},
  author={Grube, Nathan E and Mart{\'\i}n, M Pino},
  journal={Journal of Fluid Mechanics},
  volume={913},
  pages={A19},
  year={2021},
  publisher={Cambridge University Press}
}

@article{huete2021thermochemical,
  title={Thermochemical effects on hypersonic shock waves interacting with weak turbulence},
  author={Huete, C and Cuadra, A and Vera, M and Urzay, J},
  journal={Physics of Fluids},
  volume={33},
  number={8},
  year={2021},
  publisher={AIP Publishing}
}

@inproceedings{bhagwandin2019shock,
  title={Shock-Turbulent Boundary Layer Interactions in Separated Compression Corners at Mach 10.},
  author={Bhagwandin, Vishal A and Helm, Clara M and Martin, Pino},
  booktitle={AIAA Scitech 2019 Forum},
  pages={1129},
  year={2019}
}

@misc{yuting2021scaling,
  title={Scaling of interaction lengths for hypersonic shock wave/turbulent boundary layer interactions},
  author={Yuting, HONG and Zhufei, LI and Jiming, YANG},
  journal={Chinese Journal of Aeronautics},
  volume={34},
  number={5},
  pages={504--509},
  year={2021},
  publisher={Elsevier}
}

@article{rathi2024simulation,
  title={Simulation of Hypersonic Shock-Boundary Layer Interaction Using Shock-Strength Dependent Turbulence Model},
  author={Rathi, Harsha and Sinha, Krishnendu},
  journal={AIAA Journal},
  pages={1--14},
  year={2024},
  publisher={American Institute of Aeronautics and Astronautics}
}

@inproceedings{brooks2017mach,
  title={Mach 10 PIV flow field measurements of a turbulent boundary layer and shock turbulent boundary layer interaction},
  author={Brooks, Jonathan and Gupta, Ashwani and Marineau, Eric C and Smith, Michael},
  booktitle={33rd AIAA aerodynamic measurement technology and ground testing conference},
  pages={3325},
  year={2017}
}

@inproceedings{helm2016new,
  title={New LES of a hypersonic shock/turbulent boundary layer interaction},
  author={Helm, Clara M and Martin, M Pino},
  booktitle={54th AIAA Aerospace Sciences Meeting},
  pages={0346},
  year={2016}
}

@article{whalen2020hypersonic,
  title={Hypersonic fluid--structure interactions in compression corner shock-wave/boundary-layer interaction},
  author={Whalen, Thomas J and Sch{\"o}neich, Antonio Giovanni and Laurence, Stuart J and Sullivan, Bryson T and Bodony, Daniel J and Freydin, Maxim and Dowell, Earl H and Buck, Gregory M},
  journal={AIAA journal},
  volume={58},
  number={9},
  pages={4090--4105},
  year={2020},
  publisher={American Institute of Aeronautics and Astronautics}
}

@inproceedings{whalen2019unsteady,
  title={Unsteady surface and flowfield measurements in ramp-induced turbulent and transitional shock-wave boundary-layer interactions at Mach 6},
  author={Whalen, Thomas J and Kennedy, Richard E and Laurence, Stuart J and Sullivan, Bryson and Bodony, Daniel J and Buck, Gregory},
  booktitle={AIAA Scitech 2019 Forum},
  pages={1127},
  year={2019}
}

@article{martin2007direct,
  title={Direct numerical simulation of hypersonic turbulent boundary layers. Part 1. Initialization and comparison with experiments},
  author={Martin, M Pino},
  journal={Journal of Fluid Mechanics},
  volume={570},
  pages={347--364},
  year={2007},
  publisher={Cambridge University Press}
}

@inproceedings{bookey2005new,
  title={New experimental data of STBLI at DNS/LES accessible Reynolds numbers},
  author={Bookey, Patrick and Wyckham, Christopher and Smits, Alexander and Martin, Pino},
  booktitle={43rd AIAA Aerospace Sciences Meeting and Exhibit},
  pages={309},
  year={2005}
}

@article{duan2016pressure,
  title={Pressure fluctuations induced by a hypersonic turbulent boundary layer},
  author={Duan, Lian and Choudhari, Meelan M and Zhang, Chao},
  journal={Journal of Fluid Mechanics},
  volume={804},
  pages={578--607},
  year={2016},
  publisher={Cambridge University Press}
}

@article{martin1998effect,
  title={Effect of chemical reactions on decaying isotropic turbulence},
  author={Mart{\i}n, M Pino and Candler, Graham V},
  journal={Physics of Fluids},
  volume={10},
  number={7},
  pages={1715--1724},
  year={1998},
  publisher={American Institute of Physics}
}

@article{duan2011assessment,
  title={Assessment of turbulence-chemistry interaction in hypersonic turbulent boundary layers},
  author={Duan, Lian and Mart{\'\i}n, M Pino},
  journal={AIAA journal},
  volume={49},
  number={1},
  pages={172--184},
  year={2011}
}

@article{reynolds1895iv,
  title={IV. On the dynamical theory of incompressible viscous fluids and the determination of the criterion},
  author={Reynolds, Osborne},
  journal={Philosophical transactions of the royal society of london.(a.)},
  number={186},
  pages={123--164},
  year={1895},
  publisher={The Royal Society London}
}

@article{favre1965equations,
  title={Equations des gaz turbulents compressibles},
  author={Favre, A},
  journal={J. de Mecanique},
  volume={4},
  number={3},
  pages={361},
  year={1965}
}


@book{prandtl1945neues,
  title={{\"U}ber ein neues Formelsystem f{\"u}r die ausgebildete Turbulenz},
  author={Prandtl, Ludwig},
  year={1945},
  publisher={Vandenhoeck \& Ruprecht}
}


@article{zhang2022review,
  title={A review of the mathematical modeling of equilibrium and nonequilibrium hypersonic flows},
  author={Zhang, Wenqing and Zhang, Zhijun and Wang, Xiaowei and Su, Tianyi},
  journal={Advances in Aerodynamics},
  volume={4},
  number={1},
  pages={38},
  year={2022},
  publisher={Springer}
}

@article{guohua2012assessment,
  title={Assessment of two turbulence models and some compressibility corrections for hypersonic compression corners by high-order difference schemes},
  author={Guohua, TU and Xiaogang, DENG and Meiliang, MAO},
  journal={Chinese Journal of Aeronautics},
  volume={25},
  number={1},
  pages={25--32},
  year={2012},
  publisher={Elsevier}
}

@article{zhu2020analysis,
  title={Analysis of compressibility corrections for turbulence models in hypersonic boundary-layer applications},
  author={Zhu, Zhibin and Zhang, Xuejun and Wang, Xun and Zhang, Liang},
  journal={Journal of Spacecraft and Rockets},
  volume={57},
  number={2},
  pages={364--372},
  year={2020},
  publisher={American Institute of Aeronautics and Astronautics}
}

@article{barone2022internal,
  title={Internal energy balance and aerodynamic heating predictions for hypersonic turbulent boundary layers},
  author={Barone, Matthew and Nicholson, Gary L and Duan, Lian},
  journal={Physical Review Fluids},
  volume={7},
  number={8},
  pages={084604},
  year={2022},
  publisher={APS}
}

@article{marchenay2022hypersonic,
  title={Hypersonic Turbulent Flow Reynolds-Averaged Navier--Stokes Simulations with Roughness and Blowing Effects},
  author={Marchenay, Yann and Olazabal Loum{\'e}, M and Chedevergne, Fran{\c{c}}ois},
  journal={Journal of Spacecraft and Rockets},
  volume={59},
  number={5},
  pages={1686--1696},
  year={2022},
  publisher={American Institute of Aeronautics and Astronautics}
}

@inproceedings{buck2021rans,
  title={On RANS turbulence models for high-speed applications},
  author={Buck, Axel and Mundt, Christian},
  booktitle={AIAA Scitech 2021 Forum},
  pages={1744},
  year={2021}
}

@inproceedings{oliver2007assessment,
  title={Assessment of turbulent shock-boundary layer interaction computations using the OVERFLOW code},
  author={Oliver, Anthony and Lillard, Randolph and Schwing, Alan and Blaisdell, Gregory and Lyrintzis, Anastasios},
  booktitle={45th AIAA Aerospace Sciences Meeting and Exhibit},
  pages={104},
  year={2007}
}

@inproceedings{huang2017high,
  title={High-Mach-number turbulence modeling using machine learning and direct numerical simulation database},
  author={Huang, Junji and Duan, Lian and Wang, Jianxun and Sun, Rui and Xiao, Heng},
  booktitle={55th AIAA Aerospace Sciences Meeting},
  pages={0315},
  year={2017}
}

@article{wang2019prediction,
  title={{Prediction of Reynolds stresses in high-Mach-number turbulent boundary layers using physics-informed machine learning}},
  author={Wang, Jian-Xun and Huang, Junji and Duan, Lian and Xiao, Heng},
  journal={Theoretical and Computational Fluid Dynamics},
  volume={33},
  pages={1--19},
  year={2019},
  publisher={Springer}
}

@article{chowdhary2022calibrating,
  title={Calibrating hypersonic turbulence flow models with the HIFiRE-1 experiment using data-driven machine-learned models},
  author={Chowdhary, Kenny and Hoang, Chi and Lee, Kookjin and Ray, Jaideep and Weirs, V Gregory and Carnes, Brian},
  journal={Computer Methods in Applied Mechanics and Engineering},
  volume={401},
  pages={115396},
  year={2022},
  publisher={Elsevier}
}

@article{cherroud2023space,
  title={Space-dependent Aggregation of Stochastic Data-driven Turbulence Models},
  author={Cherroud, S. and Merle, X. and Cinnella, P. and Gloerfelt, X.},
  journal={arXiv preprint arXiv:2306.16996},
  year={2023}
}

@article{oulghelou2024machine,
  title={Machine-learning-assisted Blending of Data-Driven Turbulence Models},
  author={Oulghelou, M. and Cherroud, S. and Merle, X. and Cinnella, P.},
  journal={arXiv preprint arXiv:2410.14431},
  year={2024}
}

@article{rincon2023progressive,
  title={Progressive augmentation of Reynolds stress tensor models for secondary flow prediction by computational fluid dynamics driven surrogate optimisation},
  author={Rinc{\'o}n, Mario Javier and Amarloo, Ali and Reclari, Martino and Yang, Xiang IA and Abkar, Mahdi},
  journal={International Journal of Heat and Fluid Flow},
  volume={104},
  pages={109242},
  year={2023},
  publisher={Elsevier}
}

@article{fang2023toward,
  title={Toward more general turbulence models via multicase computational-fluid-dynamics-driven training},
  author={Fang, Yuan and Zhao, Yaomin and Waschkowski, Fabian and Ooi, Andrew SH and Sandberg, Richard D},
  journal={AIAA Journal},
  volume={61},
  number={5},
  pages={2100--2115},
  year={2023},
  publisher={American Institute of Aeronautics and Astronautics}
}

@article{huang2022direct,
  title={Direct numerical simulation of hypersonic turbulent boundary layers: effect of spatial evolution and Reynolds number},
  author={Huang, Junji and Duan, Lian and Choudhari, Meelan M},
  journal={Journal of Fluid Mechanics},
  volume={937},
  pages={A3},
  year={2022},
  publisher={Cambridge University Press}
}

@inproceedings{dann2007cfd,
  title={CFD designed experiments for shock wave/boundary layer interactions in hypersonic ducted flows},
  author={Dann, AG and Morgan, RG},
  booktitle={Proceedings of the 16th Australasian Fluid Mechanics Conference, 16AFMC},
  volume={1},
  pages={1304--1308},
  year={2007},
  organization={Citeseer}
}

@techreport{dilley2001evaluation,
  title={Evaluation of CFD turbulent heating prediction techniques and comparison with hypersonic experimental data},
  author={Dilley, Arthur D and McClinton, Charles R},
  year={2001}
}

@article{hejranfar2012dual,
  title={Dual-code solution procedure for efficient computing equilibrium hypersonic axisymmetric transitional/turbulent flows},
  author={Hejranfar, Kazem and Esfahanian, Vahid and Kamali-Moghadam, Ramin},
  journal={Aerospace Science and Technology},
  volume={21},
  number={1},
  pages={64--74},
  year={2012},
  publisher={Elsevier}
}

@inproceedings{horstman1987prediction,
  title={Prediction of hypersonic shock-wave/turbulent-boundary-layer interaction flows},
  author={Horstman, CC},
  booktitle={19th AIAA, Fluid Dynamics, Plasma Dynamics, and Lasers Conference},
  pages={1367},
  year={1987}
}

@inproceedings{kim1988hypersonic,
  title={Hypersonic turbulent wall boundary layer computations},
  author={KIM, S and HARLOFF, G},
  booktitle={24th Joint Propulsion Conference},
  pages={2829},
  year={1988}
}

@inproceedings{narayanswami1993numerical,
  title={Numerical simulation of crossing/turbulent boundary layer interaction at Mach 8.3 comparison of zero and two-equation turbulence models},
  author={Narayanswami, N and HORSTMAN, C and KNIGHT, D},
  booktitle={31st Aerospace Sciences Meeting},
  pages={779},
  year={1993}
}

@inproceedings{ozgun2015hypersonic,
  title={Hypersonic Flow Analysis of Re-entry Vehicles Using Three Dimensional Navier-Stokes Equations},
  author={Ozgun, Muharrem and Eyi, Sinan},
  booktitle={13th International Energy Conversion Engineering Conference},
  pages={3881},
  year={2015}
}

@article{panaras2015turbulence,
  title={Turbulence modeling of flows with extensive crossflow separation},
  author={Panaras, Argyris G},
  journal={Aerospace},
  volume={2},
  number={3},
  pages={461--481},
  year={2015},
  publisher={MDPI}
}

@article{sarkar1992pressure,
  title={The pressure--dilatation correlation in compressible flows},
  author={Sarkar, S},
  journal={Physics of Fluids A: Fluid Dynamics},
  volume={4},
  number={12},
  pages={2674--2682},
  year={1992},
  publisher={American Institute of Physics}
}

@article{so1998morkovin,
  title={Morkovin hypothesis and the modeling of wall-bounded compressible turbulent flows},
  author={So, RMC and Gatski, TB and Sommer, TP},
  journal={AIAA journal},
  volume={36},
  number={9},
  pages={1583--1592},
  year={1998}
}


@article{sinha2003modeling,
  title={Modeling shock unsteadiness in shock/turbulence interaction},
  author={Sinha, Krishnendu and Mahesh, Krishnan and Candler, Graham V},
  journal={Physics of fluids},
  volume={15},
  number={8},
  pages={2290--2297},
  year={2003},
  publisher={American Institute of Physics}
}

@article{veera2009modeling,
  title={Modeling the effect of upstream temperature fluctuations on shock/homogeneous turbulence interaction},
  author={Veera, Vijay K and Sinha, Krishnendu},
  journal={Physics of fluids},
  volume={21},
  number={2},
  year={2009},
  publisher={AIP Publishing}
}

@article{marchenay2022hypersonic,
  title={Hypersonic Turbulent Flow Reynolds-Averaged Navier--Stokes Simulations with Roughness and Blowing Effects},
  author={Marchenay, Yann and Olazabal Loum{\'e}, M and Chedevergne, Fran{\c{c}}ois},
  journal={Journal of Spacecraft and Rockets},
  volume={59},
  number={5},
  pages={1686--1696},
  year={2022},
  publisher={American Institute of Aeronautics and Astronautics}
}

@article{nicholson2024direct,
  title={Direct Numerical Simulation Database of High-Speed Flow over Parameterized Curved Walls},
  author={Nicholson, Gary L and Duan, Lian and Bisek, Nicholas J},
  journal={AIAA Journal},
  volume={62},
  number={6},
  pages={2095--2118},
  year={2024},
  publisher={American Institute of Aeronautics and Astronautics}
}

@article{marvin1983turbulence,
  title={Turbulence modeling for computational aerodynamics},
  author={Marvin, Joseph G},
  journal={AIAA Journal},
  volume={21},
  number={7},
  pages={941--955},
  year={1983}
}

@inproceedings{bardina1994three,
  title={Three-dimensional Navier-Stokes method with two-equation turbulence models for efficient numerical simulation of hypersonic flows},
  author={Bardina, J},
  booktitle={30th Joint Propulsion Conference and Exhibit},
  pages={2950},
  year={1994}
}


@techreport{settles1991,
  title={Hypersonic shock/boundary-layer interaction database},
  author={Settles, GS and Dodson, LJ},
  year={1991},
  institution={NASA CR 177577}
}

@article{bose2018wall,
  title={Wall-modeled large-eddy simulation for complex turbulent flows},
  author={Bose, Sanjeeb T and Park, George Ilhwan},
  journal={Annual review of fluid mechanics},
  volume={50},
  number={1},
  pages={535--561},
  year={2018},
  publisher={Annual Reviews}
}

@article{mettu2022wall,
  title={Wall-modeled large eddy simulation of high speed flows},
  author={Mettu, Balachandra R and Subbareddy, Pramod K},
  journal={AIAA journal},
  volume={60},
  number={7},
  pages={4302--4324},
  year={2022},
  publisher={American Institute of Aeronautics and Astronautics}
}

@inproceedings{zangeneh2021wall,
  title={Wall-modeled large-eddy simulation of hypersonic turbulent boundary-layers},
  author={Zangeneh, Rozie},
  booktitle={AIAA Scitech 2021 Forum},
  pages={1076},
  year={2021}
}
 @article{Schwarzkopf_Livescu_Baltzer_Gore_Ristorcelli_2016, title={A Two-length Scale Turbulence Model for Single-phase Multi-fluid Mixing}, abstractNote={A two-length scale, second moment turbulence model (Reynolds averaged Navier-Stokes, RANS) is proposed to capture a wide variety of single-phase flows, spanning from incompressible flows with single fluids and mixtures of different density fluids (variable density flows) to flows over shock waves. The two-length scale model was developed to address an inconsistency present in the single-length scale models, e.g. the inability to match both variable density homogeneous Rayleigh-Taylor turbulence and Rayleigh-Taylor induced turbulence, as well as the inability to match both homogeneous shear and free shear flows. The two-length scale model focuses on separating the decay and transport length scales, as the two physical processes are generally different in inhomogeneous turbulence. This allows reasonable comparisons with statistics and spreading rates over such a wide range of turbulent flows using a common set of model coefficients. The specific canonical flows considered for calibrating the model include homogeneous shear, single-phase incompressible shear driven turbulence, variable density homogeneous Rayleigh-Taylor turbulence, Rayleigh-Taylor induced turbulence, and shocked isotropic turbulence. The second moment model shows to compare reasonably well with direct numerical simulations (DNS), experiments, and theory in most cases. The model was then applied to variable density shear layer and shock tube data and shows to be in reasonable agreement with DNS and experiments. The importance of using DNS to calibrate and assess RANS type turbulence models is also highlighted.}, author={Schwarzkopf, J D and Livescu, D and Baltzer, J R and Gore, R A and Ristorcelli, J R}, year={2016}, language={en} }


 @inbook{Karl_Hickey_Lacombe_2019, address={Cham}, title={Reynolds Stress Models for Shock-Turbulence Interaction}, ISBN={978-3-319-91019-2}, url={http://link.springer.com/10.1007/978-3-319-91020-8_60}, DOI={10.1007/978-3-319-91020-8_60}, abstractNote={A systematic deﬁciency of current turbulence models which are based on the Reynolds-averaged Navier-Stokes equations (RANS) is their inability to correctly predict the interaction of turbulence with shocks. This is because RANS models do not account for the unsteady motion or fragmentation of the shock wave within the interaction zone. Typically, signiﬁcant over-prediction of the turbulent energy ampliﬁcation occurs without dedicated adjustment of the applied turbulence model.}, booktitle={31st International Symposium on Shock Waves 1}, publisher={Springer International Publishing}, author={Karl, Sebastian and Hickey, Jean-Pierre and Lacombe, Francis}, editor={Sasoh, Akihiro and Aoki, Toshiyuki and Katayama, Masahide}, year={2019}, pages={511–517}, language={en} }

@article{spalart2009detached,
  title={Detached-eddy simulation},
  author={Spalart, Philippe R},
  journal={Annual review of fluid mechanics},
  volume={41},
  number={1},
  pages={181--202},
  year={2009},
  publisher={Annual Reviews}
}

@article{moin1998direct,
  title={Direct numerical simulation: a tool in turbulence research},
  author={Moin, Parviz and Mahesh, Krishnan},
  journal={Annual review of fluid mechanics},
  volume={30},
  number={1},
  pages={539--578},
  year={1998},
  publisher={Annual Reviews 4139 El Camino Way, PO Box 10139, Palo Alto, CA 94303-0139, USA}
}

@article{zhiyin2015large,
  title={Large-eddy simulation: Past, present and the future},
  author={Zhiyin, Yang},
  journal={Chinese journal of Aeronautics},
  volume={28},
  number={1},
  pages={11--24},
  year={2015},
  publisher={Elsevier}
}

@article{horstman1972turbulent,
  title={Turbulent properties of a compressible boundary layer.},
  author={Horstman, CC and Owen, FK},
  journal={AIAA Journal},
  volume={10},
  number={11},
  pages={1418--1424},
  year={1972}
}

@article{owen1972structure,
  title={On the structure of hypersonic turbulent boundary layers},
  author={Owen, FK and Horstman, CC},
  journal={Journal of Fluid Mechanics},
  volume={53},
  number={4},
  pages={611--636},
  year={1972},
  publisher={Cambridge University Press}
}

@article{owen1975mean,
  title={Mean and fluctuating flow measurements of a fully-developed, non-adiabatic, hypersonic boundary layer},
  author={Owen, FK and Horstman, CC and Kussoy, MI},
  journal={Journal of Fluid Mechanics},
  volume={70},
  number={2},
  pages={393--413},
  year={1975},
  publisher={Cambridge University Press}
}




@article{PaChJo24,
	abstract = { We develop a neural-network-based variable turbulent Prandtl number model for the k-eps turbulence model for improved wall heating predictions in hypersonic shock--boundary-layer interactions (SBLIs). The model is developed by performing a finite-dimensional field inference for a spatially varying turbulent Prandtl number on six canonical SBLIs: three compression ramps at Mach 8 and three impinging shocks at Mach 5. The inference results identify a turbulent Prandtl number that reduces wall heating by systematically directing heat transfer away from the wall. An ensemble of Lipschitz-continuous neural networks is then trained on the inferred turbulent Prandtl number fields to develop a predictive model. We evaluate the resulting variable turbulent Prandtl number model on a suite of test cases, including the hollow cylinder flare and HIFiRE ground test experiments. The machine-learning-augmented model systematically increases Prt near the wall to reduce negative turbulent heat flux while decreasing Prt away from the wall to enhance positive turbulent heat flux, collectively reducing overall heat transfer to the surface. Results show that the learned model consistently improves peak heating predictions by 40--70\% compared to the baseline k--ϵ model, a k--ϵ model augmented with various high-speed corrections, and the shear stress transport model across a range of conditions. },
	author = {Parish, Eric and Ching, David S. and Jordan, Cyrus and Nicholson, Gary and Miller, Nathan E. and Beresh, Steven and Barone, Matthew and Gupta, Niloy and Duraisamy, Karthik},
	doi = {10.2514/1.J064745},
	eprint = {https://doi.org/10.2514/1.J064745},
	journal = {AIAA Journal},
	number = {0},
	pages = {1-22},
	title = {Data-Driven Turbulent Prandtl Number Modeling for Hypersonic Shock--Boundary-Layer Interactions},
	url = {https://doi.org/10.2514/1.J064745},
	volume = {0},
	year = {0},
	bdsk-url-1 = {https://doi.org/10.2514/1.J064745}}

@article{mikulla1976turbulence,
  title={Turbulence measurements in hypersonic shock-wave boundary-layer interaction flows},
  author={Mikulla, Vo and Horstman, Co C},
  journal={AIAA Journal},
  volume={14},
  number={5},
  pages={568--575},
  year={1976}
}

@inproceedings{sahoo2009effects,
  title={Effects of roughness on a turbulent bloundary layer in hypersonic flow},
  author={Sahoo, Dipankar and Schultze, Marco and Smits, Alexander},
  booktitle={39th AIAA Fluid Dynamics Conference},
  pages={3678},
  year={2009}
}

@book{baumgartner1997turbulence,
  title={Turbulence structure in a hypersonic boundary layer},
  author={Baumgartner, Mark Lawrence},
  year={1997},
  publisher={Princeton University}
}

@inproceedings{mcginley1994turbulence,
  title={Turbulence measurements in a Mach 11 helium boundary layer},
  author={McGinley, C and Spina, E and Sheplak, M},
  booktitle={Fluid Dynamics Conference},
  pages={2364},
  year={1994}
}

@inproceedings{nance1999turbulence,
  title={Turbulence modeling of shock-dominated flows with a k-zeta formulation},
  author={Nance, Robert and Hassan, H},
  booktitle={37th Aerospace Sciences Meeting and Exhibit},
  pages={153},
  year={1999}
}

@article{robinson1998further,
  title={Further Development of the k-?(Enstrophy) Turbulence Closure Model},
  author={Robinson, DF and Hassan, HA},
  journal={AIAA journal},
  volume={36},
  number={10},
  pages={1825--1833},
  year={1998}
}

@inproceedings{smith1995prediction,
  title={Prediction of hypersonic shock wave turbulent boundary layer interactions with the kl two equation turbulence model},
  author={Smith, Brian},
  booktitle={33rd Aerospace Sciences Meeting and Exhibit},
  pages={232},
  year={1995}
}

@inproceedings{coakley1983turbulence,
  title={Turbulence modeling methods for the compressible Navier-Stokes equations},
  author={Coakley, Ti},
  booktitle={16th fluid and plasmadynamics conference},
  pages={1693},
  year={1983}
}

@article{leschziner2006modelling,
  title={Modelling turbulent separated flow in the context of aerodynamic applications},
  author={Leschziner, MA},
  journal={Fluid dynamics research},
  volume={38},
  number={2-3},
  pages={174--210},
  year={2006},
  publisher={Elsevier}
}

@article{shih1993realisable,
  title={A Realisable Reynolds Stress Algebraic Equation Model},
  author={SHIH, TH},
  journal={NASA TM-105993},
  year={1993}
}

@inproceedings{lien1996low,
  title={Low-Reynolds-number eddy-viscosity modelling based on non-linear stress-strain/vorticity relations},
  author={Lien, Fue-Sang},
  booktitle={Proc. 3rd Symposium On Engineering Turbulence Modelling and Measurements},
  pages={1--10},
  year={1996}
}

@article{coleman1972heat,
  title={Heat transfer from hypersonic turbulent flow at a wedge compression corner},
  author={Coleman, GT and Stollery, JL},
  journal={Journal of Fluid Mechanics},
  volume={56},
  number={4},
  pages={741--752},
  year={1972},
  publisher={Cambridge University Press}
}

@article{holden1992turbulent,
  title={Turbulent boundary layer development on curved compression surfaces},
  author={Holden, MS},
  journal={Calspan Report No},
  pages={7724--1},
  year={1992}
}

@article{craft2000progress,
  title={Progress in the use of non-linear two-equation models in the computation of convective heat-transfer in impinging and separated flows},
  author={Craft, TJ and Iacovides, H and Yoon, JH},
  journal={Flow, Turbulence and Combustion},
  volume={63},
  pages={59--80},
  year={2000},
  publisher={Springer}
}

@article{craft1996development,
  title={Development and application of a cubic eddy-viscosity model of turbulence},
  author={Craft, TJ and Launder, BE and Suga, K},
  journal={International Journal of Heat and Fluid Flow},
  volume={17},
  number={2},
  pages={108--115},
  year={1996},
  publisher={Elsevier}
}


@incollection{rung1999assessment,
  title={Assessment of explicit algebraic stress models in transonic flows},
  author={Rung, Thomas and L{\"u}bcke, H and Franke, M and Xue, L and Thiele, Frank and Fu, Song},
  booktitle={Engineering Turbulence Modelling and Experiments 4},
  pages={659--668},
  year={1999},
  publisher={Elsevier}
}

@article{gatski1993explicit,
  title={On explicit algebraic stress models for complex turbulent flows},
  author={Gatski, Thomas B and Speziale, Charles G},
  journal={Journal of fluid Mechanics},
  volume={254},
  pages={59--78},
  year={1993},
  publisher={Cambridge University Press}
}

@article{speziale1991modelling,
  title={Modelling the pressure--strain correlation of turbulence: an invariant dynamical systems approach},
  author={Speziale, Charles G and Sarkar, Sutanu and Gatski, Thomas B},
  journal={Journal of fluid mechanics},
  volume={227},
  pages={245--272},
  year={1991},
  publisher={Cambridge University Press}
}

@misc{M5DNS_SBLI,
    title = {NASA TMR},
    howpublished =  {\url{https://turbmodels.larc.nasa.gov/Other_DNS_Data/highspeed_curvedwalls.html}},
    author = {Nicholson, et al.},
    year = {2024},
    note = {Last accessed on Oct --, 2024}
}

@article{prince2005experiments,
  title={Experiments on the hypersonic turbulent shock-wave/boundary-layer interaction and the effects of surface roughness},
  author={Prince, SA and Vannahme, M and Stollery, JL},
  journal={The Aeronautical Journal},
  volume={109},
  number={1094},
  pages={177--184},
  year={2005},
  publisher={Cambridge University Press}
}

@article{WaMaHo08,
  author = "T. Wadhams and M. MacLean and M. Holden and E. Mundy",
  title = "Ground test studies of the {HIFiRE-1} transition experiment {Part 1: Experimental} results",
  journal = "J. Spacecraft Rockets",
  volume = "45",
  number = 6,
  year = 2008,
  pages = "1134--1148",
  doi = "doi: 10.2514/138338",
}


@inproceedings{holden2008review,
  title={A review of experimental studies of surface roughness and blowing on the heat transfer and skin friction to nosetips and slender cones in high mach numbers flows},
  author={Holden, Michael and Wadhams, Timothy and Mundy, Erik},
  booktitle={40th Thermophysics Conference},
  pages={3907},
  year={2008}
}


@inproceedings{sahoo2009effects,
  title={Effects of roughness on a turbulent bloundary layer in hypersonic flow},
  author={Sahoo, Dipankar and Schultze, Marco and Smits, Alexander},
  booktitle={39th AIAA Fluid Dynamics Conference},
  pages={3678},
  year={2009}
}

@article{peltier2016crosshatch,
  title={Crosshatch roughness distortions on a hypersonic turbulent boundary layer},
  author={Peltier, SJ and Humble, RA and Bowersox, RDW},
  journal={Physics of Fluids},
  volume={28},
  number={4},
  year={2016},
  publisher={AIP Publishing}
}

@book{berg1977surface,
  title={SURFACE ROUGHNESS EFFECTS ON THE HYPERSONIC TURBULENT BOUNDARY LAYER.},
  author={Berg, Dale Evan},
  year={1977},
  publisher={California Institute of Technology}
}

@inproceedings{finson1980effect,
  title={The effect of surface roughness character on turbulent re-entry heating},
  author={Finson, M and Clarke, A},
  booktitle={15th Thermophysics Conference},
  pages={1459},
  year={1980}
}


@article{aupoix2015roughness,
  title={Roughness corrections for the k--$\omega$ shear stress transport model: Status and proposals},
  author={Aupoix, B},
  journal={Journal of Fluids Engineering},
  volume={137},
  number={2},
  pages={021202},
  year={2015},
  publisher={American Society of Mechanical Engineers}
}

@article{aupoix2015improved,
  title={Improved heat transfer predictions on rough surfaces},
  author={Aupoix, B},
  journal={International Journal of Heat and Fluid Flow},
  volume={56},
  pages={160--171},
  year={2015},
  publisher={Elsevier}
}

@article{marchenay2021modeling,
  title={Modeling of combined effects of surface roughness and blowing for Reynolds-averaged Navier--Stokes turbulence models},
  author={Marchenay, Yann and Chedevergne, Fran{\c{c}}ois and Olazabal Loum{\'e}, M},
  journal={Physics of Fluids},
  volume={33},
  number={4},
  year={2021},
  publisher={AIP Publishing}
}

@inproceedings{olazabal2017study,
  title={Study on k-$\omega$ shear stress transport model corrections applied to rough wall turbulent hypersonic boundary layers},
  author={Olazabal-Loum{\'e}, M and Danvin, Florian and Mathiaud, Julien and Aupoix, Bertrand and Toulouse, ONERA},
  booktitle={Seventh European Conference for Aeronautics and Space Sciences},
  year={2017}
}

@inproceedings{olazabal2019roughness,
  title={Roughness corrections applied to the simulation of turbulent hypersonic flows},
  author={Olazabal-Loum{\'e}, Marina and Chedevergne, Fran{\c{c}}ois and Danvin, Florian and Mathiaud, Julien},
  booktitle={EUCASS 2019},
  year={2019}
}

@article{wilcox1988reassessment,
  title={Reassessment of the scale-determining equation for advanced turbulence models},
  author={Wilcox, David C},
  journal={AIAA journal},
  volume={26},
  number={11},
  pages={1299--1310},
  year={1988}
}

@article{goddard1959effect,
  title={Effect of uniformly distributed roughness on turbulent skin-friction drag at supersonic speeds},
  author={Goddard Jr, Frank E},
  journal={Journal of the Aerospace Sciences},
  volume={26},
  number={1},
  pages={1--15},
  year={1959}
}

@article{lin1982turbulence,
  title={Turbulence models for high-speed, rough-wall boundary layers},
  author={Lin, TC and Bywater, RJ},
  journal={AIAA Journal},
  volume={20},
  number={3},
  pages={325--333},
  year={1982}
}

@article{choi2012grid,
  title={Grid-point requirements for large eddy simulation: Chapman’s estimates revisited},
  author={Choi, Haecheon and Moin, Parviz},
  journal={Physics of fluids},
  volume={24},
  number={1},
  year={2012},
  publisher={AIP Publishing}
}

@article{yang2021grid,
  title={Grid-point and time-step requirements for direct numerical simulation and large-eddy simulation},
  author={Yang, Xiang IA and Griffin, Kevin P},
  journal={Physics of Fluids},
  volume={33},
  number={1},
  year={2021},
  publisher={AIP Publishing}
}


@article{goyne2003skin,
  title={Skin-friction measurements in high-enthalpy hypersonic boundary layers},
  author={Goyne, CP and Stalker, RJ and Paull, A},
  journal={Journal of Fluid Mechanics},
  volume={485},
  pages={1--32},
  year={2003},
  publisher={Cambridge University Press}
}

@inproceedings{fang2015large,
  title={Large-eddy simulation of a three-dimensional hypersonic shock wave turbulent boundary layer interaction of a single-fin},
  author={Fang, Jian and Yao, Yufeng and Zheltovodov, Alexander and Lu, Lipeng},
  booktitle={53rd AIAA Aerospace Sciences Meeting},
  pages={1062},
  year={2015}
}


@article{fang2017investigation,
  title={Investigation of three-dimensional shock wave/turbulent-boundary-layer interaction initiated by a single fin},
  author={Fang, Jian and Yao, Yufeng and Zheltovodov, Alexandr A and Lu, Lipeng},
  journal={AIAA Journal},
  volume={55},
  number={2},
  pages={509--523},
  year={2017},
  publisher={American Institute of Aeronautics and Astronautics}
}

@inproceedings{kianvashrad2021large,
  title={Large Eddy Simulation of Hypersonic Cold Wall Flat Plate},
  author={Kianvashrad, Nadia and Knight, Doyle D},
  booktitle={AIAA AVIATION 2021 FORUM},
  pages={2882},
  year={2021}
}

@inproceedings{knight2024large,
  title={Large Eddy Simulation of Cold Wall Hypersonic Turbulent Boundary Layers},
  author={Knight, Doyle D},
  booktitle={AIAA AVIATION FORUM AND ASCEND 2024},
  pages={3966},
  year={2024}
}

@inproceedings{spalart1992one,
  title={A one-equation turbulence model for aerodynamic flows},
  author={Spalart, Philippe and Allmaras, Steven},
  booktitle={30th aerospace sciences meeting and exhibit},
  pages={439},
  year={1992}
}

@book{smith1967numerical,
  title={Numerical solution of the turbulent-boundary-layer equations},
  author={Smith, AMOj and Cebeci, Tuncer},
  year={1967},
  publisher={Douglas Aircraft Company, Douglas Aircraft Division}
}

@inproceedings{baldwin1978thin,
  title={Thin-layer approximation and algebraic model for separated turbulentflows},
  author={Baldwin, Barrett and Lomax, Harvard},
  booktitle={16th aerospace sciences meeting},
  pages={257},
  year={1978}
}


@article{maeder2001direct,
  title={Direct simulation of turbulent supersonic boundary layers by an extended temporal approach},
  author={Maeder, Thierry and Adams, Nikolaus A and Kleiser, Leonhard},
  journal={Journal of Fluid Mechanics},
  volume={429},
  pages={187--216},
  year={2001},
  publisher={Cambridge University Press}
}


@article{xin2006direct,
  title={Direct numerical simulation of a spatially evolving supersonic turbulent boundary layer at Ma= 6},
  author={Xin-Liang, Li and De-Xun, Fu and Yan-Wen, Ma},
  journal={Chinese Physics Letters},
  volume={23},
  number={6},
  pages={1519},
  year={2006},
  publisher={IOP Publishing}
}

@article{duan2010direct,
  title={Direct numerical simulation of hypersonic turbulent boundary layers. Part 2. Effect of wall temperature},
  author={Duan, Lian and Beekman, I and Martin, MP},
  journal={Journal of Fluid Mechanics},
  volume={655},
  pages={419--445},
  year={2010},
  publisher={Cambridge University Press}
}

@article{duan2011direct,
  title={Direct numerical simulation of hypersonic turbulent boundary layers. Part 3. Effect of Mach number},
  author={Duan, Lian and Beekman, I and Martin, MP},
  journal={Journal of Fluid Mechanics},
  volume={672},
  pages={245--267},
  year={2011},
  publisher={Cambridge University Press}
}

@article{chu2013effect,
  title={Effect of wall temperature on hypersonic turbulent boundary layer},
  author={Chu, You-Biao and Zhuang, Yue-Qing and Lu, Xi-Yun},
  journal={Journal of Turbulence},
  volume={14},
  number={12},
  pages={37--57},
  year={2013},
  publisher={Taylor \& Francis}
}

@article{liang2012dns,
  title={DNS and analysis of a spatially evolving hypersonic turbulent boundary layer over a flat plate at Mach 8},
  author={Liang, Xian},
  journal={SCIENTIA SINICA Physica, Mechanica \& Astronomica},
  volume={42},
  number={3},
  pages={282},
  year={2012}
}

@article{liang2013dns,
  title={DNS of a spatially evolving hypersonic turbulent boundary layer at Mach 8},
  author={Liang, Xian and Li, XinLiang},
  journal={Science China Physics, Mechanics and Astronomy},
  volume={56},
  pages={1408--1418},
  year={2013},
  publisher={Springer}
}

@article{duan2016pressure,
  title={Pressure fluctuations induced by a hypersonic turbulent boundary layer},
  author={Duan, Lian and Choudhari, Meelan M and Zhang, Chao},
  journal={Journal of Fluid Mechanics},
  volume={804},
  pages={578--607},
  year={2016},
  publisher={Cambridge University Press}
}

@article{zhang2017effect,
  title={Effect of wall cooling on boundary-layer-induced pressure fluctuations at Mach 6},
  author={Zhang, Chao and Duan, Lian and Choudhari, Meelan M},
  journal={Journal of Fluid Mechanics},
  volume={822},
  pages={5--30},
  year={2017},
  publisher={Cambridge University Press}
}

@article{zhang2018direct,
  title={Direct numerical simulation database for supersonic and hypersonic turbulent boundary layers},
  author={Zhang, Chao and Duan, Lian and Choudhari, Meelan M},
  journal={AIAA journal},
  volume={56},
  number={11},
  pages={4297--4311},
  year={2018},
  publisher={American Institute of Aeronautics and Astronautics}
}


@article{duan2012study,
  title={Study of turbulence-radiation interaction in hypersonic turbulent boundary layers},
  author={Duan, Lian and Martin, MP and Feldick, AM and Modest, MF and Levin, DA},
  journal={AIAA journal},
  volume={50},
  number={2},
  pages={447--453},
  year={2012}
}

@article{trettel2016mean,
  title={Mean velocity scaling for compressible wall turbulence with heat transfer},
  author={Trettel, Andrew and Larsson, Johan},
  journal={Physics of Fluids},
  volume={28},
  number={2},
  year={2016},
  publisher={AIP Publishing}
}

@article{griffin2021velocity,
  title={Velocity transformation for compressible wall-bounded turbulent flows with and without heat transfer},
  author={Griffin, Kevin Patrick and Fu, Lin and Moin, Parviz},
  journal={Proceedings of the National Academy of Sciences},
  volume={118},
  number={34},
  pages={e2111144118},
  year={2021},
  publisher={National Acad Sciences}
}

@article{zhang2014generalized,
  title={A generalized Reynolds analogy for compressible wall-bounded turbulent flows},
  author={Zhang, You-Sheng and Bi, Wei-Tao and Hussain, Fazle and She, Zhen-Su},
  journal={Journal of Fluid Mechanics},
  volume={739},
  pages={392--420},
  year={2014},
  publisher={Cambridge University Press}
}

@book{walz1969,
  title={Boundary layers of flow and temperature},
  author={Walz, Alfred},
  year={1969},
  publisher={Cambridge, MA: MIT Press}
}


@inproceedings{mcdonald1968extended,
  title={An extended mixing length approach for computing the turbulent boundary layer development},
  author={McDonald, Henry and Camarata, FJ},
  booktitle={Proceedings, Stanford Conference on Computation of Turbulent Boundary Layers},
  volume={1},
  pages={83--98},
  year={1968}
}


@article{menter1997eddy,
  title={Eddy Viscosity Transport Equations and Their Relation to the k-$\varepsilon$ Model},
  author={Menter, FR},
  journal={Journal of Fluids Engineering},
  volume={119},
  number={4},
  pages={876--884},
  year={1997},
  publisher={ASME International}
}

@inproceedings{bosco2011investigation,
  title={Investigation of a compression corner at hypersonic conditions using a reynolds stress model},
  author={Bosco, Arianna and Reinartz, Birgit and Brown, Laurie and Boyce, Russell},
  booktitle={17th AIAA International Space Planes and Hypersonic Systems and Technologies Conference},
  pages={2217},
  year={2011}
}

@article{frauholz2014investigation,
  title={Investigation of hypersonic intakes using Reynolds stress modeling and wavelet-based adaptation},
  author={Frauholz, Sarah and Bosco, Arianna and Reinartz, Birgit U and M{\"u}ller, Siegfried and Behr, Marek},
  journal={AIAA journal},
  volume={52},
  number={12},
  pages={2765--2781},
  year={2014},
  publisher={American Institute of Aeronautics and Astronautics}
}


@inproceedings{eisfeld2005advanced,
  title={Advanced turbulence modelling and stress analysis for the DLR-F6 configuration},
  author={Eisfeld, Bernhard and Brodersen, Olaf},
  booktitle={23rd AIAA Applied Aerodynamics Conference},
  pages={4727},
  year={2005}
}

@article{gerolymos2012term,
  title={Term-by-term analysis of near-wall second-moment closures},
  author={Gerolymos, GA and Lo, C and Vallet, I and Younis, BA},
  journal={AIAA journal},
  volume={50},
  number={12},
  pages={2848--2864},
  year={2012}
}

@article{gerolymos2002wall,
  title={Wall-normal-free near-wall Reynolds-stress model for 3-D turbomachinery flows},
  author={Gerolymos, GA and Vallet, I},
  journal={AIAA Journal},
  volume={40},
  number={2},
  pages={199208},
  year={2002}
}


@article{gerolymos2004contribution,
  title={Contribution to single-point closure Reynolds-stress modelling of inhomogeneous flow},
  author={Gerolymos, Georges A and Sauret, Emilie and Vallet, I},
  journal={Theoretical and Computational Fluid Dynamics},
  volume={17},
  pages={407--431},
  year={2004},
  publisher={Springer}
}

@techreport{speziale1989preliminary,
  title={A preliminary compressible second-order closure model for high speed flows},
  author={Speziale, Charles G and Sarkar, Sutanu},
  year={1989}
}

@article{grasso1993high,
  title={High-speed turbulence modeling of shock-wave/boundary-layer interaction},
  author={Grasso, F and Falconi, D},
  journal={AIAA journal},
  volume={31},
  number={7},
  pages={1199--1206},
  year={1993}
}

@article{van1956problem,
  title={The Problem of Aerodynamic Heating},
  author={VAN DRIEST, ER},
  journal={Aeronaut. Eng. Rev.},
  volume={15},
  number={10},
  pages={26--41},
  year={1956}
}

@inproceedings{neeb2015experimental,
  title={Experimental flow characterization and heat flux augmentation analysis of a hypersonic turbulent boundary layer along a rough surface},
  author={Neeb, Dominik and Saile, Dominik and G{\"u}lhan, Ali},
  booktitle={Proceedings of the 8th European Symposium on Aerothermodynamics for Space Vehicles},
  number={89873},
  pages={1--15},
  year={2015}
}

@article{helm2022large,
  title={Large eddy simulation of two separated hypersonic shock/turbulent boundary layer interactions},
  author={Helm, Clara M and Mart{\'\i}n, MP},
  journal={Physical Review Fluids},
  volume={7},
  number={7},
  pages={074601},
  year={2022},
  publisher={APS}
}

@article{spalding1964drag,
  title={The drag of a compressible turbulent boundary layer on a smooth flat plate with and without heat transfer},
  author={Spalding, DB and Chi, SW},
  journal={Journal of Fluid Mechanics},
  volume={18},
  number={1},
  pages={117--143},
  year={1964},
  publisher={Cambridge University Press}
}


@article{white1972simple,
  title={A simple theory for the two-dimensional compressible turbulent boundary layer},
  author={White, FM and Christoph, GH},
  journal={Journal of Fluids Engineering, Transactions of the ASME},
  volume={94},
  number={3},
  pages={636},
  year={1972}
}

@article{hopkins1971evaluation,
  title={An evaluation of theories for predicting turbulent skin friction andheat transfer on flat plates at supersonic and hypersonic Mach numbers},
  author={Hopkins, Edward J and Inouye, Mamoru},
  journal={AIAA Journal},
  volume={9},
  number={6},
  pages={993--1003},
  year={1971}
}


@inproceedings{holden1972shock,
  title={Shock wave-turbulent boundary layer interaction in hypersonic flow},
  author={HOLDEN, MS},
  booktitle={10th Aerospace Sciences Meeting},
  year={1972},
  organization={American Institute of Aeronautics and Astronautics}
}

@inproceedings{watson1973measurements,
  title={Measurements in a transitional/turbulent Mach 10 boundary layer at high-Reynolds numbers},
  author={Watson, R and Harris, J and ANDERS, JR, J},
  booktitle={11th Aerospace Sciences Meeting},
  pages={165},
  year={1973}
}

@book{winkler1959investigation,
  title={Investigation of flat plate hypersonic turbulent boundary layers with heat transfer at a Mach number of 5.2},
  author={Winkler, Ava M and Cha, Moon H},
  year={1959},
  publisher={NAVORD}
}

@article{winkler1961investigation,
  title={Investigation of Flat-Plate Hypersonic, Turbulent Boundary Layers With Heat Transfer},
  author={Winkler, Eva M},
  journal={Journal of Applied Mechanics},
  volume={28},
  number={3},
  pages={323},
  year={1961}
}

@techreport{young1965experimental,
  title={Experimental investigation of the effects of surface roughness on compressible turbulent boundary layer skin friction and heat transfer},
  author={Young, Frank Levi},
  year={1965},
  institution={Technical Report DLR-532, CR-
21 Defense Research Laboratory, University of Texas, Austin}
}

@article{heronimus1966hypersonic,
  title={Hypersonic shock tunnel experiments on the W7 flat plate model-expansion side, turbulent flow and leading edge transpiration data},
  author={Heronimus, GA},
  journal={CAL Report No. AA-1952-Y-2 (Contract No. AF 33 (615)-1847), Cornell Aeronautical Lab},
  year={1966}
}

@book{cary1970summary,
  title={Summary of available information on Reynolds analogy for zero-pressure-gradient, compressible, turbulent-boundary-layer flow},
  author={Cary, Aubrey M},
  year={1970},
  publisher={NASA TN D-5560}
}


@article{suraweera2006reynolds,
  title={Reynolds analogy in high-enthalpy and high-Mach-number turbulent flows},
  author={Suraweera, MV and Mee, DJ and Stalker, RJ},
  journal={AIAA journal},
  volume={44},
  number={4},
  pages={917--919},
  year={2006}
}

@techreport{neal1966study,
  title={A study of the pressure, heat transfer, and skin friction on sharp and blunt flat plates at Mach 6.8},
  author={Neal Jr, Luther},
  year={1966},
  publisher={Technical Note D-3312 (Washington, DC: National Aeronautics and Space Administration)}
}

@article{danberg1964characteristics,
  title={Characteristics of the turbulent boundary layer with heat and mass transfer at M= 6.7},
  author={Danberg, James Edward},
  year={1964},
  publisher={Technical Report NOLTR 64-99 (White Oak, Silver Spring, MD: United States Naval Ordnance Laboratory)}
}

@article{danberg1967characteristics,
  title={Characteristics of the turbulent boundary layer with heat and mass transfer: data tabulation},
  author={Danberg, James E},
  journal={NOLTR 675-6},
  year={1967}
}


@article{wallace1967hypersonic,
  title={Hypersonic Turbulent Boundary Layer Studies at Cold Wall Temperatures},
  author={Wallace, J},
  journal={Proceedings of the 1967 Heat Transfer and Fluid Mechanics Institute held at the University of California, San Diego, La Jolla, California},
  year={1967}
}

@article{keener1972measurements,
  title={Measurements of Reynolds analogy for a hypersonic turbulent boundary layer on a nonadiabatic flat plate},
  author={Keener, Earl R and Polek, Thomas E},
  journal={AIAA Journal},
  volume={10},
  number={6},
  pages={845--846},
  year={1972}
}

@techreport{voisinet1972measurements,
  title={Measurements of a Mach 4.9 zero-pressure-gradient turbulent boundary layer with heat transfer. Part 1: Data compilation},
  author={Voisinet, RLP and Lee, Roland E},
  year={1972},
  institution={Technical Report NOLTR 72-232 (White Oak, Silver Spring, MD: United States Naval Ordnance Laboratory)}
}

@article{hill1956boundary,
  title={Boundary-layer measurements in hypersonic flow},
  author={Hill, FK},
  journal={Journal of the Aeronautical Sciences},
  volume={23},
  number={1},
  pages={35--42},
  year={1956}
}

@book{lee1969velocity,
  title={Velocity Profile, Skin-friction Balance and Heat-transfer Measurements of the Turbulent Boundary Layers at Mach 5 and Zero-pressure Gradient},
  author={Lee, Roland E and Yanta, William Joseph and Leonas, Annette C},
  volume={69},
  year={1969},
  publisher={United States Naval Ordnance Laboratory}
}

@book{matting1961turbulent,
  title={Turbulent skin friction at high Mach numbers and Reynolds numbers in air and helium},
  author={Matting, Fred W},
  volume={82},
  year={1961},
  publisher={National Aeronautics and Space Administration}
}

@article{laderman1974mean,
  title={Mean and fluctuating flow measurements in the hypersonic boundary layer over a cooled wall},
  author={Laderman, AJ and Demetriades, A},
  journal={Journal of Fluid Mechanics},
  volume={63},
  number={1},
  pages={121--144},
  year={1974},
  publisher={Cambridge University Press}
}

@inproceedings{backx1973measurements,
  title={Measurements in the Mach 15 turbulent boundary layer on the wall of the longshot conical nozzle},
  booktitle={EUROMECH, Colloquium on Heat Transfer in Turbulent Boundary Layers with Variable Fluid Properties, Goettingen, West Germany},
  author={Backx, Edgard},
  year={1973}
}

@article{backx1974experimental,
  title={Experimental study of the turbulent boundary layer at Mach 15 and 19. 8 in a conical nozzle},
  author={Backx, Edgard},
  year={1974},
  publisher={Technical Note 102 von Kármán Institute for Fluid Dynamics}
}

@article{backx1976high,
  title={A high Mach Number turbulent boundary-layer study},
  author={Backx, E and Richards, BE},
  journal={AIAA Journal},
  volume={14},
  number={9},
  pages={1159--1160},
  year={1976}
}

@article{horstman1972turbulent,
  title={Turbulent properties of a compressible boundary layer.},
  author={Horstman, CC and Owen, FK},
  journal={AIAA Journal},
  volume={10},
  number={11},
  pages={1418--1424},
  year={1972}
}

@article{owen1972structure,
  title={On the structure of hypersonic turbulent boundary layers},
  author={Owen, FK and Horstman, CC},
  journal={Journal of Fluid Mechanics},
  volume={53},
  number={4},
  pages={611--636},
  year={1972},
  publisher={Cambridge University Press}
}


@article{murray2013experimental,
  title={Experimental investigation of axisymmetric hypersonic shock-wave/turbulent-boundary-layer interactions},
  author={Murray, N and Hillier, R and Williams, S},
  journal={Journal of Fluid Mechanics},
  volume={714},
  pages={152--189},
  year={2013},
  publisher={Cambridge University Press}
}


@book{samuels1967experimental,
  title={Experimental investigation of the turbulent boundary layer at a Mach number of 6 with heat transfer at high Reynolds numbers},
  author={Samuels, Richard D and Peterson, John B and Adcock, Jerry B},
  volume={3858},
  year={1967},
  publisher={National Aeronautics and Space Administration}
}

@article{sciacovelli2020numerical,
  title={Numerical investigation of high-speed turbulent boundary layers of dense gases},
  author={Sciacovelli, Luca and Gloerfelt, Xavier and Passiatore, Donatella and Cinnella, Paola and Grasso, Francesco},
  journal={Flow, Turbulence and Combustion},
  volume={105},
  pages={555--579},
  year={2020},
  publisher={Springer}
}


@techreport{kussoy1975experimental,
  title={An experimental documentation of a hypersonic shock-wave turbulent boundary layer interaction flow: With and without separation},
  author={Kussoy, MI and Horstmann, CC},
  year={1975}
}

@article{appels1973turbulent,
  title={Turbulent boundary layer separation at Mach 12},
  author={Appels, C},
  journal={VKI TN-90},
  year={1973}
}

@article{coet1993experiments,
  title={Experiments on shock wave/boundary layer interaction in hypersonic flow},
  author={COET, MARIE-},
  journal={La Recherche Aerospatiale(English Edition)},
  number={1},
  pages={61--74},
  year={1993}
}


@article{kussoy1993hypersonic,
  title={Hypersonic crossing shock-wave/turbulent-boundary-layer interactions},
  author={Kussoy, MI and Horstoman, KC and Horstman, C},
  journal={AIAA journal},
  volume={31},
  number={12},
  pages={2197--2203},
  year={1993}
}

@article{coleman1973study,
  title={A study of hypersonic boundary layers over a family of axisymmetric bodies at zero incidence: preliminary report and data tabulation},
  author={Coleman, GT},
  journal={Imperial College Aero Report 73-06 Imperial College of Science and Technology, London, UK},
  year={1973}
}

@phdthesis{coleman1973hypersonic,
  title={Hypersonic turbulent boundary layer studies},
  author={Coleman, Graham Trevor},
  year={1973},
  school={Imperial College London}
}

@article{coleman1974incipient,
  title={Incipient separation of axially symmetric hypersonic turbulent boundary layers},
  author={Coleman, GT and Stollery, JL},
  journal={AIAA Journal},
  volume={12},
  number={1},
  pages={119--120},
  year={1974}
}


@article{holden2014measurements,
  title={Measurements in Regions of Shock Wave/Turbulent Boundary Layer Interaction on Double cone and Hollow Cylinder/Flare Configurations for Open and},
  author={Holden, M and Wadhams, T and MacLean, M},
  journal={Blind” Code Evaluation/Validation,” Tech. rep., American Institute of Aeronautics and Astronautics},
  year={2014}
}


@inproceedings{holden2018measurements,
  title={Measurements in regions of shock wave/turbulent boundary layer interaction from Mach 4 to 7 at flight duplicated velocities to evaluate and improve the models of turbulence in CFD codes},
  author={Holden, Michael},
  booktitle={2018 Fluid Dynamics Conference},
  pages={3706},
  year={2018}
}

@article{running2019hypersonic,
  title={Hypersonic shock-wave/boundary-layer interactions on a cone/flare},
  author={Running, Carson L and Juliano, Thomas J and Jewell, Joseph S and Borg, Matthew P and Kimmel, Roger L},
  journal={Experimental Thermal and Fluid Science},
  volume={109},
  pages={109911},
  year={2019},
  publisher={Elsevier}
}


@article{borovoy2009laminar,
  title={Laminar-turbulent flow over wedges mounted on sharp and blunt plates},
  author={Borovoy, V Ya and Mosharov, VN and Noev, A Yu and Radchenko, VN},
  journal={Fluid Dynamics},
  volume={44},
  number={3},
  pages={382--396},
  year={2009},
  publisher={Springer}
}

@article{borovoy20123d,
  title={3D shock/turbulent boundary layer interaction on the plate near a wedge in presence of an entropy layer},
  author={Borovoy, Volf Ya and Egorov, Ivan Vladimirovich and Noev, Anton Yurievich and Radchenko, Vladimir Nikolaevich and Skuratov, Arkadii Sergeyevich and Struminskaya, Irina Vladimirovna},
  journal={TsAGI Science Journal},
  volume={43},
  number={6},
  year={2012},
  publisher={Begel House Inc.}
}

@article{borovoy2016entropy,
  title={Entropy-layer influence on single-fin and double-fin/boundary-layer interactions},
  author={Borovoy, Volf and Egorov, Ivan and Mosharov, Vladimir and Radchenko, Vladimir and Skuratov, Arkady and Struminskaya, Irina},
  journal={AIAA Journal},
  volume={54},
  number={2},
  pages={443--457},
  year={2016},
  publisher={American Institute of Aeronautics and Astronautics}
}

@book{law1975three,
  title={Three-dimensional shock wave-turbulent boundary layer interactions at Mach 6},
  author={Law, C Herbert},
  volume={75},
  number={191},
  year={1975},
  publisher={Aerospace Research Laboratories, Air Force Systems Command, United States~…}
}


@inproceedings{rodi1992experimental,
  title={An experimental/computational study of sharp fin induced shock wave/turbulent boundary layer interactions at Mach 5-Experimental results},
  author={RODI, PATRICK and DOLLING, DAVID},
  booktitle={30th Aerospace Sciences Meeting and Exhibit},
  pages={749},
  year={1992}
}


@article{rodi1995behavior,
  title={Behavior of pressure and heat transfer in sharp fin-induced turbulent interactions},
  author={Rodi, PE and Dolling, DS},
  journal={AIAA journal},
  volume={33},
  number={11},
  pages={2013--2019},
  year={1995}
}

@inproceedings{rodi1991experimental,
  title={An experimental/computational study of heat transfer in sharp fin induced turbulent interactions at Mach 5},
  author={Rodi, P and Dolling, D and KNIGHT, DOYLED},
  booktitle={22nd Fluid Dynamics, Plasma Dynamics and Lasers Conference},
  pages={1764},
  year={1991}
}


@inproceedings{holden1984experimental,
  title={Experimental studies of quasi-two-dimensional and three-dimensional viscous interaction regions induced by skewed-shock and swept-shock boundary layer interaction},
  author={Holden, M},
  booktitle={17th Fluid Dynamics, Plasma Dynamics, and Lasers Conference},
  pages={1677},
  year={1984}
}


@book{hopkins1969summary,
  title={Summary and correlation of skin-friction and heat-transfer data for a hypersonic turbulent boundary layer on simple shapes},
  author={Hopkins, Edward J},
  volume={5089},
  year={1969},
  publisher={National Aeronautics and Space Administration}
}


@book{hopkins1970direct,
  title={Direct Measurements of Turbulent Skin Friction on a Nonadiabatic Flat Plate at Mach Number 6.5 and Comparisons With Eight Theories},
  author={Hopkins, Edward J},
  year={1970},
  publisher={National Aeronautics and Space Administration}
}


@article{hopkins1972hypersonic,
  title={Hypersonic turbulent skin-friction and boundary-layer profiles on nonadiabatic flat plates},
  author={Hopkins, Edward J and Keener, Earl R and Polek, Thomas E and Dwyer, Harry A},
  journal={AIAA Journal},
  volume={10},
  number={1},
  pages={40--48},
  year={1972}
}

@article{hill1959turbulent,
  title={Turbulent boundary layer measurements at Mach numbers from 8 to 10},
  author={Hill, FK},
  journal={The Physics of Fluids},
  volume={2},
  number={6},
  pages={668--680},
  year={1959},
  publisher={American Institute of Physics}
}


@article{roy2018variable,
  title={Variable turbulent Prandtl number model for shock/boundary-layer interaction},
  author={Roy, Subhajit and Pathak, Utkarsh and Sinha, Krishnendu},
  journal={AIAA Journal},
  volume={56},
  number={1},
  pages={342--355},
  year={2018},
  publisher={American Institute of Aeronautics and Astronautics}
}


@article{guo2023amplification,
  title={Amplification of turbulent kinetic energy and temperature fluctuation in a hypersonic turbulent boundary layer over a compression ramp},
  author={Guo, Tongbiao and Zhang, Ji and Tong, Fulin and Li, Xinliang},
  journal={Physics of Fluids},
  volume={35},
  number={4},
  year={2023},
  publisher={AIP Publishing}
}


@article{collen2021development,
  title={Development and commissioning of the T6 Stalker Tunnel},
  author={Collen, Peter and Doherty, Luke J and Subiah, Suria D and Sopek, Tamara and Jahn, Ingo and Gildfind, David and Penty Geraets, Rowland and Gollan, Rowan and Hambidge, Christopher and Morgan, Richard and others},
  journal={Experiments in Fluids},
  volume={62},
  pages={1--24},
  year={2021},
  publisher={Springer}
}

@article{kimmel1997effect,
  title={The effect of pressure gradients on transition zone length in hypersonic boundary layers},
  author={KIMMEL, RL},
  journal={Journal of fluids engineering},
  volume={119},
  number={1},
  pages={36--41},
  year={1997}
}

@article{kimmel1993experimental,
  title={Experimental transition zone lengths in pressure gradient in hypersonic flow},
  author={Kimmel, RL},
  journal={ASME-PUBLICATIONS-FED},
  volume={151},
  pages={117--117},
  year={1993},
  publisher={ASME}
}


@article{babinsky1997large,
  title={Large-scale roughness influence on turbulent hypersonic boundary layers approaching compression corners},
  author={Babinsky, H and Edwards, JA},
  journal={Journal of spacecraft and rockets},
  volume={34},
  number={1},
  pages={70--75},
  year={1997}
}



@article{schulein2001documentation,
  title={Documentation of experimental data for hypersonic 3-D shock waves},
  author={Sch{\"u}lein, E and Zheltovodov, AA},
journal={Report IB 223-99 A 26 Deutsches Zentrum für Luft- und Raumfahrt e.V. (DLR), Institut für Strömungsmechanik, Göttingen,
Germany},
  year={2001}
}

@inproceedings{holden1989studies,
  title={Studies of surface roughness and blowing effects on hypersonic turbulent boundary layers over slender cones},
  author={Holden, M},
  booktitle={27th Aerospace Sciences Meeting},
  pages={458},
  year={1989}
}

@inproceedings{holden1982experimental,
  title={Experimental studies of surface roughness, entropy swallowing and boundary layer transition effects on the skin friction and heat transfer distribution in high speed flows},
  author={Holden, M},
  booktitle={20th Aerospace Sciences Meeting},
  pages={34},
  year={1982}
}

@inproceedings{hollis2014distributed,
  title={Distributed roughness effects on blunt-body transition and turbulent heating},
  author={Hollis, Brian R},
  booktitle={52nd Aerospace Sciences Meeting},
  pages={0238},
  year={2014}
}

@inproceedings{wilder2019rough,
  title={Rough-wall turbulent heat transfer experiments in hypersonic free flight},
  author={Wilder, Michael C and Prabhu, Dinesh K},
  booktitle={AIAA Aviation 2019 Forum},
  pages={3009},
  year={2019}
}

@inproceedings{aultman2024asymptotic,
  title={Asymptotic Near-Wall Behavior of a Mach 6 Cold-Wall Turbulent Boundary Layer},
  author={Aultman, Matthew T and Roy, Dhiman and Duan, Lian},
  booktitle={AIAA SCITECH 2024 Forum},
  pages={2733},
  year={2024}
}


@article{forsyth2024experimental,
  title={Experimental Assessment of Hypersonic Convective Heat Transfer Augmentation due to Surface Roughness},
  author={Forsyth, Peter R and Hambidge, Chris and McGilvray, Matthew},
  journal={Journal of Thermophysics and Heat Transfer},
  pages={1--10},
  year={2024},
  publisher={American Institute of Aeronautics and Astronautics}
}


@inproceedings{bhagwandin2021shock,
  title={LES of Shock-Turbulent Boundary Layer Interaction over a Mach 10 Hollow Cylinder with Flare.},
  author={Bhagwandin, Vishal A and Martin, Pino},
  booktitle={AIAA AVIATION 2021 FORUM},
  pages={2820},
  year={2021}
}


@book{white2006viscous,
  title={Viscous fluid flow},
  author={White, Frank M and Majdalani, Joseph},
  volume={3},
  year={2006},
  publisher={McGraw-Hill New York}
}

@article{hopkins1971evaluation,
  title={An evaluation of theories for predicting turbulent skin friction and heat transfer on flat plates at supersonic and hypersonic Mach numbers},
  author={Hopkins, Edward J and Inouye, Mamoru},
  journal={AIAA Journal},
  volume={9},
  number={6},
  pages={993--1003},
  year={1971}
}


@article{smits1983low,
  title={Low-Reynolds-number turbulent boundary layers in zero and favorable pressure gradients},
  author={Smits, AJ and Matheson, N and Joubert, PN},
  journal={Journal of ship research},
  volume={27},
  number={03},
  pages={147--157},
  year={1983},
  publisher={SNAME}
}

@article{nagib2007approach,
  title={Approach to an asymptotic state for zero pressure gradient turbulent boundary layers},
  author={Nagib, Hassan M and Chauhan, Kapil A and Monkewitz, Peter A},
  journal={Philosophical Transactions of the Royal Society A: Mathematical, Physical and Engineering Sciences},
  volume={365},
  number={1852},
  pages={755--770},
  year={2007},
  publisher={The Royal Society London}
}







%%%%%%%%%%%%%%%%%%%%%%%%%%%%%%%%%%%JPH%%%%%%%%%%%%%%%%%%%%%%%%%%%%%%


@article{Chen_Donzis_2019, title={Shock–turbulence interactions at high turbulence intensities}, volume={870}, rights={https://www.cambridge.org/core/terms}, ISSN={0022-1120, 1469-7645}, DOI={10.1017/jfm.2019.248}, abstractNote={Shock–turbulence interactions are investigated using well-resolved direct numerical simulations (DNS) and analysis at a range of Reynolds, mean and turbulent Mach numbers (Rλ, M and Mt, respectively). The simulations are shock and turbulence resolving with Rλ up to 65, Mt up to 0.54 and M up to 1.4. The focus is on the effect of strong turbulence on the jumps of mean thermodynamic variables across the shock, the shock structure and the ampliﬁcation of turbulence as it moves through the shock. Theoretical results under the so-called quasi-equilibrium (QE) assumption provide explicit laws for a number of statistics of interests which are in agreement with the new DNS data presented here as well as all the data available in the literature. While in previous studies turbulence was found to weaken jumps, it is shown here that stronger jumps are also observed depending on the regime of the interaction. Statistics of the dilatation at the shock are also investigated and found to be well represented by QE for weak turbulence but saturate at high turbulence intensities with a Reynolds number dependence also captured by the analysis. Finally, ampliﬁcation factors are found to present a universal behaviour with two limiting asymptotic regimes governed by (M − 1) and K = Mt/R1λ/2(M − 1), for weak and strong turbulence, respectively. Effect of anisotropy in the incoming ﬂow is also assessed by utilizing two different forcing mechanisms to generate turbulence.}, journal={Journal of Fluid Mechanics}, author={Chen, Chang Hsin and Donzis, Diego A.}, year={2019}, month=jul, pages={813–847}, language={en} }

 @article{Larsson_Bermejo-Moreno_Lele, title={Reynolds- and Mach-number eﬀects in canonical shock–turbulence interaction}, volume={717}, DOI={10.1017/jfm.2012.573}, journal={Journal of Fluid Mechanics}, author={Larsson, Johan and Bermejo-Moreno, Ivan and Lele, Sanjiva K.}, year={2013}, pages={293–321}}


 @book{Ribner_1954, title={Convection of a pattern of vorticity through a shock wave}, abstractNote={An arbitrary weak spatial distribution of vorticity can be represented in terms of plane sinusoidal shear waves of all orientations and wave lengths. (Fourier integral). The analysis treats the passage of a single representative weak shear wave through a plane shock and shows refraction and modijication of the shear wave with simultaneous generation of an acoustically.}, number={1164}, institution={NACA}, author={Ribner, H S}, year={1954}, language={en} }


 @article{Lele_1992, title={Shock-jump relations in a turbulent flow}, volume={4}, ISSN={0899-8213}, DOI={10.1063/1.858343}, abstractNote={The exact jump relations applying across a shock in turbulent mean flow are formulated. These are analyzed to obtain answers to the following basic questions: Is the propagation speed of a shock wave which brings about a specified compression faster or slower than its classical value when moving through a turbulent medium? What is the magnitude of the turbulence associated correction? What is the magnitude of the correction to the pressure jump? Answers to these questions are developed by combining gas-dynamic analysis with homogeneous rapid distortion approximation.}, number={12}, journal={Physics of Fluids A: Fluid Dynamics}, author={Lele, Sanjiva K.}, year={1992}, month=dec, pages={2900–2905}, language={en} }

 @article{Yang_Wang_Gao_2022, title={Studies on effects of wall temperature variation on heat transfer in hypersonic laminar boundary layer}, volume={190}, ISSN={00179310}, DOI={10.1016/j.ijheatmasstransfer.2022.122790}, journal={International Journal of Heat and Mass Transfer}, author={Yang, Zepeng and Wang, Suozhu and Gao, Zhenxun}, year={2022}, month=jul, pages={122790}, language={en} }


 @article{prasad2023turbulence,
  title={Turbulence modeling of 3D high-speed flows with upstream-informed corrections},
  author={Prasad, Chitrarth and Gaitonde, Datta V},
  journal={Shock Waves},
  volume={33},
  number={2},
  pages={99--115},
  year={2023},
  publisher={Springer}
}

 @article{Karl_Bykerk_2024, title={Sustainable space technologies—Strategies toward a predictive aerothermal design of re-useable space transportation systems}, volume={95}, ISSN={0034-6748, 1089-7623}, DOI={10.1063/5.0177075}, abstractNote={This paper presents a review of current aerothermal design and analysis methodologies for spacecraft. It briefly introduces the most important system architectures, including rockets, gliders, and capsule-based configurations, and gives an overview of the specific aerothermal and thermo-chemical effects that are encountered during their different flight phases and trajectories. Numerical and experimental design tools of different fidelity levels are reviewed and discussed, with a specific focus placed on the present limitations and uncertainty sources of models for the wide range of physical phenomena that are encountered in the analyses. This includes high temperature thermodynamics, chemical effects, turbulence, radiation, and gasdynamic effects. This is followed by a summary of current predictive capabilities and research foci, with missing capabilities identified. Finally, a future strategy toward an efficient and predictive aerothermal design of re-useable space transportation systems is proposed.}, number={2}, journal={Review of Scientific Instruments}, author={Karl, Sebastian and Bykerk, Tamas}, year={2024}, month=feb, pages={021101}, language={en} }


 @article{Griffond_Soulard_2012, title={Evolution of axisymmetric weakly turbulent mixtures interacting with shock or rarefaction waves}, volume={24}, ISSN={1070-6631, 1089-7666}, DOI={10.1063/1.4767729}, abstractNote={This paper deals with the interaction of a shock or a rarefaction wave with a weakly turbulent mixture of perfect gases. Assuming weak density-velocity fluctuations, Kovasznay decomposition applies and linear theories can be used to predict the evolution of the joint spectrum of density and velocity during the interaction. In this work, the upstream spectrum is restricted to axisymmetric fields free of acoustic perturbations, in order to comply with shock tube experimental conditions. Besides, spectral anisotropy is limited to a first order spherical harmonic decomposition. With these assumptions, transfer matrices can be obtained which relate the Reynolds stresses, turbulent mass flux and density variance after interaction to their counterparts before interaction. Results are given for both shock waves and rarefaction or compression waves; they are intended to help improve one-point statistical turbulence models applied to shock tube experiments.}, number={11}, journal={Physics of Fluids}, author={Griffond, Jérôme and Soulard, Olivier}, year={2012}, month=nov, pages={115108}, language={en} }


 @article{Tian_Gao_Jiang_Lee_2023, title={A correction for Reynolds‐averaged‐Navier–Stokes turbulence model under the effect of shock waves in hypersonic flows}, volume={95}, ISSN={0271-2091, 1097-0363}, DOI={10.1002/fld.5150}, abstractNote={Studies on the unphysical increase of turbulent quantities for RANS simulation induced by shock waves in hypersonic flows are carried out. Numerical experiments on the hypersonic flow over a blunt body reveal that the phenomenon of unphysical increase of turbulent quantities across the detached shock wave is induced by the strain-rate-based production terms of the k-�� and k-�� SST turbulence models, which leads to the over-prediction of aerothermal prediction. While this phenomenon does not occur for Spalart–Allmaras (S–A) turbulence model because of its vorticity-based production term. In order to eliminate this unphysical phenomenon, and to maintain the accuracy of the original models for boundary layer and separation flows, a new correction method for the k-�� and k-�� SST models is proposed: by comparing the orders of magnitude between the strain-rate-based and vorticity-based production terms, the vorticity-based production term is used near the shock waves, while the original strain-rate-based production term is still used in other regions. Finally, the correction method is applied to turbulence and transition flows over blunt bodies, and the numerical results show that the correction method effectively eliminates the unphysical increase of turbulent quantities across shock waves and improves the accuracy of aerothermal and transition onset location prediction.}, number={2}, journal={International Journal for Numerical Methods in Fluids}, author={Tian, Yuyan and Gao, Zhenxun and Jiang, Chongwen and Lee, Chun‐Hian}, year={2023}, month=feb, pages={313–333}, language={en} }


 @article{Zhang_Gao_Jiang_Lee_2017, title={A RANS model correction on unphysical over-prediction of turbulent quantities across shock wave}, volume={106}, ISSN={00179310}, DOI={10.1016/j.ijheatmasstransfer.2016.10.087}, abstractNote={The unphysical increase of turbulent quantities in turbulence model induced by shock wave in simulations of hypersonic ﬂows and its inﬂuence on the aerothermal prediction are studied. Using k-x SST model, numerical experiments on the hypersonic ﬂow over a blunt body indicate that the strong detached shock wave will result in serious unphysical increase of turbulent quantities and the increase magnitudes are closely related to the mesh scales near the shock wave. This phenomenon could generally cause an over-prediction of the aerothermal prediction. To eliminate this unphysical phenomenon and enhance the accuracy of aerothermal prediction, a new correction model for current k-x SST model is established, which includes two parts: a shock wave detector to identify the location of shock wave and a damping function for the production terms in turbulent transport equations. Finally, the modiﬁed k-x SST model using the correction model is applied to two types of hypersonic ﬂows. Numerical results prove that the correction model can effectively eliminate the unphysical increase of turbulent quantities around the shock wave and meanwhile does not affect the normal development of turbulent quantities inside boundary layer. Through the correction model, the accuracy of aerothermal prediction is remarkably improved.}, journal={International Journal of Heat and Mass Transfer}, author={Zhang, Zhichao and Gao, Zhenxun and Jiang, Chongwen and Lee, Chun-Hian}, year={2017}, month=mar, pages={1107–1119}, language={en} }


@article{Zank_Zhou_Matthaeus_Rice_2002, title={The interaction of turbulence with shock waves: A basic model}, volume={14}, ISSN={1070-6631, 1089-7666}, DOI={10.1063/1.1507772}, abstractNote={The interaction of turbulence and shock waves is considered self-consistently so that the back-reaction of the turbulence and its associated reaction on the turbulence is addressed. This approach differs from previous studies which considered the interaction of linear modes with a shock. The most basic model of hypersonic flow, described by the inviscid form of Burgers’ equation, is used. An energy-containing model which couples the turbulent energy density and correlation length of the flow with the mean flow is developed. Upstream turbulence interacting with a shock wave is found to mediate the shock by (1) increasing the mean shock speed, and (2) decreasing the efficiency of turbulence amplification at the shock as the upstream turbulence energy density is increased. The implication of these results is that the energy in upstream turbulent fluctuations, while being amplified at the shock, is also being converted into mean flow energy downstream. The variance in both the shock speed and position is computed, leading to the suggestion that, in an ensemble-averaged sense, the turbulence mediated shock will acquire a characteristic thickness given by the standard deviation of the shock position. Lax’s geometric entropy condition is used to show that as the upstream turbulent energy density increases, the shock is eventually destabilized, and may emit one or more shocks to produce a system of multiple shock waves. Finally, turbulence downstream of the shock is shown to decay in time t according to t−2/3.}, number={11}, journal={Physics of Fluids}, author={Zank, G. P. and Zhou, Ye and Matthaeus, W. H. and Rice, W. K. M.}, year={2002}, month=nov, pages={3766–3774}, language={en} }

 @article{Chen_Donzis_2019, title={Shock–turbulence interactions at high turbulence intensities}, volume={870}, rights={https://www.cambridge.org/core/terms}, ISSN={0022-1120, 1469-7645}, DOI={10.1017/jfm.2019.248}, abstractNote={Shock–turbulence interactions are investigated using well-resolved direct numerical simulations (DNS) and analysis at a range of Reynolds, mean and turbulent Mach numbers (Rλ, M and Mt, respectively). The simulations are shock and turbulence resolving with Rλ up to 65, Mt up to 0.54 and M up to 1.4. The focus is on the effect of strong turbulence on the jumps of mean thermodynamic variables across the shock, the shock structure and the ampliﬁcation of turbulence as it moves through the shock. Theoretical results under the so-called quasi-equilibrium (QE) assumption provide explicit laws for a number of statistics of interests which are in agreement with the new DNS data presented here as well as all the data available in the literature. While in previous studies turbulence was found to weaken jumps, it is shown here that stronger jumps are also observed depending on the regime of the interaction. Statistics of the dilatation at the shock are also investigated and found to be well represented by QE for weak turbulence but saturate at high turbulence intensities with a Reynolds number dependence also captured by the analysis. Finally, ampliﬁcation factors are found to present a universal behaviour with two limiting asymptotic regimes governed by (M − 1) and K = Mt/R1λ/2(M − 1), for weak and strong turbulence, respectively. Effect of anisotropy in the incoming ﬂow is also assessed by utilizing two different forcing mechanisms to generate turbulence.}, journal={Journal of Fluid Mechanics}, author={Chen, Chang Hsin and Donzis, Diego A.}, year={2019}, month=jul, pages={813–847}, language={en} }


 @article{Jacquin_Cambon_Blin_1993, title={Turbulence amplification by a shock wave and rapid distortion theory}, volume={5}, number={10}, journal={Phys. Fluids A}, author={Jacquin, L and Cambon, C and Blin, E}, year={1993}, language={en} }

 @article{Lacombe_Roy_Sinha_Karl_Hickey_2021, title={Characteristic Scales in Shock–Turbulence Interaction}, volume={59}, ISSN={0001-1452, 1533-385X}, DOI={10.2514/1.J059499}, number={2}, journal={AIAA Journal}, author={Lacombe, Francis and Roy, Subhajit and Sinha, Krishnendu and Karl, Sebastian and Hickey, Jean-Pierre}, year={2021}, month=feb, pages={526–532}, language={en} }


@article{SuSi19,
	author = {Roy, Subhajit and Sinha, Krishnendu},
	doi = {10.2514/1.J058334},
	eprint = {https://doi.org/10.2514/1.J058334},
	journal = {AIAA Journal},
	number = {8},
	pages = {3624-3629},
	title = {Turbulent Heat Flux Model for Hypersonic Shock--Boundary Layer Interaction},
	url = {https://doi.org/10.2514/1.J058334},
	volume = {57},
	year = {2019},
	bdsk-url-1 = {https://doi.org/10.2514/1.J058334}}



 @article{Sinha_Mahesh_Candler_2003, title={Modeling shock unsteadiness in shock/turbulence interaction}, volume={15}, ISSN={1070-6631, 1089-7666}, DOI={10.1063/1.1588306}, abstractNote={The RANS (Reynolds averaged Navier–Stokes) equations can yield significant error when applied to practical flows involving shock waves. We use the interaction of homogeneous isotropic turbulence with a normal shock to suggest improvements in the k–ε model applied to shock/turbulence interaction. Mahesh et al. [J. Fluid Mech. 334, 353 (1997)] and Lee et al. [J. Fluid Mech. 340, 22 (1997)] present direct numerical simulation (DNS) and linear analysis of the flow of isotropic turbulence through a normal shock, where it is found that mean compression, shock unsteadiness, pressure-velocity correlation, and up-stream entropy fluctuations play an important role in the interaction. Current RANS models based on the eddy viscosity assumption yield very high amplification of the turbulent kinetic energy, k, across the shock. Suppressing the eddy viscosity in a shock improves the model predictions, but is inadequate to match theoretical results at high Mach numbers. We modify the k equation to include a term due to shock unsteadiness, and model it using linear analysis. The dissipation rate equation is similarly altered based on linear analysis results. These modifications improve the model predictions considerably, and the new model is found to match the linear theory and DNS data well.}, number={8}, journal={Physics of Fluids}, author={Sinha, Krishnendu and Mahesh, Krishnan and Candler, Graham V.}, year={2003}, month=aug, pages={2290–2297}, language={en} }
 
 @article{Braun_Gore_2018, title={On primitive variable behaviour near shocks in ensemble-averaged methods}, volume={19}, abstractNote={Models for averaged shock corrugation effects and the impact of turbulent entropy or acoustic modes on the energy equation are presented, for application in Reynolds-Averaged Navier Stokes(RANS) simulations of shock-turbulence interactions. Unlike previous work that has focused the modification of turbulent statistics by the shock, the proposed models are introduced to capture the effects of the turbulence on the profiles of primitive variables - mean density, velocity, and pressure. By producing accurate profiles for the primitive variables, it is shown that the proposed models improve numerical convergence behaviour with mesh refinement about a shock, and introduce the physical effects of shock asphericity in a converging shock geometry. These effects are achieved by local closures to turbulent statistics in the averaged Navier-Stokes equations, and can be applied in conjunction with existing Reynolds stress closures that have been constructed for broader applications beyond shock-turbulence interactions.}, number={10}, journal={JOURNAL OF TURBULENCE}, author={Braun, N O and Gore, R A}, year={2018}, pages={868–888}, language={en} }




 @article{Zhang_2017, title={A RANS model correction on unphysical over-prediction of turbulent quantities across shock wave}, abstractNote={The unphysical increase of turbulent quantities in turbulence model induced by shock wave in simulations of hypersonic ﬂows and its inﬂuence on the aerothermal prediction are studied. Using k-x SST model, numerical experiments on the hypersonic ﬂow over a blunt body indicate that the strong detached shock wave will result in serious unphysical increase of turbulent quantities and the increase magnitudes are closely related to the mesh scales near the shock wave. This phenomenon could generally cause an over-prediction of the aerothermal prediction. To eliminate this unphysical phenomenon and enhance the accuracy of aerothermal prediction, a new correction model for current k-x SST model is established, which includes two parts: a shock wave detector to identify the location of shock wave and a damping function for the production terms in turbulent transport equations. Finally, the modiﬁed k-x SST model using the correction model is applied to two types of hypersonic ﬂows. Numerical results prove that the correction model can effectively eliminate the unphysical increase of turbulent quantities around the shock wave and meanwhile does not affect the normal development of turbulent quantities inside boundary layer. Through the correction model, the accuracy of aerothermal prediction is remarkably improved.}, journal={International Journal of Heat and Mass Transfer}, author={Zhang, Zhichao}, year={2017}, language={en} }


 @article{Ryu_Livescu_2014, title={Turbulence structure behind the shock in canonical shock–vortical turbulence interaction}, volume={756}, rights={https://www.cambridge.org/core/terms}, ISSN={0022-1120, 1469-7645}, DOI={10.1017/jfm.2014.477}, abstractNote={The interaction between vortical isotropic turbulence (IT) and a normal shock wave is studied using direct numerical simulation (DNS) and linear interaction analysis (LIA). In previous studies, agreement between the simulation results and the LIA predictions has been limited and, thus, the signiﬁcance of LIA has been underestimated. The results show, for the ﬁrst time, that the turbulence quantities from DNS converge to the LIA solutions as the turbulent Mach number, Mt, becomes small, even at low upstream Reynolds numbers. The classical LIA formulae are extended to compute the complete post-shock ﬂow ﬁelds using an IT database. The solutions, consistent with the DNS results, show that the shock wave signiﬁcantly changes the topology of the turbulent structures, with a symmetrization of the third invariant of the velocity gradient tensor and (Ms-mediated) of the probability density function (PDF) of the longitudinal velocity derivatives, and an Ms-dependent increase in the correlation between strain and rotation.}, journal={Journal of Fluid Mechanics}, author={Ryu, Jaiyoung and Livescu, Daniel}, year={2014}, month=oct, pages={R1}, language={en} }

@book{Howarth1953,
publisher = {Clarendon Press},
series = {The Oxford engineering science series},
title = {Modern developments in fluid dynamics : high speed flow },
booktitle = {Modern developments in fluid dynamics : high speed flow},
keywords = {Fluid dynamics},
language = {eng},
lccn = {54005834},
year = {1953},
author = {Howarth, Leslie.},
address = {Oxford},
}




 @article{Cheng_Fu_2024, title={Mean temperature scalings in compressible wall turbulence}, volume={9}, ISSN={2469-990X}, DOI={10.1103/PhysRevFluids.9.054610}, number={5}, journal={Physical Review Fluids}, author={Cheng, Cheng and Fu, Lin}, year={2024}, month=may, pages={054610}, language={en} }

 @article{Gelain2021, title={Aerothermal characterisation of a surface heat exchanger implemented in a turbofan by-pass duct}, author={Gelain, Matteo}, language={en} ,  year={2021},
  school={Universit{\'e} Paris-Saclay}}


 @inproceedings{Maheu_Moureau_Domingo_2012, title={Large-Eddy Simulations of flow and heat transfer around a low-Mach number turbine blade}, ISBN={978-1-56700-302-4}, url={http://www.dl.begellhouse.com/references/1bb331655c289a0a,7c69ce3452e09525,6e8103ad5d4ba006.html}, DOI={10.1615/ICHMT.2012.ProcSevIntSympTurbHeatTransfPal.810}, booktitle={Proceeding of THMT-12. Proceedings of the Seventh International Symposium On Turbulence, Heat and Mass Transfer Palermo, Italy, 24-27 September, 2012}, publisher={Begellhouse}, author={Maheu, Nicolas and Moureau, Vincent and Domingo, P.}, year={2012}, pages={12}, language={en} }


 @inproceedings{muller2024investigation,
  title={Investigation of Loosely-Coupled Conjugate Heat Transfer With Wall-Modeled Large Eddy Simulation in a Mach 2.5 Flow},
  author={Muller, Julia and Boisvert, Joshua and Dutta, Meghna and Oefelein, Joseph},
  booktitle={AIAA AVIATION FORUM AND ASCEND 2024},
  pages={4175},
  year={2024}
}

@inproceedings{Muller_Dutta_Boisvert_Oefelein_2024, address={London, United Kingdom}, title={Investigation of Conjugate Heat Transfer in Wall-Modeled Large Eddy Simulation of High-Speed Compressible Wall-Bounded Flows}, ISBN={978-0-7918-8809-4}, url={https://asmedigitalcollection.asme.org/GT/proceedings/GT2024/88094/V013T13A034/1204874}, DOI={10.1115/GT2024-127902}, abstractNote={A parameter study is performed in a compressible boundarylayer at supersonic conditions to investigate the accuracy of a validated loosely-coupled Conjugate Heat Transfer (CHT) model combined with an advanced wall model approach for Large Eddy Simulation (LES). Fully-coupled CHT typically introduces extreme additional cost to numerical simulations due to large timescale disparities between processes in the fluid and solid domains. Loosely-coupled CHT methods seek to mitigate this cost while simultaneously maintaining the correct boundary layer dynamics and local heat transfer processes. Likewise, wall-modeling makes LES more computationally tractable, but has not been widely used with CHT. The current investigation demonstrates the performance (accuracy and computational cost) of these approaches when implemented as a unified model. Results demonstrate that the selected wall-modeled LES, combined with a fully-integrated loosely-coupled CHT model, can reproduce the accuracy of reference DNS and wall-resolved LES data with orders of magnitude reduction in cost. Moreover, they demonstrate the implementation requirements necessary to achieve this accuracy.}, booktitle={Volume 13: Heat Transfer: General Interest / Additive Manufacturing Impacts on Heat Transfer; Wind Energy}, publisher={American Society of Mechanical Engineers}, author={Muller, Julia A. and Dutta, Meghna and Boisvert, Joshua and Oefelein, Joseph C.}, year={2024}, month=jun, pages={V013T13A034}, language={en} }

 @article{Yang_Iacovides_Craft_Apsley_2021, title={RANS Model development on temperature variance in conjugate heat transfer}, volume={22}, ISSN={1468-5248}, DOI={10.1080/14685248.2020.1860214}, abstractNote={In this study, a RANS model of turbulent conjugate heat transfer has been developed, which is applicable across a range of different combination of fluid and solid thermal properties. This is achieved by focusing on the transport equations for the temperature variance and its dissipation rate across the solid walls which bound the flow region. In this investigation we make use of a wider range of DNS data reported by other researchers, to advance our understanding of the processes involved and to revise and extend the capabilities of the model of Craft et al [12] including a more physical fluid-solid interface condition on the dissipation of thermal fluctuations and a dependence of model coefficients on Prandtl number. The resulting model is shown to successfully reproduce the penetration of thermal fluctuations into solid regions, and their subsequent decay across the solid, for a wide range of fluid to solid thermal property ratios, and Prandtl numbers, thereby bringing a step change to RANS capabilities in turbulent conjugate heat transfer analysis.}, number={3}, journal={Journal of Turbulence}, author={Yang, Gaoqiang and Iacovides, Hector and Craft, Timothy and Apsley, David}, year={2021}, month=mar, pages={180–207}, language={en} }


 @inproceedings{Yoder_2016, address={San Diego, California, USA}, title={Comparison of Turbulent Thermal Diffusivity and Scalar Variance Models}, ISBN={978-1-62410-393-3}, url={https://arc.aiaa.org/doi/10.2514/6.2016-1561}, DOI={10.2514/6.2016-1561}, booktitle={54th AIAA Aerospace Sciences Meeting}, publisher={American Institute of Aeronautics and Astronautics}, author={Yoder, Dennis A.}, year={2016}, month=jan, language={en} }


 @article{Lewis_Hickey_2023, title={Conjugate Heat Transfer in High-Speed External Flows: A Review}, volume={37}, ISSN={0887-8722, 1533-6808}, DOI={10.2514/1.T6763}, number={4}, journal={Journal of Thermophysics and Heat Transfer}, author={Lewis, Mikaela T. and Hickey, Jean-Pierre}, year={2023}, month=oct, pages={697–712}, language={en} }


%%%%%%%%%%%%%%%%%%%%%%%%%%%%%%%%%%%END JPH%%%%%%%%%%%%%%%%%%%%%%%%%%%

%%%%%%%%%%%%%%%%%%%%%%%%%%%%%%%%%%%%%%%%%%%%%%%%%%%%%%%%%%%%%%%%%%%%

\end{document}